%% file: 9307365.tex
\documentstyle[12pt,epic,eepic]{article}
\newcommand{\abs}[1]{\left\vert{#1}\right\vert}

\newcommand{\del}{\partial}
\newcommand{\delm}{\partial_\mu}

\newcommand{\tr}{\hbox{tr}}

\newcommand{\bra}[1]{\left\langle{#1}\right|}
\newcommand{\ket}[1]{\left|{#1}\right\rangle}

\newcommand{\nonum}{\nonumber\\}

\newlength{\mleng}
\newcommand{\mq}{\makebox[0.5\mleng]{ }}
\newcommand{\mqq}{\makebox[\mleng]{ }}
\newcommand{\mqqq}{\makebox[1.5\mleng]{ }}

\newcommand{\Lag}{{\cal L}}
\renewcommand{\Im}{\hbox{Im}\;}
\renewcommand{\Re}{\hbox{Re}\;}

\newcommand{\be}{\begin{equation}}
\newcommand{\ee}{\end{equation}}
\newcommand{\ba}{\begin{eqnarray}}
\newcommand{\ea}{\end{eqnarray}}
\newcommand{\bd}{\begin{displaymath}}
\newcommand{\ed}{\end{displaymath}}

\newcommand{\bfpi}{\makebox{\boldmath$\pi$}}
\newcommand{\bfrho}{\makebox{\boldmath$\rho$}}

\newcommand{\Trang}[1]{\,\tr\left\langle{#1}\right\rangle}

\newcommand{\msection}[1]{\setcounter{equation}{0}\section{#1}}
\newcommand{\ssize}{\scriptsize}
\newcommand{\fsize}{\footnotesize}

\begin{document}

\begin{titlepage}

\begin{flushright}
\begin{minipage}[t]{3.5cm}
\begin{flushleft}
KUNS-1209 \\
HE(TH) 93/06\\
hep-ph/9307365 \\
July, 1993
\end{flushleft}
\end{minipage}
\end{flushright}

\vspace{0.7cm}

\begin{center}
\Large\bf
Interpolating\\
Axial Anomaly Induced Amplitudes
\end{center}

\vfill

\begin{center}
\large
Masako {\sc Bando}\footnote{
Permanent address: Aichi University, Miyoshi Aichi 470-02, Japan} 
\ and \ 
Masayasu {\sc Harada}%
\footnote{Fellow of the
Japan Society for the Promotion of Science for Japanese Junior
Scientists} \\
\ \\
{\normalsize\it
Department of Physics,
Kyoto University,\\
Kyoto 606-01, Japan}
\end{center}

\vfill

\begin{abstract}
We propose an interpolating formula for the amplitude induced 
by the axial anomaly, 
concentrating on the $\pi^0\gamma^{\ast}\gamma^{\ast}$
transition form factor. The QCD corrections to this amplitude are 
generally described by two major contributions
coming from the $q\bar{q}$ bound state 
and the background continuous spectrum, respectively.

For the first contribution, 
we include the lowest vector bound state using the 
realization of the dynamical gauge boson of hidden local symmetry.
The second contribution 
is included as the triangle quark loop
in which a constituent mass is adopted as a internal quark mass
and the amplitudes are smeared out around the threshold. 
Using the resulting form factor, we fit the 
experimental data for the $\pi^0\gamma$ and the $\omega\pi^0$ transition 
form factors and show that our result describes 
the experimental data well.
\end{abstract}

\end{titlepage}

\setcounter{footnote}{0}

\msection{Introduction}

The non-Abelian anomaly is one of the most prominent features 
of quantum gauge theories, 
and it has provided us with important information
on the matter content of gauge theories.
In particular the well-studied $\pi^0\rightarrow\gamma\gamma$ process,
which is controlled by the axial anomaly,
gave the first clear indication of the physical existence of the 
color degrees of freedom in the quark model.

As is well known, a low energy effective chiral Lagrangian 
may be written
in terms of the Nambu Goldstone (NG) bosons:
this Lagrangian is based on
the nonlinear realization of the chiral symmetry of QCD.
All relevant anomaly terms must be taken into account 
in this low energy effective chiral Lagrangian. 
This is done by introducing Wess-Zumino terms%
\cite{WessZumino,Witten:anomaly}, $\Lag_{\rm WZ}$, 
which determine
exactly the scattering amplitudes in the zero momentum limit
(the low energy theorem%
\cite{Adler:Anom:69,BellJackiw:Anom:69,AdlerBardeen}).
The values of the anomaly terms are determined 
by calculating the 
well-known fermion triangle graph of 
Fig.~\ref{fig:trianglegraph}.
The value of the amplitude in the low energy limit 
does not depend on either 
the higher order corrections or the internal fermion masses.

If one wants to extend the Lagrangian 
to a small but nonzero energy scale,  
one should also include
the non-anomalous terms,
and incorporate the QCD corrections to this triangle graph 
into the effective Lagrangian.
The QCD effects are represented as a rich spectrum of
intermediate states in each channel. 
Those may be
quite well described by the following 
two major contributions: the first is a variety of 
quark-antiquark bound states (hadrons)
which come into the amplitudes 
as poles in each 
channel and the second is a continuous spectrum 
which contributes to the background amplitudes.

The most important candidates for the bound states are 
the non NG boson low-mass states, i.e., the  vector mesons. 
It is known that the easiest way to include these in a consistent 
and systematic formulation is to treat them in the framework of 
dynamical realization of
hidden local symmetry\cite{BandoKugoYamawaki:PRep}.
In this section
the vector mesons are introduced in the effective Lagrangian
without any mismatch to the low energy theorem.    
The second effect may be well modeled by including
the triangle quark one-loop contributions.
In calculating these, however, one should adopt 
the non-perturbative ``constituent quark masses" instead of 
the current quark masses.
The calculation should also be smeared out  
around the $q\bar{q}$ threshold because quarks are confined 
and multi-hadron (pion) threshold effects arise
instead of multi-quarks.

On the other hand, the asymptotic freedom of QCD facilitates us to 
predict the hadronic amplitudes in the extremely high energy region,
where again the triangle diagram becomes dominant.
The  contributions have been extensively 
studied\cite{LepageBrodsky,GuberinaKuhnPecceiRuckel:80}
in connection with deep scattering exclusive processes. 
The high energy behavior is obtained 
almost exactly up to an ambiguity in the coefficient factor. 
Thus it is important to notice that 
the anomaly induced amplitudes are constrained not only by the low 
energy theorem but also by the  high energy behavior.

In this paper we propose an interpolating formula for the anomaly 
induced amplitudes which matches the behaviors 
in both the low and high energy limits.
We focus on the electromagnetic pion  transition form factor 
as an example,
and check our interpolating formula reproduces the 
existing experimental data.

This paper is organized as follows.
Section~\ref{sec:definition} is devoted to the formulation and the
presentation of the basic tools for computing 
the anomaly induced decay amplitudes
and the corresponding form factors.
In section~\ref{sec:QCDloop},
the quark triangle graph is studied,
and we propose how to take account of QCD corrections.
Section~\ref{sec:hidden} is devoted to a review of the hidden local
symmetry model
and discussions on the low energy form factor including the effects of
the vector mesons.
In sect.~\ref{sec:formfactor} we propose the interpolating form
factors.
In sect.~\ref{sec:pgg} we fit our form factor with the experimental
data for the $\pi^0\gamma$ transition form factor and give the
$Z\rightarrow\pi^0\gamma$ decay width.
In sects.~\ref{sec:wpg} and \ref{wpgform} we study the $\omega\pi^0$
transition form factor and its related processes 
to check the validity of our form factor.
Section~\ref{sec:concl} contains summary and discussions.

\msection{Definition \label{sec:definition}}

Let us start with the
anomalous Ward-Takahashi identity:
\be
  \delm j_5^\mu = 2 m_0 j_5 + \frac{e^2}{16\pi^2} F \widetilde{F},
\ee
where $j_5^\mu$ is the axial-vector current%
\footnote{Here we restrict ourselves to neutral currents.}
which generally couples to NG bosons, 
and $j_5$ is the corresponding pseudoscalar density.
In particular, for the $\pi^0$ case,
$j_5^\mu = \sum_i \bar{q}_i T_i \gamma^\mu \gamma_5 q_i$,
$q_i=(u,d)$, $T_i=(1/2,-1/2)$,
and $2m_0 j_5 \equiv \sum_i 2 m_{0i} \bar{q}_i T_i i \gamma_5 q_i$,
where $m_{0i}$ is the current quark mass.
Hereafter, we shall take as an example the case in which $j_5^\mu$
couples to $\pi^0$ to demonstrate the concrete forms of the amplitudes,
although general expressions can be obtained straightforwardly.
In the $\pi^0$ case, only the $u$ and $d$ quarks come into play.

We consider the following three-point function:
\be
  T^{AB,\mu\nu\rho}(p,q)
  =
  - i \int d^4x d^4y e^{-i k\cdot y + i p\cdot x}
  \bra{0}\mbox{T} j^{A\nu}(x) j^{B\rho}(0) j_5^\mu(y) \ket{0},
\label{def:threepoint}
\ee
where $q\equiv k-p$,
and $j^{A\nu}(x)$ and $j^{B\rho}(0)$ are the vector currents,
$j^{A\nu} = \sum_i \bar{q}_i Q^A_i \gamma^\nu q_i$ 
with $Q^A_i$ being the vector
charge of the $i$th quark $q_i$.
(The suffices $A$ and $B$ run over $\gamma$, $Z$, $\cdots$, etc., 
depending on the gauge field to which the current couples.
The normalization of $Q^A$ is determined so that the coupling between
the current and the gauge field is given by $e j^{A\mu} A_\mu$.)

PCAC in the presence of the anomaly
gives the following
relation\cite{Hikasa:Anom:90,DeshpandePalOlness:90}:
\be
  k_\mu T^{AB,\mu\nu\rho}(p,q)
  =
  M^{AB,\mu\nu} (p,q) + 
  \frac{N_{\rm c} e^2}{4\pi^2} 
  \Trang{ T \left\{ Q^A , Q^B \right\} }
  \varepsilon^{\alpha\beta\nu\rho} p_\alpha q_\beta,
\label{eq:anomalousWardTakahashi}
\ee
where $N_{\rm c}$($=3$) is the number of colors,
and 
\be
  M^{AB,\nu\rho}(p,q)
  =
  - f_{\rm P} m_{\rm P}^2 \int d^4x d^4y \exp ( ip\cdot x) \exp(-ik\cdot y)
  \bra{0}\mbox{T} j^{A\nu}(x) j^{B\rho}(0) \phi_{\rm P} (y) \ket{0}
\ee
represents the anomaly induced amplitude 
for the NG boson, $\phi_{\rm P}(y)$,
to couple to vector currents $j^{A\nu}$ and $j^{B\rho}$.
If one defines the following function $T(p^2,q^2,k^2)$:
\be
  M^{AB,\nu\rho}(p,q) \equiv
  - \frac{N_{\rm c}e^2}{4\pi^2f_{\rm P}} 
  \Trang{T \left\{ Q^A ,  Q^B \right\} }
  \varepsilon^{\alpha\beta\nu\rho} p_\alpha q_\beta
  T(p^2,q^2,k^2),
\label{def:piamp}
\ee
this function $T(p^2,q^2,k^2)$ determines the axial anomaly induced
amplitude.
In the low energy limit
this is determined exactly by the low energy theorem.
The $k^2=0$ limit of the function $T(p^2,q^2,k^2)$ 
for $\pi^0$ case
determines the $\pi^0\gamma^\ast\gamma^\ast$ transition form factor:
\be
  F\left(\pi^0, \gamma(p^2) , \gamma(q^2)\right)
  =
  T(p^2,q^2,k^2=0).
\ee
The low energy theorem requires the 
following normalization of this form factor:
\be
  F\left(\pi^0,\gamma(p^2=0),\gamma(q^2=0)\right) =1 .
\label{eq:nomalcond}
\ee

It is our purpose to investigate the properties of $T(p^2,q^2,k^2)$
not only in the low energy limit
but also in the high energy region where $p^2$ or $q^2$ is large.

\msection{Triangle Anomaly Graph and QCD Corrections
\label{sec:QCDloop}}

\subsection{Triangle Quark One-Loop Graph}

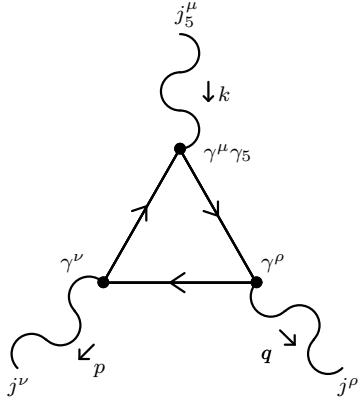
\begin{figure}[htbp]
\begin{center}
\input{fig_tri}
\ 
\end{center}
\caption{The quark triangle anomaly graph
\label{fig:trianglegraph}}
\end{figure}

The quark triangle graph shown in
Fig.~\ref{fig:trianglegraph} gives the main contribution
to the LHS of Eq.(\ref{eq:anomalousWardTakahashi}) 
This is evaluated as 
\ba
&{}&
  \left.
    k_\mu T^{AB,\mu\nu\rho}(p,q)
  \right\vert_{\rm triangle}
\nonum
&{}& \mq
  =
  \frac{N_{\rm c} e^2}{4\pi^2f_\pi} 
  2 \sum_i T_i Q^A_{i} Q^B_{i}
  \varepsilon^{\alpha\beta\nu\rho} p_\alpha q_\beta
  \left[
    1 - \int [dz] 
    \frac{m_i^2}{m_i^2 - z_2z_3k^2 - z_3z_1p^2 - z_1z_2 q^2}
  \right],
\label{res:triangle}
\ea
where $m_i$ is the $i$th quark mass,
$f_\pi=93{\rm MeV}$ is the pion decay constant
and the Feynman parameter integral is defined by
\be
  \int [dz]
  \equiv
  2 \int_0^1
  dz_1 dz_2 dz_3 \delta (1-z_1-z_2-z_3) .
\ee
Thus, we get
\be
  \left.
    T(p^2,q^2,k^2)
  \right\vert_{\rm triangle}
  =
  \frac{
    2 \sum_i T_i Q^A_i Q^B_i
  }{
    \Trang{T \left\{ Q^A,  Q^B \right\} }
  }  
  \int [dz] 
  \frac{m_i^2}{m_i^2 - z_2z_3k^2 - z_3z_1p^2 - z_1z_2 q^2}.
\label{res:functionT}
\ee
It is easy to see that in the low energy limit in which
$k^2=p^2=q^2=0$, this reduces to
\be
  \left.
    T(p^2=0,q^2=0,k^2=0)
  \right\vert_{\rm triangle}
  =
  1,
\label{eq:normalization}
\ee
independently of $m_i$.
This is of course because the anomaly term is exactly reproduced by the
lowest order triangle graph.
Here we define the triangle graph contributions to $T(p^2,q^2,k^2=0)$:
\be
  I(p^2,q^2) 
  \equiv
  \left.
    T(p^2,q^2,k^2=0)
  \right\vert_{\rm triangle} .
\label{def:functionI:general}
\ee
Let us study the high energy behavior of this function.
After performing the parameter integration we obtain
\ba
  \!\!\!\!
  I(p^2,q^2) \!\!\!
&=& \!\!\!
  \frac{
    2 \sum_i T_i Q^A_i Q^B_i
  }{
    \Trang{T \left\{ Q^A,  Q^B \right\} }
  }  
\nonum
&{}& 
  \times
  \frac{m_i^2}{p^2-q^2}
  \left[
    \left\{
      \ln
      \frac{
        \sqrt{4m_i^2-q^2} + \sqrt{-q^2}
      }{
        \sqrt{4m_i^2-q^2} - \sqrt{-q^2}
      }
    \right\}^2
    -
    \left\{
      \ln
      \frac{
        \sqrt{4m_i^2-p^2} + \sqrt{-p^2}
      }{
        \sqrt{4m_i^2-p^2} - \sqrt{-p^2}
      }
    \right\}^2
  \right] .
\label{eq:functionI}
\ea
For the $\pi^0$ case, we take $m_u\simeq m_d=m$ and get
\be
  I(p^2,q^2) 
  =
  \int [dz] 
  \frac{m^2}{m^2 - z_3z_1p^2 - z_1z_2 q^2} ,
\label{def:functionI}
\ee
whose high energy behavior is given by
\be
  I(p^2\gg m^2,q^2\gg m^2) 
  \simeq
  \frac{m^2}{p^2-q^2}
  \left[
    \left\{
      \ln \frac{m^2}{q^2} + i \pi 
    \right\}^2
    -
    \left\{
      \ln \frac{m^2}{p^2} + i \pi 
    \right\}^2
  \right] .
\label{eq:high:I}
\ee
If one photon is on its mass shell (e.g., $p^2=0$), then the function
$I(p^2,q^2)$ reduces to 
\ba
  J(q^2)
&\equiv&
  I(p^2=0,q^2) 
\nonum
&=&
  - \frac{m^2}{q^2}
  \left[
    \left\{
      \ln
      \frac{
        \sqrt{4m^2-q^2} + \sqrt{-q^2}
      }{
        \sqrt{4m^2-q^2} - \sqrt{-q^2}
      }
    \right\}^2
  \right] ,
\label{def:tildeI}
\ea
which gives the $\pi^0\gamma$ transition form factor.
The high energy behavior of this function is given by
\be
  J(q^2\gg m^2)
  \simeq
  \frac{m^2}{q^2}
  \left[
    \pi^2 - 
    \left(
      \ln \frac{m^2}{q^2}
    \right)^2
    - 2 i \pi \ln \frac{m^2}{q^2}
  \right] .
\label{eq:Itilde}
\ee
As we shall see later, this form factor, with
the constituent quark mass 
used in place of the current quark mass,
is consistent with the short range
behavior obtained from the operator product expansion (OPE)
technique\cite{LepageBrodsky,Manohar:ZWpi:90}.
In the intermediate region, however, we have to take account of
the higher order corrections, especially, the QCD effects can not
be neglected.

\subsection{QCD Corrections to Quark Triangle Anomaly Graph}

It is well known that, in the low energy limit,
the higher order corrections do not change 
the value of the anomaly induced three-point function in
Eq.(\ref{def:threepoint}), or $T(p^2=q^2=k^2=0)$ itself.
However, higher order corrections do 
contribute appreciably to its higher energy
behavior.

Generally QCD effects are represented as a rich hadron spectrum,
with the most important contributions coming from the
lowest energy bound states, i.e. the vector mesons.
This can be elegantly described 
in the framework of hidden local symmetry, 
which will be briefly summarized in the section \ref{sec:hidden}.
All other QCD effects, including the higher resonances with
broader widths,
are expressed as a background amplitude.
This background amplitude may be reasonably well described by
the triangle quark 
contributions of Fig.~\ref{fig:trianglegraph}, in which, 
however, one should include the following two kinds 
of major non-perturbative QCD effects:
1) the current quark masses should be replaced by the constituent
quark masses;
2) the amplitude should be smeared out around the
$q\bar{q}$ threshold.

1) 
Since the chiral symmetry has been spontaneously broken 
and there appear NG bosons,
the quarks necessarily acquire dynamical masses.
Thus, in this phase of broken chiral symmetry,
it is natural to replace the current quark mass
by the constituent quark mass in the function $I(p^2,q^2)$.
Note that, however, 
this replacement never changes the value of the low energy limit, 
because $I(p^2=0,q^2=0)=1$ independently of the quark mass 
(see Eq.(\ref{eq:normalization})).
Further, the asymptotic behavior of $I(p^2=0,q^2)=1$
at large $q^2$ has been found to be of order ${\cal O}(m^2/q^2)$,
which is almost consistent with the short range behavior
obtained from the OPE\cite{LepageBrodsky,Manohar:ZWpi:90},
because the constituent quark masses are of order $f_\pi$
(note that in the chiral limit use of $m=0$ gives 
completely the wrong behavior.)
It is the asymptotic freedom of QCD that guarantees 
the validity of the perturbative expansion at short distance.
As a result the triangle graph becomes dominant.

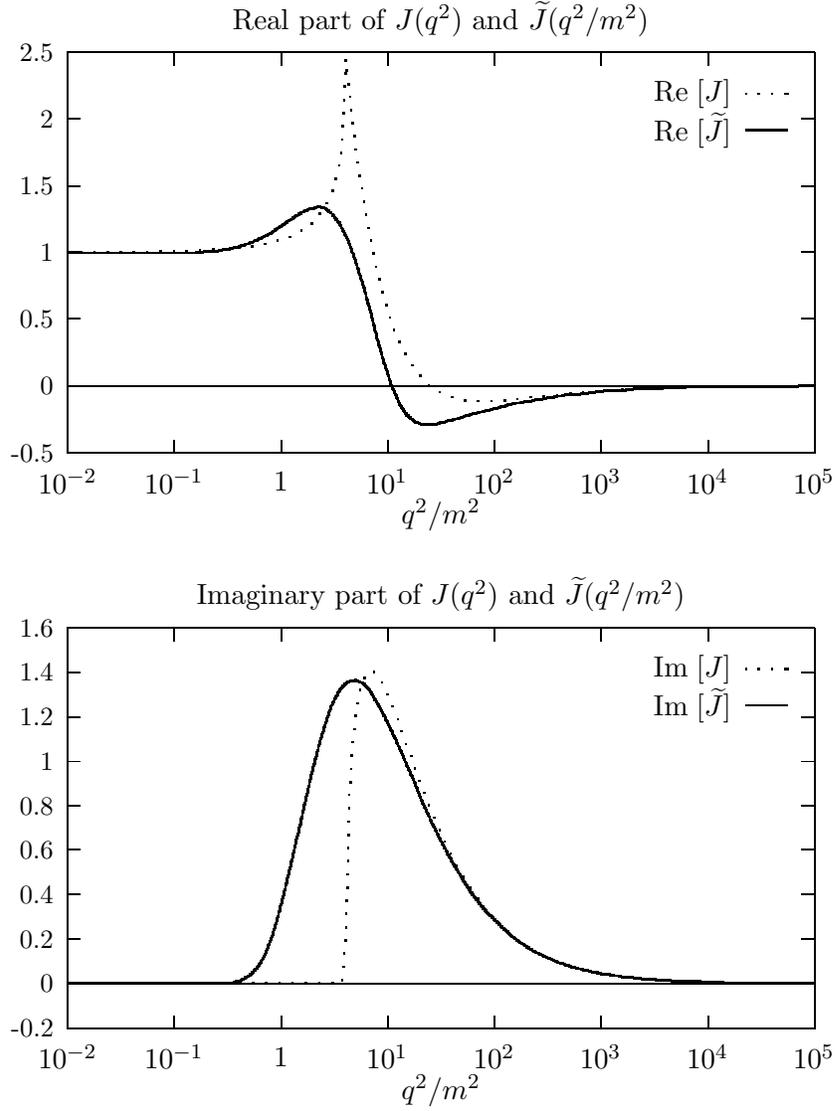
\begin{figure}[htbp]
\begin{center}
\input{modfunre}
\input{modfunim}
\end{center}
\caption[]{The QCD corrected-improved 
functions $\widetilde{J}(q^2/m^2)$ (solid lines)
comparing with the original functions $J(q^2)$ (dotted lines), 
where we take the parameter choice 
$\alpha=1.4$, $\beta=0.35$ and $\lambda=2.5$.
\label{fig:modifiedI}}
\end{figure}

2)
The graph in Fig.~\ref{fig:trianglegraph} shows the contribution of
the $q\bar{q}$ threshold at $p^2=4m^2$ ($q^2=4m^2$) 
from which an imaginary part emerges
as shown in Fig.~\ref{fig:modifiedI} (the dotted lines).
However, since the quarks are confined,
the intermediate states are not multi-quark states 
but actually multi-hadron states
(2$\pi$, 4$\pi$, ... etc.).
The above effect may be properly taken into account by smearing out
the original $J(q^2)$.
Here we adopt the following function:
\be
  \widetilde{J}(x)
  \equiv
  - \frac{\left(\ln(1+x)\right)^2}{x}
  + \frac{\ln(1+x)}{x} + \alpha x \exp(-\beta x)
  + 2 \pi i \frac{\ln(1+x)}{x} 
  \exp\left(-\frac{\lambda}{x} \right) ,
\label{smooth:tildeI}
\ee
where $x=q^2/m^2$.
Both $J$ and $\widetilde{J}$ are 
shown in Fig.~\ref{fig:modifiedI},
where we take the parameter choice 
$\alpha=1.4$, $\beta=0.35$ and $\lambda=2.5$.
{}From this figure, we see that the modified amplitudes
$\widetilde{J}$ are properly smeared out.
This form has been chosen so as to 
coincide with the original $J$ 
(the dotted lines)
both in the high and low energy limits.

\msection{Hidden Local Symmetry and Vector Mesons \label{sec:hidden}}

In order to get the full amplitude, we need further to take account
of the dominant resonances, 
which in our case are the vector meson poles 
in the relatively low energy region.
This can be done consistently and systematically in the framework of
hidden local symmetry (HLS)\cite{FujiwaraKugoTeraoUeharaYamawaki}.

\subsection{Hidden Local Symmetry}

First, following Ref.\cite{BandoKugoYamawaki:PRep},
we briefly review hidden local symmetry.
In this framework the vector mesons 
are introduced as the dynamical gauge
bosons of hidden local symmetry%
\cite{BandoKugoUeharaYamawakiYanagida}.

For the case of the $\pi^0$ form factor,
the low energy QCD system is known to be described by the model based
on the manifold 
$G/H=\mbox{U(2)}_{\rm L} \times 
\mbox{U(2)}_{\rm R}/\mbox{U(2)}_{\rm V}$%
\footnote{We have to take account of gluon anomaly
in the case of $\eta$ and $\eta'$.}.
The low energy effective chiral Lagrangian is expressed 
in terms of two SU(2) matrix valued
variables $\xi_{\rm L,R}(x)$ such that%
\footnote{To be exact, we have to consider the 
$[\mbox{U(3)}_{\rm L}\times\mbox{U(3)}_{\rm R}]_{\rm global}
\times[\mbox{U(3)}_{\rm V}]_{\rm hidden}$ model%
\cite{BandoKugoYamawaki:NP},
which contains the nonet vector mesons $\rho$, $K^{\ast}$,
$\bar{K}^{\ast}$, $\omega$ and $\phi$.
We only write the part concerning to the $\pi$ 
for reasons of mathematical simplicity, hereafter.}
\be
  \xi^{\dag}_{\rm L}(x) \xi_{\rm R}(x) 
  = e^{2i\pi/f_\pi} \left( = U(x) \right) ,
\ee
from which it is convenient to define the algebra valued
(covariantized) 1-form;
\ba
  \hat{\alpha}_{\mu\parallel}
&=&
  \left(
    D_\mu \xi_{\rm L} \cdot \xi_{\rm L}^{\dag}
    + D_\mu \xi_{\rm R} \cdot \xi_{\rm R}^{\dag}
  \right)
  / 2i ,
\nonum
  \hat{\alpha}_{\mu\perp}
&=&
  \left(
    D_\mu \xi_{\rm L} \cdot \xi_{\rm L}^{\dag}
    - D_\mu \xi_{\rm R} \cdot \xi_{\rm R}^{\dag}
  \right)
  / 2i ,
\ea
with the covariant derivatives defined by
\ba
  D_\mu \xi_{\rm L}
&=&
  \delm\xi_{\rm L} - i g V_\mu \xi_{\rm L} + i e \xi_{\rm L}{\cal L},
\nonum
  D_\mu \xi_{\rm R}
&=&
  \delm\xi_{\rm R} - i g V_\mu \xi_{\rm R} + i e \xi_{\rm R}{\cal R},
\ea
where $V_\mu$ are the vector mesons realized as the hidden local gauge
fields:
\be
  V = \frac{1}{\sqrt{2}}
  \left(
  \begin{array}{cc}
    \frac{1}{\sqrt{2}} (\rho^0 + \omega) & \rho^+ \\
    \rho^- & -\frac{1}{\sqrt{2}} (\rho^0 - \omega) 
  \end{array}
  \right) ,
\ee
${\cal L}$ and ${\cal R}$ denote the external gauge field,
$g$ the hidden local gauge coupling and $e$ the electromagnetic
coupling constant.
The external fields are rewritten into the 
electroweak gauge fields, ${\cal B}$ (photon),
$Z$ and $W$:
\ba
  {\cal L}_\mu 
&=&
  Q 
  \left[
    {\cal B}_\mu - \tan \theta_{\rm W} {\cal Z}_\mu
  \right]
  +
  \frac{1}{\sin \theta_{\rm W} \cos \theta_{\rm W}}
  T_z \cdot {\cal Z}_\mu 
  +
  \frac{1}{\sqrt{2}\sin \theta_{\rm W}} {\cal W}_\mu ,
\nonum
  {\cal R}_\mu 
&=&
  Q 
  \left[
    {\cal B}_\mu - \tan \theta_{\rm W} {\cal Z}_\mu
  \right] ,
\ea
with
\ba
&{}&
  {\cal W}_\mu
  =
  \left(
    \begin{array}{cc}
      0 & {\cal W}_\mu^+ \cos \theta_{\rm C} \\
      {\cal W}_\mu^- \cos \theta_{\rm C} & 0 
    \end{array}
  \right) ,
\\
&{}&
  Q = \frac{1}{3}
  \left(
    \begin{array}{cc}
      2 & 0 \\
      0 & -1 
    \end{array}
  \right) ,
  \mqq
  T_z = \frac{1}{2}
  \left(
    \begin{array}{cc}
      1 & 0 \\
      0 & -1 
    \end{array}
  \right) ,
\ea
where $\theta_{\rm W}$ and $\theta_{\rm C}$ are the Weinberg and
Cabibbo angles, respectively.

Since $\hat{\alpha}_{\mu\parallel}$ and $\hat{\alpha}_{\mu\perp}$ 
transform as
\be
  \hat{\alpha}_{\mu\parallel,\perp}(x)
  \rightarrow
  h(x) \hat{\alpha}_{\mu\parallel,\perp}(x) h^{\dag}(x) ,
\ee
where $h(x)$ is an element of the hidden local gauge group,
the Lagrangian is given by\cite{BandoKugoYamawaki:PRep}
\ba
  \Lag_0
&=&
  \Lag_{\rm A} + a \Lag_{\rm V} + \Lag_{\rm kin}(V_\mu) ,
\nonum
&{}&
  \Lag_{\rm V}
  =
  f_\pi^2 
  \Trang{ (\hat{\alpha}_{\mu\parallel} )^2 } ,
  \mqq
  \Lag_{\rm A}
  =
  f_\pi^2 
  \Trang{ (\hat{\alpha}_{\mu\perp} )^2 } ,
\label{Lag:hidden}
\ea
where $a$ is an arbitrary parameter
to be determined from experiment.
Experimentally, in the real QCD system 
the hidden local gauge coupling is found to be%
\footnote{
The upper (lower) value corresponds to $a = 2$ ($2.2$).
The parameter choice $a=2$ 
reproduces\cite{BandoKugoUeharaYamawakiYanagida} 
the complete $\rho$ meson
dominance for the electromagnetic form factor of the pion\cite{Sakurai}.
}
\be
  \frac{g^2}{4\pi} = 2.7 \sim 3.0 .
\label{eq:gvalue}
\ee

{}From the Lagrangian, Eq.(\ref{Lag:hidden}), we get the $V{\cal V}$
mixing (${\cal V}_\mu \equiv ( {\cal R}_\mu + {\cal L}_\mu )/2$)
as
\ba
  \Lag_{V{\cal V}}
&=&
  - 2 e g_\rho \Trang{ V_\mu {\cal V}^\mu }
\nonum
&=&
  - e
  \Biggl[
    {\cal B}^\mu
    \left\{
      g_\rho \rho_\mu^0 + g_\omega \omega_\mu 
    \right\}
    +
    {\cal Z}^\mu 
    \left\{
      g_{Z\rho} \rho_\mu^0 +
      g_{Z\omega} \omega_\mu 
    \right\}
    + \cdots
  \Biggr] ,
\label{eq:VVmixing}
\ea
where
\ba
&{}&
  g_\rho
  =
  3 g_\omega
  =
  a g f_\pi^2 ,
  \mq
  g_{Z\rho}
  =
  \frac{
    1- 2 \sin^2 \theta_{\rm W}
  }{
    2\sin\theta_{\rm W}\cos\theta_{\rm W}
  }
  g_\rho ,
  \mq
  g_{Z\omega}
  =
  - \frac{\sin\theta_{\rm W}}{3\cos\theta_{\rm W}} g_\rho .
\ea

\subsection{Hidden Local Symmetry with Anomaly Terms}

This framework of HLS
provides us with the easiest way to incorporate the vector mesons 
consistently with the non-abelian anomaly.
Following 
Ref.\cite{BandoKugoYamawaki:PRep,FujiwaraKugoTeraoUeharaYamawaki},
we introduce the non-abelian anomaly into the HLS model.
The general solution to the Wess-Zumino anomaly
equation\cite{WessZumino} possessing HLS is
given by
\be
\Gamma
\left[
  \xi_{\rm L}^{\dag} \xi_{\rm R} , V, {\cal L}, {\cal R}
\right]
=
\Gamma_{\rm WZ} 
\left[
  \xi_{\rm L}^{\dag} \xi_{\rm R} , {\cal L}, {\cal R}
\right]
+
\int_{M^4} \sum_{i=1}^4
c_i {\cal L}_i ,
\label{Lag:Anom}
\ee
where $\Gamma_{\rm WZ}$ denotes the Wess-Zumino
term\cite{WessZumino,Witten:anomaly},
$c_i$ are arbitrary constants and ${\cal L}_i$ are the
gauge invariant 4-forms 
which conserve parity and charge conjugation but 
violate intrinsic parity%
\footnote{The intrinsic parity of a particle is
defined to be even if its parity equals $(-1)^{\rm spin}$,
and odd otherwise\cite{FujiwaraKugoTeraoUeharaYamawaki}.}:
\ba
&{}&
  \Lag_1
  =
  i 
  \Trang{ 
    \hat{\alpha}_{\rm L}^3 \hat{\alpha}_{\rm R}
    - \hat{\alpha}_{\rm R}^3 \hat{\alpha}_{\rm L}
  } ,
\mqqq
  \Lag_2
  =
  i 
  \Trang{ 
    \hat{\alpha}_{\rm L} \hat{\alpha}_{\rm R}
    \hat{\alpha}_{\rm L} \hat{\alpha}_{\rm R}
  } ,
\nonum
&{}&
  \Lag_3
  =
  \Trang{ 
    F_{\rm V}
    \left(
      \hat{\alpha}_{\rm L} \hat{\alpha}_{\rm R}
      - \hat{\alpha}_{\rm R} \hat{\alpha}_{\rm L}
    \right)
  } ,
\mq
  \Lag_4
  =
  \Trang{ 
    \hat{F}_{\rm L} \hat{\alpha}_{\rm L} \hat{\alpha}_{\rm R}
    - \hat{F}_{\rm R} \hat{\alpha}_{\rm R} \hat{\alpha}_{\rm L}
  } ,
\ea
where the gauge covariant building blocks are given by
\ba
  \hat{\alpha}_{\rm L,R}
&\equiv&
  \frac{1}{i} D \xi_{\rm L,R} \cdot \xi_{\rm L,R}^{\dag}
  = \frac{1}{i} D_\mu \xi_{\rm L,R} \cdot \xi_{\rm L,R}^{\dag} dx^\mu,
\nonum
  F_{\rm V}
&\equiv&
  g (d V - i g V^2) ,
\nonum
  \hat{F}_{\rm L,R}
&=&
  \xi_{\rm L,R} F_{\rm L,R} \xi_{\rm L,R}^{\dag} ,
  \mqq
  F_{\rm L}
  =
  e (d {\cal L} - i e {\cal L}^2) ,
  \mqq
  F_{\rm R}
  =
  e (d {\cal R} - i e {\cal R}^2) .
\ea

{}From Eq.(\ref{Lag:Anom}) we find the $VV\pi$ 
vertex\cite{FujiwaraKugoTeraoUeharaYamawaki} as
\ba
  \Lag_{VV\pi}
&=&
  2 g_{\omega\rho\pi}
  \varepsilon^{\mu\nu\rho\sigma}
  \Trang{ \delm V_\nu \del_\rho V_\sigma \cdot \pi }
\nonum
&=&
  g_{\omega\rho\pi}
  \varepsilon^{\mu\nu\rho\sigma}
  \Biggl[
    \delm \omega_\nu \del_\rho
    \bfrho_\sigma \cdot \bfpi 
    + \cdots 
  \Biggr] ,
\ea
where the coupling constant $g_{\omega\rho\pi}$
is given by
\be
g_{\omega\rho\pi} = 
- \frac{3 g^2}{8 \pi^2 f_\pi} c_3 .
\ee
Similarly, the $V{\cal V}\pi$ vertex is given by
\ba
  \Lag_{V{\cal V}\pi}
&=&
  2 g_{\omega\gamma\pi} (c_4-c_3)
  \varepsilon^{\mu\nu\rho\sigma}
  \Trang{
    \left\{
      \delm V_\nu , \del_\rho {\cal V}_\sigma   
    \right\}
    \cdot \pi
  }
\nonum
&=& 
  g_{\omega\gamma\pi} (c_4-c_3)
  \varepsilon^{\mu\nu\rho\sigma}
  \Biggl[
    \delm \omega_\nu \del_\rho {\cal B}_\sigma \cdot \pi^0 
    + \frac{1}{3} \delm \rho_\nu^0 \del_\rho{\cal B}_\sigma\cdot\pi^0
\nonum
&{}& \mqq
    + \bar{r} \delm \omega_\nu \del_\rho {\cal Z}_\sigma \cdot \pi^0
    {}- \frac{1}{3} \tan\theta_{\rm W} 
    \delm \rho_\nu^0 \del_\rho{\cal Z}_\sigma \cdot \pi^0
    + \cdots 
  \Biggr] ,
\ea
where the coupling constant $g_{\omega\gamma\pi}$ and $\bar{r}$
are given by
\be
  g_{\omega\gamma\pi} = 
  - \frac{3 e g}{16 \pi^2 f_\pi} ,
  \mqq
  \bar{r} \equiv 
  \frac{
    1-2\sin^2\theta_{\rm W}
  }{
    2\sin\theta_{\rm W}\cos\theta_{\rm W}
  } .
\label{def:rbar}
\ee
Moreover, the ${\cal V}{\cal V}\pi$ vertex is given by
\ba
  \Lag_{{\cal V}{\cal V}\pi}
&=&
  3 g_{\gamma\gamma\pi} (1-c_4)
  \varepsilon^{\mu\nu\rho\sigma}
  \Trang{ 
    \delm {\cal V}_\nu \del_\rho {\cal V}_\sigma \cdot \pi
  }
\nonum
&=&
  \frac{1}{2} g_{\gamma\gamma\pi} (1-c_4)
  \varepsilon^{\mu\nu\rho\sigma}
  \left[
    \delm {\cal B}_\nu \del_\rho {\cal B}_\sigma \cdot \pi^0
    + r \delm {\cal B}_\nu \del_\rho {\cal Z}_\sigma \cdot \pi^0
    + \cdots
  \right],
\ea
where
\be
  g_{\gamma\gamma\pi} 
  =
  - \frac{e^2}{4\pi^2f_\pi} ,
  \mqq
  r \equiv
  \frac{
    1- 4 \sin^2\theta_{\rm W}
  }{
    2 \sin\theta_{\rm W} \cos \theta_{\rm W}
  }.
\label{def:r}
\ee

\subsection{Transition Form Factors Based on Hidden Local Symmetry}

Now, the total Lagrangian is given by
\ba
  \Lag 
&=&
  \Lag_0 + \Lag_{\rm WZ} + \sum_{i=1}^4 c_i \Lag_i
\nonum
&=&
  \Lag_0 + 
  \Lag_{{\cal V}{\cal V}\pi} + \Lag_{V{\cal V}\pi} 
  + \Lag_{VV\pi} + \cdots ,
\label{def:total:Lag}
\ea
and we can write down the various transition form factors
using this Lagrangian.
The $\pi^0 \gamma^{\ast} \gamma^{\ast}$ transition form factor
is given by
\ba
  F^{\rm(HLS)}
  \left(
    \pi^0 ; \gamma (p^2) ; \gamma (q^2)
  \right)
&=&
    (1-c_4) 
\nonum
&{}& 
  {}+ \frac{c_4-c_3}{4}
    \left\{
      D_\rho(p^2) + D_\rho(q^2) +
      D_\omega(p^2) + D_\omega(q^2)
    \right\}
\nonum
&{}& 
    {}+ \frac{c_3}{2} 
    \left\{
      D_\rho(p^2) D_\omega(q^2) + D_\rho(q^2) D_\omega(p^2)
    \right\} ,
\label{form:hid:pgg}
\ea
where $D_\rho(p^2)$ and $D_\omega(p^2)$ are 
the Breit-Wigner type propagators for $\rho$ and $\omega$ mesons:
\ba
  D_\rho(p^2)
&=&
  \frac{m_\rho^2}{m_\rho^2-p^2 - i \sqrt{p^2}\Gamma_\rho} ,
\\
  D_\omega(p^2)
&=&
  \frac{m_\omega^2}{m_\omega^2-p^2 - i \sqrt{p^2}\Gamma_\omega} ,
\ea
with $\Gamma_\rho$ and $\Gamma_\omega$ being
the widths of $\rho$ and $\omega$ mesons, respectively.
If we set one photon on its mass-shell ($p^2=0$), this reduces to
\ba
  F^{\rm(HLS)}
  \left(
    \pi^0 ; \gamma (p^2=0) ; \gamma  (q^2)
  \right)
&=&
  \left(
    1-\frac{c_3+c_4}{2} 
  \right)
\nonum
&{}&
  {}+ \frac{c_4+c_3}{4}
    \left\{
      D_\rho(q^2) + D_\omega(q^2)
    \right\} .
\label{eq:hid:pg}
\ea

The $Z \pi^0 \gamma$ transition
form factor is also given as
\ba
  F^{\rm(HLS)}
  \left(
    \pi^0 ; \gamma (p^2) ; {\cal Z}(q^2)
  \right) \!\!\!
&=& \!\!\!
  (1-c_4)
\nonum
&{}&\!\!\!\!\!\!
  {}+ \frac{c_4-c_3}{2}
  \left\{
    \frac{\bar{r}}{r} 
    \left(
      D_\rho(q^2) + D_\omega(p^2)
    \right)
    - \frac{\bar{r}-r}{r}
    \left(
      D_\omega(q^2) + D_\rho(p^2)
    \right)
  \right\}
\nonum
&{}&\!\!\!\!\!\!
  {}+ c_3
  \left\{
    \frac{\bar{r}}{r} D_\rho(q^2) D_\omega(p^2)
    - \frac{\bar{r}-r}{r} D_\omega(q^2) D_\rho(p^2) 
  \right\},
\label{form:hid:pzg}
\ea
where $\bar{r}$ and $r$ are already 
defined in Eqs.(\ref{def:rbar}) and 
(\ref{def:r}).
For the on-shell photon ($p^2=0$), this reduces to
\ba
  F^{\rm(HLS)}
  \left(
    \pi^0 ; \gamma (p^2=0) ; {\cal Z} (q^2)
  \right)
&=&
  \left(
    1 - \frac{c_3+c_4}{2}
  \right)
\nonum
&{}&
  {}+ \frac{c_3+c_4}{2}
    \left\{
      \frac{\bar{r}}{r} D_\rho(q^2) -
      \frac{(\bar{r}-r)}{r} D_\omega(q^2)
    \right\} .
\label{form:hid:pzg2}
\ea

The $Z$ into $\pi^0\gamma$ transition amplitude is obtained by
multiplying this form factor by $g_{Z\gamma\pi}$
(see Eq.(\ref{width:Zpg}))
\be
  g_{Z\gamma\pi} \equiv \frac{-r e^2}{8\pi^2f_\pi}.
\label{def:gZgp}
\ee

For $p^2=q^2=0$,
$D_\rho$ and $D_\omega$ all reduce to 1.
Then the RHS's of Eq.(\ref{form:hid:pgg}) and Eq.(\ref{eq:hid:pg})
become 1
independently of $c_3$ and $c_4$:
\be
  F^{\rm(HLS)}
  \left(
    \pi^0 ; \gamma (p^2=0) ; \gamma  (q^2=0)
  \right)
  =
  1 .
\ee
This together with the coefficient factor 
$g_{\gamma\gamma\pi}$ of Eq.(\ref{def:r}),
excellently reproduces the experimental value of the
$\pi^0\rightarrow\gamma\gamma$ process:
this is nothing but the expression of the low energy theorem.
On the other hand, in the high energy limit, 
since $D_\rho$ and $D_\omega$ are suppressed by order $1/q^2$, 
the constant terms
$(1-c_4)$ in Eqs.(\ref{form:hid:pgg}) and (\ref{form:hid:pzg})
and $1-(c_3+c_4)/2$ in Eqs.(\ref{eq:hid:pg}) and (\ref{form:hid:pzg2})
become dominant.
We know that 
these $q^2$ independent terms necessarily violate the unitarity bound
in the high $q^2$ ($p^2$) limit.
Thus it is necessary to modify the form factor 
of Eqs.(\ref{form:hid:pzg}) and (\ref{form:hid:pzg2}).

\msection{Expression for The Transition Form Factors
\label{sec:formfactor}}

In the previous section, we have derived the transition form factors
$F^{\rm(HLS)}$ from the effective chiral Lagrangian based 
on HLS.
The transition form factors $F^{\rm(HLS)}$ 
describe the low energy processes 
induced by the axial anomaly very well.
However, as we discussed in the previous section,
we can not extend these form factors to the high energy region 
in their original forms.
In this section, we propose a new expression for the
transition form factors which are applicable independent of 
the energy regions.
Let us pay attention to the constant anomaly terms appearing in
the first terms of the form factors
Eqs.(\ref{form:hid:pgg}), (\ref{eq:hid:pg}),
(\ref{form:hid:pzg}) and (\ref{form:hid:pzg2}),
which come from $\Lag_{\rm WZ}$ of Eq.(\ref{def:total:Lag}).
In the low energy limit, we have seen that 
the value of this anomaly term 
is given by calculating the famous triangle anomaly graph
of Fig.~\ref{fig:trianglegraph}.
In sect.~\ref{sec:QCDloop}, 
we have proposed $\widetilde{J}(q^2)$ 
as the QCD corrected anomaly amplitudes.
We now further propose to use modified form factors 
where the constant anomaly terms are replaced by the improved ones.
The concrete expressions are as follows.
The $\pi^0 \gamma^{\ast} \gamma^{\ast}$ transition form factor is
\ba
  F_{\pi^0\gamma^{\ast}\gamma^{\ast}}(p^2,q^2)
&\equiv&
  F\left(\pi^0, \gamma(p^2) , \gamma(q^2)\right)
\nonum
&=&
  (1-c_4) \widetilde{I}(p^2,q^2) 
\nonum
&{}&
  {}+ \frac{c_4-c_3}{4}
  \left[
    \left\{
      D_\rho(p^2) + D_\omega(p^2)
    \right\}
    \widetilde{J}(q^2) +
    \left\{
      D_\rho(q^2) + D_\omega(q^2)
    \right\}
    \widetilde{J}(p^2)
  \right]
\nonum
&{}& 
  {}+ \frac{c_3}{2}
  \left[
    D_\rho(p^2) D_\omega(q^2)
    + D_\rho(q^2) D_\omega(p^2)
  \right] ,
\label{def:formfactor}
\ea
in which 
the constant anomaly terms
$(1-c_4)$ and $\frac{c_4-c_3}{4}$
of Eq.(\ref{form:hid:pgg})
have been replaced by
the smeared
functions $(1-c_4)\widetilde{I}$%
\footnote{%
The function $\widetilde{I}$ is obtained by the same procedure as
$\widetilde{J}$.}
and $\frac{c_4-c_3}{4}\widetilde{J}$, respectively.

As expected, this form factor again 
satisfies the low energy theorem:
\be
  F\left(\pi^0,\gamma(p^2=0),\gamma(q^2=0)\right) =1 ,
\label{eq:nomalcond2}
\ee
which is independent of the parameters $c_3$ and $c_4$.
In the high energy region, on the other hand, we have
$D_\rho(q^2) \sim m_\rho^2/q^2$ and
$D_\omega(q^2) \sim m_\omega^2/q^2$.
If one notices that 
$m_\rho \sim m_\omega \sim {\cal O}(f_\pi)$,
it is easy to see that $\rho$ and $\omega$ pole
contributions have the same high energy properties as $\widetilde{I}$
and $\widetilde{J}$.

In much the same way
we have
\ba
  F_{\pi^0\gamma}(q^2)
&\equiv&
  F
  \left(
    \pi^0, \gamma(p^2=0), \gamma(q^2)
  \right)
\nonum
&=&
  \left(
    1-\frac{c_3+c_4}{2}
  \right)
  \widetilde{J}(q^2)
  + \frac{c_3+c_4}{4}
  \left\{
    D_\rho(q^2) + D_\omega(q^2)
  \right\} ,
\label{def:pigammaformfactor}
\\
  F_{\pi^0\gamma^{\ast}Z^{\ast}}(p^2,q^2)
&\equiv&
  F
  \left(
    \pi^0 ; \gamma (p,\mu) ; {\cal Z} (q,\nu)
  \right)
\nonum
&=&
  (1-c_4) \widetilde{I}(p^2,q^2)
\nonum
&{}&
  {}+ \frac{c_4-c_3}{2}
  \biggl\{
    \frac{\bar{r}}{r} 
    \left(
      D_\rho(q^2) \widetilde{J}(p^2) 
      + D_\omega(p^2) \widetilde{J}(q^2) 
    \right)
\nonum
&{}& \mqqq
    - \frac{\bar{r}-r}{r}
    \left(
      D_\omega(q^2) \widetilde{J}(p^2) 
      + D_\rho(p^2) \widetilde{J}(q^2) 
    \right)
  \biggr\}
\nonum
&{}&
  {}+ c_3 
  \left\{
    \frac{\bar{r}}{r} D_\rho(q^2) D_\omega(p^2)
    - \frac{\bar{r}-r}{r} D_\rho(p^2) D_\omega(q^2) 
  \right\},
\\
  F_{Z\pi^0\gamma}(q^2)
&\equiv&
  F
  \left(
     \pi^0, \gamma(p^2=0), Z(q^2)
  \right)
\nonum
&=&
  \left(
    1-\frac{c_3+c_4}{2}
  \right)
  \widetilde{J}(q^2)
  + \frac{c_3+c_4}{2}
  \left\{
    \frac{\bar{r}}{r} D_\rho(q^2) - \frac{\bar{r}-r}{r} D_\omega(q^2)
  \right\} ,
\label{def:Zpigammaformfactor}
\ea
where $\bar{r}$ and $r$ are defined in Eqs.(\ref{def:rbar}) and 
(\ref{def:r}).

The vector meson form factors can also be obtained.
For the later convenience
we give the expression for the $\omega\pi^0$ transition
form factor.
By extracting the $\omega$-pole contributions from
Eq.(\ref{def:formfactor})
and by making proper normalization of the
amplitude we find
\be
  F_\omega(q^2)
  =
  - \tilde{c} \widetilde{J}(q^2) + (1+\tilde{c}) D_\rho(q^2) ,
  \mqq
  \tilde{c} \equiv \frac{c_3-c_4}{c_3+c_4} .
\label{def:OmegaPiFormfactor}
\ee

In the following sections, 
we compare the above formulae with the experimental data.
In doing this, it is important 
to bear in mind the following two points:
\begin{enumerate}
\renewcommand{\labelenumi}{\arabic{enumi})}
\item
The complete vector meson dominance (VMD) hypothesis

Since the proposal of the VMD model\cite{Sakurai},
it has often been believed that VMD
never led to contradictions with the experimental data,
and in many analyses VMD was taken for granted.
We should stress here that all the processes so far examined 
are those which occur in either 
the low energy or the high energy limit.
As may be seen from the expressions for the form factors 
derived in this section,
realization of the VMD hypothesis 
depends on the parameters $c_i$ ($c_3$ and $c_4$ in our case).
As we have already mentioned, in the low and high energy limits 
the results are independent of the choice of these parameter.
For example, 
$\Gamma(\pi^0\rightarrow\gamma\gamma)$ 
can be evaluated correctly
either from the pure Wess-Zumino term or 
by using the VMD hypothesis.
The results are found to coincide completely.
In the high energy limit, on the other hand,
Arnellos et
al.\cite{ArnellosMarcianoParsa:ZWdecay:82}
pointed out that the $\rho$ meson dominant form factor
gives the correct high energy behavior.
So we should remark that
it is the intermediate energy phenomena 
which determine the parameters $c_i$.

So, the question is in what experimental process 
we can observe the deviation from
the complete VMD hypothesis,
in which the triangle anomaly terms are appreciable.

\item
Constant versus momentum dependent triangle anomaly terms

If nature is described by the VMD model, 
the triangle anomaly term does not contribute either 
to the form factors
or to the decay amplitudes:
for example, the parameter choice $c_3=c_4=1$ in our form factor
corresponds to VMD:
\be
  F_{\pi^0\gamma^{\ast}\gamma^{\ast}}
  (p^2,q^2) =
  \frac{1}{2}
  \left[
    D_\rho(p^2) D_\omega(q^2)
    + D_\rho(q^2) D_\omega(p^2) 
  \right] ,
\ee
and in this case
the $\widetilde{I}$ and $\widetilde{J}$ terms
disappear.
However, as we shall see below, 
complete VMD is not always realized in nature,
so we need the information from the triangle anomaly terms.
For the low $q^2$ region,
most of the analyses use the amplitude 
obtained directly from the Wess-Zumino 
term\cite{WessZumino,Witten:anomaly}.
Namely, they adopt
the form factors $F^{\rm(HLS)}$ of section~\ref{sec:hidden}.
For example, Bramon et al.\cite{BramonGrauPancheri:ChPT:92}
have made an extensive analysis for relatively high $q^2$ region
using $F^{\rm(HLS)}$.

If we further straightforwardly extend the amplitudes $F^{\rm(HLS)}$
to the higher $q^2$ region,
say, for example, $q^2=M_Z^2$,
we would get an extraordinary large decay amplitude for
$Z\rightarrow\pi^0\gamma$\cite{JacobWu:89}.
There have been many discussions concerning this process
(see, for example, Refs.
\cite{Hikasa:Anom:90}, \cite{DeshpandePalOlness:90} and
\cite{Manohar:ZWpi:90}),
which have made it clear that the form factor 
should be strongly suppressed.

Our interest here is 
in examining intermediate $q^2$ phenomena
to see how the anomaly term depends on $q^2$.

\end{enumerate}

\msection{$\pi^0\gamma\gamma^{\ast}$ Process
\label{sec:pgg}}

In this section we apply our form factors of the previous section 
to the $\pi^0\gamma$ transition form factor and also
to the $Z\rightarrow\pi^0\gamma$ decay.

\subsection{$\pi^0\gamma$ Transition Form Factor 
\label{sec:pigform}}

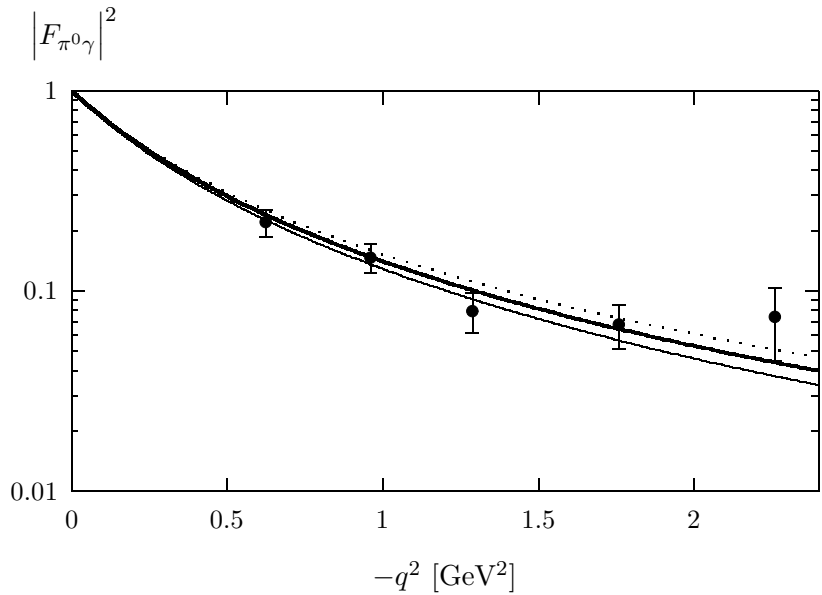
\begin{figure}[htbp]
\begin{center}
\input{pggform0}
\end{center}
\caption{
\label{fig:pggform}
The $\pi^0\gamma$ transition form factor,
where the dotted line corresponds to the parameter choice
$(c_3+c_4)/2=0.95$, the thick line, $(c_3+c_4)/2=1.0$ (VMD) 
and the thin line, $(c_3+c_4)/2=1.05$, respectively.
}
\end{figure}

The $\pi^0\gamma$ transition form factor is given in 
Eq.(\ref{def:pigammaformfactor});
\be
  F_{\pi^0\gamma} (q^2)
  =
  \left(
    1-\frac{c_3+c_4}{2}
  \right)
  \widetilde{J}(q^2)
  + \frac{c_3+c_4}{4}
  \left\{
    D_\rho(q^2) + D_\omega(q^2)
  \right\} ,
\label{def:pigammaformfactor2}
\ee
in which $(c_3+c_4)/2$ is 
to be determined from the experimental data\cite{CELLO:91}.
The experimental data points are 
shown in Fig.~\ref{fig:pggform},
together with these theoretical curves.
The thick line corresponds to the VMD:
\be
  \frac{c_3+c_4}{2}
  =
  1.0 ,
\label{fit:c3c4value}
\ee
and the thin (dotted) line corresponds to a 5\% increase (decrease)
in this value.
The result indicates that the VMD is almost realized in this process.

\subsection{$Z\rightarrow\pi^0\gamma$ Decay}

By extrapolating 
the form factor $F_{Z\pi^0\gamma}(q^2)$
of Eq.(\ref{def:Zpigammaformfactor})
to $q^2=M_Z^2$,
we obtain the decay width $\Gamma(Z\rightarrow\pi^0\gamma)$ together
with the coefficient factor $g_{Z\gamma\pi}$ of Eq.(\ref{def:gZgp}):
\be
  \Gamma(Z\rightarrow\pi^0\gamma)
  =
  \frac{\alpha^2 r^2}{48 \pi^3f_\pi^2}
  \abs{ \frac{M_Z^2 - m_\pi^2}{2M_Z} }^3
  \abs{ 
    F_{Z\pi^0\gamma}(q^2=M_Z^2)
  }^2 .
\label{width:Zpg}
\ee

We have already seen that the parameter 
$(c_3+c_4)/2$ is almost equal to $1$,
so the first term of Eq.(\ref{def:Zpigammaformfactor})
which is proportional to $\widetilde{J}$
does not contribute to the decay width.
We therefore obtain
\be
  \Gamma(Z\rightarrow\pi^0\gamma)
  =
  0.8 \times 10^{-11} \mqq \mbox{(GeV)} ,
\ee
which is of roughly the same order as has been 
predicted by other authors%
\cite{GuberinaKuhnPecceiRuckel:80,Hikasa:Anom:90,%
DeshpandePalOlness:90,Manohar:ZWpi:90,%
ArnellosMarcianoParsa:ZWdecay:82}.

\msection{Application to $\omega$ Process
\label{sec:wpg}}

The $\omega\pi^0$ transition form factor is given 
by Eq.(\ref{def:OmegaPiFormfactor}):
\be
  F_\omega(q^2)
  =
  - \tilde{c} \widetilde{J}(q^2) + (1+\tilde{c}) D_\rho(q^2) ,
  \mqq
  \tilde{c} \equiv \frac{c_3-c_4}{c_3+c_4} .
\label{def:OmegaPiFormfactor2}
\ee
where $\tilde{c}$ is to be determined.
We compare the above improved form factor
with the following unimproved form factor:
\be
  F_\omega^{\rm C}(q^2)
  =
  - \tilde{c} + (1+\tilde{c}) D_\rho(q^2) ,
\label{form:BW0}
\ee
in which the triangle anomaly term is taken to be a constant.

\subsection{$\omega\rightarrow\pi^0\gamma$ Decay}

First, we study the $\omega\rightarrow\pi^0\gamma$ decay.
Since all the external fields are on mass shell,
the form factor reduces to $1$, 
and the decay width is given by
\be
\Gamma(\omega\rightarrow\pi^0\gamma) 
=
\left(
  \frac{1}{2}(c_3+c_4) g 
\right)^2 
\frac{3\alpha}{64\pi^4f_\pi^2}
\abs{\frac{m_\omega^2-m_\pi^2}{2m_\omega}}^3 ,
\label{eq:width:wpg}
\ee
where $g$ is the gauge coupling constant of the hidden local gauge
boson.

{}From the experimental data\cite{ParticleDataGroup}
\be
  \Gamma(\omega\rightarrow\pi^0\gamma)/\Gamma_\omega
  =
  (8.5\pm0.5) \times 10^{-2} ,
\label{data:branch:omegapigamma}
\ee
or
\be
  \Gamma(\omega\rightarrow\pi^0\gamma)_{\rm exp} =
  (7.2\pm0.5) \times 10^{-1} \mq \mbox{(MeV)},
\label{omega:partialwidth}
\ee
and so we find
\be
  \left(
    \frac{1}{2}(c_3+c_4) g
  \right)^2
  = 32.6 \pm 2.3 \, .
\ee
Using the value of $g$ given in Eq.(\ref{eq:gvalue}) 
we get 
\be
  \frac{1}{2}(c_3+c_4) = 0.90 \sim 1.0 \, ,
\ee
which is to be compared with the value we independently derived 
in sect.~\ref{sec:pigform} (see Eq.(\ref{fit:c3c4value})).
Thus HLS also describes the $\omega\rightarrow\pi^0\gamma$ decay 
well\cite{FujiwaraKugoTeraoUeharaYamawaki}.
In the form factor analysis below, we take the
$\omega\rightarrow\pi^0\gamma$ decay width as an input.

\subsection{$\omega\rightarrow\pi^0\mu^+\mu^-$ Decay}

We now consider the $\omega\rightarrow\pi^0\mu^+\mu^-$ decay.
This gives information about the form factor at high $q^2$.
It is convenient to write this decay width as\cite{Dzhelyadinetal1}
\ba
  \Gamma(\omega\rightarrow\pi^0\mu^+\mu^-)
&=&
  \int_{4 m_\mu^2}^{(m_\omega-m_\pi)^2} dq^2
  \frac{\alpha}{3\pi}
  \frac{\Gamma(\omega\rightarrow\pi^0\gamma)}{q^2}
  \left(
    1 + \frac{2m_\mu^2}{q^2}
  \right)
  \sqrt{\frac{q^2-4m_\mu^2}{q^2}}
\nonum
&{}& \mq 
  \times
  \left[
    \left(
      1 + \frac{q^2}{m_\omega^2-m_\pi^2}
    \right)^2
    - \frac{4m_\omega^2q^2}{(m_\omega^2-m_\pi^2)^2}
  \right]^{3/2}
  \abs{F_\omega(q^2)}^2 ,
\ea
where $q^2$ is the intermediate photon momentum 
(or invariant mass of final muons),
and $F_\omega(q^2)$ is the $\omega\pi^0$ transition form factor
given in Eq.(\ref{def:OmegaPiFormfactor2}).

We find
\be
  \Gamma(\omega\rightarrow\pi^0\mu^+\mu^-) 
  =
  \left(
    8.97 + 1.79 \tilde{c} + 0.32 \tilde{c}^2
  \right)
  \times 10^{-4}
  \times \Gamma(\omega\rightarrow\pi^0\gamma) 
  \mq \mbox{(MeV)} .
\ee
The experimental value for the branching ratio of this mode is
given by\cite{ParticleDataGroup}
\be
  \Gamma(\omega\rightarrow\pi^0\mu^+\mu^-) / \Gamma\omega
  =
  (9.6\pm2.3) \times 10^{-5} ,
\ee
and using this and the branching ratio of
$\omega\rightarrow\pi^0\gamma$ in Eq.(\ref{data:branch:omegapigamma}),
we find
\be
  \frac{
    \Gamma(\omega\rightarrow\pi^0\mu^+\mu^-)
  }{
    \Gamma(\omega\rightarrow\pi^0\gamma)
  }
  =
  \frac{
    (9.6\pm2.3) \times 10^{-5} 
  }{
    (8.5\pm0.5) \times 10^{-2} 
  }
  \simeq
  (1.1\pm0.3) \times 10^{-3} .
\ee
{}From these we obtain%
\footnote{The another solution $\tilde{c} = -2.5 \pm 0.1$ 
is excluded by experiments.}
\be
  \tilde{c}
  =
  0.49 ^{\displaystyle +0.12}_{\displaystyle -0.13} .
\label{value:ctilde}
\ee
This clearly shows that 
complete $\rho$ meson dominance is incapable of describing
the $\omega\pi^0$ form factor\cite{Dzhelyadinetal1}
(VMD corresponds to $\tilde{c}=0$, see
Eq.(\ref{def:OmegaPiFormfactor2})).

\msection{Overall Fit of Form Factor
\label{wpgform}}

As we saw in the previous section,
the transition form factor $\omega\rightarrow\pi^0\gamma$
receives appreciable contributions from the anomaly term,
and we use this process to obtain information about the 
$q^2$ dependence from the experimental data.
The available experimental data for $\omega\pi^0\gamma^{\ast}$
lies in a relatively wide energy range 
(up to $1.4$GeV).
We compare our form factor $F_\omega$ 
with the unimproved form factor $F_\omega^{\rm C}$.

\subsection{$\omega\pi^0$ Transition Form Factor below $m_\omega$
\label{sec:omegapi}}

\begin{figure}[htbp]
\begin{center}
\input{wpgformj}
\input{wpgformw}
\caption[]{
The $\omega\pi^0$ transition form factor in the energy range below
$m_\omega$: (a) our form factor $F_\omega(q^2)$;
(b) unimproved form factor $F_\omega^{\rm C}(q^2)$.
We use the experimental data shown in Ref.\cite{Dzhelyadinetal1}.
In (a) and (b), the thick line corresponds to the parameter choice
$\tilde{c}=1$, the thin line, $\tilde{c}=0.5$ and the dotted line,
$\tilde{c}=0$ (VMD).
\label{fig:form:low}}
\end{center}
\end{figure}
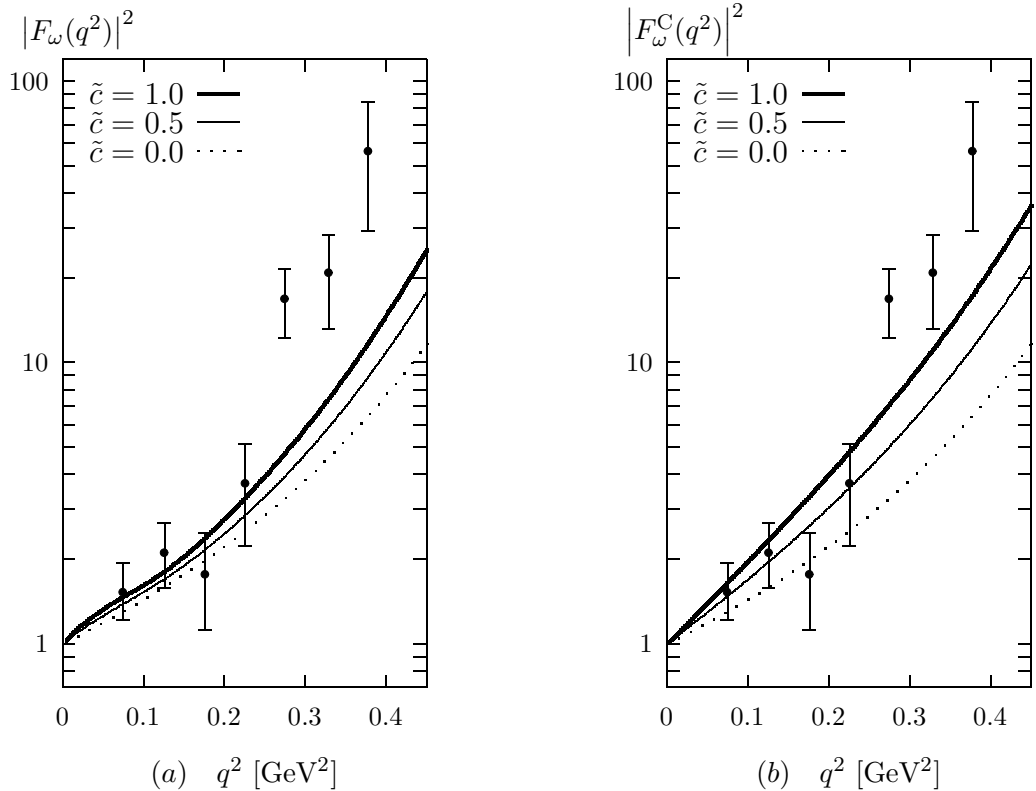

First, we fit our form factor given in
Eq.(\ref{def:OmegaPiFormfactor2}) to the experimental
data \cite{Dzhelyadinetal1}.
In the very low energy region 
the function $\widetilde{J}(q^2)$ is
approximately 1 and
so $q^2$ dependent effects are very small.
The results are shown 
in Fig.~\ref{fig:form:low},
in which the thick line corresponds to the parameter choice
$\tilde{c}=1$, the thin line, $\tilde{c}=0.5$ 
and the dotted line, $\tilde{c}=0$ (VMD).

\noindent
Although the experimental data shows some
deviation from the theoretical curve,
we can clearly exclude the VMD value $\tilde{c}\simeq0$.
A rough estimate gives\footnote{
In Ref.~\cite{BramonGrauPancheri:ChPT:92}, 
the value $\tilde{c}=0.5$ is used.}
\be
  \tilde{c} = 0.5 \sim 1.0 ,
\label{value:ctil:low}
\ee
which is to be compared with Eq.(\ref{value:ctilde}).
So far as we only use the low energy data,
we can not clearly discriminate between
our form factor $F_\omega$ 
and the unimproved one $F^{\rm C}$.

\subsection{ $e^- e^+ \rightarrow\omega\pi^0$ \label{sec:eewp}}

\begin{figure}[bthp]
\begin{center}
\input{f_eewpj}
\end{center}
\caption[]{
Total cross section of the process 
$e^- e^+ \rightarrow\omega\pi^0$ in the energy region from 1 GeV to
1.4 GeV using our form factor $F_\omega(q^2)$.
The thick line corresponds to the parameter choice
$\tilde{c}=0.8$, the thin line, $\tilde{c}=0.7$ 
and the dotted line, $\tilde{c}=0.5$.
The experimental data is given in Ref.\cite{Dolinskyetal}.
\label{fig:eewp:our}
}
\end{figure}
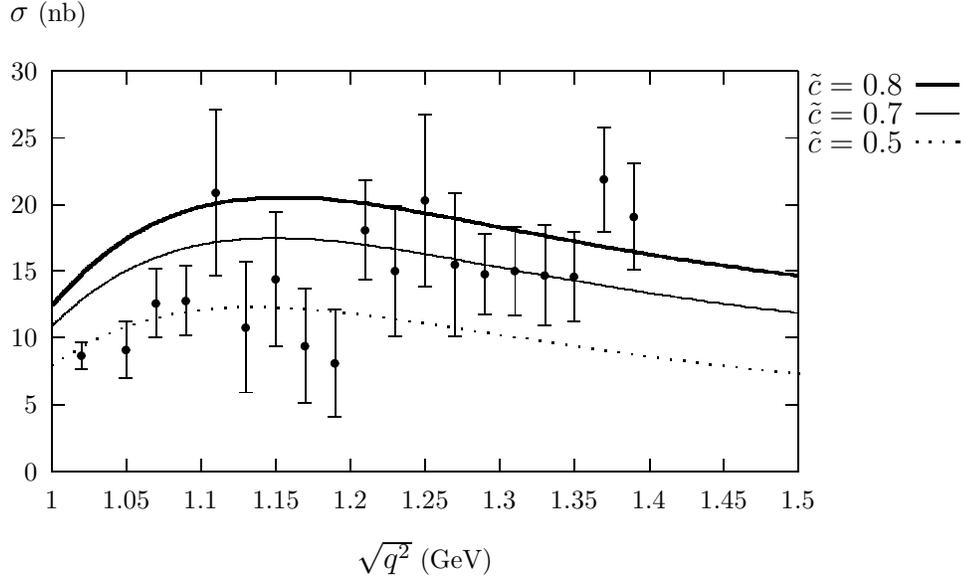
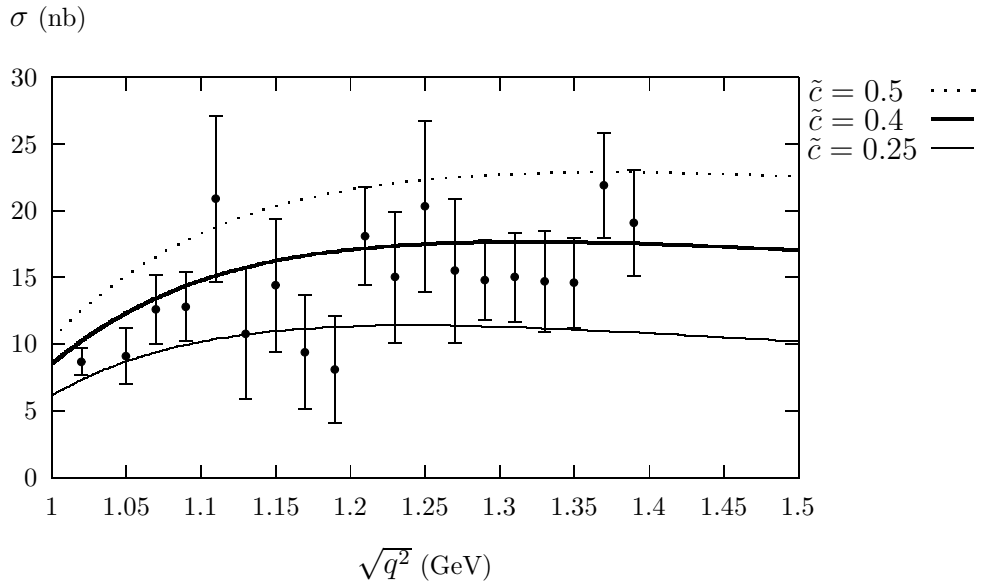
\begin{figure}[htbp]
\begin{center}
\input{f_eewp1}
\end{center}
\caption[]{
Total cross section of the process 
$e^- e^+ \rightarrow\omega\pi^0$ in the energy region from 1 GeV to
1.4 GeV using the unimproved form factor $F_\omega^{\rm C}(q^2)$.
The thick line corresponds to the parameter choice
$\tilde{c}=0.4$, the thin line, $\tilde{c}=0.25$ 
and the dotted line, $\tilde{c}=0.5$.
The experimental data is given in Ref.\cite{Dolinskyetal}.
\label{fig:eewp:hid}
}
\end{figure}

Let us go to the higher $q^2$ region and 
study the process $e^- e^+ \rightarrow\omega\pi^0$.
In this case $q^2$ is the total energy squared of 
the $e^+e^-$ pair ($q^2>(m_\omega+m_\pi)^2$).
The cross section for $e^- e^+ \rightarrow\omega\pi^0$
is given by
\be
  \sigma(e^- e^+ \rightarrow\omega\pi^0)
 =
  4 \pi \alpha \Gamma(\omega\rightarrow\pi^0\gamma)
  \frac{q^2+2m_e^2}{q^4 \sqrt{q^2-4m_e^2}}
  \frac{p_\omega^3(q^2)}{p_\gamma^3}
  \times \abs{F_\omega(q^2)}^2 \, ,
\ee
where $p_\omega(q^2)$ and $p_\gamma$ denote the $\omega$ momentum
and the photon momentum respectively.
These are given by
\ba
  p_\gamma
&=&
  \frac{m_\omega^2 - m_\pi^2}{2m_\omega} ,
\\
  p_\omega(q^2)
&=&
  \frac{1}{2\sqrt{q^2}}
  \sqrt{
    \left(
      q^2 - (m_\omega+m_\pi)^2
    \right)
    \left(
      q^2 - (m_\omega-m_\pi)^2
    \right)
  } \, .
\ea
\newline
The results using our form factor $F_\omega$
are shown in Fig.~\ref{fig:eewp:our},
from which we find that
the effect of the function $\widetilde{J}(q^2)$ suppresses 
the cross section appreciably.
This enables us to reproduce the experimental data with the value
$\tilde{c}=0.5\sim0.8$.
On the other hand, the predictions using the
unimproved form factor $F_\omega^{\rm C}$ (see Fig.~\ref{fig:eewp:hid})
prefer the value $\tilde{c}=0.25\sim0.40$,
which is smaller than that of Eq.(\ref{value:ctil:low}).

\subsection{Overall Fit in the Energy Range below 1.4~GeV}

\begin{figure}[htbp]
\begin{center}
\input{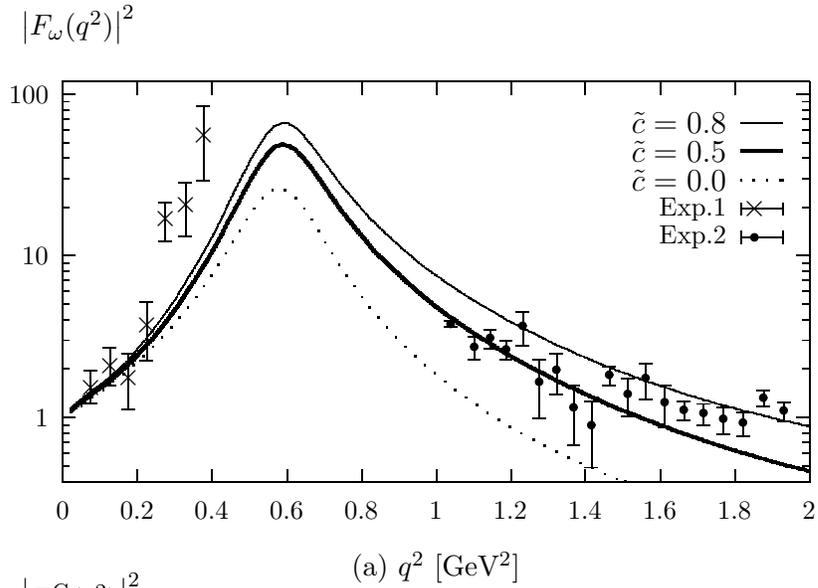}
\input{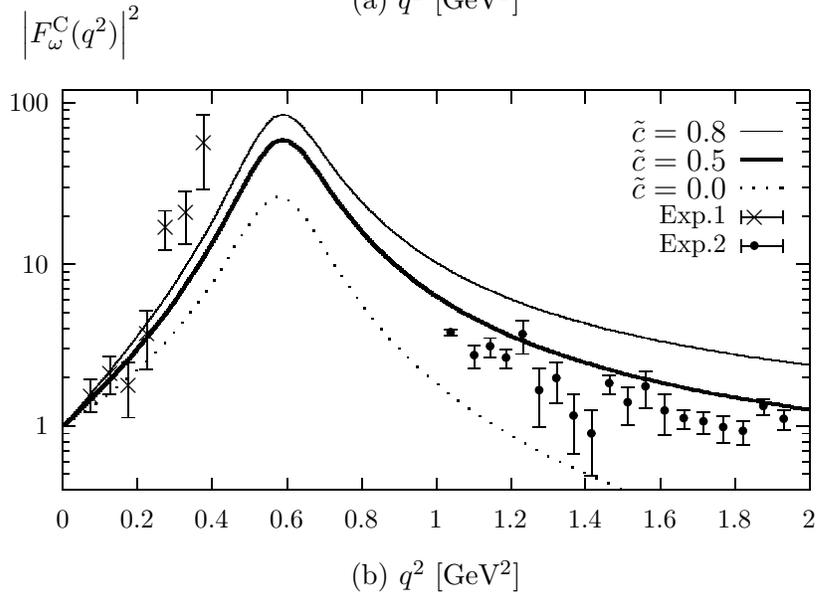}
\end{center}
\caption[]{
The form factor in the energy range below 1.4 GeV.
The low energy data (Exp.1) is given in Ref.\cite{Dzhelyadinetal1},
and the high energy data (Exp.2) is translated 
from the cross section data\cite{Dolinskyetal}.
The thick line ($\tilde{c}=0.5$), the thin line ($\tilde{c}=0.8$) and
the dotted line ($\tilde{c}=0.0$) show the values
of (a) our form factor $F_\omega(q^2)$ and 
(b) unimproved one $F_\omega^{\rm C}(q^2)$.
\label{fig:formgl}
}
\end{figure}
To summarizing the result so far examined:
Data for the
$\pi^0\gamma$ transition form factor indicates
that $(c_3+c_4)/2=1$, and so 
their experiment is consistent with VMD.
It must be noted, however, that VMD of $\pi^0\gamma$ transition form
factor does not imply $c_3=c_4=1$.
In fact,
the experimental data for the $\omega\pi^0$ transition form factor
shows $\tilde{c}\neq0$ ($\tilde{c}=0.5\sim0.8$), i.e., $c_3 \neq c_4$,
which implies some deviation from VMD.
Bearing these facts in mind we here study
our form factor $F_\omega$ 
by comparing it with the experimental data
over the whole energy range below 1.4 GeV.
To do this, it is convenient to convert
the cross section data of Fig.~\ref{fig:eewp:our}
into values for the form factor,
and combine the result with the values of Fig.~\ref{fig:form:low}.
The predicted curves are shown in Fig.~\ref{fig:formgl}.
For reference we show the fitting using the unimproved form
factor $F_\omega^{\rm C}$.

\msection{Summary and Further Outlook
\label{sec:concl}}

We have seen that the axial anomaly induced amplitudes are not always
expressed in terms of a VMD model.
So far as the $\pi^0\gamma$ transition form factor is
concerned,
the experiments indicate that VMD works well.
On the other hand, 
in the experiments for $\omega\pi^0$ form factor, it does not.
Because of this deviation from VMD,
the anomaly term contributions become important.

It should be remarked that,
even for the $\pi^0\gamma$ transition form factor, 
there may be cases where 
the $q^2$ dependence of the anomaly term becomes important.
For example, if we adopt a constant anomaly term,
even very small deviation from VMD yields a large 
$Z\rightarrow\pi^0\gamma$ decay width:
let the deviation from VMD be $\varepsilon$, 
so that $\frac{c_3+c_4}{2}=1+\varepsilon$,
and 
let us suppose that $\varepsilon$ is very small,
say for example, $10^{-1}$.
Since the suppression of 
$D_\rho(q^2)\sim m_\rho^2/M_Z^2$
is of order ${\cal O}(10^{-4})$,
the first term of Eq.(\ref{form:hid:pzg2}) dominates,
which yields
$\Gamma(Z\rightarrow\pi^0\gamma)\sim{\cal O}(10^{-5})$GeV.
The $10^7$ $Z$ events at LEP
then might include the rare decay $\pi^0\gamma$ events already.

The main task in this paper has been to investigate the $q^2$
dependence of this anomaly term.
We proposed a form factor with the improved triangle anomaly
amplitudes $\widetilde{I}$ and $\widetilde{J}$ 
of Eq.(\ref{def:formfactor})
and saw that this reproduces the existing $\omega\pi^0$ form factors
(at least up to 1.4 GeV).

This form factor is found to also be consistent with the extremely
high $q^2$ behavior.
At present, however, 
there exists no data to check whether 
we can extend our form factor to the region $q^2>(\mbox{1.4GeV})^2$.
In this sense the $Z\rightarrow\pi^0\gamma$ process may be important.
Unfortunately in addition to the suppression factor 
${\cal O}(f_\pi^2/q^2)$,
the coupling with the $Z$ is proportional
to 
\be
  Q^z = \frac{1}{\sin\theta_{\rm W}\cos\theta_{\rm W}}
  \left[
    \frac{1}{2} T_3 - Q \sin^2 \theta_{\rm W}
  \right] ,
\ee
and the ratio of the coefficient $g_{Z\gamma\pi}^2$ 
to $g_{\gamma\gamma\pi}^2$ 
(see Eqs.(\ref{def:gZgp}) and (\ref{def:r}))
is 
\be
  \left(
    \frac{r}{2} 
  \right)^2
  =
  \left(
    \frac{
      1-4\sin^2 \theta_{\rm W}
    }{
      4 \sin\theta_{\rm W}\cos\theta_{\rm W}
    } 
  \right)^2
  \simeq 10^{-3}.
\ee
This is why the $Z\rightarrow\pi^0\gamma$ decay width is predicted
very small.

In the future, the $W\rightarrow\pi\gamma$ decay processes
may give important information,
because this amplitude are not suppressed.
Two photon processes will also provide us
with the another important informations on the function $\widetilde{I}$.

Finally we make comments on the future tasks.

1)
We may expect a considerably larger amplitude for 
the high $q^2$ region for decays into NG bosons composed of
heavy quarks.
For example, the decay of $Z$ into $\gamma$ and heavy quarkonia
may be experimentally detected in the near future.
So it is interesting to extend our formula to heavy quark systems.
To do this we have to take account of 
explicit chiral symmetry breaking effects 
in the HLS framework.

2)
A more interesting possibility is to apply our formulae 
to strong interacting systems other than QCD.
For example, technicolor systems may be one of those possibilities.
Recently, there have been proposed various types of technicolor-like
dynamics whose scales are of the of order several tens of GeV.
In such cases, if the technicolor dynamics is 
of a different type from QCD,
one may expect very light vector bound
states or relatively small decay constants 
of the order of a few tens of GeV.
Then the axial anomaly induced processes might play an essential role 
and may be observed as very clear events.

\vspace{1cm}

\section*{Acknowledgement}

We would like to thank Taichiro Kugo and Nobuhiro Maekawa 
for stimulating discussions and comments and also
Ken-ichi Hikasa for the discussion about the triangle anomaly term.
We are also grateful to 
Mark Mitchard for critical reading of this manuscript.

\end{document}

%% file: fig_tri.tex
\setlength{\unitlength}{0.0050in}
\begin{picture}(385,419)(0,-10)
\thicklines
\put(180.000,275.000){\arc{40.000}{4.7124}{7.8540}}
\put(180.000,315.000){\arc{40.000}{1.5708}{4.7124}}
\put(180.000,355.000){\arc{40.000}{4.7124}{7.8540}}
\put(305.000,70.000){\arc{42.426}{3.9270}{7.0686}}
\put(335.000,40.000){\arc{42.426}{0.7854}{3.9270}}
\put(85.000,100.000){\arc{42.426}{2.3562}{5.4978}}
\put(55.000,70.000){\arc{42.426}{5.4978}{8.6394}}
\put(25.000,40.000){\arc{42.426}{2.3562}{5.4978}}
\put(275.000,100.000){\arc{42.426}{0.7854}{3.9270}}
\put(180,255){\blacken\ellipse{10}{10}}
\put(180,255){\ellipse{10}{10}}
\put(260,115){\blacken\ellipse{10}{10}}
\put(260,115){\ellipse{10}{10}}
\put(100,115){\blacken\ellipse{10}{10}}
\put(100,115){\ellipse{10}{10}}
\path(180,255)(100,115)(260,115)(180,255)
\path(210,325)(210,305)
\path(220,200)(220,185)
\path(220,185)(205,190)
\path(185,125)(170,115)(185,105)
\path(130,190)(145,195)(145,180)
\path(285,65)(300,50)
\path(300,50)(290,50)
\path(300,60)(300,50)
\path(90,50)(75,35)
\path(75,35)(85,35)
\path(75,45)(75,35)
\path(180,255)(100,115)(260,115)(180,255)
\path(180,255)(100,115)(260,115)(180,255)
\path(205,310)(210,305)
\path(215,310)(215,310)(210,305)
\put(175,390){\makebox(0,0)[lb]{\ssize$j_5^\mu$}}
\put(220,305){\makebox(0,0)[lb]{\ssize$k$}}
\put(265,35){\makebox(0,0)[lb]{\ssize$q$}}
\put(345,0){\makebox(0,0)[lb]{\ssize$j^\rho$}}
\put(0,0){\makebox(0,0)[lb]{\ssize$j^\nu$}}
\put(90,20){\makebox(0,0)[lb]{\ssize$p$}}
\put(265,35){\makebox(0,0)[lb]{\ssize$q$}}
\put(205,245){\makebox(0,0)[lb]{\ssize$\gamma^\mu\gamma_5$}}
\put(265,130){\makebox(0,0)[lb]{\ssize$\gamma^\rho$}}
\put(55,130){\makebox(0,0)[lb]{\ssize$\gamma^\nu$}}
\end{picture}

%% file: modfunre.tex
\setlength{\unitlength}{0.240900pt}
\ifx\plotpoint\undefined\newsavebox{\plotpoint}\fi
\sbox{\plotpoint}{\rule[-0.175pt]{0.350pt}{0.350pt}}%
\begin{picture}(1500,900)(0,0)
\tenrm
\sbox{\plotpoint}{\rule[-0.175pt]{0.350pt}{0.350pt}}%
\put(264,263){\rule[-0.175pt]{282.335pt}{0.350pt}}
\put(264,158){\rule[-0.175pt]{4.818pt}{0.350pt}}
\put(242,158){\makebox(0,0)[r]{-0.5}}
\put(1416,158){\rule[-0.175pt]{4.818pt}{0.350pt}}
\put(264,263){\rule[-0.175pt]{4.818pt}{0.350pt}}
\put(242,263){\makebox(0,0)[r]{0}}
\put(1416,263){\rule[-0.175pt]{4.818pt}{0.350pt}}
\put(264,368){\rule[-0.175pt]{4.818pt}{0.350pt}}
\put(242,368){\makebox(0,0)[r]{0.5}}
\put(1416,368){\rule[-0.175pt]{4.818pt}{0.350pt}}
\put(264,473){\rule[-0.175pt]{4.818pt}{0.350pt}}
\put(242,473){\makebox(0,0)[r]{1}}
\put(1416,473){\rule[-0.175pt]{4.818pt}{0.350pt}}
\put(264,577){\rule[-0.175pt]{4.818pt}{0.350pt}}
\put(242,577){\makebox(0,0)[r]{1.5}}
\put(1416,577){\rule[-0.175pt]{4.818pt}{0.350pt}}
\put(264,682){\rule[-0.175pt]{4.818pt}{0.350pt}}
\put(242,682){\makebox(0,0)[r]{2}}
\put(1416,682){\rule[-0.175pt]{4.818pt}{0.350pt}}
\put(264,787){\rule[-0.175pt]{4.818pt}{0.350pt}}
\put(242,787){\makebox(0,0)[r]{2.5}}
\put(1416,787){\rule[-0.175pt]{4.818pt}{0.350pt}}
\put(264,158){\rule[-0.175pt]{0.350pt}{4.818pt}}
\put(264,113){\makebox(0,0){\small$10^{-2}$}}
\put(264,767){\rule[-0.175pt]{0.350pt}{4.818pt}}
\put(431,158){\rule[-0.175pt]{0.350pt}{4.818pt}}
\put(431,113){\makebox(0,0){\small$10^{-1}$}}
\put(431,767){\rule[-0.175pt]{0.350pt}{4.818pt}}
\put(599,158){\rule[-0.175pt]{0.350pt}{4.818pt}}
\put(599,113){\makebox(0,0){\small$1$}}
\put(599,767){\rule[-0.175pt]{0.350pt}{4.818pt}}
\put(766,158){\rule[-0.175pt]{0.350pt}{4.818pt}}
\put(766,113){\makebox(0,0){\small$10^{1}$}}
\put(766,767){\rule[-0.175pt]{0.350pt}{4.818pt}}
\put(934,158){\rule[-0.175pt]{0.350pt}{4.818pt}}
\put(934,113){\makebox(0,0){\small$10^{2}$}}
\put(934,767){\rule[-0.175pt]{0.350pt}{4.818pt}}
\put(1101,158){\rule[-0.175pt]{0.350pt}{4.818pt}}
\put(1101,113){\makebox(0,0){\small$10^{3}$}}
\put(1101,767){\rule[-0.175pt]{0.350pt}{4.818pt}}
\put(1269,158){\rule[-0.175pt]{0.350pt}{4.818pt}}
\put(1269,113){\makebox(0,0){\small$10^{4}$}}
\put(1269,767){\rule[-0.175pt]{0.350pt}{4.818pt}}
\put(1436,158){\rule[-0.175pt]{0.350pt}{4.818pt}}
\put(1436,113){\makebox(0,0){\small$10^{5}$}}
\put(1436,767){\rule[-0.175pt]{0.350pt}{4.818pt}}
\put(264,158){\rule[-0.175pt]{282.335pt}{0.350pt}}
\put(1436,158){\rule[-0.175pt]{0.350pt}{151.526pt}}
\put(264,787){\rule[-0.175pt]{282.335pt}{0.350pt}}
\put(850,60){\makebox(0,0){\small$q^2/m^2$}}
\put(850,840){\makebox(0,0){\small Real part of $J(q^2)$ and $\widetilde{J}(q^2/m^2)$}}
\put(264,158){\rule[-0.175pt]{0.350pt}{151.526pt}}
\sbox{\plotpoint}{\rule[-0.250pt]{0.500pt}{0.500pt}}%
\put(1306,722){\makebox(0,0)[r]{\small$\Re[J]$}}
\put(1328,722){\usebox{\plotpoint}}
\put(1348,722){\usebox{\plotpoint}}
\put(1369,722){\usebox{\plotpoint}}
\put(1390,722){\usebox{\plotpoint}}
\put(1394,722){\usebox{\plotpoint}}
\put(264,473){\usebox{\plotpoint}}
\put(284,473){\usebox{\plotpoint}}
\put(305,473){\usebox{\plotpoint}}
\put(326,473){\usebox{\plotpoint}}
\put(347,473){\usebox{\plotpoint}}
\put(367,473){\usebox{\plotpoint}}
\put(388,473){\usebox{\plotpoint}}
\put(409,474){\usebox{\plotpoint}}
\put(430,474){\usebox{\plotpoint}}
\put(450,475){\usebox{\plotpoint}}
\put(471,476){\usebox{\plotpoint}}
\put(492,477){\usebox{\plotpoint}}
\put(512,478){\usebox{\plotpoint}}
\put(533,480){\usebox{\plotpoint}}
\put(553,482){\usebox{\plotpoint}}
\put(573,486){\usebox{\plotpoint}}
\put(593,491){\usebox{\plotpoint}}
\put(612,497){\usebox{\plotpoint}}
\put(630,507){\usebox{\plotpoint}}
\put(646,518){\usebox{\plotpoint}}
\put(659,533){\usebox{\plotpoint}}
\put(670,550){\usebox{\plotpoint}}
\put(679,569){\usebox{\plotpoint}}
\put(684,588){\usebox{\plotpoint}}
\put(689,609){\usebox{\plotpoint}}
\put(692,629){\usebox{\plotpoint}}
\put(694,650){\usebox{\plotpoint}}
\put(695,671){\usebox{\plotpoint}}
\put(696,691){\usebox{\plotpoint}}
\put(697,712){\usebox{\plotpoint}}
\put(698,733){\usebox{\plotpoint}}
\put(699,754){\usebox{\plotpoint}}
\put(699,774){\usebox{\plotpoint}}
\put(701,758){\usebox{\plotpoint}}
\put(704,737){\usebox{\plotpoint}}
\put(706,717){\usebox{\plotpoint}}
\put(708,696){\usebox{\plotpoint}}
\put(710,676){\usebox{\plotpoint}}
\put(713,655){\usebox{\plotpoint}}
\put(715,634){\usebox{\plotpoint}}
\put(718,614){\usebox{\plotpoint}}
\put(721,593){\usebox{\plotpoint}}
\put(725,573){\usebox{\plotpoint}}
\put(728,552){\usebox{\plotpoint}}
\put(731,532){\usebox{\plotpoint}}
\put(735,511){\usebox{\plotpoint}}
\put(739,491){\usebox{\plotpoint}}
\put(743,471){\usebox{\plotpoint}}
\put(747,450){\usebox{\plotpoint}}
\put(752,430){\usebox{\plotpoint}}
\put(757,410){\usebox{\plotpoint}}
\put(763,390){\usebox{\plotpoint}}
\put(769,370){\usebox{\plotpoint}}
\put(775,351){\usebox{\plotpoint}}
\put(784,332){\usebox{\plotpoint}}
\put(793,313){\usebox{\plotpoint}}
\put(804,295){\usebox{\plotpoint}}
\put(816,279){\usebox{\plotpoint}}
\put(830,264){\usebox{\plotpoint}}
\put(847,253){\usebox{\plotpoint}}
\put(866,245){\usebox{\plotpoint}}
\put(887,240){\usebox{\plotpoint}}
\put(907,239){\usebox{\plotpoint}}
\put(928,239){\usebox{\plotpoint}}
\put(949,240){\usebox{\plotpoint}}
\put(969,242){\usebox{\plotpoint}}
\put(990,245){\usebox{\plotpoint}}
\put(1010,247){\usebox{\plotpoint}}
\put(1031,248){\usebox{\plotpoint}}
\put(1052,251){\usebox{\plotpoint}}
\put(1072,253){\usebox{\plotpoint}}
\put(1093,254){\usebox{\plotpoint}}
\put(1113,255){\usebox{\plotpoint}}
\put(1134,257){\usebox{\plotpoint}}
\put(1155,258){\usebox{\plotpoint}}
\put(1175,259){\usebox{\plotpoint}}
\put(1196,260){\usebox{\plotpoint}}
\put(1217,260){\usebox{\plotpoint}}
\put(1238,261){\usebox{\plotpoint}}
\put(1258,261){\usebox{\plotpoint}}
\put(1279,261){\usebox{\plotpoint}}
\put(1300,262){\usebox{\plotpoint}}
\put(1320,262){\usebox{\plotpoint}}
\put(1341,262){\usebox{\plotpoint}}
\put(1362,262){\usebox{\plotpoint}}
\put(1383,262){\usebox{\plotpoint}}
\put(1404,262){\usebox{\plotpoint}}
\put(1424,263){\usebox{\plotpoint}}
\put(1436,263){\usebox{\plotpoint}}
\sbox{\plotpoint}{\rule[-0.350pt]{0.700pt}{0.700pt}}%
\put(1306,665){\makebox(0,0)[r]{\small$\Re[\widetilde{J}]$}}
\put(1328,665){\rule[-0.350pt]{15.899pt}{0.700pt}}
\put(264,472){\usebox{\plotpoint}}
\put(264,472){\rule[-0.350pt]{48.421pt}{0.700pt}}
\put(465,473){\rule[-0.350pt]{2.891pt}{0.700pt}}
\put(477,474){\rule[-0.350pt]{4.336pt}{0.700pt}}
\put(495,475){\rule[-0.350pt]{1.445pt}{0.700pt}}
\put(501,476){\rule[-0.350pt]{1.445pt}{0.700pt}}
\put(507,477){\rule[-0.350pt]{1.445pt}{0.700pt}}
\put(513,478){\rule[-0.350pt]{1.325pt}{0.700pt}}
\put(518,479){\rule[-0.350pt]{1.325pt}{0.700pt}}
\put(524,480){\rule[-0.350pt]{0.964pt}{0.700pt}}
\put(528,481){\rule[-0.350pt]{0.964pt}{0.700pt}}
\put(532,482){\rule[-0.350pt]{0.964pt}{0.700pt}}
\put(536,483){\rule[-0.350pt]{0.723pt}{0.700pt}}
\put(539,484){\rule[-0.350pt]{0.723pt}{0.700pt}}
\put(542,485){\rule[-0.350pt]{0.723pt}{0.700pt}}
\put(545,486){\rule[-0.350pt]{0.723pt}{0.700pt}}
\put(548,487){\usebox{\plotpoint}}
\put(550,488){\usebox{\plotpoint}}
\put(552,489){\usebox{\plotpoint}}
\put(555,490){\usebox{\plotpoint}}
\put(557,491){\usebox{\plotpoint}}
\put(560,492){\usebox{\plotpoint}}
\put(562,493){\usebox{\plotpoint}}
\put(564,494){\usebox{\plotpoint}}
\put(566,495){\usebox{\plotpoint}}
\put(568,496){\usebox{\plotpoint}}
\put(570,497){\usebox{\plotpoint}}
\put(572,498){\usebox{\plotpoint}}
\put(574,499){\usebox{\plotpoint}}
\put(576,500){\usebox{\plotpoint}}
\put(578,501){\usebox{\plotpoint}}
\put(580,502){\usebox{\plotpoint}}
\put(582,503){\usebox{\plotpoint}}
\put(584,504){\usebox{\plotpoint}}
\put(585,505){\usebox{\plotpoint}}
\put(586,506){\usebox{\plotpoint}}
\put(588,507){\usebox{\plotpoint}}
\put(589,508){\usebox{\plotpoint}}
\put(590,509){\usebox{\plotpoint}}
\put(592,510){\usebox{\plotpoint}}
\put(593,511){\usebox{\plotpoint}}
\put(595,512){\usebox{\plotpoint}}
\put(596,513){\usebox{\plotpoint}}
\put(598,514){\usebox{\plotpoint}}
\put(599,515){\usebox{\plotpoint}}
\put(601,516){\usebox{\plotpoint}}
\put(602,517){\usebox{\plotpoint}}
\put(604,518){\usebox{\plotpoint}}
\put(605,519){\usebox{\plotpoint}}
\put(607,520){\usebox{\plotpoint}}
\put(608,521){\usebox{\plotpoint}}
\put(610,522){\usebox{\plotpoint}}
\put(611,523){\usebox{\plotpoint}}
\put(613,524){\usebox{\plotpoint}}
\put(614,525){\usebox{\plotpoint}}
\put(616,526){\usebox{\plotpoint}}
\put(617,527){\usebox{\plotpoint}}
\put(619,528){\usebox{\plotpoint}}
\put(620,529){\usebox{\plotpoint}}
\put(622,530){\usebox{\plotpoint}}
\put(623,531){\usebox{\plotpoint}}
\put(625,532){\usebox{\plotpoint}}
\put(626,533){\usebox{\plotpoint}}
\put(628,534){\usebox{\plotpoint}}
\put(629,535){\usebox{\plotpoint}}
\put(631,536){\usebox{\plotpoint}}
\put(633,537){\usebox{\plotpoint}}
\put(635,538){\usebox{\plotpoint}}
\put(638,539){\usebox{\plotpoint}}
\put(640,540){\usebox{\plotpoint}}
\put(643,541){\rule[-0.350pt]{0.964pt}{0.700pt}}
\put(647,542){\rule[-0.350pt]{0.964pt}{0.700pt}}
\put(651,543){\rule[-0.350pt]{0.964pt}{0.700pt}}
\put(655,544){\rule[-0.350pt]{1.445pt}{0.700pt}}
\put(661,543){\rule[-0.350pt]{1.445pt}{0.700pt}}
\put(667,542){\usebox{\plotpoint}}
\put(668,541){\usebox{\plotpoint}}
\put(669,540){\usebox{\plotpoint}}
\put(671,539){\usebox{\plotpoint}}
\put(672,538){\usebox{\plotpoint}}
\put(673,537){\usebox{\plotpoint}}
\put(675,536){\usebox{\plotpoint}}
\put(676,535){\usebox{\plotpoint}}
\put(678,532){\usebox{\plotpoint}}
\put(679,531){\usebox{\plotpoint}}
\put(680,530){\usebox{\plotpoint}}
\put(681,529){\usebox{\plotpoint}}
\put(682,527){\usebox{\plotpoint}}
\put(683,526){\usebox{\plotpoint}}
\put(684,525){\usebox{\plotpoint}}
\put(685,524){\usebox{\plotpoint}}
\put(686,522){\usebox{\plotpoint}}
\put(687,521){\usebox{\plotpoint}}
\put(688,520){\usebox{\plotpoint}}
\put(689,519){\usebox{\plotpoint}}
\put(690,517){\usebox{\plotpoint}}
\put(691,515){\usebox{\plotpoint}}
\put(692,513){\usebox{\plotpoint}}
\put(693,511){\usebox{\plotpoint}}
\put(694,509){\usebox{\plotpoint}}
\put(695,507){\usebox{\plotpoint}}
\put(696,505){\usebox{\plotpoint}}
\put(697,503){\usebox{\plotpoint}}
\put(698,501){\usebox{\plotpoint}}
\put(699,499){\usebox{\plotpoint}}
\put(700,497){\usebox{\plotpoint}}
\put(701,496){\usebox{\plotpoint}}
\put(702,493){\usebox{\plotpoint}}
\put(703,490){\usebox{\plotpoint}}
\put(704,488){\usebox{\plotpoint}}
\put(705,485){\usebox{\plotpoint}}
\put(706,482){\usebox{\plotpoint}}
\put(707,480){\usebox{\plotpoint}}
\put(708,477){\usebox{\plotpoint}}
\put(709,474){\usebox{\plotpoint}}
\put(710,472){\usebox{\plotpoint}}
\put(711,469){\usebox{\plotpoint}}
\put(712,466){\usebox{\plotpoint}}
\put(713,464){\usebox{\plotpoint}}
\put(714,460){\rule[-0.350pt]{0.700pt}{0.763pt}}
\put(715,457){\rule[-0.350pt]{0.700pt}{0.763pt}}
\put(716,454){\rule[-0.350pt]{0.700pt}{0.763pt}}
\put(717,451){\rule[-0.350pt]{0.700pt}{0.763pt}}
\put(718,448){\rule[-0.350pt]{0.700pt}{0.763pt}}
\put(719,445){\rule[-0.350pt]{0.700pt}{0.763pt}}
\put(720,441){\rule[-0.350pt]{0.700pt}{0.763pt}}
\put(721,438){\rule[-0.350pt]{0.700pt}{0.763pt}}
\put(722,435){\rule[-0.350pt]{0.700pt}{0.763pt}}
\put(723,432){\rule[-0.350pt]{0.700pt}{0.763pt}}
\put(724,429){\rule[-0.350pt]{0.700pt}{0.763pt}}
\put(725,426){\rule[-0.350pt]{0.700pt}{0.763pt}}
\put(726,422){\rule[-0.350pt]{0.700pt}{0.863pt}}
\put(727,418){\rule[-0.350pt]{0.700pt}{0.863pt}}
\put(728,415){\rule[-0.350pt]{0.700pt}{0.863pt}}
\put(729,411){\rule[-0.350pt]{0.700pt}{0.863pt}}
\put(730,408){\rule[-0.350pt]{0.700pt}{0.863pt}}
\put(731,404){\rule[-0.350pt]{0.700pt}{0.863pt}}
\put(732,400){\rule[-0.350pt]{0.700pt}{0.863pt}}
\put(733,397){\rule[-0.350pt]{0.700pt}{0.863pt}}
\put(734,393){\rule[-0.350pt]{0.700pt}{0.863pt}}
\put(735,390){\rule[-0.350pt]{0.700pt}{0.863pt}}
\put(736,386){\rule[-0.350pt]{0.700pt}{0.863pt}}
\put(737,383){\rule[-0.350pt]{0.700pt}{0.863pt}}
\put(738,379){\rule[-0.350pt]{0.700pt}{0.964pt}}
\put(739,375){\rule[-0.350pt]{0.700pt}{0.964pt}}
\put(740,371){\rule[-0.350pt]{0.700pt}{0.964pt}}
\put(741,367){\rule[-0.350pt]{0.700pt}{0.964pt}}
\put(742,363){\rule[-0.350pt]{0.700pt}{0.964pt}}
\put(743,359){\rule[-0.350pt]{0.700pt}{0.964pt}}
\put(744,355){\rule[-0.350pt]{0.700pt}{0.964pt}}
\put(745,351){\rule[-0.350pt]{0.700pt}{0.964pt}}
\put(746,347){\rule[-0.350pt]{0.700pt}{0.964pt}}
\put(747,343){\rule[-0.350pt]{0.700pt}{0.964pt}}
\put(748,339){\rule[-0.350pt]{0.700pt}{0.964pt}}
\put(749,335){\rule[-0.350pt]{0.700pt}{0.843pt}}
\put(750,332){\rule[-0.350pt]{0.700pt}{0.843pt}}
\put(751,328){\rule[-0.350pt]{0.700pt}{0.843pt}}
\put(752,325){\rule[-0.350pt]{0.700pt}{0.843pt}}
\put(753,321){\rule[-0.350pt]{0.700pt}{0.843pt}}
\put(754,318){\rule[-0.350pt]{0.700pt}{0.843pt}}
\put(755,314){\rule[-0.350pt]{0.700pt}{0.843pt}}
\put(756,311){\rule[-0.350pt]{0.700pt}{0.843pt}}
\put(757,307){\rule[-0.350pt]{0.700pt}{0.843pt}}
\put(758,304){\rule[-0.350pt]{0.700pt}{0.843pt}}
\put(759,300){\rule[-0.350pt]{0.700pt}{0.843pt}}
\put(760,297){\rule[-0.350pt]{0.700pt}{0.843pt}}
\put(761,294){\rule[-0.350pt]{0.700pt}{0.703pt}}
\put(762,291){\rule[-0.350pt]{0.700pt}{0.703pt}}
\put(763,288){\rule[-0.350pt]{0.700pt}{0.703pt}}
\put(764,285){\rule[-0.350pt]{0.700pt}{0.703pt}}
\put(765,282){\rule[-0.350pt]{0.700pt}{0.703pt}}
\put(766,279){\rule[-0.350pt]{0.700pt}{0.703pt}}
\put(767,276){\rule[-0.350pt]{0.700pt}{0.703pt}}
\put(768,273){\rule[-0.350pt]{0.700pt}{0.703pt}}
\put(769,270){\rule[-0.350pt]{0.700pt}{0.703pt}}
\put(770,267){\rule[-0.350pt]{0.700pt}{0.703pt}}
\put(771,264){\rule[-0.350pt]{0.700pt}{0.703pt}}
\put(772,262){\rule[-0.350pt]{0.700pt}{0.703pt}}
\put(773,259){\usebox{\plotpoint}}
\put(774,257){\usebox{\plotpoint}}
\put(775,254){\usebox{\plotpoint}}
\put(776,252){\usebox{\plotpoint}}
\put(777,250){\usebox{\plotpoint}}
\put(778,248){\usebox{\plotpoint}}
\put(779,245){\usebox{\plotpoint}}
\put(780,243){\usebox{\plotpoint}}
\put(781,241){\usebox{\plotpoint}}
\put(782,238){\usebox{\plotpoint}}
\put(783,236){\usebox{\plotpoint}}
\put(784,234){\usebox{\plotpoint}}
\put(785,232){\usebox{\plotpoint}}
\put(786,231){\usebox{\plotpoint}}
\put(787,229){\usebox{\plotpoint}}
\put(788,228){\usebox{\plotpoint}}
\put(789,226){\usebox{\plotpoint}}
\put(790,225){\usebox{\plotpoint}}
\put(791,223){\usebox{\plotpoint}}
\put(792,222){\usebox{\plotpoint}}
\put(793,220){\usebox{\plotpoint}}
\put(794,219){\usebox{\plotpoint}}
\put(795,217){\usebox{\plotpoint}}
\put(796,216){\usebox{\plotpoint}}
\put(797,216){\usebox{\plotpoint}}
\put(798,215){\usebox{\plotpoint}}
\put(799,214){\usebox{\plotpoint}}
\put(800,213){\usebox{\plotpoint}}
\put(801,212){\usebox{\plotpoint}}
\put(803,211){\usebox{\plotpoint}}
\put(804,210){\usebox{\plotpoint}}
\put(805,209){\usebox{\plotpoint}}
\put(806,208){\usebox{\plotpoint}}
\put(807,207){\usebox{\plotpoint}}
\put(809,206){\usebox{\plotpoint}}
\put(811,205){\usebox{\plotpoint}}
\put(814,204){\usebox{\plotpoint}}
\put(817,203){\usebox{\plotpoint}}
\put(820,202){\rule[-0.350pt]{4.336pt}{0.700pt}}
\put(838,203){\rule[-0.350pt]{1.445pt}{0.700pt}}
\put(844,204){\rule[-0.350pt]{0.964pt}{0.700pt}}
\put(848,205){\rule[-0.350pt]{0.964pt}{0.700pt}}
\put(852,206){\rule[-0.350pt]{0.964pt}{0.700pt}}
\put(856,207){\rule[-0.350pt]{0.964pt}{0.700pt}}
\put(860,208){\rule[-0.350pt]{0.964pt}{0.700pt}}
\put(864,209){\rule[-0.350pt]{0.964pt}{0.700pt}}
\put(868,210){\rule[-0.350pt]{0.964pt}{0.700pt}}
\put(872,211){\rule[-0.350pt]{0.964pt}{0.700pt}}
\put(876,212){\rule[-0.350pt]{0.964pt}{0.700pt}}
\put(880,213){\usebox{\plotpoint}}
\put(882,214){\usebox{\plotpoint}}
\put(885,215){\usebox{\plotpoint}}
\put(888,216){\usebox{\plotpoint}}
\put(891,217){\rule[-0.350pt]{0.964pt}{0.700pt}}
\put(895,218){\rule[-0.350pt]{0.964pt}{0.700pt}}
\put(899,219){\rule[-0.350pt]{0.964pt}{0.700pt}}
\put(903,220){\rule[-0.350pt]{0.964pt}{0.700pt}}
\put(907,221){\rule[-0.350pt]{0.964pt}{0.700pt}}
\put(911,222){\rule[-0.350pt]{0.964pt}{0.700pt}}
\put(915,223){\rule[-0.350pt]{0.964pt}{0.700pt}}
\put(919,224){\rule[-0.350pt]{0.964pt}{0.700pt}}
\put(923,225){\rule[-0.350pt]{0.964pt}{0.700pt}}
\put(927,226){\rule[-0.350pt]{0.964pt}{0.700pt}}
\put(931,227){\rule[-0.350pt]{0.964pt}{0.700pt}}
\put(935,228){\rule[-0.350pt]{0.964pt}{0.700pt}}
\put(939,229){\rule[-0.350pt]{0.964pt}{0.700pt}}
\put(943,230){\rule[-0.350pt]{0.964pt}{0.700pt}}
\put(947,231){\rule[-0.350pt]{0.964pt}{0.700pt}}
\put(951,232){\rule[-0.350pt]{0.883pt}{0.700pt}}
\put(954,233){\rule[-0.350pt]{0.883pt}{0.700pt}}
\put(958,234){\rule[-0.350pt]{0.883pt}{0.700pt}}
\put(962,235){\rule[-0.350pt]{1.445pt}{0.700pt}}
\put(968,236){\rule[-0.350pt]{1.445pt}{0.700pt}}
\put(974,237){\rule[-0.350pt]{1.445pt}{0.700pt}}
\put(980,238){\rule[-0.350pt]{1.445pt}{0.700pt}}
\put(986,239){\rule[-0.350pt]{1.445pt}{0.700pt}}
\put(992,240){\rule[-0.350pt]{1.445pt}{0.700pt}}
\put(998,241){\rule[-0.350pt]{1.445pt}{0.700pt}}
\put(1004,242){\rule[-0.350pt]{1.445pt}{0.700pt}}
\put(1010,243){\rule[-0.350pt]{1.445pt}{0.700pt}}
\put(1016,244){\rule[-0.350pt]{1.445pt}{0.700pt}}
\put(1022,245){\rule[-0.350pt]{1.325pt}{0.700pt}}
\put(1027,246){\rule[-0.350pt]{1.325pt}{0.700pt}}
\put(1033,247){\rule[-0.350pt]{1.445pt}{0.700pt}}
\put(1039,248){\rule[-0.350pt]{1.445pt}{0.700pt}}
\put(1045,249){\rule[-0.350pt]{2.891pt}{0.700pt}}
\put(1057,250){\rule[-0.350pt]{2.891pt}{0.700pt}}
\put(1069,251){\rule[-0.350pt]{2.891pt}{0.700pt}}
\put(1081,252){\rule[-0.350pt]{1.445pt}{0.700pt}}
\put(1087,253){\rule[-0.350pt]{1.445pt}{0.700pt}}
\put(1093,254){\rule[-0.350pt]{2.891pt}{0.700pt}}
\put(1105,255){\rule[-0.350pt]{5.541pt}{0.700pt}}
\put(1128,256){\rule[-0.350pt]{2.891pt}{0.700pt}}
\put(1140,257){\rule[-0.350pt]{2.891pt}{0.700pt}}
\put(1152,258){\rule[-0.350pt]{5.782pt}{0.700pt}}
\put(1176,259){\rule[-0.350pt]{5.541pt}{0.700pt}}
\put(1199,260){\rule[-0.350pt]{8.672pt}{0.700pt}}
\put(1235,261){\rule[-0.350pt]{14.213pt}{0.700pt}}
\put(1294,262){\rule[-0.350pt]{28.426pt}{0.700pt}}
\put(1412,263){\rule[-0.350pt]{5.782pt}{0.700pt}}
\end{picture}

%% file: modfunim.tex
\setlength{\unitlength}{0.240900pt}
\ifx\plotpoint\undefined\newsavebox{\plotpoint}\fi
\sbox{\plotpoint}{\rule[-0.175pt]{0.350pt}{0.350pt}}%
\begin{picture}(1500,900)(0,0)
\tenrm
\sbox{\plotpoint}{\rule[-0.175pt]{0.350pt}{0.350pt}}%
\put(264,228){\rule[-0.175pt]{282.335pt}{0.350pt}}
\put(264,158){\rule[-0.175pt]{4.818pt}{0.350pt}}
\put(242,158){\makebox(0,0)[r]{-0.2}}
\put(1416,158){\rule[-0.175pt]{4.818pt}{0.350pt}}
\put(264,228){\rule[-0.175pt]{4.818pt}{0.350pt}}
\put(242,228){\makebox(0,0)[r]{0}}
\put(1416,228){\rule[-0.175pt]{4.818pt}{0.350pt}}
\put(264,298){\rule[-0.175pt]{4.818pt}{0.350pt}}
\put(242,298){\makebox(0,0)[r]{0.2}}
\put(1416,298){\rule[-0.175pt]{4.818pt}{0.350pt}}
\put(264,368){\rule[-0.175pt]{4.818pt}{0.350pt}}
\put(242,368){\makebox(0,0)[r]{0.4}}
\put(1416,368){\rule[-0.175pt]{4.818pt}{0.350pt}}
\put(264,438){\rule[-0.175pt]{4.818pt}{0.350pt}}
\put(242,438){\makebox(0,0)[r]{0.6}}
\put(1416,438){\rule[-0.175pt]{4.818pt}{0.350pt}}
\put(264,507){\rule[-0.175pt]{4.818pt}{0.350pt}}
\put(242,507){\makebox(0,0)[r]{0.8}}
\put(1416,507){\rule[-0.175pt]{4.818pt}{0.350pt}}
\put(264,577){\rule[-0.175pt]{4.818pt}{0.350pt}}
\put(242,577){\makebox(0,0)[r]{1}}
\put(1416,577){\rule[-0.175pt]{4.818pt}{0.350pt}}
\put(264,647){\rule[-0.175pt]{4.818pt}{0.350pt}}
\put(242,647){\makebox(0,0)[r]{1.2}}
\put(1416,647){\rule[-0.175pt]{4.818pt}{0.350pt}}
\put(264,717){\rule[-0.175pt]{4.818pt}{0.350pt}}
\put(242,717){\makebox(0,0)[r]{1.4}}
\put(1416,717){\rule[-0.175pt]{4.818pt}{0.350pt}}
\put(264,787){\rule[-0.175pt]{4.818pt}{0.350pt}}
\put(242,787){\makebox(0,0)[r]{1.6}}
\put(1416,787){\rule[-0.175pt]{4.818pt}{0.350pt}}
\put(264,158){\rule[-0.175pt]{0.350pt}{4.818pt}}
\put(264,113){\makebox(0,0){\small$10^{-2}$}}
\put(264,767){\rule[-0.175pt]{0.350pt}{4.818pt}}
\put(431,158){\rule[-0.175pt]{0.350pt}{4.818pt}}
\put(431,113){\makebox(0,0){\small$10^{-1}$}}
\put(431,767){\rule[-0.175pt]{0.350pt}{4.818pt}}
\put(599,158){\rule[-0.175pt]{0.350pt}{4.818pt}}
\put(599,113){\makebox(0,0){\small$1$}}
\put(599,767){\rule[-0.175pt]{0.350pt}{4.818pt}}
\put(766,158){\rule[-0.175pt]{0.350pt}{4.818pt}}
\put(766,113){\makebox(0,0){\small$10^{1}$}}
\put(766,767){\rule[-0.175pt]{0.350pt}{4.818pt}}
\put(934,158){\rule[-0.175pt]{0.350pt}{4.818pt}}
\put(934,113){\makebox(0,0){\small$10^{2}$}}
\put(934,767){\rule[-0.175pt]{0.350pt}{4.818pt}}
\put(1101,158){\rule[-0.175pt]{0.350pt}{4.818pt}}
\put(1101,113){\makebox(0,0){\small$10^{3}$}}
\put(1101,767){\rule[-0.175pt]{0.350pt}{4.818pt}}
\put(1269,158){\rule[-0.175pt]{0.350pt}{4.818pt}}
\put(1269,113){\makebox(0,0){\small$10^{4}$}}
\put(1269,767){\rule[-0.175pt]{0.350pt}{4.818pt}}
\put(1436,158){\rule[-0.175pt]{0.350pt}{4.818pt}}
\put(1436,113){\makebox(0,0){\small$10^{5}$}}
\put(1436,767){\rule[-0.175pt]{0.350pt}{4.818pt}}
\put(264,158){\rule[-0.175pt]{282.335pt}{0.350pt}}
\put(1436,158){\rule[-0.175pt]{0.350pt}{151.526pt}}
\put(264,787){\rule[-0.175pt]{282.335pt}{0.350pt}}
\put(850,60){\makebox(0,0){\small$q^2/m^2$}}
\put(850,840){\makebox(0,0){\small Imaginary part of $J(q^2)$ and $\widetilde{J}(q^2/m^2)$}}
\put(264,158){\rule[-0.175pt]{0.350pt}{151.526pt}}
\sbox{\plotpoint}{\rule[-0.250pt]{0.500pt}{0.500pt}}%
\put(1306,722){\makebox(0,0)[r]{\small$\Im[J]$}}
\put(1328,722){\usebox{\plotpoint}}
\put(1348,722){\usebox{\plotpoint}}
\put(1369,722){\usebox{\plotpoint}}
\put(1390,722){\usebox{\plotpoint}}
\put(1394,722){\usebox{\plotpoint}}
\put(264,228){\usebox{\plotpoint}}
\put(284,228){\usebox{\plotpoint}}
\put(305,228){\usebox{\plotpoint}}
\put(326,228){\usebox{\plotpoint}}
\put(347,228){\usebox{\plotpoint}}
\put(367,228){\usebox{\plotpoint}}
\put(388,228){\usebox{\plotpoint}}
\put(409,228){\usebox{\plotpoint}}
\put(430,228){\usebox{\plotpoint}}
\put(450,228){\usebox{\plotpoint}}
\put(471,228){\usebox{\plotpoint}}
\put(492,228){\usebox{\plotpoint}}
\put(513,228){\usebox{\plotpoint}}
\put(533,228){\usebox{\plotpoint}}
\put(554,228){\usebox{\plotpoint}}
\put(575,228){\usebox{\plotpoint}}
\put(596,228){\usebox{\plotpoint}}
\put(616,228){\usebox{\plotpoint}}
\put(637,228){\usebox{\plotpoint}}
\put(658,228){\usebox{\plotpoint}}
\put(679,228){\usebox{\plotpoint}}
\put(694,233){\usebox{\plotpoint}}
\put(696,254){\usebox{\plotpoint}}
\put(697,275){\usebox{\plotpoint}}
\put(699,295){\usebox{\plotpoint}}
\put(700,316){\usebox{\plotpoint}}
\put(700,337){\usebox{\plotpoint}}
\put(701,358){\usebox{\plotpoint}}
\put(702,378){\usebox{\plotpoint}}
\put(702,399){\usebox{\plotpoint}}
\put(703,420){\usebox{\plotpoint}}
\put(703,441){\usebox{\plotpoint}}
\put(704,461){\usebox{\plotpoint}}
\put(704,482){\usebox{\plotpoint}}
\put(705,503){\usebox{\plotpoint}}
\put(705,524){\usebox{\plotpoint}}
\put(707,544){\usebox{\plotpoint}}
\put(708,565){\usebox{\plotpoint}}
\put(709,586){\usebox{\plotpoint}}
\put(710,607){\usebox{\plotpoint}}
\put(713,627){\usebox{\plotpoint}}
\put(715,648){\usebox{\plotpoint}}
\put(718,668){\usebox{\plotpoint}}
\put(722,689){\usebox{\plotpoint}}
\put(729,708){\usebox{\plotpoint}}
\put(744,717){\usebox{\plotpoint}}
\put(756,701){\usebox{\plotpoint}}
\put(765,682){\usebox{\plotpoint}}
\put(773,663){\usebox{\plotpoint}}
\put(781,643){\usebox{\plotpoint}}
\put(787,624){\usebox{\plotpoint}}
\put(794,604){\usebox{\plotpoint}}
\put(802,585){\usebox{\plotpoint}}
\put(808,566){\usebox{\plotpoint}}
\put(815,546){\usebox{\plotpoint}}
\put(823,527){\usebox{\plotpoint}}
\put(830,508){\usebox{\plotpoint}}
\put(838,488){\usebox{\plotpoint}}
\put(847,469){\usebox{\plotpoint}}
\put(855,451){\usebox{\plotpoint}}
\put(864,432){\usebox{\plotpoint}}
\put(873,413){\usebox{\plotpoint}}
\put(884,395){\usebox{\plotpoint}}
\put(895,378){\usebox{\plotpoint}}
\put(907,361){\usebox{\plotpoint}}
\put(919,344){\usebox{\plotpoint}}
\put(933,328){\usebox{\plotpoint}}
\put(947,314){\usebox{\plotpoint}}
\put(963,300){\usebox{\plotpoint}}
\put(980,288){\usebox{\plotpoint}}
\put(997,277){\usebox{\plotpoint}}
\put(1016,268){\usebox{\plotpoint}}
\put(1035,260){\usebox{\plotpoint}}
\put(1055,253){\usebox{\plotpoint}}
\put(1075,248){\usebox{\plotpoint}}
\put(1095,243){\usebox{\plotpoint}}
\put(1115,240){\usebox{\plotpoint}}
\put(1136,238){\usebox{\plotpoint}}
\put(1156,236){\usebox{\plotpoint}}
\put(1177,234){\usebox{\plotpoint}}
\put(1198,233){\usebox{\plotpoint}}
\put(1218,232){\usebox{\plotpoint}}
\put(1239,231){\usebox{\plotpoint}}
\put(1260,230){\usebox{\plotpoint}}
\put(1280,230){\usebox{\plotpoint}}
\put(1301,229){\usebox{\plotpoint}}
\put(1322,229){\usebox{\plotpoint}}
\put(1343,229){\usebox{\plotpoint}}
\put(1363,228){\usebox{\plotpoint}}
\put(1384,228){\usebox{\plotpoint}}
\put(1405,228){\usebox{\plotpoint}}
\put(1426,228){\usebox{\plotpoint}}
\put(1436,228){\usebox{\plotpoint}}
\sbox{\plotpoint}{\rule[-0.350pt]{0.700pt}{0.700pt}}%
\put(1306,665){\makebox(0,0)[r]{\small$\Im[\widetilde{J}]$}}
\put(1328,665){\rule[-0.350pt]{15.899pt}{0.700pt}}
\put(264,228){\usebox{\plotpoint}}
\put(264,228){\rule[-0.350pt]{61.309pt}{0.700pt}}
\put(518,229){\rule[-0.350pt]{1.325pt}{0.700pt}}
\put(524,230){\rule[-0.350pt]{0.964pt}{0.700pt}}
\put(528,231){\rule[-0.350pt]{0.964pt}{0.700pt}}
\put(532,232){\rule[-0.350pt]{0.964pt}{0.700pt}}
\put(536,233){\usebox{\plotpoint}}
\put(537,234){\usebox{\plotpoint}}
\put(539,235){\usebox{\plotpoint}}
\put(541,236){\usebox{\plotpoint}}
\put(542,237){\usebox{\plotpoint}}
\put(544,238){\usebox{\plotpoint}}
\put(546,239){\usebox{\plotpoint}}
\put(548,240){\usebox{\plotpoint}}
\put(549,241){\usebox{\plotpoint}}
\put(550,242){\usebox{\plotpoint}}
\put(551,243){\usebox{\plotpoint}}
\put(552,244){\usebox{\plotpoint}}
\put(553,245){\usebox{\plotpoint}}
\put(554,246){\usebox{\plotpoint}}
\put(555,247){\usebox{\plotpoint}}
\put(556,248){\usebox{\plotpoint}}
\put(557,249){\usebox{\plotpoint}}
\put(558,250){\usebox{\plotpoint}}
\put(559,251){\usebox{\plotpoint}}
\put(560,252){\usebox{\plotpoint}}
\put(561,253){\usebox{\plotpoint}}
\put(562,255){\usebox{\plotpoint}}
\put(563,257){\usebox{\plotpoint}}
\put(564,258){\usebox{\plotpoint}}
\put(565,260){\usebox{\plotpoint}}
\put(566,261){\usebox{\plotpoint}}
\put(567,263){\usebox{\plotpoint}}
\put(568,265){\usebox{\plotpoint}}
\put(569,266){\usebox{\plotpoint}}
\put(570,268){\usebox{\plotpoint}}
\put(571,270){\usebox{\plotpoint}}
\put(572,271){\usebox{\plotpoint}}
\put(573,274){\usebox{\plotpoint}}
\put(574,277){\usebox{\plotpoint}}
\put(575,279){\usebox{\plotpoint}}
\put(576,282){\usebox{\plotpoint}}
\put(577,284){\usebox{\plotpoint}}
\put(578,287){\usebox{\plotpoint}}
\put(579,289){\usebox{\plotpoint}}
\put(580,292){\usebox{\plotpoint}}
\put(581,294){\usebox{\plotpoint}}
\put(582,297){\usebox{\plotpoint}}
\put(583,299){\usebox{\plotpoint}}
\put(584,302){\rule[-0.350pt]{0.700pt}{0.832pt}}
\put(585,305){\rule[-0.350pt]{0.700pt}{0.832pt}}
\put(586,308){\rule[-0.350pt]{0.700pt}{0.832pt}}
\put(587,312){\rule[-0.350pt]{0.700pt}{0.832pt}}
\put(588,315){\rule[-0.350pt]{0.700pt}{0.832pt}}
\put(589,319){\rule[-0.350pt]{0.700pt}{0.832pt}}
\put(590,322){\rule[-0.350pt]{0.700pt}{0.832pt}}
\put(591,326){\rule[-0.350pt]{0.700pt}{0.832pt}}
\put(592,329){\rule[-0.350pt]{0.700pt}{0.832pt}}
\put(593,333){\rule[-0.350pt]{0.700pt}{0.832pt}}
\put(594,336){\rule[-0.350pt]{0.700pt}{0.832pt}}
\put(595,340){\rule[-0.350pt]{0.700pt}{0.944pt}}
\put(596,343){\rule[-0.350pt]{0.700pt}{0.944pt}}
\put(597,347){\rule[-0.350pt]{0.700pt}{0.944pt}}
\put(598,351){\rule[-0.350pt]{0.700pt}{0.944pt}}
\put(599,355){\rule[-0.350pt]{0.700pt}{0.944pt}}
\put(600,359){\rule[-0.350pt]{0.700pt}{0.944pt}}
\put(601,363){\rule[-0.350pt]{0.700pt}{0.944pt}}
\put(602,367){\rule[-0.350pt]{0.700pt}{0.944pt}}
\put(603,371){\rule[-0.350pt]{0.700pt}{0.944pt}}
\put(604,375){\rule[-0.350pt]{0.700pt}{0.944pt}}
\put(605,379){\rule[-0.350pt]{0.700pt}{0.944pt}}
\put(606,383){\rule[-0.350pt]{0.700pt}{0.944pt}}
\put(607,386){\rule[-0.350pt]{0.700pt}{1.044pt}}
\put(608,391){\rule[-0.350pt]{0.700pt}{1.044pt}}
\put(609,395){\rule[-0.350pt]{0.700pt}{1.044pt}}
\put(610,400){\rule[-0.350pt]{0.700pt}{1.044pt}}
\put(611,404){\rule[-0.350pt]{0.700pt}{1.044pt}}
\put(612,408){\rule[-0.350pt]{0.700pt}{1.044pt}}
\put(613,413){\rule[-0.350pt]{0.700pt}{1.044pt}}
\put(614,417){\rule[-0.350pt]{0.700pt}{1.044pt}}
\put(615,421){\rule[-0.350pt]{0.700pt}{1.044pt}}
\put(616,426){\rule[-0.350pt]{0.700pt}{1.044pt}}
\put(617,430){\rule[-0.350pt]{0.700pt}{1.044pt}}
\put(618,434){\rule[-0.350pt]{0.700pt}{1.044pt}}
\put(619,439){\rule[-0.350pt]{0.700pt}{1.084pt}}
\put(620,443){\rule[-0.350pt]{0.700pt}{1.084pt}}
\put(621,448){\rule[-0.350pt]{0.700pt}{1.084pt}}
\put(622,452){\rule[-0.350pt]{0.700pt}{1.084pt}}
\put(623,457){\rule[-0.350pt]{0.700pt}{1.084pt}}
\put(624,461){\rule[-0.350pt]{0.700pt}{1.084pt}}
\put(625,466){\rule[-0.350pt]{0.700pt}{1.084pt}}
\put(626,470){\rule[-0.350pt]{0.700pt}{1.084pt}}
\put(627,475){\rule[-0.350pt]{0.700pt}{1.084pt}}
\put(628,479){\rule[-0.350pt]{0.700pt}{1.084pt}}
\put(629,484){\rule[-0.350pt]{0.700pt}{1.084pt}}
\put(630,488){\rule[-0.350pt]{0.700pt}{1.084pt}}
\put(631,493){\rule[-0.350pt]{0.700pt}{1.064pt}}
\put(632,497){\rule[-0.350pt]{0.700pt}{1.064pt}}
\put(633,501){\rule[-0.350pt]{0.700pt}{1.064pt}}
\put(634,506){\rule[-0.350pt]{0.700pt}{1.064pt}}
\put(635,510){\rule[-0.350pt]{0.700pt}{1.064pt}}
\put(636,515){\rule[-0.350pt]{0.700pt}{1.064pt}}
\put(637,519){\rule[-0.350pt]{0.700pt}{1.064pt}}
\put(638,523){\rule[-0.350pt]{0.700pt}{1.064pt}}
\put(639,528){\rule[-0.350pt]{0.700pt}{1.064pt}}
\put(640,532){\rule[-0.350pt]{0.700pt}{1.064pt}}
\put(641,537){\rule[-0.350pt]{0.700pt}{1.064pt}}
\put(642,541){\rule[-0.350pt]{0.700pt}{1.064pt}}
\put(643,546){\rule[-0.350pt]{0.700pt}{0.984pt}}
\put(644,550){\rule[-0.350pt]{0.700pt}{0.984pt}}
\put(645,554){\rule[-0.350pt]{0.700pt}{0.984pt}}
\put(646,558){\rule[-0.350pt]{0.700pt}{0.984pt}}
\put(647,562){\rule[-0.350pt]{0.700pt}{0.984pt}}
\put(648,566){\rule[-0.350pt]{0.700pt}{0.984pt}}
\put(649,570){\rule[-0.350pt]{0.700pt}{0.984pt}}
\put(650,574){\rule[-0.350pt]{0.700pt}{0.984pt}}
\put(651,578){\rule[-0.350pt]{0.700pt}{0.984pt}}
\put(652,582){\rule[-0.350pt]{0.700pt}{0.984pt}}
\put(653,586){\rule[-0.350pt]{0.700pt}{0.984pt}}
\put(654,590){\rule[-0.350pt]{0.700pt}{0.984pt}}
\put(655,594){\rule[-0.350pt]{0.700pt}{0.823pt}}
\put(656,598){\rule[-0.350pt]{0.700pt}{0.823pt}}
\put(657,601){\rule[-0.350pt]{0.700pt}{0.823pt}}
\put(658,605){\rule[-0.350pt]{0.700pt}{0.823pt}}
\put(659,608){\rule[-0.350pt]{0.700pt}{0.823pt}}
\put(660,612){\rule[-0.350pt]{0.700pt}{0.823pt}}
\put(661,615){\rule[-0.350pt]{0.700pt}{0.823pt}}
\put(662,618){\rule[-0.350pt]{0.700pt}{0.823pt}}
\put(663,622){\rule[-0.350pt]{0.700pt}{0.823pt}}
\put(664,625){\rule[-0.350pt]{0.700pt}{0.823pt}}
\put(665,629){\rule[-0.350pt]{0.700pt}{0.823pt}}
\put(666,632){\rule[-0.350pt]{0.700pt}{0.823pt}}
\put(667,636){\rule[-0.350pt]{0.700pt}{0.701pt}}
\put(668,638){\rule[-0.350pt]{0.700pt}{0.701pt}}
\put(669,641){\rule[-0.350pt]{0.700pt}{0.701pt}}
\put(670,644){\rule[-0.350pt]{0.700pt}{0.701pt}}
\put(671,647){\rule[-0.350pt]{0.700pt}{0.701pt}}
\put(672,650){\rule[-0.350pt]{0.700pt}{0.701pt}}
\put(673,653){\rule[-0.350pt]{0.700pt}{0.701pt}}
\put(674,656){\rule[-0.350pt]{0.700pt}{0.701pt}}
\put(675,659){\rule[-0.350pt]{0.700pt}{0.701pt}}
\put(676,662){\rule[-0.350pt]{0.700pt}{0.701pt}}
\put(677,665){\rule[-0.350pt]{0.700pt}{0.701pt}}
\put(678,668){\usebox{\plotpoint}}
\put(679,669){\usebox{\plotpoint}}
\put(680,671){\usebox{\plotpoint}}
\put(681,673){\usebox{\plotpoint}}
\put(682,675){\usebox{\plotpoint}}
\put(683,677){\usebox{\plotpoint}}
\put(684,678){\usebox{\plotpoint}}
\put(685,680){\usebox{\plotpoint}}
\put(686,682){\usebox{\plotpoint}}
\put(687,684){\usebox{\plotpoint}}
\put(688,686){\usebox{\plotpoint}}
\put(689,688){\usebox{\plotpoint}}
\put(690,689){\usebox{\plotpoint}}
\put(690,690){\usebox{\plotpoint}}
\put(691,691){\usebox{\plotpoint}}
\put(692,692){\usebox{\plotpoint}}
\put(693,693){\usebox{\plotpoint}}
\put(694,694){\usebox{\plotpoint}}
\put(695,695){\usebox{\plotpoint}}
\put(696,696){\usebox{\plotpoint}}
\put(697,697){\usebox{\plotpoint}}
\put(698,698){\usebox{\plotpoint}}
\put(699,699){\usebox{\plotpoint}}
\put(700,700){\usebox{\plotpoint}}
\put(701,701){\usebox{\plotpoint}}
\put(702,702){\rule[-0.350pt]{0.964pt}{0.700pt}}
\put(706,703){\rule[-0.350pt]{0.964pt}{0.700pt}}
\put(710,704){\rule[-0.350pt]{0.964pt}{0.700pt}}
\put(714,705){\usebox{\plotpoint}}
\put(716,704){\usebox{\plotpoint}}
\put(718,703){\usebox{\plotpoint}}
\put(721,702){\usebox{\plotpoint}}
\put(723,701){\usebox{\plotpoint}}
\put(726,700){\usebox{\plotpoint}}
\put(727,699){\usebox{\plotpoint}}
\put(728,698){\usebox{\plotpoint}}
\put(729,697){\usebox{\plotpoint}}
\put(730,696){\usebox{\plotpoint}}
\put(731,695){\usebox{\plotpoint}}
\put(732,694){\usebox{\plotpoint}}
\put(733,693){\usebox{\plotpoint}}
\put(734,692){\usebox{\plotpoint}}
\put(735,691){\usebox{\plotpoint}}
\put(736,690){\usebox{\plotpoint}}
\put(737,689){\usebox{\plotpoint}}
\put(738,686){\usebox{\plotpoint}}
\put(739,684){\usebox{\plotpoint}}
\put(740,683){\usebox{\plotpoint}}
\put(741,681){\usebox{\plotpoint}}
\put(742,679){\usebox{\plotpoint}}
\put(743,678){\usebox{\plotpoint}}
\put(744,676){\usebox{\plotpoint}}
\put(745,674){\usebox{\plotpoint}}
\put(746,673){\usebox{\plotpoint}}
\put(747,671){\usebox{\plotpoint}}
\put(748,670){\usebox{\plotpoint}}
\put(749,668){\usebox{\plotpoint}}
\put(750,666){\usebox{\plotpoint}}
\put(751,664){\usebox{\plotpoint}}
\put(752,662){\usebox{\plotpoint}}
\put(753,660){\usebox{\plotpoint}}
\put(754,659){\usebox{\plotpoint}}
\put(755,657){\usebox{\plotpoint}}
\put(756,655){\usebox{\plotpoint}}
\put(757,653){\usebox{\plotpoint}}
\put(758,651){\usebox{\plotpoint}}
\put(759,649){\usebox{\plotpoint}}
\put(760,648){\usebox{\plotpoint}}
\put(761,645){\usebox{\plotpoint}}
\put(762,643){\usebox{\plotpoint}}
\put(763,641){\usebox{\plotpoint}}
\put(764,639){\usebox{\plotpoint}}
\put(765,637){\usebox{\plotpoint}}
\put(766,635){\usebox{\plotpoint}}
\put(767,633){\usebox{\plotpoint}}
\put(768,631){\usebox{\plotpoint}}
\put(769,629){\usebox{\plotpoint}}
\put(770,627){\usebox{\plotpoint}}
\put(771,625){\usebox{\plotpoint}}
\put(772,623){\usebox{\plotpoint}}
\put(773,620){\usebox{\plotpoint}}
\put(774,618){\usebox{\plotpoint}}
\put(775,616){\usebox{\plotpoint}}
\put(776,614){\usebox{\plotpoint}}
\put(777,612){\usebox{\plotpoint}}
\put(778,609){\usebox{\plotpoint}}
\put(779,607){\usebox{\plotpoint}}
\put(780,605){\usebox{\plotpoint}}
\put(781,603){\usebox{\plotpoint}}
\put(782,601){\usebox{\plotpoint}}
\put(783,599){\usebox{\plotpoint}}
\put(784,597){\usebox{\plotpoint}}
\put(785,594){\usebox{\plotpoint}}
\put(786,592){\usebox{\plotpoint}}
\put(787,590){\usebox{\plotpoint}}
\put(788,587){\usebox{\plotpoint}}
\put(789,585){\usebox{\plotpoint}}
\put(790,583){\usebox{\plotpoint}}
\put(791,580){\usebox{\plotpoint}}
\put(792,578){\usebox{\plotpoint}}
\put(793,576){\usebox{\plotpoint}}
\put(794,573){\usebox{\plotpoint}}
\put(795,571){\usebox{\plotpoint}}
\put(796,569){\usebox{\plotpoint}}
\put(797,566){\usebox{\plotpoint}}
\put(798,564){\usebox{\plotpoint}}
\put(799,562){\usebox{\plotpoint}}
\put(800,560){\usebox{\plotpoint}}
\put(801,557){\usebox{\plotpoint}}
\put(802,555){\usebox{\plotpoint}}
\put(803,553){\usebox{\plotpoint}}
\put(804,551){\usebox{\plotpoint}}
\put(805,548){\usebox{\plotpoint}}
\put(806,546){\usebox{\plotpoint}}
\put(807,544){\usebox{\plotpoint}}
\put(808,542){\usebox{\plotpoint}}
\put(809,539){\usebox{\plotpoint}}
\put(810,536){\usebox{\plotpoint}}
\put(811,534){\usebox{\plotpoint}}
\put(812,531){\usebox{\plotpoint}}
\put(813,529){\usebox{\plotpoint}}
\put(814,526){\usebox{\plotpoint}}
\put(815,524){\usebox{\plotpoint}}
\put(816,521){\usebox{\plotpoint}}
\put(817,519){\usebox{\plotpoint}}
\put(818,516){\usebox{\plotpoint}}
\put(819,514){\usebox{\plotpoint}}
\put(820,511){\usebox{\plotpoint}}
\put(821,509){\usebox{\plotpoint}}
\put(822,507){\usebox{\plotpoint}}
\put(823,505){\usebox{\plotpoint}}
\put(824,503){\usebox{\plotpoint}}
\put(825,501){\usebox{\plotpoint}}
\put(826,498){\usebox{\plotpoint}}
\put(827,496){\usebox{\plotpoint}}
\put(828,494){\usebox{\plotpoint}}
\put(829,492){\usebox{\plotpoint}}
\put(830,490){\usebox{\plotpoint}}
\put(831,488){\usebox{\plotpoint}}
\put(832,485){\usebox{\plotpoint}}
\put(833,483){\usebox{\plotpoint}}
\put(834,481){\usebox{\plotpoint}}
\put(835,479){\usebox{\plotpoint}}
\put(836,477){\usebox{\plotpoint}}
\put(837,475){\usebox{\plotpoint}}
\put(838,473){\usebox{\plotpoint}}
\put(839,471){\usebox{\plotpoint}}
\put(840,469){\usebox{\plotpoint}}
\put(841,467){\usebox{\plotpoint}}
\put(842,465){\usebox{\plotpoint}}
\put(843,463){\usebox{\plotpoint}}
\put(844,461){\usebox{\plotpoint}}
\put(845,459){\usebox{\plotpoint}}
\put(846,457){\usebox{\plotpoint}}
\put(847,455){\usebox{\plotpoint}}
\put(848,453){\usebox{\plotpoint}}
\put(849,451){\usebox{\plotpoint}}
\put(850,449){\usebox{\plotpoint}}
\put(851,447){\usebox{\plotpoint}}
\put(852,445){\usebox{\plotpoint}}
\put(853,443){\usebox{\plotpoint}}
\put(854,441){\usebox{\plotpoint}}
\put(855,440){\usebox{\plotpoint}}
\put(856,438){\usebox{\plotpoint}}
\put(857,436){\usebox{\plotpoint}}
\put(858,434){\usebox{\plotpoint}}
\put(859,432){\usebox{\plotpoint}}
\put(860,430){\usebox{\plotpoint}}
\put(861,428){\usebox{\plotpoint}}
\put(862,427){\usebox{\plotpoint}}
\put(863,425){\usebox{\plotpoint}}
\put(864,423){\usebox{\plotpoint}}
\put(865,421){\usebox{\plotpoint}}
\put(866,419){\usebox{\plotpoint}}
\put(867,418){\usebox{\plotpoint}}
\put(868,416){\usebox{\plotpoint}}
\put(869,414){\usebox{\plotpoint}}
\put(870,413){\usebox{\plotpoint}}
\put(871,411){\usebox{\plotpoint}}
\put(872,409){\usebox{\plotpoint}}
\put(873,408){\usebox{\plotpoint}}
\put(874,406){\usebox{\plotpoint}}
\put(875,404){\usebox{\plotpoint}}
\put(876,403){\usebox{\plotpoint}}
\put(877,401){\usebox{\plotpoint}}
\put(878,399){\usebox{\plotpoint}}
\put(879,398){\usebox{\plotpoint}}
\put(880,396){\usebox{\plotpoint}}
\put(881,394){\usebox{\plotpoint}}
\put(882,393){\usebox{\plotpoint}}
\put(883,391){\usebox{\plotpoint}}
\put(884,389){\usebox{\plotpoint}}
\put(885,388){\usebox{\plotpoint}}
\put(886,386){\usebox{\plotpoint}}
\put(887,384){\usebox{\plotpoint}}
\put(888,383){\usebox{\plotpoint}}
\put(889,381){\usebox{\plotpoint}}
\put(890,380){\usebox{\plotpoint}}
\put(891,378){\usebox{\plotpoint}}
\put(892,377){\usebox{\plotpoint}}
\put(893,375){\usebox{\plotpoint}}
\put(894,374){\usebox{\plotpoint}}
\put(895,372){\usebox{\plotpoint}}
\put(896,371){\usebox{\plotpoint}}
\put(897,370){\usebox{\plotpoint}}
\put(898,368){\usebox{\plotpoint}}
\put(899,367){\usebox{\plotpoint}}
\put(900,365){\usebox{\plotpoint}}
\put(901,364){\usebox{\plotpoint}}
\put(902,363){\usebox{\plotpoint}}
\put(903,361){\usebox{\plotpoint}}
\put(904,360){\usebox{\plotpoint}}
\put(905,359){\usebox{\plotpoint}}
\put(906,358){\usebox{\plotpoint}}
\put(907,356){\usebox{\plotpoint}}
\put(908,355){\usebox{\plotpoint}}
\put(909,354){\usebox{\plotpoint}}
\put(910,353){\usebox{\plotpoint}}
\put(911,351){\usebox{\plotpoint}}
\put(912,350){\usebox{\plotpoint}}
\put(913,349){\usebox{\plotpoint}}
\put(914,348){\usebox{\plotpoint}}
\put(915,346){\usebox{\plotpoint}}
\put(916,345){\usebox{\plotpoint}}
\put(917,344){\usebox{\plotpoint}}
\put(918,343){\usebox{\plotpoint}}
\put(919,342){\usebox{\plotpoint}}
\put(920,341){\usebox{\plotpoint}}
\put(921,339){\usebox{\plotpoint}}
\put(922,338){\usebox{\plotpoint}}
\put(923,337){\usebox{\plotpoint}}
\put(924,336){\usebox{\plotpoint}}
\put(925,335){\usebox{\plotpoint}}
\put(926,334){\usebox{\plotpoint}}
\put(927,334){\usebox{\plotpoint}}
\put(927,334){\usebox{\plotpoint}}
\put(928,333){\usebox{\plotpoint}}
\put(929,332){\usebox{\plotpoint}}
\put(930,331){\usebox{\plotpoint}}
\put(931,330){\usebox{\plotpoint}}
\put(932,329){\usebox{\plotpoint}}
\put(933,328){\usebox{\plotpoint}}
\put(934,327){\usebox{\plotpoint}}
\put(935,326){\usebox{\plotpoint}}
\put(936,325){\usebox{\plotpoint}}
\put(937,324){\usebox{\plotpoint}}
\put(938,323){\usebox{\plotpoint}}
\put(939,322){\usebox{\plotpoint}}
\put(940,321){\usebox{\plotpoint}}
\put(941,320){\usebox{\plotpoint}}
\put(942,319){\usebox{\plotpoint}}
\put(943,318){\usebox{\plotpoint}}
\put(944,317){\usebox{\plotpoint}}
\put(945,316){\usebox{\plotpoint}}
\put(946,315){\usebox{\plotpoint}}
\put(947,314){\usebox{\plotpoint}}
\put(948,313){\usebox{\plotpoint}}
\put(949,312){\usebox{\plotpoint}}
\put(950,311){\usebox{\plotpoint}}
\put(952,310){\usebox{\plotpoint}}
\put(953,309){\usebox{\plotpoint}}
\put(954,308){\usebox{\plotpoint}}
\put(955,307){\usebox{\plotpoint}}
\put(956,306){\usebox{\plotpoint}}
\put(957,305){\usebox{\plotpoint}}
\put(958,304){\usebox{\plotpoint}}
\put(959,303){\usebox{\plotpoint}}
\put(960,302){\usebox{\plotpoint}}
\put(961,301){\usebox{\plotpoint}}
\put(963,300){\usebox{\plotpoint}}
\put(964,299){\usebox{\plotpoint}}
\put(965,298){\usebox{\plotpoint}}
\put(967,297){\usebox{\plotpoint}}
\put(968,296){\usebox{\plotpoint}}
\put(969,295){\usebox{\plotpoint}}
\put(971,294){\usebox{\plotpoint}}
\put(972,293){\usebox{\plotpoint}}
\put(973,292){\usebox{\plotpoint}}
\put(975,291){\usebox{\plotpoint}}
\put(977,290){\usebox{\plotpoint}}
\put(978,289){\usebox{\plotpoint}}
\put(980,288){\usebox{\plotpoint}}
\put(981,287){\usebox{\plotpoint}}
\put(983,286){\usebox{\plotpoint}}
\put(984,285){\usebox{\plotpoint}}
\put(986,284){\usebox{\plotpoint}}
\put(987,283){\usebox{\plotpoint}}
\put(989,282){\usebox{\plotpoint}}
\put(991,281){\usebox{\plotpoint}}
\put(992,280){\usebox{\plotpoint}}
\put(994,279){\usebox{\plotpoint}}
\put(996,278){\usebox{\plotpoint}}
\put(998,277){\usebox{\plotpoint}}
\put(1000,276){\usebox{\plotpoint}}
\put(1002,275){\usebox{\plotpoint}}
\put(1004,274){\usebox{\plotpoint}}
\put(1006,273){\usebox{\plotpoint}}
\put(1008,272){\usebox{\plotpoint}}
\put(1010,271){\usebox{\plotpoint}}
\put(1012,270){\usebox{\plotpoint}}
\put(1014,269){\usebox{\plotpoint}}
\put(1017,268){\usebox{\plotpoint}}
\put(1019,267){\usebox{\plotpoint}}
\put(1022,266){\usebox{\plotpoint}}
\put(1024,265){\usebox{\plotpoint}}
\put(1026,264){\usebox{\plotpoint}}
\put(1028,263){\usebox{\plotpoint}}
\put(1030,262){\usebox{\plotpoint}}
\put(1032,261){\rule[-0.350pt]{0.723pt}{0.700pt}}
\put(1036,260){\rule[-0.350pt]{0.723pt}{0.700pt}}
\put(1039,259){\rule[-0.350pt]{0.723pt}{0.700pt}}
\put(1042,258){\rule[-0.350pt]{0.723pt}{0.700pt}}
\put(1045,257){\rule[-0.350pt]{0.723pt}{0.700pt}}
\put(1048,256){\rule[-0.350pt]{0.723pt}{0.700pt}}
\put(1051,255){\rule[-0.350pt]{0.723pt}{0.700pt}}
\put(1054,254){\rule[-0.350pt]{0.723pt}{0.700pt}}
\put(1057,253){\rule[-0.350pt]{0.964pt}{0.700pt}}
\put(1061,252){\rule[-0.350pt]{0.964pt}{0.700pt}}
\put(1065,251){\rule[-0.350pt]{0.964pt}{0.700pt}}
\put(1069,250){\rule[-0.350pt]{0.964pt}{0.700pt}}
\put(1073,249){\rule[-0.350pt]{0.964pt}{0.700pt}}
\put(1077,248){\rule[-0.350pt]{0.964pt}{0.700pt}}
\put(1081,247){\rule[-0.350pt]{1.445pt}{0.700pt}}
\put(1087,246){\rule[-0.350pt]{1.445pt}{0.700pt}}
\put(1093,245){\rule[-0.350pt]{0.964pt}{0.700pt}}
\put(1097,244){\rule[-0.350pt]{0.964pt}{0.700pt}}
\put(1101,243){\rule[-0.350pt]{0.964pt}{0.700pt}}
\put(1105,242){\rule[-0.350pt]{2.650pt}{0.700pt}}
\put(1116,241){\rule[-0.350pt]{1.445pt}{0.700pt}}
\put(1122,240){\rule[-0.350pt]{1.445pt}{0.700pt}}
\put(1128,239){\rule[-0.350pt]{1.445pt}{0.700pt}}
\put(1134,238){\rule[-0.350pt]{1.445pt}{0.700pt}}
\put(1140,237){\rule[-0.350pt]{2.891pt}{0.700pt}}
\put(1152,236){\rule[-0.350pt]{2.891pt}{0.700pt}}
\put(1164,235){\rule[-0.350pt]{2.891pt}{0.700pt}}
\put(1176,234){\rule[-0.350pt]{2.650pt}{0.700pt}}
\put(1187,233){\rule[-0.350pt]{5.782pt}{0.700pt}}
\put(1211,232){\rule[-0.350pt]{2.891pt}{0.700pt}}
\put(1223,231){\rule[-0.350pt]{8.431pt}{0.700pt}}
\put(1258,230){\rule[-0.350pt]{8.672pt}{0.700pt}}
\put(1294,229){\rule[-0.350pt]{19.995pt}{0.700pt}}
\put(1377,228){\rule[-0.350pt]{14.213pt}{0.700pt}}
\end{picture}

%% file: pggform0.tex
\setlength{\unitlength}{0.240900pt}
\ifx\plotpoint\undefined\newsavebox{\plotpoint}\fi
\begin{picture}(1500,900)(0,0)
\tenrm
\sbox{\plotpoint}{\rule[-0.175pt]{0.350pt}{0.350pt}}%
\put(264,158){\rule[-0.175pt]{0.350pt}{151.526pt}}
\put(264,158){\rule[-0.175pt]{4.818pt}{0.350pt}}
\put(242,158){\makebox(0,0)[r]{0.01}}
\put(1416,158){\rule[-0.175pt]{4.818pt}{0.350pt}}
\put(264,253){\rule[-0.175pt]{2.409pt}{0.350pt}}
\put(1426,253){\rule[-0.175pt]{2.409pt}{0.350pt}}
\put(264,308){\rule[-0.175pt]{2.409pt}{0.350pt}}
\put(1426,308){\rule[-0.175pt]{2.409pt}{0.350pt}}
\put(264,347){\rule[-0.175pt]{2.409pt}{0.350pt}}
\put(1426,347){\rule[-0.175pt]{2.409pt}{0.350pt}}
\put(264,378){\rule[-0.175pt]{2.409pt}{0.350pt}}
\put(1426,378){\rule[-0.175pt]{2.409pt}{0.350pt}}
\put(264,403){\rule[-0.175pt]{2.409pt}{0.350pt}}
\put(1426,403){\rule[-0.175pt]{2.409pt}{0.350pt}}
\put(264,424){\rule[-0.175pt]{2.409pt}{0.350pt}}
\put(1426,424){\rule[-0.175pt]{2.409pt}{0.350pt}}
\put(264,442){\rule[-0.175pt]{2.409pt}{0.350pt}}
\put(1426,442){\rule[-0.175pt]{2.409pt}{0.350pt}}
\put(264,458){\rule[-0.175pt]{2.409pt}{0.350pt}}
\put(1426,458){\rule[-0.175pt]{2.409pt}{0.350pt}}
\put(264,473){\rule[-0.175pt]{4.818pt}{0.350pt}}
\put(242,473){\makebox(0,0)[r]{0.1}}
\put(1416,473){\rule[-0.175pt]{4.818pt}{0.350pt}}
\put(264,567){\rule[-0.175pt]{2.409pt}{0.350pt}}
\put(1426,567){\rule[-0.175pt]{2.409pt}{0.350pt}}
\put(264,623){\rule[-0.175pt]{2.409pt}{0.350pt}}
\put(1426,623){\rule[-0.175pt]{2.409pt}{0.350pt}}
\put(264,662){\rule[-0.175pt]{2.409pt}{0.350pt}}
\put(1426,662){\rule[-0.175pt]{2.409pt}{0.350pt}}
\put(264,692){\rule[-0.175pt]{2.409pt}{0.350pt}}
\put(1426,692){\rule[-0.175pt]{2.409pt}{0.350pt}}
\put(264,717){\rule[-0.175pt]{2.409pt}{0.350pt}}
\put(1426,717){\rule[-0.175pt]{2.409pt}{0.350pt}}
\put(264,738){\rule[-0.175pt]{2.409pt}{0.350pt}}
\put(1426,738){\rule[-0.175pt]{2.409pt}{0.350pt}}
\put(264,757){\rule[-0.175pt]{2.409pt}{0.350pt}}
\put(1426,757){\rule[-0.175pt]{2.409pt}{0.350pt}}
\put(264,773){\rule[-0.175pt]{2.409pt}{0.350pt}}
\put(1426,773){\rule[-0.175pt]{2.409pt}{0.350pt}}
\put(264,787){\rule[-0.175pt]{4.818pt}{0.350pt}}
\put(242,787){\makebox(0,0)[r]{1}}
\put(1416,787){\rule[-0.175pt]{4.818pt}{0.350pt}}
\put(264,158){\rule[-0.175pt]{0.350pt}{4.818pt}}
\put(264,113){\makebox(0,0){0}}
\put(264,767){\rule[-0.175pt]{0.350pt}{4.818pt}}
\put(508,158){\rule[-0.175pt]{0.350pt}{4.818pt}}
\put(508,113){\makebox(0,0){0.5}}
\put(508,767){\rule[-0.175pt]{0.350pt}{4.818pt}}
\put(752,158){\rule[-0.175pt]{0.350pt}{4.818pt}}
\put(752,113){\makebox(0,0){1}}
\put(752,767){\rule[-0.175pt]{0.350pt}{4.818pt}}
\put(997,158){\rule[-0.175pt]{0.350pt}{4.818pt}}
\put(997,113){\makebox(0,0){1.5}}
\put(997,767){\rule[-0.175pt]{0.350pt}{4.818pt}}
\put(1241,158){\rule[-0.175pt]{0.350pt}{4.818pt}}
\put(1241,113){\makebox(0,0){2}}
\put(1241,767){\rule[-0.175pt]{0.350pt}{4.818pt}}
\put(264,158){\rule[-0.175pt]{282.335pt}{0.350pt}}
\put(1436,158){\rule[-0.175pt]{0.350pt}{151.526pt}}
\put(264,787){\rule[-0.175pt]{282.335pt}{0.350pt}}
\put(199,877){\makebox(0,0)[l]{\shortstack{\small$\abs{F_{\pi^0\gamma}}^2$}}}
\put(850,23){\makebox(0,0){\small$-q^2$ [GeV${}^2$]}}
\put(264,158){\rule[-0.175pt]{0.350pt}{151.526pt}}
\sbox{\plotpoint}{\rule[-0.250pt]{0.500pt}{0.500pt}}%
\put(1436,369){\usebox{\plotpoint}}
\put(1415,372){\usebox{\plotpoint}}
\put(1395,375){\usebox{\plotpoint}}
\put(1374,379){\usebox{\plotpoint}}
\put(1354,383){\usebox{\plotpoint}}
\put(1333,386){\usebox{\plotpoint}}
\put(1313,391){\usebox{\plotpoint}}
\put(1293,395){\usebox{\plotpoint}}
\put(1272,398){\usebox{\plotpoint}}
\put(1252,403){\usebox{\plotpoint}}
\put(1232,407){\usebox{\plotpoint}}
\put(1211,411){\usebox{\plotpoint}}
\put(1191,415){\usebox{\plotpoint}}
\put(1171,419){\usebox{\plotpoint}}
\put(1150,424){\usebox{\plotpoint}}
\put(1130,428){\usebox{\plotpoint}}
\put(1110,432){\usebox{\plotpoint}}
\put(1090,437){\usebox{\plotpoint}}
\put(1069,442){\usebox{\plotpoint}}
\put(1049,446){\usebox{\plotpoint}}
\put(1029,451){\usebox{\plotpoint}}
\put(1009,456){\usebox{\plotpoint}}
\put(989,461){\usebox{\plotpoint}}
\put(969,466){\usebox{\plotpoint}}
\put(949,472){\usebox{\plotpoint}}
\put(929,477){\usebox{\plotpoint}}
\put(909,483){\usebox{\plotpoint}}
\put(889,488){\usebox{\plotpoint}}
\put(869,494){\usebox{\plotpoint}}
\put(849,500){\usebox{\plotpoint}}
\put(829,506){\usebox{\plotpoint}}
\put(809,512){\usebox{\plotpoint}}
\put(789,518){\usebox{\plotpoint}}
\put(769,524){\usebox{\plotpoint}}
\put(750,530){\usebox{\plotpoint}}
\put(730,538){\usebox{\plotpoint}}
\put(711,544){\usebox{\plotpoint}}
\put(691,551){\usebox{\plotpoint}}
\put(672,558){\usebox{\plotpoint}}
\put(652,565){\usebox{\plotpoint}}
\put(633,573){\usebox{\plotpoint}}
\put(613,580){\usebox{\plotpoint}}
\put(594,588){\usebox{\plotpoint}}
\put(575,597){\usebox{\plotpoint}}
\put(556,605){\usebox{\plotpoint}}
\put(537,614){\usebox{\plotpoint}}
\put(518,622){\usebox{\plotpoint}}
\put(500,631){\usebox{\plotpoint}}
\put(481,641){\usebox{\plotpoint}}
\put(463,650){\usebox{\plotpoint}}
\put(444,660){\usebox{\plotpoint}}
\put(426,670){\usebox{\plotpoint}}
\put(409,681){\usebox{\plotpoint}}
\put(391,691){\usebox{\plotpoint}}
\put(373,703){\usebox{\plotpoint}}
\put(356,714){\usebox{\plotpoint}}
\put(339,726){\usebox{\plotpoint}}
\put(322,738){\usebox{\plotpoint}}
\put(306,751){\usebox{\plotpoint}}
\put(290,764){\usebox{\plotpoint}}
\put(274,777){\usebox{\plotpoint}}
\put(264,787){\usebox{\plotpoint}}
\sbox{\plotpoint}{\rule[-0.500pt]{1.000pt}{1.000pt}}%
\put(1430,348){\rule[-0.500pt]{1.445pt}{1.000pt}}
\put(1424,349){\rule[-0.500pt]{1.445pt}{1.000pt}}
\put(1418,350){\rule[-0.500pt]{1.445pt}{1.000pt}}
\put(1412,351){\rule[-0.500pt]{1.445pt}{1.000pt}}
\put(1407,352){\rule[-0.500pt]{1.044pt}{1.000pt}}
\put(1403,353){\rule[-0.500pt]{1.044pt}{1.000pt}}
\put(1399,354){\rule[-0.500pt]{1.044pt}{1.000pt}}
\put(1393,355){\rule[-0.500pt]{1.445pt}{1.000pt}}
\put(1387,356){\rule[-0.500pt]{1.445pt}{1.000pt}}
\put(1381,357){\rule[-0.500pt]{1.445pt}{1.000pt}}
\put(1375,358){\rule[-0.500pt]{1.445pt}{1.000pt}}
\put(1371,359){\usebox{\plotpoint}}
\put(1367,360){\usebox{\plotpoint}}
\put(1363,361){\usebox{\plotpoint}}
\put(1357,362){\rule[-0.500pt]{1.445pt}{1.000pt}}
\put(1351,363){\rule[-0.500pt]{1.445pt}{1.000pt}}
\put(1346,364){\rule[-0.500pt]{1.044pt}{1.000pt}}
\put(1342,365){\rule[-0.500pt]{1.044pt}{1.000pt}}
\put(1338,366){\rule[-0.500pt]{1.044pt}{1.000pt}}
\put(1332,367){\rule[-0.500pt]{1.445pt}{1.000pt}}
\put(1326,368){\rule[-0.500pt]{1.445pt}{1.000pt}}
\put(1320,369){\rule[-0.500pt]{1.445pt}{1.000pt}}
\put(1314,370){\rule[-0.500pt]{1.445pt}{1.000pt}}
\put(1310,371){\usebox{\plotpoint}}
\put(1306,372){\usebox{\plotpoint}}
\put(1302,373){\usebox{\plotpoint}}
\put(1295,374){\rule[-0.500pt]{1.566pt}{1.000pt}}
\put(1289,375){\rule[-0.500pt]{1.566pt}{1.000pt}}
\put(1285,376){\usebox{\plotpoint}}
\put(1281,377){\usebox{\plotpoint}}
\put(1277,378){\usebox{\plotpoint}}
\put(1273,379){\usebox{\plotpoint}}
\put(1269,380){\usebox{\plotpoint}}
\put(1265,381){\usebox{\plotpoint}}
\put(1259,382){\rule[-0.500pt]{1.445pt}{1.000pt}}
\put(1253,383){\rule[-0.500pt]{1.445pt}{1.000pt}}
\put(1249,384){\usebox{\plotpoint}}
\put(1245,385){\usebox{\plotpoint}}
\put(1241,386){\usebox{\plotpoint}}
\put(1234,387){\rule[-0.500pt]{1.566pt}{1.000pt}}
\put(1228,388){\rule[-0.500pt]{1.566pt}{1.000pt}}
\put(1224,389){\usebox{\plotpoint}}
\put(1220,390){\usebox{\plotpoint}}
\put(1216,391){\usebox{\plotpoint}}
\put(1212,392){\usebox{\plotpoint}}
\put(1208,393){\usebox{\plotpoint}}
\put(1204,394){\usebox{\plotpoint}}
\put(1198,395){\rule[-0.500pt]{1.445pt}{1.000pt}}
\put(1192,396){\rule[-0.500pt]{1.445pt}{1.000pt}}
\put(1188,397){\usebox{\plotpoint}}
\put(1184,398){\usebox{\plotpoint}}
\put(1180,399){\usebox{\plotpoint}}
\put(1175,400){\rule[-0.500pt]{1.044pt}{1.000pt}}
\put(1171,401){\rule[-0.500pt]{1.044pt}{1.000pt}}
\put(1167,402){\rule[-0.500pt]{1.044pt}{1.000pt}}
\put(1163,403){\usebox{\plotpoint}}
\put(1159,404){\usebox{\plotpoint}}
\put(1155,405){\usebox{\plotpoint}}
\put(1151,406){\usebox{\plotpoint}}
\put(1147,407){\usebox{\plotpoint}}
\put(1143,408){\usebox{\plotpoint}}
\put(1137,409){\rule[-0.500pt]{1.445pt}{1.000pt}}
\put(1131,410){\rule[-0.500pt]{1.445pt}{1.000pt}}
\put(1127,411){\usebox{\plotpoint}}
\put(1123,412){\usebox{\plotpoint}}
\put(1119,413){\usebox{\plotpoint}}
\put(1114,414){\rule[-0.500pt]{1.044pt}{1.000pt}}
\put(1110,415){\rule[-0.500pt]{1.044pt}{1.000pt}}
\put(1106,416){\rule[-0.500pt]{1.044pt}{1.000pt}}
\put(1102,417){\usebox{\plotpoint}}
\put(1098,418){\usebox{\plotpoint}}
\put(1094,419){\usebox{\plotpoint}}
\put(1090,420){\usebox{\plotpoint}}
\put(1086,421){\usebox{\plotpoint}}
\put(1082,422){\usebox{\plotpoint}}
\put(1078,423){\usebox{\plotpoint}}
\put(1074,424){\usebox{\plotpoint}}
\put(1070,425){\usebox{\plotpoint}}
\put(1066,426){\usebox{\plotpoint}}
\put(1062,427){\usebox{\plotpoint}}
\put(1058,428){\usebox{\plotpoint}}
\put(1053,429){\rule[-0.500pt]{1.044pt}{1.000pt}}
\put(1049,430){\rule[-0.500pt]{1.044pt}{1.000pt}}
\put(1045,431){\rule[-0.500pt]{1.044pt}{1.000pt}}
\put(1041,432){\usebox{\plotpoint}}
\put(1037,433){\usebox{\plotpoint}}
\put(1033,434){\usebox{\plotpoint}}
\put(1030,435){\usebox{\plotpoint}}
\put(1027,436){\usebox{\plotpoint}}
\put(1024,437){\usebox{\plotpoint}}
\put(1021,438){\usebox{\plotpoint}}
\put(1017,439){\usebox{\plotpoint}}
\put(1013,440){\usebox{\plotpoint}}
\put(1009,441){\usebox{\plotpoint}}
\put(1005,442){\usebox{\plotpoint}}
\put(1001,443){\usebox{\plotpoint}}
\put(997,444){\usebox{\plotpoint}}
\put(992,445){\rule[-0.500pt]{1.044pt}{1.000pt}}
\put(988,446){\rule[-0.500pt]{1.044pt}{1.000pt}}
\put(984,447){\rule[-0.500pt]{1.044pt}{1.000pt}}
\put(981,448){\usebox{\plotpoint}}
\put(978,449){\usebox{\plotpoint}}
\put(975,450){\usebox{\plotpoint}}
\put(972,451){\usebox{\plotpoint}}
\put(968,452){\usebox{\plotpoint}}
\put(964,453){\usebox{\plotpoint}}
\put(960,454){\usebox{\plotpoint}}
\put(956,455){\usebox{\plotpoint}}
\put(952,456){\usebox{\plotpoint}}
\put(948,457){\usebox{\plotpoint}}
\put(944,458){\usebox{\plotpoint}}
\put(941,459){\usebox{\plotpoint}}
\put(938,460){\usebox{\plotpoint}}
\put(935,461){\usebox{\plotpoint}}
\put(931,462){\usebox{\plotpoint}}
\put(927,463){\usebox{\plotpoint}}
\put(923,464){\usebox{\plotpoint}}
\put(920,465){\usebox{\plotpoint}}
\put(917,466){\usebox{\plotpoint}}
\put(914,467){\usebox{\plotpoint}}
\put(911,468){\usebox{\plotpoint}}
\put(907,469){\usebox{\plotpoint}}
\put(903,470){\usebox{\plotpoint}}
\put(899,471){\usebox{\plotpoint}}
\put(896,472){\usebox{\plotpoint}}
\put(893,473){\usebox{\plotpoint}}
\put(890,474){\usebox{\plotpoint}}
\put(887,475){\usebox{\plotpoint}}
\put(883,476){\usebox{\plotpoint}}
\put(880,477){\usebox{\plotpoint}}
\put(877,478){\usebox{\plotpoint}}
\put(874,479){\usebox{\plotpoint}}
\put(870,480){\usebox{\plotpoint}}
\put(866,481){\usebox{\plotpoint}}
\put(862,482){\usebox{\plotpoint}}
\put(859,483){\usebox{\plotpoint}}
\put(856,484){\usebox{\plotpoint}}
\put(853,485){\usebox{\plotpoint}}
\put(850,486){\usebox{\plotpoint}}
\put(847,487){\usebox{\plotpoint}}
\put(844,488){\usebox{\plotpoint}}
\put(841,489){\usebox{\plotpoint}}
\put(838,490){\usebox{\plotpoint}}
\put(835,491){\usebox{\plotpoint}}
\put(832,492){\usebox{\plotpoint}}
\put(829,493){\usebox{\plotpoint}}
\put(826,494){\usebox{\plotpoint}}
\put(822,495){\usebox{\plotpoint}}
\put(819,496){\usebox{\plotpoint}}
\put(816,497){\usebox{\plotpoint}}
\put(813,498){\usebox{\plotpoint}}
\put(810,499){\usebox{\plotpoint}}
\put(807,500){\usebox{\plotpoint}}
\put(804,501){\usebox{\plotpoint}}
\put(801,502){\usebox{\plotpoint}}
\put(798,503){\usebox{\plotpoint}}
\put(795,504){\usebox{\plotpoint}}
\put(792,505){\usebox{\plotpoint}}
\put(789,506){\usebox{\plotpoint}}
\put(786,507){\usebox{\plotpoint}}
\put(783,508){\usebox{\plotpoint}}
\put(780,509){\usebox{\plotpoint}}
\put(777,510){\usebox{\plotpoint}}
\put(774,511){\usebox{\plotpoint}}
\put(771,512){\usebox{\plotpoint}}
\put(768,513){\usebox{\plotpoint}}
\put(765,514){\usebox{\plotpoint}}
\put(761,515){\usebox{\plotpoint}}
\put(758,516){\usebox{\plotpoint}}
\put(755,517){\usebox{\plotpoint}}
\put(752,518){\usebox{\plotpoint}}
\put(749,519){\usebox{\plotpoint}}
\put(747,520){\usebox{\plotpoint}}
\put(744,521){\usebox{\plotpoint}}
\put(742,522){\usebox{\plotpoint}}
\put(740,523){\usebox{\plotpoint}}
\put(737,524){\usebox{\plotpoint}}
\put(734,525){\usebox{\plotpoint}}
\put(731,526){\usebox{\plotpoint}}
\put(728,527){\usebox{\plotpoint}}
\put(725,528){\usebox{\plotpoint}}
\put(722,529){\usebox{\plotpoint}}
\put(719,530){\usebox{\plotpoint}}
\put(716,531){\usebox{\plotpoint}}
\put(713,532){\usebox{\plotpoint}}
\put(711,533){\usebox{\plotpoint}}
\put(708,534){\usebox{\plotpoint}}
\put(706,535){\usebox{\plotpoint}}
\put(704,536){\usebox{\plotpoint}}
\put(701,537){\usebox{\plotpoint}}
\put(698,538){\usebox{\plotpoint}}
\put(696,539){\usebox{\plotpoint}}
\put(693,540){\usebox{\plotpoint}}
\put(691,541){\usebox{\plotpoint}}
\put(688,542){\usebox{\plotpoint}}
\put(685,543){\usebox{\plotpoint}}
\put(682,544){\usebox{\plotpoint}}
\put(679,545){\usebox{\plotpoint}}
\put(676,546){\usebox{\plotpoint}}
\put(674,547){\usebox{\plotpoint}}
\put(671,548){\usebox{\plotpoint}}
\put(669,549){\usebox{\plotpoint}}
\put(667,550){\usebox{\plotpoint}}
\put(664,551){\usebox{\plotpoint}}
\put(662,552){\usebox{\plotpoint}}
\put(659,553){\usebox{\plotpoint}}
\put(657,554){\usebox{\plotpoint}}
\put(655,555){\usebox{\plotpoint}}
\put(652,556){\usebox{\plotpoint}}
\put(649,557){\usebox{\plotpoint}}
\put(647,558){\usebox{\plotpoint}}
\put(644,559){\usebox{\plotpoint}}
\put(642,560){\usebox{\plotpoint}}
\put(639,561){\usebox{\plotpoint}}
\put(637,562){\usebox{\plotpoint}}
\put(634,563){\usebox{\plotpoint}}
\put(632,564){\usebox{\plotpoint}}
\put(630,565){\usebox{\plotpoint}}
\put(627,566){\usebox{\plotpoint}}
\put(625,567){\usebox{\plotpoint}}
\put(622,568){\usebox{\plotpoint}}
\put(620,569){\usebox{\plotpoint}}
\put(618,570){\usebox{\plotpoint}}
\put(615,571){\usebox{\plotpoint}}
\put(613,572){\usebox{\plotpoint}}
\put(610,573){\usebox{\plotpoint}}
\put(608,574){\usebox{\plotpoint}}
\put(606,575){\usebox{\plotpoint}}
\put(603,576){\usebox{\plotpoint}}
\put(601,577){\usebox{\plotpoint}}
\put(598,578){\usebox{\plotpoint}}
\put(596,579){\usebox{\plotpoint}}
\put(594,580){\usebox{\plotpoint}}
\put(591,581){\usebox{\plotpoint}}
\put(589,582){\usebox{\plotpoint}}
\put(587,583){\usebox{\plotpoint}}
\put(585,584){\usebox{\plotpoint}}
\put(583,585){\usebox{\plotpoint}}
\put(581,586){\usebox{\plotpoint}}
\put(578,587){\usebox{\plotpoint}}
\put(576,588){\usebox{\plotpoint}}
\put(573,589){\usebox{\plotpoint}}
\put(571,590){\usebox{\plotpoint}}
\put(569,591){\usebox{\plotpoint}}
\put(567,592){\usebox{\plotpoint}}
\put(565,593){\usebox{\plotpoint}}
\put(563,594){\usebox{\plotpoint}}
\put(561,595){\usebox{\plotpoint}}
\put(559,596){\usebox{\plotpoint}}
\put(557,597){\usebox{\plotpoint}}
\put(555,598){\usebox{\plotpoint}}
\put(553,599){\usebox{\plotpoint}}
\put(551,600){\usebox{\plotpoint}}
\put(549,601){\usebox{\plotpoint}}
\put(547,602){\usebox{\plotpoint}}
\put(545,603){\usebox{\plotpoint}}
\put(542,604){\usebox{\plotpoint}}
\put(540,605){\usebox{\plotpoint}}
\put(537,606){\usebox{\plotpoint}}
\put(535,607){\usebox{\plotpoint}}
\put(533,608){\usebox{\plotpoint}}
\put(530,609){\usebox{\plotpoint}}
\put(528,610){\usebox{\plotpoint}}
\put(526,611){\usebox{\plotpoint}}
\put(524,612){\usebox{\plotpoint}}
\put(522,613){\usebox{\plotpoint}}
\put(520,614){\usebox{\plotpoint}}
\put(518,615){\usebox{\plotpoint}}
\put(516,616){\usebox{\plotpoint}}
\put(514,617){\usebox{\plotpoint}}
\put(513,618){\usebox{\plotpoint}}
\put(511,619){\usebox{\plotpoint}}
\put(509,620){\usebox{\plotpoint}}
\put(508,621){\usebox{\plotpoint}}
\put(506,622){\usebox{\plotpoint}}
\put(504,623){\usebox{\plotpoint}}
\put(502,624){\usebox{\plotpoint}}
\put(500,625){\usebox{\plotpoint}}
\put(498,626){\usebox{\plotpoint}}
\put(496,627){\usebox{\plotpoint}}
\put(494,628){\usebox{\plotpoint}}
\put(492,629){\usebox{\plotpoint}}
\put(490,630){\usebox{\plotpoint}}
\put(488,631){\usebox{\plotpoint}}
\put(486,632){\usebox{\plotpoint}}
\put(484,633){\usebox{\plotpoint}}
\put(482,634){\usebox{\plotpoint}}
\put(480,635){\usebox{\plotpoint}}
\put(478,636){\usebox{\plotpoint}}
\put(477,637){\usebox{\plotpoint}}
\put(475,638){\usebox{\plotpoint}}
\put(473,639){\usebox{\plotpoint}}
\put(472,640){\usebox{\plotpoint}}
\put(470,641){\usebox{\plotpoint}}
\put(468,642){\usebox{\plotpoint}}
\put(466,643){\usebox{\plotpoint}}
\put(464,644){\usebox{\plotpoint}}
\put(462,645){\usebox{\plotpoint}}
\put(460,646){\usebox{\plotpoint}}
\put(459,647){\usebox{\plotpoint}}
\put(457,648){\usebox{\plotpoint}}
\put(455,649){\usebox{\plotpoint}}
\put(453,650){\usebox{\plotpoint}}
\put(452,651){\usebox{\plotpoint}}
\put(450,652){\usebox{\plotpoint}}
\put(448,653){\usebox{\plotpoint}}
\put(447,654){\usebox{\plotpoint}}
\put(445,655){\usebox{\plotpoint}}
\put(443,656){\usebox{\plotpoint}}
\put(441,657){\usebox{\plotpoint}}
\put(440,658){\usebox{\plotpoint}}
\put(438,659){\usebox{\plotpoint}}
\put(436,660){\usebox{\plotpoint}}
\put(435,661){\usebox{\plotpoint}}
\put(433,662){\usebox{\plotpoint}}
\put(431,663){\usebox{\plotpoint}}
\put(429,664){\usebox{\plotpoint}}
\put(428,665){\usebox{\plotpoint}}
\put(426,666){\usebox{\plotpoint}}
\put(424,667){\usebox{\plotpoint}}
\put(423,668){\usebox{\plotpoint}}
\put(421,669){\usebox{\plotpoint}}
\put(419,670){\usebox{\plotpoint}}
\put(417,671){\usebox{\plotpoint}}
\put(416,672){\usebox{\plotpoint}}
\put(414,673){\usebox{\plotpoint}}
\put(412,674){\usebox{\plotpoint}}
\put(411,675){\usebox{\plotpoint}}
\put(409,676){\usebox{\plotpoint}}
\put(407,677){\usebox{\plotpoint}}
\put(406,678){\usebox{\plotpoint}}
\put(404,679){\usebox{\plotpoint}}
\put(402,680){\usebox{\plotpoint}}
\put(401,681){\usebox{\plotpoint}}
\put(399,682){\usebox{\plotpoint}}
\put(398,683){\usebox{\plotpoint}}
\put(396,684){\usebox{\plotpoint}}
\put(395,685){\usebox{\plotpoint}}
\put(393,686){\usebox{\plotpoint}}
\put(392,687){\usebox{\plotpoint}}
\put(390,688){\usebox{\plotpoint}}
\put(389,689){\usebox{\plotpoint}}
\put(387,690){\usebox{\plotpoint}}
\put(386,691){\usebox{\plotpoint}}
\put(384,692){\usebox{\plotpoint}}
\put(383,693){\usebox{\plotpoint}}
\put(381,694){\usebox{\plotpoint}}
\put(380,695){\usebox{\plotpoint}}
\put(378,696){\usebox{\plotpoint}}
\put(377,697){\usebox{\plotpoint}}
\put(375,698){\usebox{\plotpoint}}
\put(374,699){\usebox{\plotpoint}}
\put(372,700){\usebox{\plotpoint}}
\put(371,701){\usebox{\plotpoint}}
\put(369,702){\usebox{\plotpoint}}
\put(368,703){\usebox{\plotpoint}}
\put(367,704){\usebox{\plotpoint}}
\put(365,705){\usebox{\plotpoint}}
\put(364,706){\usebox{\plotpoint}}
\put(363,707){\usebox{\plotpoint}}
\put(362,708){\usebox{\plotpoint}}
\put(360,709){\usebox{\plotpoint}}
\put(358,710){\usebox{\plotpoint}}
\put(357,711){\usebox{\plotpoint}}
\put(355,712){\usebox{\plotpoint}}
\put(353,713){\usebox{\plotpoint}}
\put(352,714){\usebox{\plotpoint}}
\put(350,715){\usebox{\plotpoint}}
\put(349,716){\usebox{\plotpoint}}
\put(347,717){\usebox{\plotpoint}}
\put(346,718){\usebox{\plotpoint}}
\put(344,719){\usebox{\plotpoint}}
\put(343,720){\usebox{\plotpoint}}
\put(342,721){\usebox{\plotpoint}}
\put(340,722){\usebox{\plotpoint}}
\put(339,723){\usebox{\plotpoint}}
\put(338,724){\usebox{\plotpoint}}
\put(337,725){\usebox{\plotpoint}}
\put(335,726){\usebox{\plotpoint}}
\put(334,727){\usebox{\plotpoint}}
\put(332,728){\usebox{\plotpoint}}
\put(331,729){\usebox{\plotpoint}}
\put(330,730){\usebox{\plotpoint}}
\put(328,731){\usebox{\plotpoint}}
\put(327,732){\usebox{\plotpoint}}
\put(326,733){\usebox{\plotpoint}}
\put(325,734){\usebox{\plotpoint}}
\put(323,735){\usebox{\plotpoint}}
\put(322,736){\usebox{\plotpoint}}
\put(321,737){\usebox{\plotpoint}}
\put(320,738){\usebox{\plotpoint}}
\put(318,739){\usebox{\plotpoint}}
\put(317,740){\usebox{\plotpoint}}
\put(316,741){\usebox{\plotpoint}}
\put(315,742){\usebox{\plotpoint}}
\put(314,743){\usebox{\plotpoint}}
\put(313,744){\usebox{\plotpoint}}
\put(311,745){\usebox{\plotpoint}}
\put(310,746){\usebox{\plotpoint}}
\put(309,747){\usebox{\plotpoint}}
\put(308,748){\usebox{\plotpoint}}
\put(306,749){\usebox{\plotpoint}}
\put(305,750){\usebox{\plotpoint}}
\put(304,751){\usebox{\plotpoint}}
\put(303,752){\usebox{\plotpoint}}
\put(302,753){\usebox{\plotpoint}}
\put(301,754){\usebox{\plotpoint}}
\put(299,755){\usebox{\plotpoint}}
\put(298,756){\usebox{\plotpoint}}
\put(297,757){\usebox{\plotpoint}}
\put(295,758){\usebox{\plotpoint}}
\put(294,759){\usebox{\plotpoint}}
\put(293,760){\usebox{\plotpoint}}
\put(291,761){\usebox{\plotpoint}}
\put(290,762){\usebox{\plotpoint}}
\put(289,763){\usebox{\plotpoint}}
\put(288,764){\usebox{\plotpoint}}
\put(286,765){\usebox{\plotpoint}}
\put(285,766){\usebox{\plotpoint}}
\put(284,767){\usebox{\plotpoint}}
\put(283,768){\usebox{\plotpoint}}
\put(282,769){\usebox{\plotpoint}}
\put(281,770){\usebox{\plotpoint}}
\put(280,771){\usebox{\plotpoint}}
\put(279,772){\usebox{\plotpoint}}
\put(278,773){\usebox{\plotpoint}}
\put(277,774){\usebox{\plotpoint}}
\put(276,775){\usebox{\plotpoint}}
\put(274,776){\usebox{\plotpoint}}
\put(273,777){\usebox{\plotpoint}}
\put(272,778){\usebox{\plotpoint}}
\put(271,779){\usebox{\plotpoint}}
\put(270,780){\usebox{\plotpoint}}
\put(269,781){\usebox{\plotpoint}}
\put(268,782){\usebox{\plotpoint}}
\put(267,783){\usebox{\plotpoint}}
\put(266,784){\usebox{\plotpoint}}
\put(265,785){\usebox{\plotpoint}}
\put(264,786){\usebox{\plotpoint}}
\sbox{\plotpoint}{\rule[-0.175pt]{0.350pt}{0.350pt}}%
\put(1432,325){\rule[-0.175pt]{0.964pt}{0.350pt}}
\put(1428,326){\rule[-0.175pt]{0.964pt}{0.350pt}}
\put(1424,327){\rule[-0.175pt]{0.964pt}{0.350pt}}
\put(1418,328){\rule[-0.175pt]{1.445pt}{0.350pt}}
\put(1412,329){\rule[-0.175pt]{1.445pt}{0.350pt}}
\put(1407,330){\rule[-0.175pt]{1.044pt}{0.350pt}}
\put(1403,331){\rule[-0.175pt]{1.044pt}{0.350pt}}
\put(1399,332){\rule[-0.175pt]{1.044pt}{0.350pt}}
\put(1393,333){\rule[-0.175pt]{1.445pt}{0.350pt}}
\put(1387,334){\rule[-0.175pt]{1.445pt}{0.350pt}}
\put(1383,335){\rule[-0.175pt]{0.964pt}{0.350pt}}
\put(1379,336){\rule[-0.175pt]{0.964pt}{0.350pt}}
\put(1375,337){\rule[-0.175pt]{0.964pt}{0.350pt}}
\put(1369,338){\rule[-0.175pt]{1.445pt}{0.350pt}}
\put(1363,339){\rule[-0.175pt]{1.445pt}{0.350pt}}
\put(1359,340){\rule[-0.175pt]{0.964pt}{0.350pt}}
\put(1355,341){\rule[-0.175pt]{0.964pt}{0.350pt}}
\put(1351,342){\rule[-0.175pt]{0.964pt}{0.350pt}}
\put(1344,343){\rule[-0.175pt]{1.566pt}{0.350pt}}
\put(1338,344){\rule[-0.175pt]{1.566pt}{0.350pt}}
\put(1334,345){\rule[-0.175pt]{0.964pt}{0.350pt}}
\put(1330,346){\rule[-0.175pt]{0.964pt}{0.350pt}}
\put(1326,347){\rule[-0.175pt]{0.964pt}{0.350pt}}
\put(1322,348){\rule[-0.175pt]{0.964pt}{0.350pt}}
\put(1318,349){\rule[-0.175pt]{0.964pt}{0.350pt}}
\put(1314,350){\rule[-0.175pt]{0.964pt}{0.350pt}}
\put(1308,351){\rule[-0.175pt]{1.445pt}{0.350pt}}
\put(1302,352){\rule[-0.175pt]{1.445pt}{0.350pt}}
\put(1297,353){\rule[-0.175pt]{1.044pt}{0.350pt}}
\put(1293,354){\rule[-0.175pt]{1.044pt}{0.350pt}}
\put(1289,355){\rule[-0.175pt]{1.044pt}{0.350pt}}
\put(1285,356){\rule[-0.175pt]{0.964pt}{0.350pt}}
\put(1281,357){\rule[-0.175pt]{0.964pt}{0.350pt}}
\put(1277,358){\rule[-0.175pt]{0.964pt}{0.350pt}}
\put(1271,359){\rule[-0.175pt]{1.445pt}{0.350pt}}
\put(1265,360){\rule[-0.175pt]{1.445pt}{0.350pt}}
\put(1261,361){\rule[-0.175pt]{0.964pt}{0.350pt}}
\put(1257,362){\rule[-0.175pt]{0.964pt}{0.350pt}}
\put(1253,363){\rule[-0.175pt]{0.964pt}{0.350pt}}
\put(1249,364){\rule[-0.175pt]{0.964pt}{0.350pt}}
\put(1245,365){\rule[-0.175pt]{0.964pt}{0.350pt}}
\put(1241,366){\rule[-0.175pt]{0.964pt}{0.350pt}}
\put(1236,367){\rule[-0.175pt]{1.044pt}{0.350pt}}
\put(1232,368){\rule[-0.175pt]{1.044pt}{0.350pt}}
\put(1228,369){\rule[-0.175pt]{1.044pt}{0.350pt}}
\put(1224,370){\rule[-0.175pt]{0.964pt}{0.350pt}}
\put(1220,371){\rule[-0.175pt]{0.964pt}{0.350pt}}
\put(1216,372){\rule[-0.175pt]{0.964pt}{0.350pt}}
\put(1210,373){\rule[-0.175pt]{1.445pt}{0.350pt}}
\put(1204,374){\rule[-0.175pt]{1.445pt}{0.350pt}}
\put(1200,375){\rule[-0.175pt]{0.964pt}{0.350pt}}
\put(1196,376){\rule[-0.175pt]{0.964pt}{0.350pt}}
\put(1192,377){\rule[-0.175pt]{0.964pt}{0.350pt}}
\put(1188,378){\rule[-0.175pt]{0.964pt}{0.350pt}}
\put(1184,379){\rule[-0.175pt]{0.964pt}{0.350pt}}
\put(1180,380){\rule[-0.175pt]{0.964pt}{0.350pt}}
\put(1175,381){\rule[-0.175pt]{1.044pt}{0.350pt}}
\put(1171,382){\rule[-0.175pt]{1.044pt}{0.350pt}}
\put(1167,383){\rule[-0.175pt]{1.044pt}{0.350pt}}
\put(1163,384){\rule[-0.175pt]{0.964pt}{0.350pt}}
\put(1159,385){\rule[-0.175pt]{0.964pt}{0.350pt}}
\put(1155,386){\rule[-0.175pt]{0.964pt}{0.350pt}}
\put(1151,387){\rule[-0.175pt]{0.964pt}{0.350pt}}
\put(1147,388){\rule[-0.175pt]{0.964pt}{0.350pt}}
\put(1143,389){\rule[-0.175pt]{0.964pt}{0.350pt}}
\put(1139,390){\rule[-0.175pt]{0.964pt}{0.350pt}}
\put(1135,391){\rule[-0.175pt]{0.964pt}{0.350pt}}
\put(1131,392){\rule[-0.175pt]{0.964pt}{0.350pt}}
\put(1127,393){\rule[-0.175pt]{0.964pt}{0.350pt}}
\put(1123,394){\rule[-0.175pt]{0.964pt}{0.350pt}}
\put(1119,395){\rule[-0.175pt]{0.964pt}{0.350pt}}
\put(1115,396){\rule[-0.175pt]{0.783pt}{0.350pt}}
\put(1112,397){\rule[-0.175pt]{0.783pt}{0.350pt}}
\put(1109,398){\rule[-0.175pt]{0.783pt}{0.350pt}}
\put(1106,399){\rule[-0.175pt]{0.783pt}{0.350pt}}
\put(1102,400){\rule[-0.175pt]{0.964pt}{0.350pt}}
\put(1098,401){\rule[-0.175pt]{0.964pt}{0.350pt}}
\put(1094,402){\rule[-0.175pt]{0.964pt}{0.350pt}}
\put(1090,403){\rule[-0.175pt]{0.964pt}{0.350pt}}
\put(1086,404){\rule[-0.175pt]{0.964pt}{0.350pt}}
\put(1082,405){\rule[-0.175pt]{0.964pt}{0.350pt}}
\put(1078,406){\rule[-0.175pt]{0.964pt}{0.350pt}}
\put(1074,407){\rule[-0.175pt]{0.964pt}{0.350pt}}
\put(1070,408){\rule[-0.175pt]{0.964pt}{0.350pt}}
\put(1066,409){\rule[-0.175pt]{0.964pt}{0.350pt}}
\put(1062,410){\rule[-0.175pt]{0.964pt}{0.350pt}}
\put(1058,411){\rule[-0.175pt]{0.964pt}{0.350pt}}
\put(1054,412){\rule[-0.175pt]{0.783pt}{0.350pt}}
\put(1051,413){\rule[-0.175pt]{0.783pt}{0.350pt}}
\put(1048,414){\rule[-0.175pt]{0.783pt}{0.350pt}}
\put(1045,415){\rule[-0.175pt]{0.783pt}{0.350pt}}
\put(1041,416){\rule[-0.175pt]{0.964pt}{0.350pt}}
\put(1037,417){\rule[-0.175pt]{0.964pt}{0.350pt}}
\put(1033,418){\rule[-0.175pt]{0.964pt}{0.350pt}}
\put(1029,419){\rule[-0.175pt]{0.964pt}{0.350pt}}
\put(1025,420){\rule[-0.175pt]{0.964pt}{0.350pt}}
\put(1021,421){\rule[-0.175pt]{0.964pt}{0.350pt}}
\put(1018,422){\rule[-0.175pt]{0.723pt}{0.350pt}}
\put(1015,423){\rule[-0.175pt]{0.723pt}{0.350pt}}
\put(1012,424){\rule[-0.175pt]{0.723pt}{0.350pt}}
\put(1009,425){\rule[-0.175pt]{0.723pt}{0.350pt}}
\put(1005,426){\rule[-0.175pt]{0.964pt}{0.350pt}}
\put(1001,427){\rule[-0.175pt]{0.964pt}{0.350pt}}
\put(997,428){\rule[-0.175pt]{0.964pt}{0.350pt}}
\put(993,429){\rule[-0.175pt]{0.783pt}{0.350pt}}
\put(990,430){\rule[-0.175pt]{0.783pt}{0.350pt}}
\put(987,431){\rule[-0.175pt]{0.783pt}{0.350pt}}
\put(984,432){\rule[-0.175pt]{0.783pt}{0.350pt}}
\put(980,433){\rule[-0.175pt]{0.964pt}{0.350pt}}
\put(976,434){\rule[-0.175pt]{0.964pt}{0.350pt}}
\put(972,435){\rule[-0.175pt]{0.964pt}{0.350pt}}
\put(969,436){\rule[-0.175pt]{0.723pt}{0.350pt}}
\put(966,437){\rule[-0.175pt]{0.723pt}{0.350pt}}
\put(963,438){\rule[-0.175pt]{0.723pt}{0.350pt}}
\put(960,439){\rule[-0.175pt]{0.723pt}{0.350pt}}
\put(956,440){\rule[-0.175pt]{0.964pt}{0.350pt}}
\put(952,441){\rule[-0.175pt]{0.964pt}{0.350pt}}
\put(948,442){\rule[-0.175pt]{0.964pt}{0.350pt}}
\put(944,443){\rule[-0.175pt]{0.783pt}{0.350pt}}
\put(941,444){\rule[-0.175pt]{0.783pt}{0.350pt}}
\put(938,445){\rule[-0.175pt]{0.783pt}{0.350pt}}
\put(935,446){\rule[-0.175pt]{0.783pt}{0.350pt}}
\put(932,447){\rule[-0.175pt]{0.723pt}{0.350pt}}
\put(929,448){\rule[-0.175pt]{0.723pt}{0.350pt}}
\put(926,449){\rule[-0.175pt]{0.723pt}{0.350pt}}
\put(923,450){\rule[-0.175pt]{0.723pt}{0.350pt}}
\put(919,451){\rule[-0.175pt]{0.964pt}{0.350pt}}
\put(915,452){\rule[-0.175pt]{0.964pt}{0.350pt}}
\put(911,453){\rule[-0.175pt]{0.964pt}{0.350pt}}
\put(908,454){\rule[-0.175pt]{0.723pt}{0.350pt}}
\put(905,455){\rule[-0.175pt]{0.723pt}{0.350pt}}
\put(902,456){\rule[-0.175pt]{0.723pt}{0.350pt}}
\put(899,457){\rule[-0.175pt]{0.723pt}{0.350pt}}
\put(896,458){\rule[-0.175pt]{0.723pt}{0.350pt}}
\put(893,459){\rule[-0.175pt]{0.723pt}{0.350pt}}
\put(890,460){\rule[-0.175pt]{0.723pt}{0.350pt}}
\put(887,461){\rule[-0.175pt]{0.723pt}{0.350pt}}
\put(883,462){\rule[-0.175pt]{0.783pt}{0.350pt}}
\put(880,463){\rule[-0.175pt]{0.783pt}{0.350pt}}
\put(877,464){\rule[-0.175pt]{0.783pt}{0.350pt}}
\put(874,465){\rule[-0.175pt]{0.783pt}{0.350pt}}
\put(871,466){\rule[-0.175pt]{0.723pt}{0.350pt}}
\put(868,467){\rule[-0.175pt]{0.723pt}{0.350pt}}
\put(865,468){\rule[-0.175pt]{0.723pt}{0.350pt}}
\put(862,469){\rule[-0.175pt]{0.723pt}{0.350pt}}
\put(859,470){\rule[-0.175pt]{0.723pt}{0.350pt}}
\put(856,471){\rule[-0.175pt]{0.723pt}{0.350pt}}
\put(853,472){\rule[-0.175pt]{0.723pt}{0.350pt}}
\put(850,473){\rule[-0.175pt]{0.723pt}{0.350pt}}
\put(847,474){\rule[-0.175pt]{0.723pt}{0.350pt}}
\put(844,475){\rule[-0.175pt]{0.723pt}{0.350pt}}
\put(841,476){\rule[-0.175pt]{0.723pt}{0.350pt}}
\put(838,477){\rule[-0.175pt]{0.723pt}{0.350pt}}
\put(835,478){\rule[-0.175pt]{0.723pt}{0.350pt}}
\put(832,479){\rule[-0.175pt]{0.723pt}{0.350pt}}
\put(829,480){\rule[-0.175pt]{0.723pt}{0.350pt}}
\put(826,481){\rule[-0.175pt]{0.723pt}{0.350pt}}
\put(822,482){\rule[-0.175pt]{0.783pt}{0.350pt}}
\put(819,483){\rule[-0.175pt]{0.783pt}{0.350pt}}
\put(816,484){\rule[-0.175pt]{0.783pt}{0.350pt}}
\put(813,485){\rule[-0.175pt]{0.783pt}{0.350pt}}
\put(810,486){\rule[-0.175pt]{0.723pt}{0.350pt}}
\put(807,487){\rule[-0.175pt]{0.723pt}{0.350pt}}
\put(804,488){\rule[-0.175pt]{0.723pt}{0.350pt}}
\put(801,489){\rule[-0.175pt]{0.723pt}{0.350pt}}
\put(798,490){\rule[-0.175pt]{0.723pt}{0.350pt}}
\put(795,491){\rule[-0.175pt]{0.723pt}{0.350pt}}
\put(792,492){\rule[-0.175pt]{0.723pt}{0.350pt}}
\put(789,493){\rule[-0.175pt]{0.723pt}{0.350pt}}
\put(786,494){\rule[-0.175pt]{0.578pt}{0.350pt}}
\put(784,495){\rule[-0.175pt]{0.578pt}{0.350pt}}
\put(781,496){\rule[-0.175pt]{0.578pt}{0.350pt}}
\put(779,497){\rule[-0.175pt]{0.578pt}{0.350pt}}
\put(777,498){\rule[-0.175pt]{0.578pt}{0.350pt}}
\put(774,499){\rule[-0.175pt]{0.723pt}{0.350pt}}
\put(771,500){\rule[-0.175pt]{0.723pt}{0.350pt}}
\put(768,501){\rule[-0.175pt]{0.723pt}{0.350pt}}
\put(765,502){\rule[-0.175pt]{0.723pt}{0.350pt}}
\put(762,503){\rule[-0.175pt]{0.626pt}{0.350pt}}
\put(759,504){\rule[-0.175pt]{0.626pt}{0.350pt}}
\put(757,505){\rule[-0.175pt]{0.626pt}{0.350pt}}
\put(754,506){\rule[-0.175pt]{0.626pt}{0.350pt}}
\put(752,507){\rule[-0.175pt]{0.626pt}{0.350pt}}
\put(749,508){\rule[-0.175pt]{0.723pt}{0.350pt}}
\put(746,509){\rule[-0.175pt]{0.723pt}{0.350pt}}
\put(743,510){\rule[-0.175pt]{0.723pt}{0.350pt}}
\put(740,511){\rule[-0.175pt]{0.723pt}{0.350pt}}
\put(737,512){\rule[-0.175pt]{0.578pt}{0.350pt}}
\put(735,513){\rule[-0.175pt]{0.578pt}{0.350pt}}
\put(732,514){\rule[-0.175pt]{0.578pt}{0.350pt}}
\put(730,515){\rule[-0.175pt]{0.578pt}{0.350pt}}
\put(728,516){\rule[-0.175pt]{0.578pt}{0.350pt}}
\put(725,517){\rule[-0.175pt]{0.578pt}{0.350pt}}
\put(723,518){\rule[-0.175pt]{0.578pt}{0.350pt}}
\put(720,519){\rule[-0.175pt]{0.578pt}{0.350pt}}
\put(718,520){\rule[-0.175pt]{0.578pt}{0.350pt}}
\put(716,521){\rule[-0.175pt]{0.578pt}{0.350pt}}
\put(713,522){\rule[-0.175pt]{0.723pt}{0.350pt}}
\put(710,523){\rule[-0.175pt]{0.723pt}{0.350pt}}
\put(707,524){\rule[-0.175pt]{0.723pt}{0.350pt}}
\put(704,525){\rule[-0.175pt]{0.723pt}{0.350pt}}
\put(701,526){\rule[-0.175pt]{0.626pt}{0.350pt}}
\put(698,527){\rule[-0.175pt]{0.626pt}{0.350pt}}
\put(696,528){\rule[-0.175pt]{0.626pt}{0.350pt}}
\put(693,529){\rule[-0.175pt]{0.626pt}{0.350pt}}
\put(691,530){\rule[-0.175pt]{0.626pt}{0.350pt}}
\put(688,531){\rule[-0.175pt]{0.578pt}{0.350pt}}
\put(686,532){\rule[-0.175pt]{0.578pt}{0.350pt}}
\put(683,533){\rule[-0.175pt]{0.578pt}{0.350pt}}
\put(681,534){\rule[-0.175pt]{0.578pt}{0.350pt}}
\put(679,535){\rule[-0.175pt]{0.578pt}{0.350pt}}
\put(676,536){\rule[-0.175pt]{0.578pt}{0.350pt}}
\put(674,537){\rule[-0.175pt]{0.578pt}{0.350pt}}
\put(671,538){\rule[-0.175pt]{0.578pt}{0.350pt}}
\put(669,539){\rule[-0.175pt]{0.578pt}{0.350pt}}
\put(667,540){\rule[-0.175pt]{0.578pt}{0.350pt}}
\put(664,541){\rule[-0.175pt]{0.578pt}{0.350pt}}
\put(662,542){\rule[-0.175pt]{0.578pt}{0.350pt}}
\put(659,543){\rule[-0.175pt]{0.578pt}{0.350pt}}
\put(657,544){\rule[-0.175pt]{0.578pt}{0.350pt}}
\put(655,545){\rule[-0.175pt]{0.578pt}{0.350pt}}
\put(652,546){\rule[-0.175pt]{0.626pt}{0.350pt}}
\put(649,547){\rule[-0.175pt]{0.626pt}{0.350pt}}
\put(647,548){\rule[-0.175pt]{0.626pt}{0.350pt}}
\put(644,549){\rule[-0.175pt]{0.626pt}{0.350pt}}
\put(642,550){\rule[-0.175pt]{0.626pt}{0.350pt}}
\put(640,551){\rule[-0.175pt]{0.482pt}{0.350pt}}
\put(638,552){\rule[-0.175pt]{0.482pt}{0.350pt}}
\put(636,553){\rule[-0.175pt]{0.482pt}{0.350pt}}
\put(634,554){\rule[-0.175pt]{0.482pt}{0.350pt}}
\put(632,555){\rule[-0.175pt]{0.482pt}{0.350pt}}
\put(630,556){\rule[-0.175pt]{0.482pt}{0.350pt}}
\put(627,557){\rule[-0.175pt]{0.578pt}{0.350pt}}
\put(625,558){\rule[-0.175pt]{0.578pt}{0.350pt}}
\put(622,559){\rule[-0.175pt]{0.578pt}{0.350pt}}
\put(620,560){\rule[-0.175pt]{0.578pt}{0.350pt}}
\put(618,561){\rule[-0.175pt]{0.578pt}{0.350pt}}
\put(615,562){\rule[-0.175pt]{0.578pt}{0.350pt}}
\put(613,563){\rule[-0.175pt]{0.578pt}{0.350pt}}
\put(610,564){\rule[-0.175pt]{0.578pt}{0.350pt}}
\put(608,565){\rule[-0.175pt]{0.578pt}{0.350pt}}
\put(606,566){\rule[-0.175pt]{0.578pt}{0.350pt}}
\put(604,567){\rule[-0.175pt]{0.482pt}{0.350pt}}
\put(602,568){\rule[-0.175pt]{0.482pt}{0.350pt}}
\put(600,569){\rule[-0.175pt]{0.482pt}{0.350pt}}
\put(598,570){\rule[-0.175pt]{0.482pt}{0.350pt}}
\put(596,571){\rule[-0.175pt]{0.482pt}{0.350pt}}
\put(594,572){\rule[-0.175pt]{0.482pt}{0.350pt}}
\put(591,573){\rule[-0.175pt]{0.522pt}{0.350pt}}
\put(589,574){\rule[-0.175pt]{0.522pt}{0.350pt}}
\put(587,575){\rule[-0.175pt]{0.522pt}{0.350pt}}
\put(585,576){\rule[-0.175pt]{0.522pt}{0.350pt}}
\put(583,577){\rule[-0.175pt]{0.522pt}{0.350pt}}
\put(581,578){\rule[-0.175pt]{0.522pt}{0.350pt}}
\put(578,579){\rule[-0.175pt]{0.578pt}{0.350pt}}
\put(576,580){\rule[-0.175pt]{0.578pt}{0.350pt}}
\put(573,581){\rule[-0.175pt]{0.578pt}{0.350pt}}
\put(571,582){\rule[-0.175pt]{0.578pt}{0.350pt}}
\put(569,583){\rule[-0.175pt]{0.578pt}{0.350pt}}
\put(567,584){\rule[-0.175pt]{0.482pt}{0.350pt}}
\put(565,585){\rule[-0.175pt]{0.482pt}{0.350pt}}
\put(563,586){\rule[-0.175pt]{0.482pt}{0.350pt}}
\put(561,587){\rule[-0.175pt]{0.482pt}{0.350pt}}
\put(559,588){\rule[-0.175pt]{0.482pt}{0.350pt}}
\put(557,589){\rule[-0.175pt]{0.482pt}{0.350pt}}
\put(555,590){\rule[-0.175pt]{0.482pt}{0.350pt}}
\put(553,591){\rule[-0.175pt]{0.482pt}{0.350pt}}
\put(551,592){\rule[-0.175pt]{0.482pt}{0.350pt}}
\put(549,593){\rule[-0.175pt]{0.482pt}{0.350pt}}
\put(547,594){\rule[-0.175pt]{0.482pt}{0.350pt}}
\put(545,595){\rule[-0.175pt]{0.482pt}{0.350pt}}
\put(543,596){\rule[-0.175pt]{0.482pt}{0.350pt}}
\put(541,597){\rule[-0.175pt]{0.482pt}{0.350pt}}
\put(539,598){\rule[-0.175pt]{0.482pt}{0.350pt}}
\put(537,599){\rule[-0.175pt]{0.482pt}{0.350pt}}
\put(535,600){\rule[-0.175pt]{0.482pt}{0.350pt}}
\put(533,601){\rule[-0.175pt]{0.482pt}{0.350pt}}
\put(531,602){\rule[-0.175pt]{0.447pt}{0.350pt}}
\put(529,603){\rule[-0.175pt]{0.447pt}{0.350pt}}
\put(527,604){\rule[-0.175pt]{0.447pt}{0.350pt}}
\put(525,605){\rule[-0.175pt]{0.447pt}{0.350pt}}
\put(523,606){\rule[-0.175pt]{0.447pt}{0.350pt}}
\put(521,607){\rule[-0.175pt]{0.447pt}{0.350pt}}
\put(520,608){\rule[-0.175pt]{0.447pt}{0.350pt}}
\put(518,609){\rule[-0.175pt]{0.482pt}{0.350pt}}
\put(516,610){\rule[-0.175pt]{0.482pt}{0.350pt}}
\put(514,611){\rule[-0.175pt]{0.482pt}{0.350pt}}
\put(512,612){\rule[-0.175pt]{0.482pt}{0.350pt}}
\put(510,613){\rule[-0.175pt]{0.482pt}{0.350pt}}
\put(508,614){\rule[-0.175pt]{0.482pt}{0.350pt}}
\put(506,615){\rule[-0.175pt]{0.413pt}{0.350pt}}
\put(504,616){\rule[-0.175pt]{0.413pt}{0.350pt}}
\put(502,617){\rule[-0.175pt]{0.413pt}{0.350pt}}
\put(501,618){\rule[-0.175pt]{0.413pt}{0.350pt}}
\put(499,619){\rule[-0.175pt]{0.413pt}{0.350pt}}
\put(497,620){\rule[-0.175pt]{0.413pt}{0.350pt}}
\put(496,621){\rule[-0.175pt]{0.413pt}{0.350pt}}
\put(494,622){\rule[-0.175pt]{0.482pt}{0.350pt}}
\put(492,623){\rule[-0.175pt]{0.482pt}{0.350pt}}
\put(490,624){\rule[-0.175pt]{0.482pt}{0.350pt}}
\put(488,625){\rule[-0.175pt]{0.482pt}{0.350pt}}
\put(486,626){\rule[-0.175pt]{0.482pt}{0.350pt}}
\put(484,627){\rule[-0.175pt]{0.482pt}{0.350pt}}
\put(482,628){\rule[-0.175pt]{0.413pt}{0.350pt}}
\put(480,629){\rule[-0.175pt]{0.413pt}{0.350pt}}
\put(478,630){\rule[-0.175pt]{0.413pt}{0.350pt}}
\put(477,631){\rule[-0.175pt]{0.413pt}{0.350pt}}
\put(475,632){\rule[-0.175pt]{0.413pt}{0.350pt}}
\put(473,633){\rule[-0.175pt]{0.413pt}{0.350pt}}
\put(472,634){\rule[-0.175pt]{0.413pt}{0.350pt}}
\put(470,635){\rule[-0.175pt]{0.447pt}{0.350pt}}
\put(468,636){\rule[-0.175pt]{0.447pt}{0.350pt}}
\put(466,637){\rule[-0.175pt]{0.447pt}{0.350pt}}
\put(464,638){\rule[-0.175pt]{0.447pt}{0.350pt}}
\put(462,639){\rule[-0.175pt]{0.447pt}{0.350pt}}
\put(460,640){\rule[-0.175pt]{0.447pt}{0.350pt}}
\put(459,641){\rule[-0.175pt]{0.447pt}{0.350pt}}
\put(457,642){\rule[-0.175pt]{0.361pt}{0.350pt}}
\put(456,643){\rule[-0.175pt]{0.361pt}{0.350pt}}
\put(454,644){\rule[-0.175pt]{0.361pt}{0.350pt}}
\put(453,645){\rule[-0.175pt]{0.361pt}{0.350pt}}
\put(451,646){\rule[-0.175pt]{0.361pt}{0.350pt}}
\put(450,647){\rule[-0.175pt]{0.361pt}{0.350pt}}
\put(448,648){\rule[-0.175pt]{0.361pt}{0.350pt}}
\put(447,649){\rule[-0.175pt]{0.361pt}{0.350pt}}
\put(445,650){\rule[-0.175pt]{0.413pt}{0.350pt}}
\put(443,651){\rule[-0.175pt]{0.413pt}{0.350pt}}
\put(441,652){\rule[-0.175pt]{0.413pt}{0.350pt}}
\put(440,653){\rule[-0.175pt]{0.413pt}{0.350pt}}
\put(438,654){\rule[-0.175pt]{0.413pt}{0.350pt}}
\put(436,655){\rule[-0.175pt]{0.413pt}{0.350pt}}
\put(435,656){\rule[-0.175pt]{0.413pt}{0.350pt}}
\put(433,657){\rule[-0.175pt]{0.413pt}{0.350pt}}
\put(431,658){\rule[-0.175pt]{0.413pt}{0.350pt}}
\put(429,659){\rule[-0.175pt]{0.413pt}{0.350pt}}
\put(428,660){\rule[-0.175pt]{0.413pt}{0.350pt}}
\put(426,661){\rule[-0.175pt]{0.413pt}{0.350pt}}
\put(424,662){\rule[-0.175pt]{0.413pt}{0.350pt}}
\put(423,663){\rule[-0.175pt]{0.413pt}{0.350pt}}
\put(421,664){\rule[-0.175pt]{0.361pt}{0.350pt}}
\put(420,665){\rule[-0.175pt]{0.361pt}{0.350pt}}
\put(418,666){\rule[-0.175pt]{0.361pt}{0.350pt}}
\put(417,667){\rule[-0.175pt]{0.361pt}{0.350pt}}
\put(415,668){\rule[-0.175pt]{0.361pt}{0.350pt}}
\put(414,669){\rule[-0.175pt]{0.361pt}{0.350pt}}
\put(412,670){\rule[-0.175pt]{0.361pt}{0.350pt}}
\put(411,671){\rule[-0.175pt]{0.361pt}{0.350pt}}
\put(409,672){\rule[-0.175pt]{0.391pt}{0.350pt}}
\put(407,673){\rule[-0.175pt]{0.391pt}{0.350pt}}
\put(406,674){\rule[-0.175pt]{0.391pt}{0.350pt}}
\put(404,675){\rule[-0.175pt]{0.391pt}{0.350pt}}
\put(402,676){\rule[-0.175pt]{0.391pt}{0.350pt}}
\put(401,677){\rule[-0.175pt]{0.391pt}{0.350pt}}
\put(399,678){\rule[-0.175pt]{0.391pt}{0.350pt}}
\put(398,679){\rule[-0.175pt]{0.391pt}{0.350pt}}
\put(396,680){\rule[-0.175pt]{0.361pt}{0.350pt}}
\put(395,681){\rule[-0.175pt]{0.361pt}{0.350pt}}
\put(393,682){\rule[-0.175pt]{0.361pt}{0.350pt}}
\put(392,683){\rule[-0.175pt]{0.361pt}{0.350pt}}
\put(390,684){\rule[-0.175pt]{0.361pt}{0.350pt}}
\put(389,685){\rule[-0.175pt]{0.361pt}{0.350pt}}
\put(387,686){\rule[-0.175pt]{0.361pt}{0.350pt}}
\put(386,687){\rule[-0.175pt]{0.361pt}{0.350pt}}
\put(384,688){\usebox{\plotpoint}}
\put(383,689){\usebox{\plotpoint}}
\put(381,690){\usebox{\plotpoint}}
\put(380,691){\usebox{\plotpoint}}
\put(379,692){\usebox{\plotpoint}}
\put(377,693){\usebox{\plotpoint}}
\put(376,694){\usebox{\plotpoint}}
\put(375,695){\usebox{\plotpoint}}
\put(374,696){\usebox{\plotpoint}}
\put(372,697){\usebox{\plotpoint}}
\put(371,698){\usebox{\plotpoint}}
\put(369,699){\usebox{\plotpoint}}
\put(368,700){\usebox{\plotpoint}}
\put(367,701){\usebox{\plotpoint}}
\put(365,702){\usebox{\plotpoint}}
\put(364,703){\usebox{\plotpoint}}
\put(363,704){\usebox{\plotpoint}}
\put(362,705){\usebox{\plotpoint}}
\put(360,706){\usebox{\plotpoint}}
\put(359,707){\usebox{\plotpoint}}
\put(357,708){\usebox{\plotpoint}}
\put(356,709){\usebox{\plotpoint}}
\put(354,710){\usebox{\plotpoint}}
\put(353,711){\usebox{\plotpoint}}
\put(351,712){\usebox{\plotpoint}}
\put(350,713){\usebox{\plotpoint}}
\put(349,714){\usebox{\plotpoint}}
\put(347,715){\usebox{\plotpoint}}
\put(346,716){\usebox{\plotpoint}}
\put(344,717){\usebox{\plotpoint}}
\put(343,718){\usebox{\plotpoint}}
\put(342,719){\usebox{\plotpoint}}
\put(340,720){\usebox{\plotpoint}}
\put(339,721){\usebox{\plotpoint}}
\put(338,722){\usebox{\plotpoint}}
\put(337,723){\usebox{\plotpoint}}
\put(335,724){\usebox{\plotpoint}}
\put(334,725){\usebox{\plotpoint}}
\put(333,726){\usebox{\plotpoint}}
\put(332,727){\usebox{\plotpoint}}
\put(330,728){\usebox{\plotpoint}}
\put(329,729){\usebox{\plotpoint}}
\put(328,730){\usebox{\plotpoint}}
\put(327,731){\usebox{\plotpoint}}
\put(326,732){\usebox{\plotpoint}}
\put(325,733){\usebox{\plotpoint}}
\put(323,734){\usebox{\plotpoint}}
\put(322,735){\usebox{\plotpoint}}
\put(320,736){\usebox{\plotpoint}}
\put(319,737){\usebox{\plotpoint}}
\put(318,738){\usebox{\plotpoint}}
\put(316,739){\usebox{\plotpoint}}
\put(315,740){\usebox{\plotpoint}}
\put(314,741){\usebox{\plotpoint}}
\put(313,742){\usebox{\plotpoint}}
\put(311,743){\usebox{\plotpoint}}
\put(310,744){\usebox{\plotpoint}}
\put(309,745){\usebox{\plotpoint}}
\put(308,746){\usebox{\plotpoint}}
\put(307,747){\usebox{\plotpoint}}
\put(306,748){\usebox{\plotpoint}}
\put(305,749){\usebox{\plotpoint}}
\put(304,750){\usebox{\plotpoint}}
\put(303,751){\usebox{\plotpoint}}
\put(302,752){\usebox{\plotpoint}}
\put(301,753){\usebox{\plotpoint}}
\put(299,754){\usebox{\plotpoint}}
\put(298,755){\usebox{\plotpoint}}
\put(297,756){\usebox{\plotpoint}}
\put(295,757){\usebox{\plotpoint}}
\put(294,758){\usebox{\plotpoint}}
\put(293,759){\usebox{\plotpoint}}
\put(291,760){\usebox{\plotpoint}}
\put(290,761){\usebox{\plotpoint}}
\put(289,762){\usebox{\plotpoint}}
\put(288,763){\usebox{\plotpoint}}
\put(286,764){\usebox{\plotpoint}}
\put(285,765){\usebox{\plotpoint}}
\put(284,766){\usebox{\plotpoint}}
\put(283,767){\usebox{\plotpoint}}
\put(282,768){\usebox{\plotpoint}}
\put(281,769){\usebox{\plotpoint}}
\put(280,770){\usebox{\plotpoint}}
\put(279,771){\usebox{\plotpoint}}
\put(278,772){\usebox{\plotpoint}}
\put(277,773){\usebox{\plotpoint}}
\put(276,774){\usebox{\plotpoint}}
\put(275,775){\usebox{\plotpoint}}
\put(274,776){\usebox{\plotpoint}}
\put(273,777){\usebox{\plotpoint}}
\put(272,778){\usebox{\plotpoint}}
\put(271,779){\usebox{\plotpoint}}
\put(270,780){\usebox{\plotpoint}}
\put(269,781){\usebox{\plotpoint}}
\put(268,782){\usebox{\plotpoint}}
\put(267,783){\usebox{\plotpoint}}
\put(266,784){\usebox{\plotpoint}}
\put(265,785){\usebox{\plotpoint}}
\put(264,786){\usebox{\plotpoint}}
\put(569,581){\circle*{18}}
\put(733,525){\circle*{18}}
\put(893,441){\circle*{18}}
\put(1123,420){\circle*{18}}
\put(1368,432){\circle*{18}}
\put(569,558){\rule[-0.175pt]{0.350pt}{10.118pt}}
\put(559,558){\rule[-0.175pt]{4.818pt}{0.350pt}}
\put(559,600){\rule[-0.175pt]{4.818pt}{0.350pt}}
\put(733,500){\rule[-0.175pt]{0.350pt}{11.081pt}}
\put(723,500){\rule[-0.175pt]{4.818pt}{0.350pt}}
\put(723,546){\rule[-0.175pt]{4.818pt}{0.350pt}}
\put(893,407){\rule[-0.175pt]{0.350pt}{15.177pt}}
\put(883,407){\rule[-0.175pt]{4.818pt}{0.350pt}}
\put(883,470){\rule[-0.175pt]{4.818pt}{0.350pt}}
\put(1123,381){\rule[-0.175pt]{0.350pt}{16.863pt}}
\put(1113,381){\rule[-0.175pt]{4.818pt}{0.350pt}}
\put(1113,451){\rule[-0.175pt]{4.818pt}{0.350pt}}
\put(1368,362){\rule[-0.175pt]{0.350pt}{27.703pt}}
\put(1358,362){\rule[-0.175pt]{4.818pt}{0.350pt}}
\put(1358,477){\rule[-0.175pt]{4.818pt}{0.350pt}}
\end{picture}

%% file: wpgformj.tex
\setlength{\unitlength}{0.240900pt}
\ifx\plotpoint\undefined\newsavebox{\plotpoint}\fi
\begin{picture}(900,1259)(0,0)
\tenrm
\sbox{\plotpoint}{\rule[-0.175pt]{0.350pt}{0.350pt}}%
\put(264,158){\rule[-0.175pt]{0.350pt}{238.009pt}}
\put(264,184){\rule[-0.175pt]{4.818pt}{0.350pt}}
\put(816,184){\rule[-0.175pt]{4.818pt}{0.350pt}}
\put(264,206){\rule[-0.175pt]{4.818pt}{0.350pt}}
\put(816,206){\rule[-0.175pt]{4.818pt}{0.350pt}}
\put(264,227){\rule[-0.175pt]{4.818pt}{0.350pt}}
\put(242,227){\makebox(0,0)[r]{\fsize1}}
\put(816,227){\rule[-0.175pt]{4.818pt}{0.350pt}}
\put(264,360){\rule[-0.175pt]{4.818pt}{0.350pt}}
\put(816,360){\rule[-0.175pt]{4.818pt}{0.350pt}}
\put(264,438){\rule[-0.175pt]{4.818pt}{0.350pt}}
\put(816,438){\rule[-0.175pt]{4.818pt}{0.350pt}}
\put(264,493){\rule[-0.175pt]{4.818pt}{0.350pt}}
\put(816,493){\rule[-0.175pt]{4.818pt}{0.350pt}}
\put(264,536){\rule[-0.175pt]{4.818pt}{0.350pt}}
\put(816,536){\rule[-0.175pt]{4.818pt}{0.350pt}}
\put(264,571){\rule[-0.175pt]{4.818pt}{0.350pt}}
\put(816,571){\rule[-0.175pt]{4.818pt}{0.350pt}}
\put(264,600){\rule[-0.175pt]{4.818pt}{0.350pt}}
\put(816,600){\rule[-0.175pt]{4.818pt}{0.350pt}}
\put(264,626){\rule[-0.175pt]{4.818pt}{0.350pt}}
\put(816,626){\rule[-0.175pt]{4.818pt}{0.350pt}}
\put(264,649){\rule[-0.175pt]{4.818pt}{0.350pt}}
\put(816,649){\rule[-0.175pt]{4.818pt}{0.350pt}}
\put(264,669){\rule[-0.175pt]{4.818pt}{0.350pt}}
\put(242,669){\makebox(0,0)[r]{\fsize10}}
\put(816,669){\rule[-0.175pt]{4.818pt}{0.350pt}}
\put(264,802){\rule[-0.175pt]{4.818pt}{0.350pt}}
\put(816,802){\rule[-0.175pt]{4.818pt}{0.350pt}}
\put(264,880){\rule[-0.175pt]{4.818pt}{0.350pt}}
\put(816,880){\rule[-0.175pt]{4.818pt}{0.350pt}}
\put(264,935){\rule[-0.175pt]{4.818pt}{0.350pt}}
\put(816,935){\rule[-0.175pt]{4.818pt}{0.350pt}}
\put(264,978){\rule[-0.175pt]{4.818pt}{0.350pt}}
\put(816,978){\rule[-0.175pt]{4.818pt}{0.350pt}}
\put(264,1013){\rule[-0.175pt]{4.818pt}{0.350pt}}
\put(816,1013){\rule[-0.175pt]{4.818pt}{0.350pt}}
\put(264,1042){\rule[-0.175pt]{4.818pt}{0.350pt}}
\put(816,1042){\rule[-0.175pt]{4.818pt}{0.350pt}}
\put(264,1068){\rule[-0.175pt]{4.818pt}{0.350pt}}
\put(816,1068){\rule[-0.175pt]{4.818pt}{0.350pt}}
\put(264,1091){\rule[-0.175pt]{4.818pt}{0.350pt}}
\put(816,1091){\rule[-0.175pt]{4.818pt}{0.350pt}}
\put(264,1111){\rule[-0.175pt]{4.818pt}{0.350pt}}
\put(242,1111){\makebox(0,0)[r]{\fsize100}}
\put(816,1111){\rule[-0.175pt]{4.818pt}{0.350pt}}
\put(264,158){\rule[-0.175pt]{0.350pt}{4.818pt}}
\put(264,113){\makebox(0,0){\fsize0}}
\put(264,1126){\rule[-0.175pt]{0.350pt}{4.818pt}}
\put(391,158){\rule[-0.175pt]{0.350pt}{4.818pt}}
\put(391,113){\makebox(0,0){\fsize0.1}}
\put(391,1126){\rule[-0.175pt]{0.350pt}{4.818pt}}
\put(518,158){\rule[-0.175pt]{0.350pt}{4.818pt}}
\put(518,113){\makebox(0,0){\fsize0.2}}
\put(518,1126){\rule[-0.175pt]{0.350pt}{4.818pt}}
\put(645,158){\rule[-0.175pt]{0.350pt}{4.818pt}}
\put(645,113){\makebox(0,0){\fsize0.3}}
\put(645,1126){\rule[-0.175pt]{0.350pt}{4.818pt}}
\put(772,158){\rule[-0.175pt]{0.350pt}{4.818pt}}
\put(772,113){\makebox(0,0){\fsize0.4}}
\put(772,1126){\rule[-0.175pt]{0.350pt}{4.818pt}}
\put(264,158){\rule[-0.175pt]{137.795pt}{0.350pt}}
\put(836,158){\rule[-0.175pt]{0.350pt}{238.009pt}}
\put(264,1146){\rule[-0.175pt]{137.795pt}{0.350pt}}
\put(199,1192){\makebox(0,0)[l]{\shortstack{\small$\abs{F_\omega(q^2)}^2$}}}
\put(550,23){\makebox(0,0){\small$(a)\mq q^2$ [GeV${}^2$]}}
\put(264,158){\rule[-0.175pt]{0.350pt}{238.009pt}}
\sbox{\plotpoint}{\rule[-0.500pt]{1.000pt}{1.000pt}}%
\put(455,1091){\makebox(0,0)[r]{$\tilde{c}=1.0$}}
\put(477,1091){\rule[-0.500pt]{15.899pt}{1.000pt}}
\put(270,233){\usebox{\plotpoint}}
\put(270,233){\usebox{\plotpoint}}
\put(271,234){\usebox{\plotpoint}}
\put(272,235){\usebox{\plotpoint}}
\put(273,236){\usebox{\plotpoint}}
\put(274,237){\usebox{\plotpoint}}
\put(275,238){\usebox{\plotpoint}}
\put(276,240){\usebox{\plotpoint}}
\put(277,241){\usebox{\plotpoint}}
\put(278,242){\usebox{\plotpoint}}
\put(279,243){\usebox{\plotpoint}}
\put(280,244){\usebox{\plotpoint}}
\put(281,245){\usebox{\plotpoint}}
\put(282,246){\usebox{\plotpoint}}
\put(283,247){\usebox{\plotpoint}}
\put(284,248){\usebox{\plotpoint}}
\put(285,249){\usebox{\plotpoint}}
\put(286,250){\usebox{\plotpoint}}
\put(287,251){\usebox{\plotpoint}}
\put(288,252){\usebox{\plotpoint}}
\put(289,253){\usebox{\plotpoint}}
\put(290,254){\usebox{\plotpoint}}
\put(291,255){\usebox{\plotpoint}}
\put(293,256){\usebox{\plotpoint}}
\put(294,257){\usebox{\plotpoint}}
\put(296,258){\usebox{\plotpoint}}
\put(297,259){\usebox{\plotpoint}}
\put(299,260){\usebox{\plotpoint}}
\put(300,261){\usebox{\plotpoint}}
\put(301,262){\usebox{\plotpoint}}
\put(302,263){\usebox{\plotpoint}}
\put(303,264){\usebox{\plotpoint}}
\put(304,265){\usebox{\plotpoint}}
\put(305,266){\usebox{\plotpoint}}
\put(307,267){\usebox{\plotpoint}}
\put(308,268){\usebox{\plotpoint}}
\put(310,269){\usebox{\plotpoint}}
\put(311,270){\usebox{\plotpoint}}
\put(313,271){\usebox{\plotpoint}}
\put(314,272){\usebox{\plotpoint}}
\put(316,273){\usebox{\plotpoint}}
\put(317,274){\usebox{\plotpoint}}
\put(319,275){\usebox{\plotpoint}}
\put(320,276){\usebox{\plotpoint}}
\put(322,277){\usebox{\plotpoint}}
\put(323,278){\usebox{\plotpoint}}
\put(325,279){\usebox{\plotpoint}}
\put(326,280){\usebox{\plotpoint}}
\put(328,281){\usebox{\plotpoint}}
\put(329,282){\usebox{\plotpoint}}
\put(330,283){\usebox{\plotpoint}}
\put(331,284){\usebox{\plotpoint}}
\put(333,285){\usebox{\plotpoint}}
\put(334,286){\usebox{\plotpoint}}
\put(336,287){\usebox{\plotpoint}}
\put(337,288){\usebox{\plotpoint}}
\put(339,289){\usebox{\plotpoint}}
\put(341,290){\usebox{\plotpoint}}
\put(343,291){\usebox{\plotpoint}}
\put(345,292){\usebox{\plotpoint}}
\put(346,293){\usebox{\plotpoint}}
\put(348,294){\usebox{\plotpoint}}
\put(349,295){\usebox{\plotpoint}}
\put(351,296){\usebox{\plotpoint}}
\put(352,297){\usebox{\plotpoint}}
\put(354,298){\usebox{\plotpoint}}
\put(355,299){\usebox{\plotpoint}}
\put(358,300){\usebox{\plotpoint}}
\put(360,301){\usebox{\plotpoint}}
\put(362,302){\usebox{\plotpoint}}
\put(363,303){\usebox{\plotpoint}}
\put(365,304){\usebox{\plotpoint}}
\put(366,305){\usebox{\plotpoint}}
\put(368,306){\usebox{\plotpoint}}
\put(370,307){\usebox{\plotpoint}}
\put(372,308){\usebox{\plotpoint}}
\put(374,309){\usebox{\plotpoint}}
\put(376,310){\usebox{\plotpoint}}
\put(378,311){\usebox{\plotpoint}}
\put(380,312){\usebox{\plotpoint}}
\put(381,313){\usebox{\plotpoint}}
\put(382,314){\usebox{\plotpoint}}
\put(383,315){\usebox{\plotpoint}}
\put(385,316){\usebox{\plotpoint}}
\put(387,317){\usebox{\plotpoint}}
\put(389,318){\usebox{\plotpoint}}
\put(391,319){\usebox{\plotpoint}}
\put(392,320){\usebox{\plotpoint}}
\put(394,321){\usebox{\plotpoint}}
\put(395,322){\usebox{\plotpoint}}
\put(397,323){\usebox{\plotpoint}}
\put(399,324){\usebox{\plotpoint}}
\put(401,325){\usebox{\plotpoint}}
\put(403,326){\usebox{\plotpoint}}
\put(404,327){\usebox{\plotpoint}}
\put(405,328){\usebox{\plotpoint}}
\put(406,329){\usebox{\plotpoint}}
\put(408,330){\usebox{\plotpoint}}
\put(409,331){\usebox{\plotpoint}}
\put(411,332){\usebox{\plotpoint}}
\put(412,333){\usebox{\plotpoint}}
\put(414,334){\usebox{\plotpoint}}
\put(415,335){\usebox{\plotpoint}}
\put(417,336){\usebox{\plotpoint}}
\put(418,337){\usebox{\plotpoint}}
\put(420,338){\usebox{\plotpoint}}
\put(421,339){\usebox{\plotpoint}}
\put(423,340){\usebox{\plotpoint}}
\put(424,341){\usebox{\plotpoint}}
\put(426,342){\usebox{\plotpoint}}
\put(427,343){\usebox{\plotpoint}}
\put(429,344){\usebox{\plotpoint}}
\put(430,345){\usebox{\plotpoint}}
\put(432,346){\usebox{\plotpoint}}
\put(433,347){\usebox{\plotpoint}}
\put(434,348){\usebox{\plotpoint}}
\put(435,349){\usebox{\plotpoint}}
\put(436,350){\usebox{\plotpoint}}
\put(437,351){\usebox{\plotpoint}}
\put(438,352){\usebox{\plotpoint}}
\put(440,353){\usebox{\plotpoint}}
\put(441,354){\usebox{\plotpoint}}
\put(443,355){\usebox{\plotpoint}}
\put(444,356){\usebox{\plotpoint}}
\put(445,357){\usebox{\plotpoint}}
\put(446,358){\usebox{\plotpoint}}
\put(447,359){\usebox{\plotpoint}}
\put(449,360){\usebox{\plotpoint}}
\put(450,361){\usebox{\plotpoint}}
\put(452,362){\usebox{\plotpoint}}
\put(453,363){\usebox{\plotpoint}}
\put(455,364){\usebox{\plotpoint}}
\put(456,365){\usebox{\plotpoint}}
\put(457,366){\usebox{\plotpoint}}
\put(458,367){\usebox{\plotpoint}}
\put(459,368){\usebox{\plotpoint}}
\put(460,369){\usebox{\plotpoint}}
\put(461,370){\usebox{\plotpoint}}
\put(462,371){\usebox{\plotpoint}}
\put(463,372){\usebox{\plotpoint}}
\put(464,373){\usebox{\plotpoint}}
\put(466,374){\usebox{\plotpoint}}
\put(467,375){\usebox{\plotpoint}}
\put(468,376){\usebox{\plotpoint}}
\put(469,377){\usebox{\plotpoint}}
\put(470,378){\usebox{\plotpoint}}
\put(472,379){\usebox{\plotpoint}}
\put(473,380){\usebox{\plotpoint}}
\put(474,381){\usebox{\plotpoint}}
\put(475,382){\usebox{\plotpoint}}
\put(476,383){\usebox{\plotpoint}}
\put(477,384){\usebox{\plotpoint}}
\put(478,385){\usebox{\plotpoint}}
\put(479,386){\usebox{\plotpoint}}
\put(480,387){\usebox{\plotpoint}}
\put(481,388){\usebox{\plotpoint}}
\put(482,389){\usebox{\plotpoint}}
\put(484,390){\usebox{\plotpoint}}
\put(485,391){\usebox{\plotpoint}}
\put(486,392){\usebox{\plotpoint}}
\put(487,393){\usebox{\plotpoint}}
\put(488,394){\usebox{\plotpoint}}
\put(489,395){\usebox{\plotpoint}}
\put(490,396){\usebox{\plotpoint}}
\put(491,397){\usebox{\plotpoint}}
\put(492,398){\usebox{\plotpoint}}
\put(493,399){\usebox{\plotpoint}}
\put(494,400){\usebox{\plotpoint}}
\put(495,401){\usebox{\plotpoint}}
\put(496,402){\usebox{\plotpoint}}
\put(497,403){\usebox{\plotpoint}}
\put(498,404){\usebox{\plotpoint}}
\put(499,405){\usebox{\plotpoint}}
\put(501,406){\usebox{\plotpoint}}
\put(502,407){\usebox{\plotpoint}}
\put(503,408){\usebox{\plotpoint}}
\put(504,409){\usebox{\plotpoint}}
\put(505,410){\usebox{\plotpoint}}
\put(506,411){\usebox{\plotpoint}}
\put(507,412){\usebox{\plotpoint}}
\put(508,413){\usebox{\plotpoint}}
\put(509,414){\usebox{\plotpoint}}
\put(510,415){\usebox{\plotpoint}}
\put(511,416){\usebox{\plotpoint}}
\put(512,418){\usebox{\plotpoint}}
\put(513,419){\usebox{\plotpoint}}
\put(514,420){\usebox{\plotpoint}}
\put(515,421){\usebox{\plotpoint}}
\put(516,422){\usebox{\plotpoint}}
\put(518,423){\usebox{\plotpoint}}
\put(519,424){\usebox{\plotpoint}}
\put(520,425){\usebox{\plotpoint}}
\put(521,426){\usebox{\plotpoint}}
\put(522,427){\usebox{\plotpoint}}
\put(523,428){\usebox{\plotpoint}}
\put(524,429){\usebox{\plotpoint}}
\put(525,430){\usebox{\plotpoint}}
\put(526,431){\usebox{\plotpoint}}
\put(527,432){\usebox{\plotpoint}}
\put(528,433){\usebox{\plotpoint}}
\put(529,434){\usebox{\plotpoint}}
\put(530,435){\usebox{\plotpoint}}
\put(531,436){\usebox{\plotpoint}}
\put(532,437){\usebox{\plotpoint}}
\put(533,438){\usebox{\plotpoint}}
\put(534,439){\usebox{\plotpoint}}
\put(535,440){\usebox{\plotpoint}}
\put(536,441){\usebox{\plotpoint}}
\put(537,442){\usebox{\plotpoint}}
\put(538,443){\usebox{\plotpoint}}
\put(539,444){\usebox{\plotpoint}}
\put(540,445){\usebox{\plotpoint}}
\put(541,447){\usebox{\plotpoint}}
\put(542,448){\usebox{\plotpoint}}
\put(543,449){\usebox{\plotpoint}}
\put(544,450){\usebox{\plotpoint}}
\put(545,451){\usebox{\plotpoint}}
\put(546,452){\usebox{\plotpoint}}
\put(547,453){\usebox{\plotpoint}}
\put(548,454){\usebox{\plotpoint}}
\put(549,455){\usebox{\plotpoint}}
\put(550,456){\usebox{\plotpoint}}
\put(551,457){\usebox{\plotpoint}}
\put(552,458){\usebox{\plotpoint}}
\put(553,459){\usebox{\plotpoint}}
\put(553,460){\usebox{\plotpoint}}
\put(554,461){\usebox{\plotpoint}}
\put(555,462){\usebox{\plotpoint}}
\put(556,463){\usebox{\plotpoint}}
\put(557,464){\usebox{\plotpoint}}
\put(558,465){\usebox{\plotpoint}}
\put(559,466){\usebox{\plotpoint}}
\put(560,467){\usebox{\plotpoint}}
\put(561,468){\usebox{\plotpoint}}
\put(562,469){\usebox{\plotpoint}}
\put(563,470){\usebox{\plotpoint}}
\put(564,472){\usebox{\plotpoint}}
\put(565,473){\usebox{\plotpoint}}
\put(566,474){\usebox{\plotpoint}}
\put(567,475){\usebox{\plotpoint}}
\put(568,476){\usebox{\plotpoint}}
\put(569,477){\usebox{\plotpoint}}
\put(570,478){\usebox{\plotpoint}}
\put(570,479){\usebox{\plotpoint}}
\put(571,480){\usebox{\plotpoint}}
\put(572,481){\usebox{\plotpoint}}
\put(573,482){\usebox{\plotpoint}}
\put(574,483){\usebox{\plotpoint}}
\put(575,484){\usebox{\plotpoint}}
\put(576,485){\usebox{\plotpoint}}
\put(577,486){\usebox{\plotpoint}}
\put(578,487){\usebox{\plotpoint}}
\put(579,488){\usebox{\plotpoint}}
\put(580,489){\usebox{\plotpoint}}
\put(581,490){\usebox{\plotpoint}}
\put(582,491){\usebox{\plotpoint}}
\put(583,492){\usebox{\plotpoint}}
\put(584,493){\usebox{\plotpoint}}
\put(585,494){\usebox{\plotpoint}}
\put(586,495){\usebox{\plotpoint}}
\put(587,496){\usebox{\plotpoint}}
\put(588,497){\usebox{\plotpoint}}
\put(589,499){\usebox{\plotpoint}}
\put(590,500){\usebox{\plotpoint}}
\put(591,502){\usebox{\plotpoint}}
\put(592,503){\usebox{\plotpoint}}
\put(593,504){\usebox{\plotpoint}}
\put(593,505){\usebox{\plotpoint}}
\put(594,506){\usebox{\plotpoint}}
\put(595,507){\usebox{\plotpoint}}
\put(596,508){\usebox{\plotpoint}}
\put(597,509){\usebox{\plotpoint}}
\put(598,510){\usebox{\plotpoint}}
\put(599,511){\usebox{\plotpoint}}
\put(600,512){\usebox{\plotpoint}}
\put(601,513){\usebox{\plotpoint}}
\put(602,514){\usebox{\plotpoint}}
\put(603,515){\usebox{\plotpoint}}
\put(604,516){\usebox{\plotpoint}}
\put(605,518){\usebox{\plotpoint}}
\put(606,519){\usebox{\plotpoint}}
\put(607,520){\usebox{\plotpoint}}
\put(608,521){\usebox{\plotpoint}}
\put(609,522){\usebox{\plotpoint}}
\put(610,523){\usebox{\plotpoint}}
\put(611,525){\usebox{\plotpoint}}
\put(612,526){\usebox{\plotpoint}}
\put(613,527){\usebox{\plotpoint}}
\put(614,529){\usebox{\plotpoint}}
\put(615,530){\usebox{\plotpoint}}
\put(616,532){\usebox{\plotpoint}}
\put(617,533){\usebox{\plotpoint}}
\put(618,534){\usebox{\plotpoint}}
\put(619,535){\usebox{\plotpoint}}
\put(620,536){\usebox{\plotpoint}}
\put(621,537){\usebox{\plotpoint}}
\put(622,538){\usebox{\plotpoint}}
\put(623,539){\usebox{\plotpoint}}
\put(624,540){\usebox{\plotpoint}}
\put(625,541){\usebox{\plotpoint}}
\put(626,542){\usebox{\plotpoint}}
\put(627,543){\usebox{\plotpoint}}
\put(628,545){\usebox{\plotpoint}}
\put(629,546){\usebox{\plotpoint}}
\put(630,547){\usebox{\plotpoint}}
\put(631,548){\usebox{\plotpoint}}
\put(632,549){\usebox{\plotpoint}}
\put(633,550){\usebox{\plotpoint}}
\put(634,552){\usebox{\plotpoint}}
\put(635,553){\usebox{\plotpoint}}
\put(636,554){\usebox{\plotpoint}}
\put(637,555){\usebox{\plotpoint}}
\put(638,557){\usebox{\plotpoint}}
\put(639,558){\usebox{\plotpoint}}
\put(640,559){\usebox{\plotpoint}}
\put(641,561){\usebox{\plotpoint}}
\put(642,562){\usebox{\plotpoint}}
\put(643,564){\usebox{\plotpoint}}
\put(644,565){\usebox{\plotpoint}}
\put(645,567){\usebox{\plotpoint}}
\put(646,568){\usebox{\plotpoint}}
\put(647,569){\usebox{\plotpoint}}
\put(648,570){\usebox{\plotpoint}}
\put(649,571){\usebox{\plotpoint}}
\put(650,572){\usebox{\plotpoint}}
\put(651,574){\usebox{\plotpoint}}
\put(652,575){\usebox{\plotpoint}}
\put(653,576){\usebox{\plotpoint}}
\put(654,577){\usebox{\plotpoint}}
\put(655,578){\usebox{\plotpoint}}
\put(656,579){\usebox{\plotpoint}}
\put(657,581){\usebox{\plotpoint}}
\put(658,582){\usebox{\plotpoint}}
\put(659,583){\usebox{\plotpoint}}
\put(660,584){\usebox{\plotpoint}}
\put(661,586){\usebox{\plotpoint}}
\put(662,587){\usebox{\plotpoint}}
\put(663,588){\usebox{\plotpoint}}
\put(664,590){\usebox{\plotpoint}}
\put(665,591){\usebox{\plotpoint}}
\put(666,593){\usebox{\plotpoint}}
\put(667,594){\usebox{\plotpoint}}
\put(668,596){\usebox{\plotpoint}}
\put(669,597){\usebox{\plotpoint}}
\put(670,598){\usebox{\plotpoint}}
\put(671,599){\usebox{\plotpoint}}
\put(672,601){\usebox{\plotpoint}}
\put(673,602){\usebox{\plotpoint}}
\put(674,603){\usebox{\plotpoint}}
\put(675,605){\usebox{\plotpoint}}
\put(676,606){\usebox{\plotpoint}}
\put(677,607){\usebox{\plotpoint}}
\put(678,608){\usebox{\plotpoint}}
\put(679,609){\usebox{\plotpoint}}
\put(680,611){\usebox{\plotpoint}}
\put(681,612){\usebox{\plotpoint}}
\put(682,613){\usebox{\plotpoint}}
\put(683,614){\usebox{\plotpoint}}
\put(684,616){\usebox{\plotpoint}}
\put(685,617){\usebox{\plotpoint}}
\put(686,618){\usebox{\plotpoint}}
\put(687,620){\usebox{\plotpoint}}
\put(688,621){\usebox{\plotpoint}}
\put(689,622){\usebox{\plotpoint}}
\put(690,624){\usebox{\plotpoint}}
\put(691,625){\usebox{\plotpoint}}
\put(692,626){\usebox{\plotpoint}}
\put(693,628){\usebox{\plotpoint}}
\put(694,630){\usebox{\plotpoint}}
\put(695,631){\usebox{\plotpoint}}
\put(696,633){\usebox{\plotpoint}}
\put(697,634){\usebox{\plotpoint}}
\put(698,636){\usebox{\plotpoint}}
\put(699,637){\usebox{\plotpoint}}
\put(700,638){\usebox{\plotpoint}}
\put(701,640){\usebox{\plotpoint}}
\put(702,641){\usebox{\plotpoint}}
\put(703,642){\usebox{\plotpoint}}
\put(704,644){\usebox{\plotpoint}}
\put(705,645){\usebox{\plotpoint}}
\put(706,646){\usebox{\plotpoint}}
\put(707,648){\usebox{\plotpoint}}
\put(708,649){\usebox{\plotpoint}}
\put(709,650){\usebox{\plotpoint}}
\put(710,652){\usebox{\plotpoint}}
\put(711,653){\usebox{\plotpoint}}
\put(712,654){\usebox{\plotpoint}}
\put(713,656){\usebox{\plotpoint}}
\put(714,657){\usebox{\plotpoint}}
\put(715,658){\usebox{\plotpoint}}
\put(716,660){\usebox{\plotpoint}}
\put(717,662){\usebox{\plotpoint}}
\put(718,663){\usebox{\plotpoint}}
\put(719,665){\usebox{\plotpoint}}
\put(720,666){\usebox{\plotpoint}}
\put(721,668){\usebox{\plotpoint}}
\put(722,669){\usebox{\plotpoint}}
\put(723,670){\usebox{\plotpoint}}
\put(724,672){\usebox{\plotpoint}}
\put(725,673){\usebox{\plotpoint}}
\put(726,674){\usebox{\plotpoint}}
\put(727,676){\usebox{\plotpoint}}
\put(728,678){\usebox{\plotpoint}}
\put(729,679){\usebox{\plotpoint}}
\put(730,681){\usebox{\plotpoint}}
\put(731,682){\usebox{\plotpoint}}
\put(732,684){\usebox{\plotpoint}}
\put(733,685){\usebox{\plotpoint}}
\put(734,686){\usebox{\plotpoint}}
\put(735,687){\usebox{\plotpoint}}
\put(736,689){\usebox{\plotpoint}}
\put(737,690){\usebox{\plotpoint}}
\put(738,691){\usebox{\plotpoint}}
\put(739,693){\usebox{\plotpoint}}
\put(740,695){\usebox{\plotpoint}}
\put(741,696){\usebox{\plotpoint}}
\put(742,698){\usebox{\plotpoint}}
\put(743,699){\usebox{\plotpoint}}
\put(744,701){\usebox{\plotpoint}}
\put(745,702){\usebox{\plotpoint}}
\put(746,704){\usebox{\plotpoint}}
\put(747,705){\usebox{\plotpoint}}
\put(748,707){\usebox{\plotpoint}}
\put(749,708){\usebox{\plotpoint}}
\put(750,710){\usebox{\plotpoint}}
\put(751,712){\usebox{\plotpoint}}
\put(752,713){\usebox{\plotpoint}}
\put(753,715){\usebox{\plotpoint}}
\put(754,716){\usebox{\plotpoint}}
\put(755,718){\usebox{\plotpoint}}
\put(756,719){\usebox{\plotpoint}}
\put(757,721){\usebox{\plotpoint}}
\put(758,722){\usebox{\plotpoint}}
\put(759,724){\usebox{\plotpoint}}
\put(760,725){\usebox{\plotpoint}}
\put(761,727){\usebox{\plotpoint}}
\put(762,728){\usebox{\plotpoint}}
\put(763,730){\usebox{\plotpoint}}
\put(764,731){\usebox{\plotpoint}}
\put(765,733){\usebox{\plotpoint}}
\put(766,734){\usebox{\plotpoint}}
\put(767,736){\usebox{\plotpoint}}
\put(768,737){\usebox{\plotpoint}}
\put(769,739){\usebox{\plotpoint}}
\put(770,741){\usebox{\plotpoint}}
\put(771,743){\usebox{\plotpoint}}
\put(772,744){\usebox{\plotpoint}}
\put(773,746){\usebox{\plotpoint}}
\put(774,748){\usebox{\plotpoint}}
\put(775,749){\usebox{\plotpoint}}
\put(776,751){\usebox{\plotpoint}}
\put(777,752){\usebox{\plotpoint}}
\put(778,754){\usebox{\plotpoint}}
\put(779,755){\usebox{\plotpoint}}
\put(780,757){\usebox{\plotpoint}}
\put(781,758){\usebox{\plotpoint}}
\put(782,760){\usebox{\plotpoint}}
\put(783,761){\usebox{\plotpoint}}
\put(784,763){\usebox{\plotpoint}}
\put(785,764){\usebox{\plotpoint}}
\put(786,766){\usebox{\plotpoint}}
\put(787,767){\usebox{\plotpoint}}
\put(788,769){\usebox{\plotpoint}}
\put(789,770){\usebox{\plotpoint}}
\put(790,772){\usebox{\plotpoint}}
\put(791,773){\usebox{\plotpoint}}
\put(792,775){\usebox{\plotpoint}}
\put(793,776){\usebox{\plotpoint}}
\put(794,778){\usebox{\plotpoint}}
\put(795,779){\usebox{\plotpoint}}
\put(796,781){\usebox{\plotpoint}}
\put(797,783){\usebox{\plotpoint}}
\put(798,785){\usebox{\plotpoint}}
\put(799,787){\usebox{\plotpoint}}
\put(800,789){\usebox{\plotpoint}}
\put(801,791){\usebox{\plotpoint}}
\put(802,792){\usebox{\plotpoint}}
\put(803,794){\usebox{\plotpoint}}
\put(804,795){\usebox{\plotpoint}}
\put(805,797){\usebox{\plotpoint}}
\put(806,798){\usebox{\plotpoint}}
\put(807,800){\usebox{\plotpoint}}
\put(808,801){\usebox{\plotpoint}}
\put(809,803){\usebox{\plotpoint}}
\put(810,805){\usebox{\plotpoint}}
\put(811,806){\usebox{\plotpoint}}
\put(812,808){\usebox{\plotpoint}}
\put(813,810){\usebox{\plotpoint}}
\put(814,811){\usebox{\plotpoint}}
\put(815,813){\usebox{\plotpoint}}
\put(816,814){\usebox{\plotpoint}}
\put(817,816){\usebox{\plotpoint}}
\put(818,817){\usebox{\plotpoint}}
\put(819,819){\usebox{\plotpoint}}
\put(820,821){\usebox{\plotpoint}}
\put(821,823){\usebox{\plotpoint}}
\put(822,825){\usebox{\plotpoint}}
\put(823,827){\usebox{\plotpoint}}
\put(824,829){\usebox{\plotpoint}}
\put(825,830){\usebox{\plotpoint}}
\put(826,832){\usebox{\plotpoint}}
\put(827,833){\usebox{\plotpoint}}
\put(828,835){\usebox{\plotpoint}}
\put(829,836){\usebox{\plotpoint}}
\put(830,838){\usebox{\plotpoint}}
\put(831,839){\usebox{\plotpoint}}
\put(832,841){\usebox{\plotpoint}}
\put(833,843){\usebox{\plotpoint}}
\put(834,844){\usebox{\plotpoint}}
\put(835,846){\usebox{\plotpoint}}
\sbox{\plotpoint}{\rule[-0.175pt]{0.350pt}{0.350pt}}%
\put(455,1046){\makebox(0,0)[r]{$\tilde{c}=0.5$}}
\put(477,1046){\rule[-0.175pt]{15.899pt}{0.350pt}}
\put(270,231){\usebox{\plotpoint}}
\put(270,231){\usebox{\plotpoint}}
\put(271,232){\usebox{\plotpoint}}
\put(272,233){\usebox{\plotpoint}}
\put(273,234){\usebox{\plotpoint}}
\put(274,235){\usebox{\plotpoint}}
\put(276,236){\usebox{\plotpoint}}
\put(277,237){\usebox{\plotpoint}}
\put(278,238){\usebox{\plotpoint}}
\put(279,239){\usebox{\plotpoint}}
\put(281,240){\rule[-0.175pt]{0.361pt}{0.350pt}}
\put(282,241){\rule[-0.175pt]{0.361pt}{0.350pt}}
\put(284,242){\rule[-0.175pt]{0.361pt}{0.350pt}}
\put(285,243){\rule[-0.175pt]{0.361pt}{0.350pt}}
\put(287,244){\rule[-0.175pt]{0.361pt}{0.350pt}}
\put(288,245){\rule[-0.175pt]{0.361pt}{0.350pt}}
\put(290,246){\rule[-0.175pt]{0.361pt}{0.350pt}}
\put(291,247){\rule[-0.175pt]{0.361pt}{0.350pt}}
\put(293,248){\rule[-0.175pt]{0.361pt}{0.350pt}}
\put(294,249){\rule[-0.175pt]{0.361pt}{0.350pt}}
\put(296,250){\rule[-0.175pt]{0.361pt}{0.350pt}}
\put(297,251){\rule[-0.175pt]{0.361pt}{0.350pt}}
\put(299,252){\usebox{\plotpoint}}
\put(300,253){\usebox{\plotpoint}}
\put(301,254){\usebox{\plotpoint}}
\put(302,255){\usebox{\plotpoint}}
\put(304,256){\rule[-0.175pt]{0.361pt}{0.350pt}}
\put(305,257){\rule[-0.175pt]{0.361pt}{0.350pt}}
\put(307,258){\rule[-0.175pt]{0.361pt}{0.350pt}}
\put(308,259){\rule[-0.175pt]{0.361pt}{0.350pt}}
\put(310,260){\rule[-0.175pt]{0.482pt}{0.350pt}}
\put(312,261){\rule[-0.175pt]{0.482pt}{0.350pt}}
\put(314,262){\rule[-0.175pt]{0.482pt}{0.350pt}}
\put(316,263){\rule[-0.175pt]{0.361pt}{0.350pt}}
\put(317,264){\rule[-0.175pt]{0.361pt}{0.350pt}}
\put(319,265){\rule[-0.175pt]{0.361pt}{0.350pt}}
\put(320,266){\rule[-0.175pt]{0.361pt}{0.350pt}}
\put(322,267){\rule[-0.175pt]{0.482pt}{0.350pt}}
\put(324,268){\rule[-0.175pt]{0.482pt}{0.350pt}}
\put(326,269){\rule[-0.175pt]{0.482pt}{0.350pt}}
\put(328,270){\usebox{\plotpoint}}
\put(329,271){\usebox{\plotpoint}}
\put(330,272){\usebox{\plotpoint}}
\put(331,273){\usebox{\plotpoint}}
\put(333,274){\rule[-0.175pt]{0.482pt}{0.350pt}}
\put(335,275){\rule[-0.175pt]{0.482pt}{0.350pt}}
\put(337,276){\rule[-0.175pt]{0.482pt}{0.350pt}}
\put(339,277){\rule[-0.175pt]{0.361pt}{0.350pt}}
\put(340,278){\rule[-0.175pt]{0.361pt}{0.350pt}}
\put(342,279){\rule[-0.175pt]{0.361pt}{0.350pt}}
\put(343,280){\rule[-0.175pt]{0.361pt}{0.350pt}}
\put(345,281){\rule[-0.175pt]{0.482pt}{0.350pt}}
\put(347,282){\rule[-0.175pt]{0.482pt}{0.350pt}}
\put(349,283){\rule[-0.175pt]{0.482pt}{0.350pt}}
\put(351,284){\usebox{\plotpoint}}
\put(352,285){\usebox{\plotpoint}}
\put(353,286){\usebox{\plotpoint}}
\put(354,287){\usebox{\plotpoint}}
\put(356,288){\rule[-0.175pt]{0.482pt}{0.350pt}}
\put(358,289){\rule[-0.175pt]{0.482pt}{0.350pt}}
\put(360,290){\rule[-0.175pt]{0.482pt}{0.350pt}}
\put(362,291){\rule[-0.175pt]{0.482pt}{0.350pt}}
\put(364,292){\rule[-0.175pt]{0.482pt}{0.350pt}}
\put(366,293){\rule[-0.175pt]{0.482pt}{0.350pt}}
\put(368,294){\rule[-0.175pt]{0.361pt}{0.350pt}}
\put(369,295){\rule[-0.175pt]{0.361pt}{0.350pt}}
\put(371,296){\rule[-0.175pt]{0.361pt}{0.350pt}}
\put(372,297){\rule[-0.175pt]{0.361pt}{0.350pt}}
\put(374,298){\rule[-0.175pt]{0.482pt}{0.350pt}}
\put(376,299){\rule[-0.175pt]{0.482pt}{0.350pt}}
\put(378,300){\rule[-0.175pt]{0.482pt}{0.350pt}}
\put(380,301){\rule[-0.175pt]{0.401pt}{0.350pt}}
\put(381,302){\rule[-0.175pt]{0.401pt}{0.350pt}}
\put(383,303){\rule[-0.175pt]{0.401pt}{0.350pt}}
\put(384,304){\rule[-0.175pt]{0.361pt}{0.350pt}}
\put(386,305){\rule[-0.175pt]{0.361pt}{0.350pt}}
\put(388,306){\rule[-0.175pt]{0.361pt}{0.350pt}}
\put(389,307){\rule[-0.175pt]{0.361pt}{0.350pt}}
\put(391,308){\rule[-0.175pt]{0.482pt}{0.350pt}}
\put(393,309){\rule[-0.175pt]{0.482pt}{0.350pt}}
\put(395,310){\rule[-0.175pt]{0.482pt}{0.350pt}}
\put(397,311){\rule[-0.175pt]{0.361pt}{0.350pt}}
\put(398,312){\rule[-0.175pt]{0.361pt}{0.350pt}}
\put(400,313){\rule[-0.175pt]{0.361pt}{0.350pt}}
\put(401,314){\rule[-0.175pt]{0.361pt}{0.350pt}}
\put(403,315){\rule[-0.175pt]{0.401pt}{0.350pt}}
\put(404,316){\rule[-0.175pt]{0.401pt}{0.350pt}}
\put(406,317){\rule[-0.175pt]{0.401pt}{0.350pt}}
\put(407,318){\rule[-0.175pt]{0.361pt}{0.350pt}}
\put(409,319){\rule[-0.175pt]{0.361pt}{0.350pt}}
\put(411,320){\rule[-0.175pt]{0.361pt}{0.350pt}}
\put(412,321){\rule[-0.175pt]{0.361pt}{0.350pt}}
\put(414,322){\rule[-0.175pt]{0.361pt}{0.350pt}}
\put(415,323){\rule[-0.175pt]{0.361pt}{0.350pt}}
\put(417,324){\rule[-0.175pt]{0.361pt}{0.350pt}}
\put(418,325){\rule[-0.175pt]{0.361pt}{0.350pt}}
\put(420,326){\rule[-0.175pt]{0.482pt}{0.350pt}}
\put(422,327){\rule[-0.175pt]{0.482pt}{0.350pt}}
\put(424,328){\rule[-0.175pt]{0.482pt}{0.350pt}}
\put(426,329){\rule[-0.175pt]{0.361pt}{0.350pt}}
\put(427,330){\rule[-0.175pt]{0.361pt}{0.350pt}}
\put(429,331){\rule[-0.175pt]{0.361pt}{0.350pt}}
\put(430,332){\rule[-0.175pt]{0.361pt}{0.350pt}}
\put(432,333){\usebox{\plotpoint}}
\put(433,334){\usebox{\plotpoint}}
\put(434,335){\usebox{\plotpoint}}
\put(435,336){\usebox{\plotpoint}}
\put(437,337){\rule[-0.175pt]{0.361pt}{0.350pt}}
\put(438,338){\rule[-0.175pt]{0.361pt}{0.350pt}}
\put(440,339){\rule[-0.175pt]{0.361pt}{0.350pt}}
\put(441,340){\rule[-0.175pt]{0.361pt}{0.350pt}}
\put(443,341){\rule[-0.175pt]{0.361pt}{0.350pt}}
\put(444,342){\rule[-0.175pt]{0.361pt}{0.350pt}}
\put(446,343){\rule[-0.175pt]{0.361pt}{0.350pt}}
\put(447,344){\rule[-0.175pt]{0.361pt}{0.350pt}}
\put(449,345){\rule[-0.175pt]{0.361pt}{0.350pt}}
\put(450,346){\rule[-0.175pt]{0.361pt}{0.350pt}}
\put(452,347){\rule[-0.175pt]{0.361pt}{0.350pt}}
\put(453,348){\rule[-0.175pt]{0.361pt}{0.350pt}}
\put(455,349){\usebox{\plotpoint}}
\put(456,350){\usebox{\plotpoint}}
\put(457,351){\usebox{\plotpoint}}
\put(458,352){\usebox{\plotpoint}}
\put(460,353){\usebox{\plotpoint}}
\put(461,354){\usebox{\plotpoint}}
\put(462,355){\usebox{\plotpoint}}
\put(463,356){\usebox{\plotpoint}}
\put(464,357){\usebox{\plotpoint}}
\put(466,358){\rule[-0.175pt]{0.361pt}{0.350pt}}
\put(467,359){\rule[-0.175pt]{0.361pt}{0.350pt}}
\put(469,360){\rule[-0.175pt]{0.361pt}{0.350pt}}
\put(470,361){\rule[-0.175pt]{0.361pt}{0.350pt}}
\put(472,362){\rule[-0.175pt]{0.361pt}{0.350pt}}
\put(473,363){\rule[-0.175pt]{0.361pt}{0.350pt}}
\put(475,364){\rule[-0.175pt]{0.361pt}{0.350pt}}
\put(476,365){\rule[-0.175pt]{0.361pt}{0.350pt}}
\put(478,366){\usebox{\plotpoint}}
\put(479,367){\usebox{\plotpoint}}
\put(480,368){\usebox{\plotpoint}}
\put(481,369){\usebox{\plotpoint}}
\put(482,370){\usebox{\plotpoint}}
\put(484,371){\usebox{\plotpoint}}
\put(485,372){\usebox{\plotpoint}}
\put(486,373){\usebox{\plotpoint}}
\put(487,374){\usebox{\plotpoint}}
\put(488,375){\usebox{\plotpoint}}
\put(489,376){\rule[-0.175pt]{0.361pt}{0.350pt}}
\put(490,377){\rule[-0.175pt]{0.361pt}{0.350pt}}
\put(492,378){\rule[-0.175pt]{0.361pt}{0.350pt}}
\put(493,379){\rule[-0.175pt]{0.361pt}{0.350pt}}
\put(495,380){\usebox{\plotpoint}}
\put(496,381){\usebox{\plotpoint}}
\put(497,382){\usebox{\plotpoint}}
\put(498,383){\usebox{\plotpoint}}
\put(499,384){\usebox{\plotpoint}}
\put(501,385){\usebox{\plotpoint}}
\put(502,386){\usebox{\plotpoint}}
\put(503,387){\usebox{\plotpoint}}
\put(504,388){\usebox{\plotpoint}}
\put(505,389){\usebox{\plotpoint}}
\put(507,390){\usebox{\plotpoint}}
\put(508,391){\usebox{\plotpoint}}
\put(509,392){\usebox{\plotpoint}}
\put(510,393){\usebox{\plotpoint}}
\put(512,394){\usebox{\plotpoint}}
\put(513,395){\usebox{\plotpoint}}
\put(514,396){\usebox{\plotpoint}}
\put(515,397){\usebox{\plotpoint}}
\put(516,398){\usebox{\plotpoint}}
\put(518,399){\usebox{\plotpoint}}
\put(519,400){\usebox{\plotpoint}}
\put(520,401){\usebox{\plotpoint}}
\put(521,402){\usebox{\plotpoint}}
\put(522,403){\usebox{\plotpoint}}
\put(524,404){\usebox{\plotpoint}}
\put(525,405){\usebox{\plotpoint}}
\put(526,406){\usebox{\plotpoint}}
\put(527,407){\usebox{\plotpoint}}
\put(528,408){\usebox{\plotpoint}}
\put(530,409){\usebox{\plotpoint}}
\put(531,410){\usebox{\plotpoint}}
\put(532,411){\usebox{\plotpoint}}
\put(533,412){\usebox{\plotpoint}}
\put(534,413){\usebox{\plotpoint}}
\put(535,414){\usebox{\plotpoint}}
\put(536,415){\usebox{\plotpoint}}
\put(537,416){\usebox{\plotpoint}}
\put(538,417){\usebox{\plotpoint}}
\put(539,418){\usebox{\plotpoint}}
\put(540,419){\usebox{\plotpoint}}
\put(541,420){\usebox{\plotpoint}}
\put(542,421){\usebox{\plotpoint}}
\put(543,422){\usebox{\plotpoint}}
\put(544,423){\usebox{\plotpoint}}
\put(545,424){\usebox{\plotpoint}}
\put(547,425){\usebox{\plotpoint}}
\put(548,426){\usebox{\plotpoint}}
\put(549,427){\usebox{\plotpoint}}
\put(550,428){\usebox{\plotpoint}}
\put(551,429){\usebox{\plotpoint}}
\put(553,430){\usebox{\plotpoint}}
\put(554,431){\usebox{\plotpoint}}
\put(555,432){\usebox{\plotpoint}}
\put(556,433){\usebox{\plotpoint}}
\put(557,434){\usebox{\plotpoint}}
\put(558,435){\usebox{\plotpoint}}
\put(559,436){\usebox{\plotpoint}}
\put(560,437){\usebox{\plotpoint}}
\put(561,438){\usebox{\plotpoint}}
\put(562,439){\usebox{\plotpoint}}
\put(563,440){\usebox{\plotpoint}}
\put(564,441){\usebox{\plotpoint}}
\put(565,442){\usebox{\plotpoint}}
\put(566,443){\usebox{\plotpoint}}
\put(567,444){\usebox{\plotpoint}}
\put(568,445){\usebox{\plotpoint}}
\put(569,446){\usebox{\plotpoint}}
\put(570,447){\usebox{\plotpoint}}
\put(571,448){\usebox{\plotpoint}}
\put(572,449){\usebox{\plotpoint}}
\put(573,450){\usebox{\plotpoint}}
\put(574,451){\usebox{\plotpoint}}
\put(576,452){\usebox{\plotpoint}}
\put(577,453){\usebox{\plotpoint}}
\put(578,454){\usebox{\plotpoint}}
\put(579,455){\usebox{\plotpoint}}
\put(580,456){\usebox{\plotpoint}}
\put(581,457){\usebox{\plotpoint}}
\put(582,458){\usebox{\plotpoint}}
\put(583,459){\usebox{\plotpoint}}
\put(584,460){\usebox{\plotpoint}}
\put(585,461){\usebox{\plotpoint}}
\put(586,462){\usebox{\plotpoint}}
\put(587,463){\usebox{\plotpoint}}
\put(588,464){\usebox{\plotpoint}}
\put(589,465){\usebox{\plotpoint}}
\put(590,466){\usebox{\plotpoint}}
\put(591,467){\usebox{\plotpoint}}
\put(592,468){\usebox{\plotpoint}}
\put(593,469){\usebox{\plotpoint}}
\put(594,470){\usebox{\plotpoint}}
\put(595,471){\usebox{\plotpoint}}
\put(596,472){\usebox{\plotpoint}}
\put(597,473){\usebox{\plotpoint}}
\put(598,474){\usebox{\plotpoint}}
\put(599,475){\usebox{\plotpoint}}
\put(600,476){\usebox{\plotpoint}}
\put(601,477){\usebox{\plotpoint}}
\put(602,478){\usebox{\plotpoint}}
\put(603,479){\usebox{\plotpoint}}
\put(604,480){\usebox{\plotpoint}}
\put(605,481){\usebox{\plotpoint}}
\put(606,482){\usebox{\plotpoint}}
\put(607,483){\usebox{\plotpoint}}
\put(608,484){\usebox{\plotpoint}}
\put(609,485){\usebox{\plotpoint}}
\put(610,486){\usebox{\plotpoint}}
\put(611,487){\usebox{\plotpoint}}
\put(612,488){\usebox{\plotpoint}}
\put(613,489){\usebox{\plotpoint}}
\put(614,490){\usebox{\plotpoint}}
\put(615,491){\usebox{\plotpoint}}
\put(616,493){\usebox{\plotpoint}}
\put(617,494){\usebox{\plotpoint}}
\put(618,495){\usebox{\plotpoint}}
\put(619,496){\usebox{\plotpoint}}
\put(620,497){\usebox{\plotpoint}}
\put(621,498){\usebox{\plotpoint}}
\put(622,499){\usebox{\plotpoint}}
\put(623,500){\usebox{\plotpoint}}
\put(624,501){\usebox{\plotpoint}}
\put(625,502){\usebox{\plotpoint}}
\put(626,503){\usebox{\plotpoint}}
\put(627,504){\usebox{\plotpoint}}
\put(628,505){\usebox{\plotpoint}}
\put(628,506){\usebox{\plotpoint}}
\put(629,507){\usebox{\plotpoint}}
\put(630,508){\usebox{\plotpoint}}
\put(631,509){\usebox{\plotpoint}}
\put(632,510){\usebox{\plotpoint}}
\put(633,511){\usebox{\plotpoint}}
\put(634,512){\usebox{\plotpoint}}
\put(635,513){\usebox{\plotpoint}}
\put(636,514){\usebox{\plotpoint}}
\put(637,515){\usebox{\plotpoint}}
\put(638,516){\usebox{\plotpoint}}
\put(639,517){\usebox{\plotpoint}}
\put(640,518){\usebox{\plotpoint}}
\put(641,519){\usebox{\plotpoint}}
\put(642,520){\usebox{\plotpoint}}
\put(643,522){\usebox{\plotpoint}}
\put(644,523){\usebox{\plotpoint}}
\put(645,525){\usebox{\plotpoint}}
\put(646,526){\usebox{\plotpoint}}
\put(647,527){\usebox{\plotpoint}}
\put(648,528){\usebox{\plotpoint}}
\put(649,529){\usebox{\plotpoint}}
\put(650,530){\usebox{\plotpoint}}
\put(651,531){\usebox{\plotpoint}}
\put(652,532){\usebox{\plotpoint}}
\put(653,533){\usebox{\plotpoint}}
\put(654,534){\usebox{\plotpoint}}
\put(655,535){\usebox{\plotpoint}}
\put(656,536){\usebox{\plotpoint}}
\put(657,538){\usebox{\plotpoint}}
\put(658,539){\usebox{\plotpoint}}
\put(659,540){\usebox{\plotpoint}}
\put(660,541){\usebox{\plotpoint}}
\put(661,542){\usebox{\plotpoint}}
\put(662,543){\usebox{\plotpoint}}
\put(663,544){\usebox{\plotpoint}}
\put(664,545){\usebox{\plotpoint}}
\put(665,546){\usebox{\plotpoint}}
\put(666,548){\usebox{\plotpoint}}
\put(667,549){\usebox{\plotpoint}}
\put(668,551){\usebox{\plotpoint}}
\put(669,552){\usebox{\plotpoint}}
\put(670,553){\usebox{\plotpoint}}
\put(671,554){\usebox{\plotpoint}}
\put(672,555){\usebox{\plotpoint}}
\put(673,556){\usebox{\plotpoint}}
\put(674,558){\usebox{\plotpoint}}
\put(675,559){\usebox{\plotpoint}}
\put(676,560){\usebox{\plotpoint}}
\put(677,561){\usebox{\plotpoint}}
\put(678,562){\usebox{\plotpoint}}
\put(679,563){\usebox{\plotpoint}}
\put(680,565){\usebox{\plotpoint}}
\put(681,566){\usebox{\plotpoint}}
\put(682,567){\usebox{\plotpoint}}
\put(683,568){\usebox{\plotpoint}}
\put(684,569){\usebox{\plotpoint}}
\put(685,570){\usebox{\plotpoint}}
\put(686,572){\usebox{\plotpoint}}
\put(687,573){\usebox{\plotpoint}}
\put(688,574){\usebox{\plotpoint}}
\put(689,575){\usebox{\plotpoint}}
\put(690,576){\usebox{\plotpoint}}
\put(691,577){\usebox{\plotpoint}}
\put(692,579){\usebox{\plotpoint}}
\put(693,580){\usebox{\plotpoint}}
\put(694,581){\usebox{\plotpoint}}
\put(695,583){\usebox{\plotpoint}}
\put(696,584){\usebox{\plotpoint}}
\put(697,586){\usebox{\plotpoint}}
\put(698,587){\usebox{\plotpoint}}
\put(699,588){\usebox{\plotpoint}}
\put(700,589){\usebox{\plotpoint}}
\put(701,590){\usebox{\plotpoint}}
\put(702,591){\usebox{\plotpoint}}
\put(703,593){\usebox{\plotpoint}}
\put(704,594){\usebox{\plotpoint}}
\put(705,595){\usebox{\plotpoint}}
\put(706,596){\usebox{\plotpoint}}
\put(707,597){\usebox{\plotpoint}}
\put(708,598){\usebox{\plotpoint}}
\put(709,600){\usebox{\plotpoint}}
\put(710,601){\usebox{\plotpoint}}
\put(711,602){\usebox{\plotpoint}}
\put(712,603){\usebox{\plotpoint}}
\put(713,605){\usebox{\plotpoint}}
\put(714,606){\usebox{\plotpoint}}
\put(715,607){\usebox{\plotpoint}}
\put(716,609){\usebox{\plotpoint}}
\put(717,610){\usebox{\plotpoint}}
\put(718,612){\usebox{\plotpoint}}
\put(719,613){\usebox{\plotpoint}}
\put(720,615){\usebox{\plotpoint}}
\put(721,616){\usebox{\plotpoint}}
\put(722,617){\usebox{\plotpoint}}
\put(723,618){\usebox{\plotpoint}}
\put(724,620){\usebox{\plotpoint}}
\put(725,621){\usebox{\plotpoint}}
\put(726,622){\usebox{\plotpoint}}
\put(727,624){\usebox{\plotpoint}}
\put(728,625){\usebox{\plotpoint}}
\put(729,626){\usebox{\plotpoint}}
\put(730,627){\usebox{\plotpoint}}
\put(731,628){\usebox{\plotpoint}}
\put(732,630){\usebox{\plotpoint}}
\put(733,631){\usebox{\plotpoint}}
\put(734,632){\usebox{\plotpoint}}
\put(735,633){\usebox{\plotpoint}}
\put(736,635){\usebox{\plotpoint}}
\put(737,636){\usebox{\plotpoint}}
\put(738,637){\usebox{\plotpoint}}
\put(739,639){\usebox{\plotpoint}}
\put(740,640){\usebox{\plotpoint}}
\put(741,641){\usebox{\plotpoint}}
\put(742,643){\usebox{\plotpoint}}
\put(743,644){\usebox{\plotpoint}}
\put(744,645){\rule[-0.175pt]{0.350pt}{0.385pt}}
\put(745,647){\rule[-0.175pt]{0.350pt}{0.385pt}}
\put(746,649){\rule[-0.175pt]{0.350pt}{0.385pt}}
\put(747,650){\rule[-0.175pt]{0.350pt}{0.385pt}}
\put(748,652){\rule[-0.175pt]{0.350pt}{0.385pt}}
\put(749,653){\usebox{\plotpoint}}
\put(750,655){\usebox{\plotpoint}}
\put(751,656){\usebox{\plotpoint}}
\put(752,657){\usebox{\plotpoint}}
\put(753,659){\usebox{\plotpoint}}
\put(754,660){\usebox{\plotpoint}}
\put(755,661){\usebox{\plotpoint}}
\put(756,663){\usebox{\plotpoint}}
\put(757,664){\usebox{\plotpoint}}
\put(758,665){\usebox{\plotpoint}}
\put(759,667){\usebox{\plotpoint}}
\put(760,668){\usebox{\plotpoint}}
\put(761,669){\usebox{\plotpoint}}
\put(762,671){\usebox{\plotpoint}}
\put(763,672){\usebox{\plotpoint}}
\put(764,673){\usebox{\plotpoint}}
\put(765,675){\usebox{\plotpoint}}
\put(766,676){\usebox{\plotpoint}}
\put(767,677){\rule[-0.175pt]{0.350pt}{0.385pt}}
\put(768,679){\rule[-0.175pt]{0.350pt}{0.385pt}}
\put(769,681){\rule[-0.175pt]{0.350pt}{0.385pt}}
\put(770,682){\rule[-0.175pt]{0.350pt}{0.385pt}}
\put(771,684){\rule[-0.175pt]{0.350pt}{0.385pt}}
\put(772,685){\usebox{\plotpoint}}
\put(773,687){\usebox{\plotpoint}}
\put(774,688){\usebox{\plotpoint}}
\put(775,689){\usebox{\plotpoint}}
\put(776,691){\usebox{\plotpoint}}
\put(777,692){\usebox{\plotpoint}}
\put(778,693){\usebox{\plotpoint}}
\put(779,695){\usebox{\plotpoint}}
\put(780,696){\usebox{\plotpoint}}
\put(781,697){\usebox{\plotpoint}}
\put(782,699){\usebox{\plotpoint}}
\put(783,700){\usebox{\plotpoint}}
\put(784,701){\rule[-0.175pt]{0.350pt}{0.361pt}}
\put(785,703){\rule[-0.175pt]{0.350pt}{0.361pt}}
\put(786,705){\rule[-0.175pt]{0.350pt}{0.361pt}}
\put(787,706){\rule[-0.175pt]{0.350pt}{0.361pt}}
\put(788,708){\rule[-0.175pt]{0.350pt}{0.361pt}}
\put(789,709){\rule[-0.175pt]{0.350pt}{0.361pt}}
\put(790,711){\usebox{\plotpoint}}
\put(791,712){\usebox{\plotpoint}}
\put(792,713){\usebox{\plotpoint}}
\put(793,714){\usebox{\plotpoint}}
\put(794,716){\usebox{\plotpoint}}
\put(795,717){\usebox{\plotpoint}}
\put(796,718){\rule[-0.175pt]{0.350pt}{0.434pt}}
\put(797,720){\rule[-0.175pt]{0.350pt}{0.434pt}}
\put(798,722){\rule[-0.175pt]{0.350pt}{0.434pt}}
\put(799,724){\rule[-0.175pt]{0.350pt}{0.434pt}}
\put(800,726){\rule[-0.175pt]{0.350pt}{0.434pt}}
\put(801,727){\usebox{\plotpoint}}
\put(802,729){\usebox{\plotpoint}}
\put(803,730){\usebox{\plotpoint}}
\put(804,731){\usebox{\plotpoint}}
\put(805,733){\usebox{\plotpoint}}
\put(806,734){\usebox{\plotpoint}}
\put(807,735){\rule[-0.175pt]{0.350pt}{0.361pt}}
\put(808,737){\rule[-0.175pt]{0.350pt}{0.361pt}}
\put(809,739){\rule[-0.175pt]{0.350pt}{0.361pt}}
\put(810,740){\rule[-0.175pt]{0.350pt}{0.361pt}}
\put(811,742){\rule[-0.175pt]{0.350pt}{0.361pt}}
\put(812,743){\rule[-0.175pt]{0.350pt}{0.361pt}}
\put(813,745){\rule[-0.175pt]{0.350pt}{0.361pt}}
\put(814,746){\rule[-0.175pt]{0.350pt}{0.361pt}}
\put(815,748){\rule[-0.175pt]{0.350pt}{0.361pt}}
\put(816,749){\rule[-0.175pt]{0.350pt}{0.361pt}}
\put(817,751){\rule[-0.175pt]{0.350pt}{0.361pt}}
\put(818,752){\rule[-0.175pt]{0.350pt}{0.361pt}}
\put(819,754){\rule[-0.175pt]{0.350pt}{0.385pt}}
\put(820,755){\rule[-0.175pt]{0.350pt}{0.385pt}}
\put(821,757){\rule[-0.175pt]{0.350pt}{0.385pt}}
\put(822,758){\rule[-0.175pt]{0.350pt}{0.385pt}}
\put(823,760){\rule[-0.175pt]{0.350pt}{0.385pt}}
\put(824,761){\rule[-0.175pt]{0.350pt}{0.361pt}}
\put(825,763){\rule[-0.175pt]{0.350pt}{0.361pt}}
\put(826,765){\rule[-0.175pt]{0.350pt}{0.361pt}}
\put(827,766){\rule[-0.175pt]{0.350pt}{0.361pt}}
\put(828,768){\rule[-0.175pt]{0.350pt}{0.361pt}}
\put(829,769){\rule[-0.175pt]{0.350pt}{0.361pt}}
\put(830,771){\rule[-0.175pt]{0.350pt}{0.361pt}}
\put(831,772){\rule[-0.175pt]{0.350pt}{0.361pt}}
\put(832,774){\rule[-0.175pt]{0.350pt}{0.361pt}}
\put(833,775){\rule[-0.175pt]{0.350pt}{0.361pt}}
\put(834,777){\rule[-0.175pt]{0.350pt}{0.361pt}}
\put(835,778){\rule[-0.175pt]{0.350pt}{0.361pt}}
\sbox{\plotpoint}{\rule[-0.250pt]{0.500pt}{0.500pt}}%
\put(455,1001){\makebox(0,0)[r]{$\tilde{c}=0.0$}}
\put(477,1001){\usebox{\plotpoint}}
\put(497,1001){\usebox{\plotpoint}}
\put(518,1001){\usebox{\plotpoint}}
\put(539,1001){\usebox{\plotpoint}}
\put(543,1001){\usebox{\plotpoint}}
\put(270,229){\usebox{\plotpoint}}
\put(270,229){\usebox{\plotpoint}}
\put(288,238){\usebox{\plotpoint}}
\put(306,248){\usebox{\plotpoint}}
\put(324,258){\usebox{\plotpoint}}
\put(343,268){\usebox{\plotpoint}}
\put(361,278){\usebox{\plotpoint}}
\put(379,288){\usebox{\plotpoint}}
\put(397,299){\usebox{\plotpoint}}
\put(414,310){\usebox{\plotpoint}}
\put(432,321){\usebox{\plotpoint}}
\put(449,332){\usebox{\plotpoint}}
\put(466,343){\usebox{\plotpoint}}
\put(484,355){\usebox{\plotpoint}}
\put(501,367){\usebox{\plotpoint}}
\put(517,379){\usebox{\plotpoint}}
\put(534,391){\usebox{\plotpoint}}
\put(550,404){\usebox{\plotpoint}}
\put(567,417){\usebox{\plotpoint}}
\put(583,430){\usebox{\plotpoint}}
\put(599,443){\usebox{\plotpoint}}
\put(614,456){\usebox{\plotpoint}}
\put(630,470){\usebox{\plotpoint}}
\put(645,484){\usebox{\plotpoint}}
\put(661,498){\usebox{\plotpoint}}
\put(675,513){\usebox{\plotpoint}}
\put(690,527){\usebox{\plotpoint}}
\put(704,542){\usebox{\plotpoint}}
\put(718,557){\usebox{\plotpoint}}
\put(733,573){\usebox{\plotpoint}}
\put(746,588){\usebox{\plotpoint}}
\put(760,603){\usebox{\plotpoint}}
\put(773,620){\usebox{\plotpoint}}
\put(787,635){\usebox{\plotpoint}}
\put(800,651){\usebox{\plotpoint}}
\put(813,668){\usebox{\plotpoint}}
\put(825,684){\usebox{\plotpoint}}
\put(836,698){\usebox{\plotpoint}}
\sbox{\plotpoint}{\rule[-0.175pt]{0.350pt}{0.350pt}}%
\put(359,308){\circle*{12}}
\put(424,370){\circle*{12}}
\put(488,336){\circle*{12}}
\put(551,479){\circle*{12}}
\put(613,769){\circle*{12}}
\put(682,810){\circle*{12}}
\put(744,1001){\circle*{12}}
\put(359,264){\rule[-0.175pt]{0.350pt}{21.681pt}}
\put(349,264){\rule[-0.175pt]{4.818pt}{0.350pt}}
\put(349,354){\rule[-0.175pt]{4.818pt}{0.350pt}}
\put(424,314){\rule[-0.175pt]{0.350pt}{24.572pt}}
\put(414,314){\rule[-0.175pt]{4.818pt}{0.350pt}}
\put(414,416){\rule[-0.175pt]{4.818pt}{0.350pt}}
\put(488,249){\rule[-0.175pt]{0.350pt}{36.617pt}}
\put(478,249){\rule[-0.175pt]{4.818pt}{0.350pt}}
\put(478,401){\rule[-0.175pt]{4.818pt}{0.350pt}}
\put(551,381){\rule[-0.175pt]{0.350pt}{38.544pt}}
\put(541,381){\rule[-0.175pt]{4.818pt}{0.350pt}}
\put(541,541){\rule[-0.175pt]{4.818pt}{0.350pt}}
\put(613,707){\rule[-0.175pt]{0.350pt}{26.258pt}}
\put(603,707){\rule[-0.175pt]{4.818pt}{0.350pt}}
\put(603,816){\rule[-0.175pt]{4.818pt}{0.350pt}}
\put(682,722){\rule[-0.175pt]{0.350pt}{35.412pt}}
\put(672,722){\rule[-0.175pt]{4.818pt}{0.350pt}}
\put(672,869){\rule[-0.175pt]{4.818pt}{0.350pt}}
\put(744,875){\rule[-0.175pt]{0.350pt}{48.903pt}}
\put(734,875){\rule[-0.175pt]{4.818pt}{0.350pt}}
\put(734,1078){\rule[-0.175pt]{4.818pt}{0.350pt}}
\end{picture}

%% file: wpgformw.tex
\setlength{\unitlength}{0.240900pt}
\ifx\plotpoint\undefined\newsavebox{\plotpoint}\fi
\begin{picture}(900,1259)(0,0)
\tenrm
\sbox{\plotpoint}{\rule[-0.175pt]{0.350pt}{0.350pt}}%
\put(264,158){\rule[-0.175pt]{0.350pt}{238.009pt}}
\put(264,184){\rule[-0.175pt]{4.818pt}{0.350pt}}
\put(816,184){\rule[-0.175pt]{4.818pt}{0.350pt}}
\put(264,206){\rule[-0.175pt]{4.818pt}{0.350pt}}
\put(816,206){\rule[-0.175pt]{4.818pt}{0.350pt}}
\put(264,227){\rule[-0.175pt]{4.818pt}{0.350pt}}
\put(242,227){\makebox(0,0)[r]{\fsize1}}
\put(816,227){\rule[-0.175pt]{4.818pt}{0.350pt}}
\put(264,360){\rule[-0.175pt]{4.818pt}{0.350pt}}
\put(816,360){\rule[-0.175pt]{4.818pt}{0.350pt}}
\put(264,438){\rule[-0.175pt]{4.818pt}{0.350pt}}
\put(816,438){\rule[-0.175pt]{4.818pt}{0.350pt}}
\put(264,493){\rule[-0.175pt]{4.818pt}{0.350pt}}
\put(816,493){\rule[-0.175pt]{4.818pt}{0.350pt}}
\put(264,536){\rule[-0.175pt]{4.818pt}{0.350pt}}
\put(816,536){\rule[-0.175pt]{4.818pt}{0.350pt}}
\put(264,571){\rule[-0.175pt]{4.818pt}{0.350pt}}
\put(816,571){\rule[-0.175pt]{4.818pt}{0.350pt}}
\put(264,600){\rule[-0.175pt]{4.818pt}{0.350pt}}
\put(816,600){\rule[-0.175pt]{4.818pt}{0.350pt}}
\put(264,626){\rule[-0.175pt]{4.818pt}{0.350pt}}
\put(816,626){\rule[-0.175pt]{4.818pt}{0.350pt}}
\put(264,649){\rule[-0.175pt]{4.818pt}{0.350pt}}
\put(816,649){\rule[-0.175pt]{4.818pt}{0.350pt}}
\put(264,669){\rule[-0.175pt]{4.818pt}{0.350pt}}
\put(242,669){\makebox(0,0)[r]{\fsize10}}
\put(816,669){\rule[-0.175pt]{4.818pt}{0.350pt}}
\put(264,802){\rule[-0.175pt]{4.818pt}{0.350pt}}
\put(816,802){\rule[-0.175pt]{4.818pt}{0.350pt}}
\put(264,880){\rule[-0.175pt]{4.818pt}{0.350pt}}
\put(816,880){\rule[-0.175pt]{4.818pt}{0.350pt}}
\put(264,935){\rule[-0.175pt]{4.818pt}{0.350pt}}
\put(816,935){\rule[-0.175pt]{4.818pt}{0.350pt}}
\put(264,978){\rule[-0.175pt]{4.818pt}{0.350pt}}
\put(816,978){\rule[-0.175pt]{4.818pt}{0.350pt}}
\put(264,1013){\rule[-0.175pt]{4.818pt}{0.350pt}}
\put(816,1013){\rule[-0.175pt]{4.818pt}{0.350pt}}
\put(264,1042){\rule[-0.175pt]{4.818pt}{0.350pt}}
\put(816,1042){\rule[-0.175pt]{4.818pt}{0.350pt}}
\put(264,1068){\rule[-0.175pt]{4.818pt}{0.350pt}}
\put(816,1068){\rule[-0.175pt]{4.818pt}{0.350pt}}
\put(264,1091){\rule[-0.175pt]{4.818pt}{0.350pt}}
\put(816,1091){\rule[-0.175pt]{4.818pt}{0.350pt}}
\put(264,1111){\rule[-0.175pt]{4.818pt}{0.350pt}}
\put(242,1111){\makebox(0,0)[r]{\fsize100}}
\put(816,1111){\rule[-0.175pt]{4.818pt}{0.350pt}}
\put(264,158){\rule[-0.175pt]{0.350pt}{4.818pt}}
\put(264,113){\makebox(0,0){\fsize0}}
\put(264,1126){\rule[-0.175pt]{0.350pt}{4.818pt}}
\put(391,158){\rule[-0.175pt]{0.350pt}{4.818pt}}
\put(391,113){\makebox(0,0){\fsize0.1}}
\put(391,1126){\rule[-0.175pt]{0.350pt}{4.818pt}}
\put(518,158){\rule[-0.175pt]{0.350pt}{4.818pt}}
\put(518,113){\makebox(0,0){\fsize0.2}}
\put(518,1126){\rule[-0.175pt]{0.350pt}{4.818pt}}
\put(645,158){\rule[-0.175pt]{0.350pt}{4.818pt}}
\put(645,113){\makebox(0,0){\fsize0.3}}
\put(645,1126){\rule[-0.175pt]{0.350pt}{4.818pt}}
\put(772,158){\rule[-0.175pt]{0.350pt}{4.818pt}}
\put(772,113){\makebox(0,0){\fsize0.4}}
\put(772,1126){\rule[-0.175pt]{0.350pt}{4.818pt}}
\put(264,158){\rule[-0.175pt]{137.795pt}{0.350pt}}
\put(836,158){\rule[-0.175pt]{0.350pt}{238.009pt}}
\put(264,1146){\rule[-0.175pt]{137.795pt}{0.350pt}}
\put(199,1192){\makebox(0,0)[l]{\shortstack{\small$\abs{F_\omega^{\rm C}(q^2)}^2$}}}
\put(550,23){\makebox(0,0){\small$(b)\mq q^2$ [GeV${}^2$]}}
\put(264,158){\rule[-0.175pt]{0.350pt}{238.009pt}}
\sbox{\plotpoint}{\rule[-0.500pt]{1.000pt}{1.000pt}}%
\put(455,1091){\makebox(0,0)[r]{$\tilde{c}=1.0$}}
\put(477,1091){\rule[-0.500pt]{15.899pt}{1.000pt}}
\put(264,227){\usebox{\plotpoint}}
\put(264,227){\usebox{\plotpoint}}
\put(265,228){\usebox{\plotpoint}}
\put(266,229){\usebox{\plotpoint}}
\put(267,230){\usebox{\plotpoint}}
\put(268,231){\usebox{\plotpoint}}
\put(270,232){\usebox{\plotpoint}}
\put(271,233){\usebox{\plotpoint}}
\put(272,234){\usebox{\plotpoint}}
\put(273,235){\usebox{\plotpoint}}
\put(274,236){\usebox{\plotpoint}}
\put(275,237){\usebox{\plotpoint}}
\put(276,238){\usebox{\plotpoint}}
\put(277,239){\usebox{\plotpoint}}
\put(278,240){\usebox{\plotpoint}}
\put(279,241){\usebox{\plotpoint}}
\put(280,242){\usebox{\plotpoint}}
\put(281,243){\usebox{\plotpoint}}
\put(281,244){\usebox{\plotpoint}}
\put(282,245){\usebox{\plotpoint}}
\put(283,246){\usebox{\plotpoint}}
\put(284,247){\usebox{\plotpoint}}
\put(285,248){\usebox{\plotpoint}}
\put(286,249){\usebox{\plotpoint}}
\put(287,250){\usebox{\plotpoint}}
\put(288,251){\usebox{\plotpoint}}
\put(289,252){\usebox{\plotpoint}}
\put(290,253){\usebox{\plotpoint}}
\put(291,254){\usebox{\plotpoint}}
\put(292,255){\usebox{\plotpoint}}
\put(293,256){\usebox{\plotpoint}}
\put(294,257){\usebox{\plotpoint}}
\put(295,258){\usebox{\plotpoint}}
\put(296,259){\usebox{\plotpoint}}
\put(297,260){\usebox{\plotpoint}}
\put(298,261){\usebox{\plotpoint}}
\put(299,262){\usebox{\plotpoint}}
\put(300,263){\usebox{\plotpoint}}
\put(301,264){\usebox{\plotpoint}}
\put(302,265){\usebox{\plotpoint}}
\put(303,266){\usebox{\plotpoint}}
\put(304,268){\usebox{\plotpoint}}
\put(305,269){\usebox{\plotpoint}}
\put(306,270){\usebox{\plotpoint}}
\put(307,271){\usebox{\plotpoint}}
\put(308,272){\usebox{\plotpoint}}
\put(309,273){\usebox{\plotpoint}}
\put(310,274){\usebox{\plotpoint}}
\put(311,275){\usebox{\plotpoint}}
\put(312,276){\usebox{\plotpoint}}
\put(313,277){\usebox{\plotpoint}}
\put(314,278){\usebox{\plotpoint}}
\put(315,279){\usebox{\plotpoint}}
\put(316,280){\usebox{\plotpoint}}
\put(317,281){\usebox{\plotpoint}}
\put(318,282){\usebox{\plotpoint}}
\put(319,283){\usebox{\plotpoint}}
\put(320,284){\usebox{\plotpoint}}
\put(321,285){\usebox{\plotpoint}}
\put(322,286){\usebox{\plotpoint}}
\put(323,287){\usebox{\plotpoint}}
\put(324,288){\usebox{\plotpoint}}
\put(325,289){\usebox{\plotpoint}}
\put(326,290){\usebox{\plotpoint}}
\put(327,291){\usebox{\plotpoint}}
\put(328,292){\usebox{\plotpoint}}
\put(329,293){\usebox{\plotpoint}}
\put(330,294){\usebox{\plotpoint}}
\put(331,295){\usebox{\plotpoint}}
\put(332,296){\usebox{\plotpoint}}
\put(333,297){\usebox{\plotpoint}}
\put(334,298){\usebox{\plotpoint}}
\put(335,299){\usebox{\plotpoint}}
\put(336,300){\usebox{\plotpoint}}
\put(337,301){\usebox{\plotpoint}}
\put(338,302){\usebox{\plotpoint}}
\put(339,303){\usebox{\plotpoint}}
\put(340,304){\usebox{\plotpoint}}
\put(341,305){\usebox{\plotpoint}}
\put(342,306){\usebox{\plotpoint}}
\put(343,307){\usebox{\plotpoint}}
\put(344,308){\usebox{\plotpoint}}
\put(345,309){\usebox{\plotpoint}}
\put(346,310){\usebox{\plotpoint}}
\put(347,311){\usebox{\plotpoint}}
\put(348,312){\usebox{\plotpoint}}
\put(349,313){\usebox{\plotpoint}}
\put(350,314){\usebox{\plotpoint}}
\put(351,315){\usebox{\plotpoint}}
\put(352,316){\usebox{\plotpoint}}
\put(353,317){\usebox{\plotpoint}}
\put(354,318){\usebox{\plotpoint}}
\put(355,319){\usebox{\plotpoint}}
\put(356,321){\usebox{\plotpoint}}
\put(357,322){\usebox{\plotpoint}}
\put(358,323){\usebox{\plotpoint}}
\put(359,324){\usebox{\plotpoint}}
\put(360,325){\usebox{\plotpoint}}
\put(361,326){\usebox{\plotpoint}}
\put(362,327){\usebox{\plotpoint}}
\put(363,328){\usebox{\plotpoint}}
\put(364,329){\usebox{\plotpoint}}
\put(365,330){\usebox{\plotpoint}}
\put(366,331){\usebox{\plotpoint}}
\put(367,332){\usebox{\plotpoint}}
\put(368,333){\usebox{\plotpoint}}
\put(369,334){\usebox{\plotpoint}}
\put(370,335){\usebox{\plotpoint}}
\put(371,336){\usebox{\plotpoint}}
\put(372,337){\usebox{\plotpoint}}
\put(373,338){\usebox{\plotpoint}}
\put(374,339){\usebox{\plotpoint}}
\put(375,340){\usebox{\plotpoint}}
\put(376,341){\usebox{\plotpoint}}
\put(377,342){\usebox{\plotpoint}}
\put(378,343){\usebox{\plotpoint}}
\put(379,344){\usebox{\plotpoint}}
\put(380,345){\usebox{\plotpoint}}
\put(381,346){\usebox{\plotpoint}}
\put(382,347){\usebox{\plotpoint}}
\put(383,348){\usebox{\plotpoint}}
\put(384,349){\usebox{\plotpoint}}
\put(385,351){\usebox{\plotpoint}}
\put(386,352){\usebox{\plotpoint}}
\put(387,353){\usebox{\plotpoint}}
\put(388,354){\usebox{\plotpoint}}
\put(389,355){\usebox{\plotpoint}}
\put(390,356){\usebox{\plotpoint}}
\put(391,357){\usebox{\plotpoint}}
\put(392,358){\usebox{\plotpoint}}
\put(393,359){\usebox{\plotpoint}}
\put(394,360){\usebox{\plotpoint}}
\put(395,361){\usebox{\plotpoint}}
\put(396,362){\usebox{\plotpoint}}
\put(397,363){\usebox{\plotpoint}}
\put(398,364){\usebox{\plotpoint}}
\put(399,365){\usebox{\plotpoint}}
\put(400,366){\usebox{\plotpoint}}
\put(401,367){\usebox{\plotpoint}}
\put(402,368){\usebox{\plotpoint}}
\put(403,369){\usebox{\plotpoint}}
\put(404,370){\usebox{\plotpoint}}
\put(405,371){\usebox{\plotpoint}}
\put(406,372){\usebox{\plotpoint}}
\put(407,373){\usebox{\plotpoint}}
\put(408,375){\usebox{\plotpoint}}
\put(409,376){\usebox{\plotpoint}}
\put(410,377){\usebox{\plotpoint}}
\put(411,378){\usebox{\plotpoint}}
\put(412,379){\usebox{\plotpoint}}
\put(413,380){\usebox{\plotpoint}}
\put(414,381){\usebox{\plotpoint}}
\put(415,382){\usebox{\plotpoint}}
\put(416,383){\usebox{\plotpoint}}
\put(417,384){\usebox{\plotpoint}}
\put(418,385){\usebox{\plotpoint}}
\put(419,386){\usebox{\plotpoint}}
\put(420,387){\usebox{\plotpoint}}
\put(421,388){\usebox{\plotpoint}}
\put(422,389){\usebox{\plotpoint}}
\put(423,390){\usebox{\plotpoint}}
\put(424,391){\usebox{\plotpoint}}
\put(425,392){\usebox{\plotpoint}}
\put(426,393){\usebox{\plotpoint}}
\put(427,394){\usebox{\plotpoint}}
\put(428,395){\usebox{\plotpoint}}
\put(429,396){\usebox{\plotpoint}}
\put(430,397){\usebox{\plotpoint}}
\put(431,398){\usebox{\plotpoint}}
\put(432,399){\usebox{\plotpoint}}
\put(433,400){\usebox{\plotpoint}}
\put(434,401){\usebox{\plotpoint}}
\put(435,402){\usebox{\plotpoint}}
\put(436,403){\usebox{\plotpoint}}
\put(437,405){\usebox{\plotpoint}}
\put(438,406){\usebox{\plotpoint}}
\put(439,407){\usebox{\plotpoint}}
\put(440,408){\usebox{\plotpoint}}
\put(441,409){\usebox{\plotpoint}}
\put(442,410){\usebox{\plotpoint}}
\put(443,411){\usebox{\plotpoint}}
\put(443,412){\usebox{\plotpoint}}
\put(444,413){\usebox{\plotpoint}}
\put(445,414){\usebox{\plotpoint}}
\put(446,415){\usebox{\plotpoint}}
\put(447,416){\usebox{\plotpoint}}
\put(448,417){\usebox{\plotpoint}}
\put(449,418){\usebox{\plotpoint}}
\put(450,419){\usebox{\plotpoint}}
\put(451,420){\usebox{\plotpoint}}
\put(452,421){\usebox{\plotpoint}}
\put(453,422){\usebox{\plotpoint}}
\put(454,423){\usebox{\plotpoint}}
\put(455,424){\usebox{\plotpoint}}
\put(456,425){\usebox{\plotpoint}}
\put(457,426){\usebox{\plotpoint}}
\put(458,427){\usebox{\plotpoint}}
\put(459,428){\usebox{\plotpoint}}
\put(460,430){\usebox{\plotpoint}}
\put(461,431){\usebox{\plotpoint}}
\put(462,432){\usebox{\plotpoint}}
\put(463,433){\usebox{\plotpoint}}
\put(464,434){\usebox{\plotpoint}}
\put(465,435){\usebox{\plotpoint}}
\put(466,436){\usebox{\plotpoint}}
\put(467,437){\usebox{\plotpoint}}
\put(468,438){\usebox{\plotpoint}}
\put(469,439){\usebox{\plotpoint}}
\put(470,440){\usebox{\plotpoint}}
\put(471,441){\usebox{\plotpoint}}
\put(472,442){\usebox{\plotpoint}}
\put(472,443){\usebox{\plotpoint}}
\put(473,444){\usebox{\plotpoint}}
\put(474,445){\usebox{\plotpoint}}
\put(475,446){\usebox{\plotpoint}}
\put(476,447){\usebox{\plotpoint}}
\put(477,448){\usebox{\plotpoint}}
\put(478,449){\usebox{\plotpoint}}
\put(479,450){\usebox{\plotpoint}}
\put(480,451){\usebox{\plotpoint}}
\put(481,452){\usebox{\plotpoint}}
\put(482,453){\usebox{\plotpoint}}
\put(483,454){\usebox{\plotpoint}}
\put(484,455){\usebox{\plotpoint}}
\put(485,456){\usebox{\plotpoint}}
\put(486,457){\usebox{\plotpoint}}
\put(487,458){\usebox{\plotpoint}}
\put(488,459){\usebox{\plotpoint}}
\put(489,461){\usebox{\plotpoint}}
\put(490,462){\usebox{\plotpoint}}
\put(491,463){\usebox{\plotpoint}}
\put(492,464){\usebox{\plotpoint}}
\put(493,465){\usebox{\plotpoint}}
\put(494,466){\usebox{\plotpoint}}
\put(495,467){\usebox{\plotpoint}}
\put(495,468){\usebox{\plotpoint}}
\put(496,469){\usebox{\plotpoint}}
\put(497,470){\usebox{\plotpoint}}
\put(498,471){\usebox{\plotpoint}}
\put(499,472){\usebox{\plotpoint}}
\put(500,473){\usebox{\plotpoint}}
\put(501,474){\usebox{\plotpoint}}
\put(502,475){\usebox{\plotpoint}}
\put(503,476){\usebox{\plotpoint}}
\put(504,477){\usebox{\plotpoint}}
\put(505,478){\usebox{\plotpoint}}
\put(506,479){\usebox{\plotpoint}}
\put(507,480){\usebox{\plotpoint}}
\put(508,481){\usebox{\plotpoint}}
\put(509,482){\usebox{\plotpoint}}
\put(510,484){\usebox{\plotpoint}}
\put(511,485){\usebox{\plotpoint}}
\put(512,486){\usebox{\plotpoint}}
\put(512,487){\usebox{\plotpoint}}
\put(513,488){\usebox{\plotpoint}}
\put(514,489){\usebox{\plotpoint}}
\put(515,490){\usebox{\plotpoint}}
\put(516,491){\usebox{\plotpoint}}
\put(517,492){\usebox{\plotpoint}}
\put(518,493){\usebox{\plotpoint}}
\put(519,494){\usebox{\plotpoint}}
\put(520,495){\usebox{\plotpoint}}
\put(521,496){\usebox{\plotpoint}}
\put(522,497){\usebox{\plotpoint}}
\put(523,498){\usebox{\plotpoint}}
\put(524,499){\usebox{\plotpoint}}
\put(524,500){\usebox{\plotpoint}}
\put(525,501){\usebox{\plotpoint}}
\put(526,502){\usebox{\plotpoint}}
\put(527,503){\usebox{\plotpoint}}
\put(528,504){\usebox{\plotpoint}}
\put(529,505){\usebox{\plotpoint}}
\put(530,506){\usebox{\plotpoint}}
\put(531,507){\usebox{\plotpoint}}
\put(532,508){\usebox{\plotpoint}}
\put(533,509){\usebox{\plotpoint}}
\put(534,510){\usebox{\plotpoint}}
\put(535,511){\usebox{\plotpoint}}
\put(536,512){\usebox{\plotpoint}}
\put(537,514){\usebox{\plotpoint}}
\put(538,515){\usebox{\plotpoint}}
\put(539,516){\usebox{\plotpoint}}
\put(540,517){\usebox{\plotpoint}}
\put(541,519){\usebox{\plotpoint}}
\put(542,520){\usebox{\plotpoint}}
\put(543,521){\usebox{\plotpoint}}
\put(544,522){\usebox{\plotpoint}}
\put(545,523){\usebox{\plotpoint}}
\put(546,524){\usebox{\plotpoint}}
\put(547,526){\usebox{\plotpoint}}
\put(548,527){\usebox{\plotpoint}}
\put(549,528){\usebox{\plotpoint}}
\put(550,529){\usebox{\plotpoint}}
\put(551,530){\usebox{\plotpoint}}
\put(552,531){\usebox{\plotpoint}}
\put(553,532){\usebox{\plotpoint}}
\put(554,533){\usebox{\plotpoint}}
\put(555,534){\usebox{\plotpoint}}
\put(556,535){\usebox{\plotpoint}}
\put(557,536){\usebox{\plotpoint}}
\put(558,537){\usebox{\plotpoint}}
\put(559,539){\usebox{\plotpoint}}
\put(560,540){\usebox{\plotpoint}}
\put(561,541){\usebox{\plotpoint}}
\put(562,543){\usebox{\plotpoint}}
\put(563,544){\usebox{\plotpoint}}
\put(564,546){\usebox{\plotpoint}}
\put(565,547){\usebox{\plotpoint}}
\put(566,548){\usebox{\plotpoint}}
\put(567,549){\usebox{\plotpoint}}
\put(568,550){\usebox{\plotpoint}}
\put(569,551){\usebox{\plotpoint}}
\put(570,552){\usebox{\plotpoint}}
\put(571,553){\usebox{\plotpoint}}
\put(572,554){\usebox{\plotpoint}}
\put(573,555){\usebox{\plotpoint}}
\put(574,556){\usebox{\plotpoint}}
\put(575,557){\usebox{\plotpoint}}
\put(576,559){\usebox{\plotpoint}}
\put(577,560){\usebox{\plotpoint}}
\put(578,561){\usebox{\plotpoint}}
\put(579,562){\usebox{\plotpoint}}
\put(580,563){\usebox{\plotpoint}}
\put(581,564){\usebox{\plotpoint}}
\put(582,566){\usebox{\plotpoint}}
\put(583,567){\usebox{\plotpoint}}
\put(584,568){\usebox{\plotpoint}}
\put(585,569){\usebox{\plotpoint}}
\put(586,570){\usebox{\plotpoint}}
\put(587,571){\usebox{\plotpoint}}
\put(588,573){\usebox{\plotpoint}}
\put(589,574){\usebox{\plotpoint}}
\put(590,575){\usebox{\plotpoint}}
\put(591,577){\usebox{\plotpoint}}
\put(592,578){\usebox{\plotpoint}}
\put(593,580){\usebox{\plotpoint}}
\put(594,581){\usebox{\plotpoint}}
\put(595,582){\usebox{\plotpoint}}
\put(596,583){\usebox{\plotpoint}}
\put(597,584){\usebox{\plotpoint}}
\put(598,585){\usebox{\plotpoint}}
\put(599,587){\usebox{\plotpoint}}
\put(600,588){\usebox{\plotpoint}}
\put(601,589){\usebox{\plotpoint}}
\put(602,590){\usebox{\plotpoint}}
\put(603,591){\usebox{\plotpoint}}
\put(604,592){\usebox{\plotpoint}}
\put(605,594){\usebox{\plotpoint}}
\put(606,595){\usebox{\plotpoint}}
\put(607,596){\usebox{\plotpoint}}
\put(608,597){\usebox{\plotpoint}}
\put(609,598){\usebox{\plotpoint}}
\put(610,599){\usebox{\plotpoint}}
\put(611,601){\usebox{\plotpoint}}
\put(612,602){\usebox{\plotpoint}}
\put(613,603){\usebox{\plotpoint}}
\put(614,605){\usebox{\plotpoint}}
\put(615,606){\usebox{\plotpoint}}
\put(616,608){\usebox{\plotpoint}}
\put(617,609){\usebox{\plotpoint}}
\put(618,610){\usebox{\plotpoint}}
\put(619,611){\usebox{\plotpoint}}
\put(620,612){\usebox{\plotpoint}}
\put(621,613){\usebox{\plotpoint}}
\put(622,615){\usebox{\plotpoint}}
\put(623,616){\usebox{\plotpoint}}
\put(624,617){\usebox{\plotpoint}}
\put(625,618){\usebox{\plotpoint}}
\put(626,619){\usebox{\plotpoint}}
\put(627,620){\usebox{\plotpoint}}
\put(628,622){\usebox{\plotpoint}}
\put(629,623){\usebox{\plotpoint}}
\put(630,624){\usebox{\plotpoint}}
\put(631,625){\usebox{\plotpoint}}
\put(632,626){\usebox{\plotpoint}}
\put(633,627){\usebox{\plotpoint}}
\put(634,629){\usebox{\plotpoint}}
\put(635,630){\usebox{\plotpoint}}
\put(636,631){\usebox{\plotpoint}}
\put(637,632){\usebox{\plotpoint}}
\put(638,633){\usebox{\plotpoint}}
\put(639,634){\usebox{\plotpoint}}
\put(640,636){\usebox{\plotpoint}}
\put(641,637){\usebox{\plotpoint}}
\put(642,639){\usebox{\plotpoint}}
\put(643,640){\usebox{\plotpoint}}
\put(644,642){\usebox{\plotpoint}}
\put(645,643){\usebox{\plotpoint}}
\put(646,645){\usebox{\plotpoint}}
\put(647,646){\usebox{\plotpoint}}
\put(648,647){\usebox{\plotpoint}}
\put(649,648){\usebox{\plotpoint}}
\put(650,649){\usebox{\plotpoint}}
\put(651,651){\usebox{\plotpoint}}
\put(652,652){\usebox{\plotpoint}}
\put(653,653){\usebox{\plotpoint}}
\put(654,654){\usebox{\plotpoint}}
\put(655,655){\usebox{\plotpoint}}
\put(656,656){\usebox{\plotpoint}}
\put(657,658){\usebox{\plotpoint}}
\put(658,659){\usebox{\plotpoint}}
\put(659,660){\usebox{\plotpoint}}
\put(660,661){\usebox{\plotpoint}}
\put(661,663){\usebox{\plotpoint}}
\put(662,664){\usebox{\plotpoint}}
\put(663,665){\usebox{\plotpoint}}
\put(664,667){\usebox{\plotpoint}}
\put(665,668){\usebox{\plotpoint}}
\put(666,670){\usebox{\plotpoint}}
\put(667,671){\usebox{\plotpoint}}
\put(668,673){\usebox{\plotpoint}}
\put(669,674){\usebox{\plotpoint}}
\put(670,675){\usebox{\plotpoint}}
\put(671,676){\usebox{\plotpoint}}
\put(672,678){\usebox{\plotpoint}}
\put(673,679){\usebox{\plotpoint}}
\put(674,680){\usebox{\plotpoint}}
\put(675,682){\usebox{\plotpoint}}
\put(676,683){\usebox{\plotpoint}}
\put(677,684){\usebox{\plotpoint}}
\put(678,685){\usebox{\plotpoint}}
\put(679,686){\usebox{\plotpoint}}
\put(680,688){\usebox{\plotpoint}}
\put(681,689){\usebox{\plotpoint}}
\put(682,690){\usebox{\plotpoint}}
\put(683,691){\usebox{\plotpoint}}
\put(684,693){\usebox{\plotpoint}}
\put(685,694){\usebox{\plotpoint}}
\put(686,695){\usebox{\plotpoint}}
\put(687,697){\usebox{\plotpoint}}
\put(688,698){\usebox{\plotpoint}}
\put(689,699){\usebox{\plotpoint}}
\put(690,701){\usebox{\plotpoint}}
\put(691,702){\usebox{\plotpoint}}
\put(692,703){\usebox{\plotpoint}}
\put(693,705){\usebox{\plotpoint}}
\put(694,707){\usebox{\plotpoint}}
\put(695,708){\usebox{\plotpoint}}
\put(696,710){\usebox{\plotpoint}}
\put(697,711){\usebox{\plotpoint}}
\put(698,713){\usebox{\plotpoint}}
\put(699,714){\usebox{\plotpoint}}
\put(700,715){\usebox{\plotpoint}}
\put(701,716){\usebox{\plotpoint}}
\put(702,717){\usebox{\plotpoint}}
\put(703,719){\usebox{\plotpoint}}
\put(704,720){\usebox{\plotpoint}}
\put(705,721){\usebox{\plotpoint}}
\put(706,722){\usebox{\plotpoint}}
\put(707,724){\usebox{\plotpoint}}
\put(708,725){\usebox{\plotpoint}}
\put(709,726){\usebox{\plotpoint}}
\put(710,728){\usebox{\plotpoint}}
\put(711,729){\usebox{\plotpoint}}
\put(712,730){\usebox{\plotpoint}}
\put(713,732){\usebox{\plotpoint}}
\put(714,733){\usebox{\plotpoint}}
\put(715,734){\usebox{\plotpoint}}
\put(716,736){\usebox{\plotpoint}}
\put(717,738){\usebox{\plotpoint}}
\put(718,739){\usebox{\plotpoint}}
\put(719,741){\usebox{\plotpoint}}
\put(720,742){\usebox{\plotpoint}}
\put(721,744){\usebox{\plotpoint}}
\put(722,745){\usebox{\plotpoint}}
\put(723,746){\usebox{\plotpoint}}
\put(724,748){\usebox{\plotpoint}}
\put(725,749){\usebox{\plotpoint}}
\put(726,750){\usebox{\plotpoint}}
\put(727,752){\usebox{\plotpoint}}
\put(728,754){\usebox{\plotpoint}}
\put(729,755){\usebox{\plotpoint}}
\put(730,757){\usebox{\plotpoint}}
\put(731,758){\usebox{\plotpoint}}
\put(732,760){\usebox{\plotpoint}}
\put(733,761){\usebox{\plotpoint}}
\put(734,762){\usebox{\plotpoint}}
\put(735,763){\usebox{\plotpoint}}
\put(736,765){\usebox{\plotpoint}}
\put(737,766){\usebox{\plotpoint}}
\put(738,767){\usebox{\plotpoint}}
\put(739,769){\usebox{\plotpoint}}
\put(740,770){\usebox{\plotpoint}}
\put(741,771){\usebox{\plotpoint}}
\put(742,773){\usebox{\plotpoint}}
\put(743,774){\usebox{\plotpoint}}
\put(744,775){\usebox{\plotpoint}}
\put(745,777){\usebox{\plotpoint}}
\put(746,779){\usebox{\plotpoint}}
\put(747,780){\usebox{\plotpoint}}
\put(748,782){\usebox{\plotpoint}}
\put(749,783){\usebox{\plotpoint}}
\put(750,785){\usebox{\plotpoint}}
\put(751,787){\usebox{\plotpoint}}
\put(752,788){\usebox{\plotpoint}}
\put(753,790){\usebox{\plotpoint}}
\put(754,791){\usebox{\plotpoint}}
\put(755,793){\usebox{\plotpoint}}
\put(756,794){\usebox{\plotpoint}}
\put(757,795){\usebox{\plotpoint}}
\put(758,796){\usebox{\plotpoint}}
\put(759,798){\usebox{\plotpoint}}
\put(760,799){\usebox{\plotpoint}}
\put(761,800){\usebox{\plotpoint}}
\put(762,802){\usebox{\plotpoint}}
\put(763,804){\usebox{\plotpoint}}
\put(764,805){\usebox{\plotpoint}}
\put(765,807){\usebox{\plotpoint}}
\put(766,808){\usebox{\plotpoint}}
\put(767,810){\usebox{\plotpoint}}
\put(768,811){\usebox{\plotpoint}}
\put(769,813){\usebox{\plotpoint}}
\put(770,815){\usebox{\plotpoint}}
\put(771,817){\usebox{\plotpoint}}
\put(772,818){\usebox{\plotpoint}}
\put(773,820){\usebox{\plotpoint}}
\put(774,821){\usebox{\plotpoint}}
\put(775,822){\usebox{\plotpoint}}
\put(776,824){\usebox{\plotpoint}}
\put(777,825){\usebox{\plotpoint}}
\put(778,826){\usebox{\plotpoint}}
\put(779,828){\usebox{\plotpoint}}
\put(780,830){\usebox{\plotpoint}}
\put(781,831){\usebox{\plotpoint}}
\put(782,833){\usebox{\plotpoint}}
\put(783,834){\usebox{\plotpoint}}
\put(784,836){\usebox{\plotpoint}}
\put(785,837){\usebox{\plotpoint}}
\put(786,839){\usebox{\plotpoint}}
\put(787,840){\usebox{\plotpoint}}
\put(788,842){\usebox{\plotpoint}}
\put(789,843){\usebox{\plotpoint}}
\put(790,845){\usebox{\plotpoint}}
\put(791,846){\usebox{\plotpoint}}
\put(792,848){\usebox{\plotpoint}}
\put(793,849){\usebox{\plotpoint}}
\put(794,851){\usebox{\plotpoint}}
\put(795,852){\usebox{\plotpoint}}
\put(796,854){\usebox{\plotpoint}}
\put(797,855){\usebox{\plotpoint}}
\put(798,857){\usebox{\plotpoint}}
\put(799,859){\usebox{\plotpoint}}
\put(800,861){\usebox{\plotpoint}}
\put(801,862){\usebox{\plotpoint}}
\put(802,864){\usebox{\plotpoint}}
\put(803,866){\usebox{\plotpoint}}
\put(804,867){\usebox{\plotpoint}}
\put(805,869){\usebox{\plotpoint}}
\put(806,870){\usebox{\plotpoint}}
\put(807,872){\usebox{\plotpoint}}
\put(808,873){\usebox{\plotpoint}}
\put(809,875){\usebox{\plotpoint}}
\put(810,876){\usebox{\plotpoint}}
\put(811,878){\usebox{\plotpoint}}
\put(812,879){\usebox{\plotpoint}}
\put(813,881){\usebox{\plotpoint}}
\put(814,882){\usebox{\plotpoint}}
\put(815,884){\usebox{\plotpoint}}
\put(816,885){\usebox{\plotpoint}}
\put(817,887){\usebox{\plotpoint}}
\put(818,888){\usebox{\plotpoint}}
\put(819,890){\usebox{\plotpoint}}
\put(820,891){\usebox{\plotpoint}}
\put(821,893){\usebox{\plotpoint}}
\put(822,895){\usebox{\plotpoint}}
\put(823,897){\usebox{\plotpoint}}
\put(824,898){\usebox{\plotpoint}}
\put(825,900){\usebox{\plotpoint}}
\put(826,902){\usebox{\plotpoint}}
\put(827,903){\usebox{\plotpoint}}
\put(828,905){\usebox{\plotpoint}}
\put(829,906){\usebox{\plotpoint}}
\put(830,908){\usebox{\plotpoint}}
\put(831,909){\usebox{\plotpoint}}
\put(832,911){\usebox{\plotpoint}}
\put(833,912){\usebox{\plotpoint}}
\put(834,914){\usebox{\plotpoint}}
\put(835,915){\usebox{\plotpoint}}
\sbox{\plotpoint}{\rule[-0.175pt]{0.350pt}{0.350pt}}%
\put(455,1046){\makebox(0,0)[r]{$\tilde{c}=0.5$}}
\put(477,1046){\rule[-0.175pt]{15.899pt}{0.350pt}}
\put(264,227){\usebox{\plotpoint}}
\put(264,227){\rule[-0.175pt]{0.361pt}{0.350pt}}
\put(265,228){\rule[-0.175pt]{0.361pt}{0.350pt}}
\put(267,229){\rule[-0.175pt]{0.361pt}{0.350pt}}
\put(268,230){\rule[-0.175pt]{0.361pt}{0.350pt}}
\put(270,231){\rule[-0.175pt]{0.361pt}{0.350pt}}
\put(271,232){\rule[-0.175pt]{0.361pt}{0.350pt}}
\put(273,233){\rule[-0.175pt]{0.361pt}{0.350pt}}
\put(274,234){\rule[-0.175pt]{0.361pt}{0.350pt}}
\put(276,235){\usebox{\plotpoint}}
\put(277,236){\usebox{\plotpoint}}
\put(278,237){\usebox{\plotpoint}}
\put(279,238){\usebox{\plotpoint}}
\put(280,239){\usebox{\plotpoint}}
\put(281,240){\rule[-0.175pt]{0.361pt}{0.350pt}}
\put(282,241){\rule[-0.175pt]{0.361pt}{0.350pt}}
\put(284,242){\rule[-0.175pt]{0.361pt}{0.350pt}}
\put(285,243){\rule[-0.175pt]{0.361pt}{0.350pt}}
\put(287,244){\usebox{\plotpoint}}
\put(288,245){\usebox{\plotpoint}}
\put(289,246){\usebox{\plotpoint}}
\put(290,247){\usebox{\plotpoint}}
\put(291,248){\usebox{\plotpoint}}
\put(293,249){\rule[-0.175pt]{0.361pt}{0.350pt}}
\put(294,250){\rule[-0.175pt]{0.361pt}{0.350pt}}
\put(296,251){\rule[-0.175pt]{0.361pt}{0.350pt}}
\put(297,252){\rule[-0.175pt]{0.361pt}{0.350pt}}
\put(299,253){\usebox{\plotpoint}}
\put(300,254){\usebox{\plotpoint}}
\put(301,255){\usebox{\plotpoint}}
\put(302,256){\usebox{\plotpoint}}
\put(303,257){\usebox{\plotpoint}}
\put(304,258){\rule[-0.175pt]{0.361pt}{0.350pt}}
\put(305,259){\rule[-0.175pt]{0.361pt}{0.350pt}}
\put(307,260){\rule[-0.175pt]{0.361pt}{0.350pt}}
\put(308,261){\rule[-0.175pt]{0.361pt}{0.350pt}}
\put(310,262){\usebox{\plotpoint}}
\put(311,263){\usebox{\plotpoint}}
\put(312,264){\usebox{\plotpoint}}
\put(313,265){\usebox{\plotpoint}}
\put(314,266){\usebox{\plotpoint}}
\put(316,267){\rule[-0.175pt]{0.361pt}{0.350pt}}
\put(317,268){\rule[-0.175pt]{0.361pt}{0.350pt}}
\put(319,269){\rule[-0.175pt]{0.361pt}{0.350pt}}
\put(320,270){\rule[-0.175pt]{0.361pt}{0.350pt}}
\put(322,271){\usebox{\plotpoint}}
\put(323,272){\usebox{\plotpoint}}
\put(324,273){\usebox{\plotpoint}}
\put(325,274){\usebox{\plotpoint}}
\put(326,275){\usebox{\plotpoint}}
\put(328,276){\usebox{\plotpoint}}
\put(329,277){\usebox{\plotpoint}}
\put(330,278){\usebox{\plotpoint}}
\put(331,279){\usebox{\plotpoint}}
\put(333,280){\usebox{\plotpoint}}
\put(334,281){\usebox{\plotpoint}}
\put(335,282){\usebox{\plotpoint}}
\put(336,283){\usebox{\plotpoint}}
\put(337,284){\usebox{\plotpoint}}
\put(339,285){\usebox{\plotpoint}}
\put(340,286){\usebox{\plotpoint}}
\put(341,287){\usebox{\plotpoint}}
\put(342,288){\usebox{\plotpoint}}
\put(343,289){\usebox{\plotpoint}}
\put(345,290){\rule[-0.175pt]{0.361pt}{0.350pt}}
\put(346,291){\rule[-0.175pt]{0.361pt}{0.350pt}}
\put(348,292){\rule[-0.175pt]{0.361pt}{0.350pt}}
\put(349,293){\rule[-0.175pt]{0.361pt}{0.350pt}}
\put(351,294){\usebox{\plotpoint}}
\put(352,295){\usebox{\plotpoint}}
\put(353,296){\usebox{\plotpoint}}
\put(354,297){\usebox{\plotpoint}}
\put(355,298){\usebox{\plotpoint}}
\put(356,299){\usebox{\plotpoint}}
\put(357,300){\usebox{\plotpoint}}
\put(358,301){\usebox{\plotpoint}}
\put(359,302){\usebox{\plotpoint}}
\put(360,303){\usebox{\plotpoint}}
\put(362,304){\rule[-0.175pt]{0.361pt}{0.350pt}}
\put(363,305){\rule[-0.175pt]{0.361pt}{0.350pt}}
\put(365,306){\rule[-0.175pt]{0.361pt}{0.350pt}}
\put(366,307){\rule[-0.175pt]{0.361pt}{0.350pt}}
\put(368,308){\usebox{\plotpoint}}
\put(369,309){\usebox{\plotpoint}}
\put(370,310){\usebox{\plotpoint}}
\put(371,311){\usebox{\plotpoint}}
\put(372,312){\usebox{\plotpoint}}
\put(374,313){\usebox{\plotpoint}}
\put(375,314){\usebox{\plotpoint}}
\put(376,315){\usebox{\plotpoint}}
\put(377,316){\usebox{\plotpoint}}
\put(378,317){\usebox{\plotpoint}}
\put(380,318){\usebox{\plotpoint}}
\put(381,319){\usebox{\plotpoint}}
\put(382,320){\usebox{\plotpoint}}
\put(383,321){\usebox{\plotpoint}}
\put(384,322){\usebox{\plotpoint}}
\put(385,323){\rule[-0.175pt]{0.361pt}{0.350pt}}
\put(386,324){\rule[-0.175pt]{0.361pt}{0.350pt}}
\put(388,325){\rule[-0.175pt]{0.361pt}{0.350pt}}
\put(389,326){\rule[-0.175pt]{0.361pt}{0.350pt}}
\put(391,327){\usebox{\plotpoint}}
\put(392,328){\usebox{\plotpoint}}
\put(393,329){\usebox{\plotpoint}}
\put(394,330){\usebox{\plotpoint}}
\put(395,331){\usebox{\plotpoint}}
\put(397,332){\usebox{\plotpoint}}
\put(398,333){\usebox{\plotpoint}}
\put(399,334){\usebox{\plotpoint}}
\put(400,335){\usebox{\plotpoint}}
\put(401,336){\usebox{\plotpoint}}
\put(403,337){\usebox{\plotpoint}}
\put(404,338){\usebox{\plotpoint}}
\put(405,339){\usebox{\plotpoint}}
\put(406,340){\usebox{\plotpoint}}
\put(407,341){\usebox{\plotpoint}}
\put(408,342){\usebox{\plotpoint}}
\put(409,343){\usebox{\plotpoint}}
\put(410,344){\usebox{\plotpoint}}
\put(411,345){\usebox{\plotpoint}}
\put(412,346){\usebox{\plotpoint}}
\put(414,347){\usebox{\plotpoint}}
\put(415,348){\usebox{\plotpoint}}
\put(416,349){\usebox{\plotpoint}}
\put(417,350){\usebox{\plotpoint}}
\put(418,351){\usebox{\plotpoint}}
\put(420,352){\usebox{\plotpoint}}
\put(421,353){\usebox{\plotpoint}}
\put(422,354){\usebox{\plotpoint}}
\put(423,355){\usebox{\plotpoint}}
\put(424,356){\usebox{\plotpoint}}
\put(426,357){\usebox{\plotpoint}}
\put(427,358){\usebox{\plotpoint}}
\put(428,359){\usebox{\plotpoint}}
\put(429,360){\usebox{\plotpoint}}
\put(430,361){\usebox{\plotpoint}}
\put(432,362){\usebox{\plotpoint}}
\put(433,363){\usebox{\plotpoint}}
\put(434,364){\usebox{\plotpoint}}
\put(435,365){\usebox{\plotpoint}}
\put(436,366){\usebox{\plotpoint}}
\put(437,367){\usebox{\plotpoint}}
\put(438,368){\usebox{\plotpoint}}
\put(439,369){\usebox{\plotpoint}}
\put(440,370){\usebox{\plotpoint}}
\put(441,371){\usebox{\plotpoint}}
\put(443,372){\usebox{\plotpoint}}
\put(444,373){\usebox{\plotpoint}}
\put(445,374){\usebox{\plotpoint}}
\put(446,375){\usebox{\plotpoint}}
\put(447,376){\usebox{\plotpoint}}
\put(449,377){\usebox{\plotpoint}}
\put(450,378){\usebox{\plotpoint}}
\put(451,379){\usebox{\plotpoint}}
\put(452,380){\usebox{\plotpoint}}
\put(453,381){\usebox{\plotpoint}}
\put(455,382){\usebox{\plotpoint}}
\put(456,383){\usebox{\plotpoint}}
\put(457,384){\usebox{\plotpoint}}
\put(458,385){\usebox{\plotpoint}}
\put(459,386){\usebox{\plotpoint}}
\put(460,387){\usebox{\plotpoint}}
\put(461,388){\usebox{\plotpoint}}
\put(462,389){\usebox{\plotpoint}}
\put(463,390){\usebox{\plotpoint}}
\put(464,391){\usebox{\plotpoint}}
\put(466,392){\usebox{\plotpoint}}
\put(467,393){\usebox{\plotpoint}}
\put(468,394){\usebox{\plotpoint}}
\put(469,395){\usebox{\plotpoint}}
\put(470,396){\usebox{\plotpoint}}
\put(471,397){\usebox{\plotpoint}}
\put(472,398){\usebox{\plotpoint}}
\put(473,399){\usebox{\plotpoint}}
\put(474,400){\usebox{\plotpoint}}
\put(475,401){\usebox{\plotpoint}}
\put(476,402){\usebox{\plotpoint}}
\put(478,403){\usebox{\plotpoint}}
\put(479,404){\usebox{\plotpoint}}
\put(480,405){\usebox{\plotpoint}}
\put(481,406){\usebox{\plotpoint}}
\put(482,407){\usebox{\plotpoint}}
\put(484,408){\usebox{\plotpoint}}
\put(485,409){\usebox{\plotpoint}}
\put(486,410){\usebox{\plotpoint}}
\put(487,411){\usebox{\plotpoint}}
\put(488,412){\usebox{\plotpoint}}
\put(489,413){\usebox{\plotpoint}}
\put(490,414){\usebox{\plotpoint}}
\put(491,415){\usebox{\plotpoint}}
\put(492,416){\usebox{\plotpoint}}
\put(493,417){\usebox{\plotpoint}}
\put(494,418){\usebox{\plotpoint}}
\put(495,419){\usebox{\plotpoint}}
\put(496,420){\usebox{\plotpoint}}
\put(497,421){\usebox{\plotpoint}}
\put(498,422){\usebox{\plotpoint}}
\put(499,423){\usebox{\plotpoint}}
\put(501,424){\usebox{\plotpoint}}
\put(502,425){\usebox{\plotpoint}}
\put(503,426){\usebox{\plotpoint}}
\put(504,427){\usebox{\plotpoint}}
\put(505,428){\usebox{\plotpoint}}
\put(507,429){\usebox{\plotpoint}}
\put(508,430){\usebox{\plotpoint}}
\put(509,431){\usebox{\plotpoint}}
\put(510,432){\usebox{\plotpoint}}
\put(511,433){\usebox{\plotpoint}}
\put(512,435){\usebox{\plotpoint}}
\put(513,436){\usebox{\plotpoint}}
\put(514,437){\usebox{\plotpoint}}
\put(515,438){\usebox{\plotpoint}}
\put(516,439){\usebox{\plotpoint}}
\put(518,440){\usebox{\plotpoint}}
\put(519,441){\usebox{\plotpoint}}
\put(520,442){\usebox{\plotpoint}}
\put(521,443){\usebox{\plotpoint}}
\put(522,444){\usebox{\plotpoint}}
\put(523,445){\usebox{\plotpoint}}
\put(524,446){\usebox{\plotpoint}}
\put(525,447){\usebox{\plotpoint}}
\put(526,448){\usebox{\plotpoint}}
\put(527,449){\usebox{\plotpoint}}
\put(528,450){\usebox{\plotpoint}}
\put(530,451){\usebox{\plotpoint}}
\put(531,452){\usebox{\plotpoint}}
\put(532,453){\usebox{\plotpoint}}
\put(533,454){\usebox{\plotpoint}}
\put(534,455){\usebox{\plotpoint}}
\put(535,456){\usebox{\plotpoint}}
\put(536,457){\usebox{\plotpoint}}
\put(537,458){\usebox{\plotpoint}}
\put(538,459){\usebox{\plotpoint}}
\put(539,460){\usebox{\plotpoint}}
\put(540,461){\usebox{\plotpoint}}
\put(541,463){\usebox{\plotpoint}}
\put(542,464){\usebox{\plotpoint}}
\put(543,465){\usebox{\plotpoint}}
\put(544,466){\usebox{\plotpoint}}
\put(545,467){\usebox{\plotpoint}}
\put(547,468){\usebox{\plotpoint}}
\put(548,469){\usebox{\plotpoint}}
\put(549,470){\usebox{\plotpoint}}
\put(550,471){\usebox{\plotpoint}}
\put(551,472){\usebox{\plotpoint}}
\put(552,473){\usebox{\plotpoint}}
\put(553,474){\usebox{\plotpoint}}
\put(554,475){\usebox{\plotpoint}}
\put(555,476){\usebox{\plotpoint}}
\put(556,477){\usebox{\plotpoint}}
\put(557,478){\usebox{\plotpoint}}
\put(558,479){\usebox{\plotpoint}}
\put(559,480){\usebox{\plotpoint}}
\put(560,481){\usebox{\plotpoint}}
\put(561,482){\usebox{\plotpoint}}
\put(562,483){\usebox{\plotpoint}}
\put(563,484){\usebox{\plotpoint}}
\put(564,485){\usebox{\plotpoint}}
\put(565,486){\usebox{\plotpoint}}
\put(566,487){\usebox{\plotpoint}}
\put(567,488){\usebox{\plotpoint}}
\put(568,489){\usebox{\plotpoint}}
\put(569,490){\usebox{\plotpoint}}
\put(570,491){\usebox{\plotpoint}}
\put(571,492){\usebox{\plotpoint}}
\put(572,493){\usebox{\plotpoint}}
\put(573,494){\usebox{\plotpoint}}
\put(574,495){\usebox{\plotpoint}}
\put(575,496){\usebox{\plotpoint}}
\put(576,497){\usebox{\plotpoint}}
\put(577,498){\usebox{\plotpoint}}
\put(578,499){\usebox{\plotpoint}}
\put(579,500){\usebox{\plotpoint}}
\put(580,501){\usebox{\plotpoint}}
\put(581,502){\usebox{\plotpoint}}
\put(582,503){\usebox{\plotpoint}}
\put(583,504){\usebox{\plotpoint}}
\put(584,505){\usebox{\plotpoint}}
\put(585,506){\usebox{\plotpoint}}
\put(586,507){\usebox{\plotpoint}}
\put(587,508){\usebox{\plotpoint}}
\put(588,509){\usebox{\plotpoint}}
\put(589,510){\usebox{\plotpoint}}
\put(590,511){\usebox{\plotpoint}}
\put(591,512){\usebox{\plotpoint}}
\put(592,513){\usebox{\plotpoint}}
\put(593,515){\usebox{\plotpoint}}
\put(594,516){\usebox{\plotpoint}}
\put(595,517){\usebox{\plotpoint}}
\put(596,518){\usebox{\plotpoint}}
\put(597,519){\usebox{\plotpoint}}
\put(598,520){\usebox{\plotpoint}}
\put(599,521){\usebox{\plotpoint}}
\put(600,522){\usebox{\plotpoint}}
\put(601,523){\usebox{\plotpoint}}
\put(602,524){\usebox{\plotpoint}}
\put(603,525){\usebox{\plotpoint}}
\put(604,526){\usebox{\plotpoint}}
\put(605,527){\usebox{\plotpoint}}
\put(606,528){\usebox{\plotpoint}}
\put(607,529){\usebox{\plotpoint}}
\put(608,530){\usebox{\plotpoint}}
\put(609,531){\usebox{\plotpoint}}
\put(610,532){\usebox{\plotpoint}}
\put(611,533){\usebox{\plotpoint}}
\put(612,534){\usebox{\plotpoint}}
\put(613,535){\usebox{\plotpoint}}
\put(614,537){\usebox{\plotpoint}}
\put(615,538){\usebox{\plotpoint}}
\put(616,540){\usebox{\plotpoint}}
\put(617,541){\usebox{\plotpoint}}
\put(618,542){\usebox{\plotpoint}}
\put(619,543){\usebox{\plotpoint}}
\put(620,544){\usebox{\plotpoint}}
\put(621,545){\usebox{\plotpoint}}
\put(622,546){\usebox{\plotpoint}}
\put(623,547){\usebox{\plotpoint}}
\put(624,548){\usebox{\plotpoint}}
\put(625,549){\usebox{\plotpoint}}
\put(626,550){\usebox{\plotpoint}}
\put(627,551){\usebox{\plotpoint}}
\put(628,552){\usebox{\plotpoint}}
\put(629,553){\usebox{\plotpoint}}
\put(630,554){\usebox{\plotpoint}}
\put(631,555){\usebox{\plotpoint}}
\put(632,556){\usebox{\plotpoint}}
\put(633,557){\usebox{\plotpoint}}
\put(634,559){\usebox{\plotpoint}}
\put(635,560){\usebox{\plotpoint}}
\put(636,561){\usebox{\plotpoint}}
\put(637,562){\usebox{\plotpoint}}
\put(638,563){\usebox{\plotpoint}}
\put(639,564){\usebox{\plotpoint}}
\put(640,565){\usebox{\plotpoint}}
\put(641,566){\usebox{\plotpoint}}
\put(642,567){\usebox{\plotpoint}}
\put(643,569){\usebox{\plotpoint}}
\put(644,570){\usebox{\plotpoint}}
\put(645,572){\usebox{\plotpoint}}
\put(646,573){\usebox{\plotpoint}}
\put(647,574){\usebox{\plotpoint}}
\put(648,575){\usebox{\plotpoint}}
\put(649,576){\usebox{\plotpoint}}
\put(650,577){\usebox{\plotpoint}}
\put(651,578){\usebox{\plotpoint}}
\put(652,579){\usebox{\plotpoint}}
\put(653,580){\usebox{\plotpoint}}
\put(654,581){\usebox{\plotpoint}}
\put(655,582){\usebox{\plotpoint}}
\put(656,583){\usebox{\plotpoint}}
\put(657,585){\usebox{\plotpoint}}
\put(658,586){\usebox{\plotpoint}}
\put(659,587){\usebox{\plotpoint}}
\put(660,588){\usebox{\plotpoint}}
\put(661,589){\usebox{\plotpoint}}
\put(662,590){\usebox{\plotpoint}}
\put(663,591){\usebox{\plotpoint}}
\put(664,592){\usebox{\plotpoint}}
\put(665,593){\usebox{\plotpoint}}
\put(666,595){\usebox{\plotpoint}}
\put(667,596){\usebox{\plotpoint}}
\put(668,598){\usebox{\plotpoint}}
\put(669,599){\usebox{\plotpoint}}
\put(670,600){\usebox{\plotpoint}}
\put(671,601){\usebox{\plotpoint}}
\put(672,602){\usebox{\plotpoint}}
\put(673,603){\usebox{\plotpoint}}
\put(674,605){\usebox{\plotpoint}}
\put(675,606){\usebox{\plotpoint}}
\put(676,607){\usebox{\plotpoint}}
\put(677,608){\usebox{\plotpoint}}
\put(678,609){\usebox{\plotpoint}}
\put(679,610){\usebox{\plotpoint}}
\put(680,612){\usebox{\plotpoint}}
\put(681,613){\usebox{\plotpoint}}
\put(682,614){\usebox{\plotpoint}}
\put(683,615){\usebox{\plotpoint}}
\put(684,616){\usebox{\plotpoint}}
\put(685,617){\usebox{\plotpoint}}
\put(686,619){\usebox{\plotpoint}}
\put(687,620){\usebox{\plotpoint}}
\put(688,621){\usebox{\plotpoint}}
\put(689,622){\usebox{\plotpoint}}
\put(690,623){\usebox{\plotpoint}}
\put(691,624){\usebox{\plotpoint}}
\put(692,626){\usebox{\plotpoint}}
\put(693,627){\usebox{\plotpoint}}
\put(694,628){\usebox{\plotpoint}}
\put(695,630){\usebox{\plotpoint}}
\put(696,631){\usebox{\plotpoint}}
\put(697,633){\usebox{\plotpoint}}
\put(698,634){\usebox{\plotpoint}}
\put(699,635){\usebox{\plotpoint}}
\put(700,636){\usebox{\plotpoint}}
\put(701,637){\usebox{\plotpoint}}
\put(702,638){\usebox{\plotpoint}}
\put(703,640){\usebox{\plotpoint}}
\put(704,641){\usebox{\plotpoint}}
\put(705,642){\usebox{\plotpoint}}
\put(706,643){\usebox{\plotpoint}}
\put(707,644){\usebox{\plotpoint}}
\put(708,645){\usebox{\plotpoint}}
\put(709,647){\usebox{\plotpoint}}
\put(710,648){\usebox{\plotpoint}}
\put(711,649){\usebox{\plotpoint}}
\put(712,650){\usebox{\plotpoint}}
\put(713,651){\usebox{\plotpoint}}
\put(714,652){\usebox{\plotpoint}}
\put(715,654){\rule[-0.175pt]{0.350pt}{0.385pt}}
\put(716,655){\rule[-0.175pt]{0.350pt}{0.385pt}}
\put(717,657){\rule[-0.175pt]{0.350pt}{0.385pt}}
\put(718,658){\rule[-0.175pt]{0.350pt}{0.385pt}}
\put(719,660){\rule[-0.175pt]{0.350pt}{0.385pt}}
\put(720,661){\usebox{\plotpoint}}
\put(721,663){\usebox{\plotpoint}}
\put(722,664){\usebox{\plotpoint}}
\put(723,665){\usebox{\plotpoint}}
\put(724,666){\usebox{\plotpoint}}
\put(725,667){\usebox{\plotpoint}}
\put(726,669){\usebox{\plotpoint}}
\put(727,670){\usebox{\plotpoint}}
\put(728,671){\usebox{\plotpoint}}
\put(729,672){\usebox{\plotpoint}}
\put(730,674){\usebox{\plotpoint}}
\put(731,675){\usebox{\plotpoint}}
\put(732,676){\usebox{\plotpoint}}
\put(733,678){\usebox{\plotpoint}}
\put(734,679){\usebox{\plotpoint}}
\put(735,680){\usebox{\plotpoint}}
\put(736,681){\usebox{\plotpoint}}
\put(737,682){\usebox{\plotpoint}}
\put(738,684){\usebox{\plotpoint}}
\put(739,685){\usebox{\plotpoint}}
\put(740,686){\usebox{\plotpoint}}
\put(741,687){\usebox{\plotpoint}}
\put(742,689){\usebox{\plotpoint}}
\put(743,690){\usebox{\plotpoint}}
\put(744,691){\usebox{\plotpoint}}
\put(745,693){\usebox{\plotpoint}}
\put(746,694){\usebox{\plotpoint}}
\put(747,696){\usebox{\plotpoint}}
\put(748,697){\usebox{\plotpoint}}
\put(749,699){\usebox{\plotpoint}}
\put(750,700){\usebox{\plotpoint}}
\put(751,701){\usebox{\plotpoint}}
\put(752,702){\usebox{\plotpoint}}
\put(753,704){\usebox{\plotpoint}}
\put(754,705){\usebox{\plotpoint}}
\put(755,706){\usebox{\plotpoint}}
\put(756,708){\usebox{\plotpoint}}
\put(757,709){\usebox{\plotpoint}}
\put(758,710){\usebox{\plotpoint}}
\put(759,712){\usebox{\plotpoint}}
\put(760,713){\usebox{\plotpoint}}
\put(761,714){\usebox{\plotpoint}}
\put(762,716){\usebox{\plotpoint}}
\put(763,717){\usebox{\plotpoint}}
\put(764,718){\usebox{\plotpoint}}
\put(765,720){\usebox{\plotpoint}}
\put(766,721){\usebox{\plotpoint}}
\put(767,722){\rule[-0.175pt]{0.350pt}{0.385pt}}
\put(768,724){\rule[-0.175pt]{0.350pt}{0.385pt}}
\put(769,726){\rule[-0.175pt]{0.350pt}{0.385pt}}
\put(770,727){\rule[-0.175pt]{0.350pt}{0.385pt}}
\put(771,729){\rule[-0.175pt]{0.350pt}{0.385pt}}
\put(772,730){\usebox{\plotpoint}}
\put(773,732){\usebox{\plotpoint}}
\put(774,733){\usebox{\plotpoint}}
\put(775,734){\usebox{\plotpoint}}
\put(776,736){\usebox{\plotpoint}}
\put(777,737){\usebox{\plotpoint}}
\put(778,738){\usebox{\plotpoint}}
\put(779,740){\usebox{\plotpoint}}
\put(780,741){\usebox{\plotpoint}}
\put(781,742){\usebox{\plotpoint}}
\put(782,744){\usebox{\plotpoint}}
\put(783,745){\usebox{\plotpoint}}
\put(784,746){\usebox{\plotpoint}}
\put(785,748){\usebox{\plotpoint}}
\put(786,749){\usebox{\plotpoint}}
\put(787,750){\usebox{\plotpoint}}
\put(788,752){\usebox{\plotpoint}}
\put(789,753){\usebox{\plotpoint}}
\put(790,754){\usebox{\plotpoint}}
\put(791,756){\usebox{\plotpoint}}
\put(792,757){\usebox{\plotpoint}}
\put(793,758){\usebox{\plotpoint}}
\put(794,760){\usebox{\plotpoint}}
\put(795,761){\usebox{\plotpoint}}
\put(796,762){\rule[-0.175pt]{0.350pt}{0.434pt}}
\put(797,764){\rule[-0.175pt]{0.350pt}{0.434pt}}
\put(798,766){\rule[-0.175pt]{0.350pt}{0.434pt}}
\put(799,768){\rule[-0.175pt]{0.350pt}{0.434pt}}
\put(800,770){\rule[-0.175pt]{0.350pt}{0.434pt}}
\put(801,771){\usebox{\plotpoint}}
\put(802,773){\usebox{\plotpoint}}
\put(803,774){\usebox{\plotpoint}}
\put(804,775){\usebox{\plotpoint}}
\put(805,777){\usebox{\plotpoint}}
\put(806,778){\usebox{\plotpoint}}
\put(807,779){\usebox{\plotpoint}}
\put(808,781){\usebox{\plotpoint}}
\put(809,782){\usebox{\plotpoint}}
\put(810,783){\usebox{\plotpoint}}
\put(811,785){\usebox{\plotpoint}}
\put(812,786){\usebox{\plotpoint}}
\put(813,787){\rule[-0.175pt]{0.350pt}{0.361pt}}
\put(814,789){\rule[-0.175pt]{0.350pt}{0.361pt}}
\put(815,791){\rule[-0.175pt]{0.350pt}{0.361pt}}
\put(816,792){\rule[-0.175pt]{0.350pt}{0.361pt}}
\put(817,794){\rule[-0.175pt]{0.350pt}{0.361pt}}
\put(818,795){\rule[-0.175pt]{0.350pt}{0.361pt}}
\put(819,797){\rule[-0.175pt]{0.350pt}{0.385pt}}
\put(820,798){\rule[-0.175pt]{0.350pt}{0.385pt}}
\put(821,800){\rule[-0.175pt]{0.350pt}{0.385pt}}
\put(822,801){\rule[-0.175pt]{0.350pt}{0.385pt}}
\put(823,803){\rule[-0.175pt]{0.350pt}{0.385pt}}
\put(824,804){\rule[-0.175pt]{0.350pt}{0.361pt}}
\put(825,806){\rule[-0.175pt]{0.350pt}{0.361pt}}
\put(826,808){\rule[-0.175pt]{0.350pt}{0.361pt}}
\put(827,809){\rule[-0.175pt]{0.350pt}{0.361pt}}
\put(828,811){\rule[-0.175pt]{0.350pt}{0.361pt}}
\put(829,812){\rule[-0.175pt]{0.350pt}{0.361pt}}
\put(830,814){\rule[-0.175pt]{0.350pt}{0.361pt}}
\put(831,815){\rule[-0.175pt]{0.350pt}{0.361pt}}
\put(832,817){\rule[-0.175pt]{0.350pt}{0.361pt}}
\put(833,818){\rule[-0.175pt]{0.350pt}{0.361pt}}
\put(834,820){\rule[-0.175pt]{0.350pt}{0.361pt}}
\put(835,821){\rule[-0.175pt]{0.350pt}{0.361pt}}
\sbox{\plotpoint}{\rule[-0.250pt]{0.500pt}{0.500pt}}%
\put(455,1001){\makebox(0,0)[r]{$\tilde{c}=0.0$}}
\put(477,1001){\usebox{\plotpoint}}
\put(497,1001){\usebox{\plotpoint}}
\put(518,1001){\usebox{\plotpoint}}
\put(539,1001){\usebox{\plotpoint}}
\put(543,1001){\usebox{\plotpoint}}
\put(264,227){\usebox{\plotpoint}}
\put(264,227){\usebox{\plotpoint}}
\put(282,235){\usebox{\plotpoint}}
\put(301,245){\usebox{\plotpoint}}
\put(319,255){\usebox{\plotpoint}}
\put(337,265){\usebox{\plotpoint}}
\put(355,275){\usebox{\plotpoint}}
\put(373,285){\usebox{\plotpoint}}
\put(391,296){\usebox{\plotpoint}}
\put(409,306){\usebox{\plotpoint}}
\put(426,317){\usebox{\plotpoint}}
\put(444,328){\usebox{\plotpoint}}
\put(461,340){\usebox{\plotpoint}}
\put(478,351){\usebox{\plotpoint}}
\put(495,363){\usebox{\plotpoint}}
\put(512,376){\usebox{\plotpoint}}
\put(529,387){\usebox{\plotpoint}}
\put(545,400){\usebox{\plotpoint}}
\put(562,413){\usebox{\plotpoint}}
\put(578,425){\usebox{\plotpoint}}
\put(594,439){\usebox{\plotpoint}}
\put(610,452){\usebox{\plotpoint}}
\put(625,466){\usebox{\plotpoint}}
\put(641,480){\usebox{\plotpoint}}
\put(656,494){\usebox{\plotpoint}}
\put(670,508){\usebox{\plotpoint}}
\put(686,523){\usebox{\plotpoint}}
\put(700,538){\usebox{\plotpoint}}
\put(714,552){\usebox{\plotpoint}}
\put(728,568){\usebox{\plotpoint}}
\put(742,583){\usebox{\plotpoint}}
\put(756,599){\usebox{\plotpoint}}
\put(769,615){\usebox{\plotpoint}}
\put(783,630){\usebox{\plotpoint}}
\put(796,646){\usebox{\plotpoint}}
\put(809,663){\usebox{\plotpoint}}
\put(821,679){\usebox{\plotpoint}}
\put(834,696){\usebox{\plotpoint}}
\put(836,698){\usebox{\plotpoint}}
\sbox{\plotpoint}{\rule[-0.175pt]{0.350pt}{0.350pt}}%
\put(359,308){\circle*{12}}
\put(424,370){\circle*{12}}
\put(488,336){\circle*{12}}
\put(551,479){\circle*{12}}
\put(613,769){\circle*{12}}
\put(682,810){\circle*{12}}
\put(744,1001){\circle*{12}}
\put(359,264){\rule[-0.175pt]{0.350pt}{21.681pt}}
\put(349,264){\rule[-0.175pt]{4.818pt}{0.350pt}}
\put(349,354){\rule[-0.175pt]{4.818pt}{0.350pt}}
\put(424,314){\rule[-0.175pt]{0.350pt}{24.572pt}}
\put(414,314){\rule[-0.175pt]{4.818pt}{0.350pt}}
\put(414,416){\rule[-0.175pt]{4.818pt}{0.350pt}}
\put(488,249){\rule[-0.175pt]{0.350pt}{36.617pt}}
\put(478,249){\rule[-0.175pt]{4.818pt}{0.350pt}}
\put(478,401){\rule[-0.175pt]{4.818pt}{0.350pt}}
\put(551,381){\rule[-0.175pt]{0.350pt}{38.544pt}}
\put(541,381){\rule[-0.175pt]{4.818pt}{0.350pt}}
\put(541,541){\rule[-0.175pt]{4.818pt}{0.350pt}}
\put(613,707){\rule[-0.175pt]{0.350pt}{26.258pt}}
\put(603,707){\rule[-0.175pt]{4.818pt}{0.350pt}}
\put(603,816){\rule[-0.175pt]{4.818pt}{0.350pt}}
\put(682,722){\rule[-0.175pt]{0.350pt}{35.412pt}}
\put(672,722){\rule[-0.175pt]{4.818pt}{0.350pt}}
\put(672,869){\rule[-0.175pt]{4.818pt}{0.350pt}}
\put(744,875){\rule[-0.175pt]{0.350pt}{48.903pt}}
\put(734,875){\rule[-0.175pt]{4.818pt}{0.350pt}}
\put(734,1078){\rule[-0.175pt]{4.818pt}{0.350pt}}
\end{picture}

%% file: f_eewpj.tex
\setlength{\unitlength}{0.240900pt}
\ifx\plotpoint\undefined\newsavebox{\plotpoint}\fi
\begin{picture}(1720,900)(0,0)
\tenrm
\sbox{\plotpoint}{\rule[-0.175pt]{0.350pt}{0.350pt}}%
\put(264,158){\rule[-0.175pt]{282.335pt}{0.350pt}}
\put(264,158){\rule[-0.175pt]{4.818pt}{0.350pt}}
\put(242,158){\makebox(0,0)[r]{0}}
\put(1416,158){\rule[-0.175pt]{4.818pt}{0.350pt}}
\put(264,263){\rule[-0.175pt]{4.818pt}{0.350pt}}
\put(242,263){\makebox(0,0)[r]{5}}
\put(1416,263){\rule[-0.175pt]{4.818pt}{0.350pt}}
\put(264,368){\rule[-0.175pt]{4.818pt}{0.350pt}}
\put(242,368){\makebox(0,0)[r]{10}}
\put(1416,368){\rule[-0.175pt]{4.818pt}{0.350pt}}
\put(264,473){\rule[-0.175pt]{4.818pt}{0.350pt}}
\put(242,473){\makebox(0,0)[r]{15}}
\put(1416,473){\rule[-0.175pt]{4.818pt}{0.350pt}}
\put(264,577){\rule[-0.175pt]{4.818pt}{0.350pt}}
\put(242,577){\makebox(0,0)[r]{20}}
\put(1416,577){\rule[-0.175pt]{4.818pt}{0.350pt}}
\put(264,682){\rule[-0.175pt]{4.818pt}{0.350pt}}
\put(242,682){\makebox(0,0)[r]{25}}
\put(1416,682){\rule[-0.175pt]{4.818pt}{0.350pt}}
\put(264,787){\rule[-0.175pt]{4.818pt}{0.350pt}}
\put(242,787){\makebox(0,0)[r]{30}}
\put(1416,787){\rule[-0.175pt]{4.818pt}{0.350pt}}
\put(264,158){\rule[-0.175pt]{0.350pt}{4.818pt}}
\put(264,113){\makebox(0,0){1}}
\put(264,767){\rule[-0.175pt]{0.350pt}{4.818pt}}
\put(381,158){\rule[-0.175pt]{0.350pt}{4.818pt}}
\put(381,113){\makebox(0,0){1.05}}
\put(381,767){\rule[-0.175pt]{0.350pt}{4.818pt}}
\put(498,158){\rule[-0.175pt]{0.350pt}{4.818pt}}
\put(498,113){\makebox(0,0){1.1}}
\put(498,767){\rule[-0.175pt]{0.350pt}{4.818pt}}
\put(616,158){\rule[-0.175pt]{0.350pt}{4.818pt}}
\put(616,113){\makebox(0,0){1.15}}
\put(616,767){\rule[-0.175pt]{0.350pt}{4.818pt}}
\put(733,158){\rule[-0.175pt]{0.350pt}{4.818pt}}
\put(733,113){\makebox(0,0){1.2}}
\put(733,767){\rule[-0.175pt]{0.350pt}{4.818pt}}
\put(850,158){\rule[-0.175pt]{0.350pt}{4.818pt}}
\put(850,113){\makebox(0,0){1.25}}
\put(850,767){\rule[-0.175pt]{0.350pt}{4.818pt}}
\put(967,158){\rule[-0.175pt]{0.350pt}{4.818pt}}
\put(967,113){\makebox(0,0){1.3}}
\put(967,767){\rule[-0.175pt]{0.350pt}{4.818pt}}
\put(1084,158){\rule[-0.175pt]{0.350pt}{4.818pt}}
\put(1084,113){\makebox(0,0){1.35}}
\put(1084,767){\rule[-0.175pt]{0.350pt}{4.818pt}}
\put(1202,158){\rule[-0.175pt]{0.350pt}{4.818pt}}
\put(1202,113){\makebox(0,0){1.4}}
\put(1202,767){\rule[-0.175pt]{0.350pt}{4.818pt}}
\put(1319,158){\rule[-0.175pt]{0.350pt}{4.818pt}}
\put(1319,113){\makebox(0,0){1.45}}
\put(1319,767){\rule[-0.175pt]{0.350pt}{4.818pt}}
\put(1436,158){\rule[-0.175pt]{0.350pt}{4.818pt}}
\put(1436,113){\makebox(0,0){1.5}}
\put(1436,767){\rule[-0.175pt]{0.350pt}{4.818pt}}
\put(264,158){\rule[-0.175pt]{282.335pt}{0.350pt}}
\put(1436,158){\rule[-0.175pt]{0.350pt}{151.526pt}}
\put(264,787){\rule[-0.175pt]{282.335pt}{0.350pt}}
\put(199,877){\makebox(0,0)[l]{\shortstack{$\sigma$ (nb)}}}
\put(850,23){\makebox(0,0){$\sqrt{q^2}$ (GeV)}}
\put(264,158){\rule[-0.175pt]{0.350pt}{151.526pt}}
\sbox{\plotpoint}{\rule[-0.500pt]{1.000pt}{1.000pt}}%
\put(1600,766){\makebox(0,0)[r]{$\tilde{c}=0.8$}}
\put(1622,766){\rule[-0.500pt]{15.899pt}{1.000pt}}
\put(264,420){\usebox{\plotpoint}}
\put(264,420){\usebox{\plotpoint}}
\put(265,421){\usebox{\plotpoint}}
\put(266,422){\usebox{\plotpoint}}
\put(267,423){\usebox{\plotpoint}}
\put(268,424){\usebox{\plotpoint}}
\put(269,425){\usebox{\plotpoint}}
\put(270,426){\usebox{\plotpoint}}
\put(271,428){\usebox{\plotpoint}}
\put(272,429){\usebox{\plotpoint}}
\put(273,430){\usebox{\plotpoint}}
\put(274,431){\usebox{\plotpoint}}
\put(275,432){\usebox{\plotpoint}}
\put(276,433){\usebox{\plotpoint}}
\put(276,434){\usebox{\plotpoint}}
\put(277,435){\usebox{\plotpoint}}
\put(278,436){\usebox{\plotpoint}}
\put(279,437){\usebox{\plotpoint}}
\put(280,438){\usebox{\plotpoint}}
\put(281,439){\usebox{\plotpoint}}
\put(282,440){\usebox{\plotpoint}}
\put(283,441){\usebox{\plotpoint}}
\put(284,442){\usebox{\plotpoint}}
\put(285,443){\usebox{\plotpoint}}
\put(286,444){\usebox{\plotpoint}}
\put(287,445){\usebox{\plotpoint}}
\put(288,446){\usebox{\plotpoint}}
\put(289,447){\usebox{\plotpoint}}
\put(290,448){\usebox{\plotpoint}}
\put(291,449){\usebox{\plotpoint}}
\put(292,450){\usebox{\plotpoint}}
\put(293,451){\usebox{\plotpoint}}
\put(294,452){\usebox{\plotpoint}}
\put(295,453){\usebox{\plotpoint}}
\put(296,454){\usebox{\plotpoint}}
\put(297,455){\usebox{\plotpoint}}
\put(298,456){\usebox{\plotpoint}}
\put(299,457){\usebox{\plotpoint}}
\put(300,458){\usebox{\plotpoint}}
\put(301,459){\usebox{\plotpoint}}
\put(302,460){\usebox{\plotpoint}}
\put(303,461){\usebox{\plotpoint}}
\put(304,462){\usebox{\plotpoint}}
\put(305,463){\usebox{\plotpoint}}
\put(306,464){\usebox{\plotpoint}}
\put(307,465){\usebox{\plotpoint}}
\put(308,466){\usebox{\plotpoint}}
\put(309,467){\usebox{\plotpoint}}
\put(310,468){\usebox{\plotpoint}}
\put(311,470){\usebox{\plotpoint}}
\put(312,471){\usebox{\plotpoint}}
\put(313,472){\usebox{\plotpoint}}
\put(314,473){\usebox{\plotpoint}}
\put(315,474){\usebox{\plotpoint}}
\put(317,475){\usebox{\plotpoint}}
\put(318,476){\usebox{\plotpoint}}
\put(319,477){\usebox{\plotpoint}}
\put(320,478){\usebox{\plotpoint}}
\put(321,479){\usebox{\plotpoint}}
\put(323,480){\usebox{\plotpoint}}
\put(324,481){\usebox{\plotpoint}}
\put(325,482){\usebox{\plotpoint}}
\put(326,483){\usebox{\plotpoint}}
\put(327,484){\usebox{\plotpoint}}
\put(329,485){\usebox{\plotpoint}}
\put(330,486){\usebox{\plotpoint}}
\put(331,487){\usebox{\plotpoint}}
\put(332,488){\usebox{\plotpoint}}
\put(333,489){\usebox{\plotpoint}}
\put(335,490){\usebox{\plotpoint}}
\put(336,491){\usebox{\plotpoint}}
\put(337,492){\usebox{\plotpoint}}
\put(338,493){\usebox{\plotpoint}}
\put(339,494){\usebox{\plotpoint}}
\put(341,495){\usebox{\plotpoint}}
\put(342,496){\usebox{\plotpoint}}
\put(343,497){\usebox{\plotpoint}}
\put(344,498){\usebox{\plotpoint}}
\put(345,499){\usebox{\plotpoint}}
\put(347,500){\usebox{\plotpoint}}
\put(348,501){\usebox{\plotpoint}}
\put(349,502){\usebox{\plotpoint}}
\put(351,503){\usebox{\plotpoint}}
\put(352,504){\usebox{\plotpoint}}
\put(353,505){\usebox{\plotpoint}}
\put(355,506){\usebox{\plotpoint}}
\put(356,507){\usebox{\plotpoint}}
\put(357,508){\usebox{\plotpoint}}
\put(359,509){\usebox{\plotpoint}}
\put(360,510){\usebox{\plotpoint}}
\put(362,511){\usebox{\plotpoint}}
\put(363,512){\usebox{\plotpoint}}
\put(365,513){\usebox{\plotpoint}}
\put(366,514){\usebox{\plotpoint}}
\put(368,515){\usebox{\plotpoint}}
\put(369,516){\usebox{\plotpoint}}
\put(371,517){\usebox{\plotpoint}}
\put(372,518){\usebox{\plotpoint}}
\put(373,519){\usebox{\plotpoint}}
\put(375,520){\usebox{\plotpoint}}
\put(376,521){\usebox{\plotpoint}}
\put(377,522){\usebox{\plotpoint}}
\put(379,523){\usebox{\plotpoint}}
\put(380,524){\usebox{\plotpoint}}
\put(382,525){\usebox{\plotpoint}}
\put(383,526){\usebox{\plotpoint}}
\put(385,527){\usebox{\plotpoint}}
\put(387,528){\usebox{\plotpoint}}
\put(388,529){\usebox{\plotpoint}}
\put(390,530){\usebox{\plotpoint}}
\put(392,531){\usebox{\plotpoint}}
\put(394,532){\usebox{\plotpoint}}
\put(396,533){\usebox{\plotpoint}}
\put(398,534){\usebox{\plotpoint}}
\put(400,535){\usebox{\plotpoint}}
\put(402,536){\usebox{\plotpoint}}
\put(404,537){\usebox{\plotpoint}}
\put(406,538){\usebox{\plotpoint}}
\put(407,539){\usebox{\plotpoint}}
\put(409,540){\usebox{\plotpoint}}
\put(411,541){\usebox{\plotpoint}}
\put(412,542){\usebox{\plotpoint}}
\put(414,543){\usebox{\plotpoint}}
\put(416,544){\usebox{\plotpoint}}
\put(418,545){\usebox{\plotpoint}}
\put(420,546){\usebox{\plotpoint}}
\put(422,547){\usebox{\plotpoint}}
\put(425,548){\usebox{\plotpoint}}
\put(427,549){\usebox{\plotpoint}}
\put(429,550){\usebox{\plotpoint}}
\put(432,551){\usebox{\plotpoint}}
\put(434,552){\usebox{\plotpoint}}
\put(437,553){\usebox{\plotpoint}}
\put(439,554){\usebox{\plotpoint}}
\put(441,555){\usebox{\plotpoint}}
\put(444,556){\usebox{\plotpoint}}
\put(446,557){\usebox{\plotpoint}}
\put(448,558){\usebox{\plotpoint}}
\put(450,559){\usebox{\plotpoint}}
\put(453,560){\usebox{\plotpoint}}
\put(456,561){\usebox{\plotpoint}}
\put(459,562){\usebox{\plotpoint}}
\put(462,563){\usebox{\plotpoint}}
\put(465,564){\usebox{\plotpoint}}
\put(468,565){\usebox{\plotpoint}}
\put(471,566){\usebox{\plotpoint}}
\put(474,567){\usebox{\plotpoint}}
\put(477,568){\usebox{\plotpoint}}
\put(480,569){\usebox{\plotpoint}}
\put(483,570){\usebox{\plotpoint}}
\put(486,571){\usebox{\plotpoint}}
\put(489,572){\usebox{\plotpoint}}
\put(493,573){\usebox{\plotpoint}}
\put(497,574){\usebox{\plotpoint}}
\put(501,575){\rule[-0.500pt]{1.445pt}{1.000pt}}
\put(507,576){\rule[-0.500pt]{1.445pt}{1.000pt}}
\put(513,577){\usebox{\plotpoint}}
\put(516,578){\usebox{\plotpoint}}
\put(520,579){\usebox{\plotpoint}}
\put(524,580){\rule[-0.500pt]{1.445pt}{1.000pt}}
\put(530,581){\rule[-0.500pt]{1.445pt}{1.000pt}}
\put(536,582){\rule[-0.500pt]{1.445pt}{1.000pt}}
\put(542,583){\rule[-0.500pt]{1.445pt}{1.000pt}}
\put(548,584){\rule[-0.500pt]{2.891pt}{1.000pt}}
\put(560,585){\rule[-0.500pt]{2.891pt}{1.000pt}}
\put(572,586){\rule[-0.500pt]{2.891pt}{1.000pt}}
\put(584,587){\rule[-0.500pt]{2.650pt}{1.000pt}}
\put(595,588){\rule[-0.500pt]{17.345pt}{1.000pt}}
\put(667,587){\rule[-0.500pt]{5.541pt}{1.000pt}}
\put(690,586){\rule[-0.500pt]{2.891pt}{1.000pt}}
\put(702,585){\rule[-0.500pt]{1.445pt}{1.000pt}}
\put(708,584){\rule[-0.500pt]{1.445pt}{1.000pt}}
\put(714,583){\rule[-0.500pt]{2.891pt}{1.000pt}}
\put(726,582){\rule[-0.500pt]{2.891pt}{1.000pt}}
\put(738,581){\rule[-0.500pt]{1.325pt}{1.000pt}}
\put(743,580){\rule[-0.500pt]{1.325pt}{1.000pt}}
\put(749,579){\rule[-0.500pt]{2.891pt}{1.000pt}}
\put(761,578){\rule[-0.500pt]{1.445pt}{1.000pt}}
\put(767,577){\rule[-0.500pt]{1.445pt}{1.000pt}}
\put(773,576){\rule[-0.500pt]{1.445pt}{1.000pt}}
\put(779,575){\rule[-0.500pt]{1.445pt}{1.000pt}}
\put(785,574){\rule[-0.500pt]{1.445pt}{1.000pt}}
\put(791,573){\rule[-0.500pt]{1.445pt}{1.000pt}}
\put(797,572){\rule[-0.500pt]{1.445pt}{1.000pt}}
\put(803,571){\rule[-0.500pt]{1.445pt}{1.000pt}}
\put(809,570){\rule[-0.500pt]{2.650pt}{1.000pt}}
\put(820,569){\usebox{\plotpoint}}
\put(824,568){\usebox{\plotpoint}}
\put(828,567){\usebox{\plotpoint}}
\put(832,566){\rule[-0.500pt]{1.445pt}{1.000pt}}
\put(838,565){\rule[-0.500pt]{1.445pt}{1.000pt}}
\put(844,564){\rule[-0.500pt]{1.445pt}{1.000pt}}
\put(850,563){\rule[-0.500pt]{1.445pt}{1.000pt}}
\put(856,562){\rule[-0.500pt]{1.445pt}{1.000pt}}
\put(862,561){\rule[-0.500pt]{1.445pt}{1.000pt}}
\put(868,560){\rule[-0.500pt]{1.445pt}{1.000pt}}
\put(874,559){\rule[-0.500pt]{1.445pt}{1.000pt}}
\put(880,558){\rule[-0.500pt]{1.325pt}{1.000pt}}
\put(885,557){\rule[-0.500pt]{1.325pt}{1.000pt}}
\put(891,556){\rule[-0.500pt]{1.445pt}{1.000pt}}
\put(897,555){\rule[-0.500pt]{1.445pt}{1.000pt}}
\put(903,554){\usebox{\plotpoint}}
\put(907,553){\usebox{\plotpoint}}
\put(911,552){\usebox{\plotpoint}}
\put(915,551){\rule[-0.500pt]{1.445pt}{1.000pt}}
\put(921,550){\rule[-0.500pt]{1.445pt}{1.000pt}}
\put(927,549){\rule[-0.500pt]{1.445pt}{1.000pt}}
\put(933,548){\rule[-0.500pt]{1.445pt}{1.000pt}}
\put(939,547){\usebox{\plotpoint}}
\put(943,546){\usebox{\plotpoint}}
\put(947,545){\usebox{\plotpoint}}
\put(951,544){\rule[-0.500pt]{1.325pt}{1.000pt}}
\put(956,543){\rule[-0.500pt]{1.325pt}{1.000pt}}
\put(962,542){\rule[-0.500pt]{1.445pt}{1.000pt}}
\put(968,541){\rule[-0.500pt]{1.445pt}{1.000pt}}
\put(974,540){\rule[-0.500pt]{1.445pt}{1.000pt}}
\put(980,539){\rule[-0.500pt]{1.445pt}{1.000pt}}
\put(986,538){\usebox{\plotpoint}}
\put(990,537){\usebox{\plotpoint}}
\put(994,536){\usebox{\plotpoint}}
\put(998,535){\rule[-0.500pt]{1.445pt}{1.000pt}}
\put(1004,534){\rule[-0.500pt]{1.445pt}{1.000pt}}
\put(1010,533){\rule[-0.500pt]{1.445pt}{1.000pt}}
\put(1016,532){\rule[-0.500pt]{1.445pt}{1.000pt}}
\put(1022,531){\rule[-0.500pt]{1.325pt}{1.000pt}}
\put(1027,530){\rule[-0.500pt]{1.325pt}{1.000pt}}
\put(1033,529){\usebox{\plotpoint}}
\put(1037,528){\usebox{\plotpoint}}
\put(1041,527){\usebox{\plotpoint}}
\put(1045,526){\rule[-0.500pt]{1.445pt}{1.000pt}}
\put(1051,525){\rule[-0.500pt]{1.445pt}{1.000pt}}
\put(1057,524){\rule[-0.500pt]{1.445pt}{1.000pt}}
\put(1063,523){\rule[-0.500pt]{1.445pt}{1.000pt}}
\put(1069,522){\rule[-0.500pt]{1.445pt}{1.000pt}}
\put(1075,521){\rule[-0.500pt]{1.445pt}{1.000pt}}
\put(1081,520){\rule[-0.500pt]{1.445pt}{1.000pt}}
\put(1087,519){\rule[-0.500pt]{1.445pt}{1.000pt}}
\put(1093,518){\usebox{\plotpoint}}
\put(1097,517){\usebox{\plotpoint}}
\put(1101,516){\usebox{\plotpoint}}
\put(1105,515){\rule[-0.500pt]{1.325pt}{1.000pt}}
\put(1110,514){\rule[-0.500pt]{1.325pt}{1.000pt}}
\put(1116,513){\rule[-0.500pt]{1.445pt}{1.000pt}}
\put(1122,512){\rule[-0.500pt]{1.445pt}{1.000pt}}
\put(1128,511){\rule[-0.500pt]{1.445pt}{1.000pt}}
\put(1134,510){\rule[-0.500pt]{1.445pt}{1.000pt}}
\put(1140,509){\rule[-0.500pt]{1.445pt}{1.000pt}}
\put(1146,508){\rule[-0.500pt]{1.445pt}{1.000pt}}
\put(1152,507){\rule[-0.500pt]{1.445pt}{1.000pt}}
\put(1158,506){\rule[-0.500pt]{1.445pt}{1.000pt}}
\put(1164,505){\rule[-0.500pt]{1.445pt}{1.000pt}}
\put(1170,504){\rule[-0.500pt]{1.445pt}{1.000pt}}
\put(1176,503){\rule[-0.500pt]{1.325pt}{1.000pt}}
\put(1181,502){\rule[-0.500pt]{1.325pt}{1.000pt}}
\put(1187,501){\rule[-0.500pt]{1.445pt}{1.000pt}}
\put(1193,500){\rule[-0.500pt]{1.445pt}{1.000pt}}
\put(1199,499){\rule[-0.500pt]{1.445pt}{1.000pt}}
\put(1205,498){\rule[-0.500pt]{1.445pt}{1.000pt}}
\put(1211,497){\rule[-0.500pt]{1.445pt}{1.000pt}}
\put(1217,496){\rule[-0.500pt]{1.445pt}{1.000pt}}
\put(1223,495){\rule[-0.500pt]{1.445pt}{1.000pt}}
\put(1229,494){\rule[-0.500pt]{1.445pt}{1.000pt}}
\put(1235,493){\rule[-0.500pt]{1.445pt}{1.000pt}}
\put(1241,492){\rule[-0.500pt]{1.445pt}{1.000pt}}
\put(1247,491){\rule[-0.500pt]{2.650pt}{1.000pt}}
\put(1258,490){\rule[-0.500pt]{1.445pt}{1.000pt}}
\put(1264,489){\rule[-0.500pt]{1.445pt}{1.000pt}}
\put(1270,488){\rule[-0.500pt]{1.445pt}{1.000pt}}
\put(1276,487){\rule[-0.500pt]{1.445pt}{1.000pt}}
\put(1282,486){\rule[-0.500pt]{1.445pt}{1.000pt}}
\put(1288,485){\rule[-0.500pt]{1.445pt}{1.000pt}}
\put(1294,484){\rule[-0.500pt]{2.891pt}{1.000pt}}
\put(1306,483){\rule[-0.500pt]{1.445pt}{1.000pt}}
\put(1312,482){\rule[-0.500pt]{1.445pt}{1.000pt}}
\put(1318,481){\rule[-0.500pt]{1.325pt}{1.000pt}}
\put(1323,480){\rule[-0.500pt]{1.325pt}{1.000pt}}
\put(1329,479){\rule[-0.500pt]{2.891pt}{1.000pt}}
\put(1341,478){\rule[-0.500pt]{1.445pt}{1.000pt}}
\put(1347,477){\rule[-0.500pt]{1.445pt}{1.000pt}}
\put(1353,476){\rule[-0.500pt]{1.445pt}{1.000pt}}
\put(1359,475){\rule[-0.500pt]{1.445pt}{1.000pt}}
\put(1365,474){\rule[-0.500pt]{2.891pt}{1.000pt}}
\put(1377,473){\rule[-0.500pt]{1.445pt}{1.000pt}}
\put(1383,472){\rule[-0.500pt]{1.445pt}{1.000pt}}
\put(1389,471){\rule[-0.500pt]{2.650pt}{1.000pt}}
\put(1400,470){\rule[-0.500pt]{1.445pt}{1.000pt}}
\put(1406,469){\rule[-0.500pt]{1.445pt}{1.000pt}}
\put(1412,468){\rule[-0.500pt]{2.891pt}{1.000pt}}
\put(1424,467){\rule[-0.500pt]{1.445pt}{1.000pt}}
\put(1430,466){\rule[-0.500pt]{1.445pt}{1.000pt}}
\sbox{\plotpoint}{\rule[-0.175pt]{0.350pt}{0.350pt}}%
\put(1600,721){\makebox(0,0)[r]{$\tilde{c}=0.7$}}
\put(1622,721){\rule[-0.175pt]{15.899pt}{0.350pt}}
\put(264,386){\usebox{\plotpoint}}
\put(264,386){\usebox{\plotpoint}}
\put(265,387){\usebox{\plotpoint}}
\put(266,388){\usebox{\plotpoint}}
\put(267,389){\usebox{\plotpoint}}
\put(268,390){\usebox{\plotpoint}}
\put(269,391){\usebox{\plotpoint}}
\put(270,392){\usebox{\plotpoint}}
\put(271,393){\usebox{\plotpoint}}
\put(272,394){\usebox{\plotpoint}}
\put(273,395){\usebox{\plotpoint}}
\put(274,396){\usebox{\plotpoint}}
\put(276,397){\usebox{\plotpoint}}
\put(277,398){\usebox{\plotpoint}}
\put(278,399){\usebox{\plotpoint}}
\put(279,400){\usebox{\plotpoint}}
\put(280,401){\usebox{\plotpoint}}
\put(281,402){\usebox{\plotpoint}}
\put(282,403){\usebox{\plotpoint}}
\put(283,404){\usebox{\plotpoint}}
\put(284,405){\usebox{\plotpoint}}
\put(285,406){\usebox{\plotpoint}}
\put(286,407){\usebox{\plotpoint}}
\put(288,408){\usebox{\plotpoint}}
\put(289,409){\usebox{\plotpoint}}
\put(290,410){\usebox{\plotpoint}}
\put(291,411){\usebox{\plotpoint}}
\put(292,412){\usebox{\plotpoint}}
\put(294,413){\usebox{\plotpoint}}
\put(295,414){\usebox{\plotpoint}}
\put(296,415){\usebox{\plotpoint}}
\put(297,416){\usebox{\plotpoint}}
\put(298,417){\usebox{\plotpoint}}
\put(300,418){\usebox{\plotpoint}}
\put(301,419){\usebox{\plotpoint}}
\put(302,420){\usebox{\plotpoint}}
\put(303,421){\usebox{\plotpoint}}
\put(304,422){\usebox{\plotpoint}}
\put(305,423){\usebox{\plotpoint}}
\put(306,424){\usebox{\plotpoint}}
\put(307,425){\usebox{\plotpoint}}
\put(308,426){\usebox{\plotpoint}}
\put(309,427){\usebox{\plotpoint}}
\put(311,428){\usebox{\plotpoint}}
\put(312,429){\usebox{\plotpoint}}
\put(313,430){\usebox{\plotpoint}}
\put(315,431){\usebox{\plotpoint}}
\put(316,432){\usebox{\plotpoint}}
\put(317,433){\usebox{\plotpoint}}
\put(319,434){\usebox{\plotpoint}}
\put(320,435){\usebox{\plotpoint}}
\put(321,436){\usebox{\plotpoint}}
\put(323,437){\rule[-0.175pt]{0.361pt}{0.350pt}}
\put(324,438){\rule[-0.175pt]{0.361pt}{0.350pt}}
\put(326,439){\rule[-0.175pt]{0.361pt}{0.350pt}}
\put(327,440){\rule[-0.175pt]{0.361pt}{0.350pt}}
\put(329,441){\rule[-0.175pt]{0.361pt}{0.350pt}}
\put(330,442){\rule[-0.175pt]{0.361pt}{0.350pt}}
\put(332,443){\rule[-0.175pt]{0.361pt}{0.350pt}}
\put(333,444){\rule[-0.175pt]{0.361pt}{0.350pt}}
\put(335,445){\rule[-0.175pt]{0.361pt}{0.350pt}}
\put(336,446){\rule[-0.175pt]{0.361pt}{0.350pt}}
\put(338,447){\rule[-0.175pt]{0.361pt}{0.350pt}}
\put(339,448){\rule[-0.175pt]{0.361pt}{0.350pt}}
\put(341,449){\rule[-0.175pt]{0.361pt}{0.350pt}}
\put(342,450){\rule[-0.175pt]{0.361pt}{0.350pt}}
\put(344,451){\rule[-0.175pt]{0.361pt}{0.350pt}}
\put(345,452){\rule[-0.175pt]{0.361pt}{0.350pt}}
\put(347,453){\rule[-0.175pt]{0.361pt}{0.350pt}}
\put(348,454){\rule[-0.175pt]{0.361pt}{0.350pt}}
\put(350,455){\rule[-0.175pt]{0.361pt}{0.350pt}}
\put(351,456){\rule[-0.175pt]{0.361pt}{0.350pt}}
\put(353,457){\rule[-0.175pt]{0.361pt}{0.350pt}}
\put(354,458){\rule[-0.175pt]{0.361pt}{0.350pt}}
\put(356,459){\rule[-0.175pt]{0.361pt}{0.350pt}}
\put(357,460){\rule[-0.175pt]{0.361pt}{0.350pt}}
\put(359,461){\rule[-0.175pt]{0.413pt}{0.350pt}}
\put(360,462){\rule[-0.175pt]{0.413pt}{0.350pt}}
\put(362,463){\rule[-0.175pt]{0.413pt}{0.350pt}}
\put(364,464){\rule[-0.175pt]{0.413pt}{0.350pt}}
\put(365,465){\rule[-0.175pt]{0.413pt}{0.350pt}}
\put(367,466){\rule[-0.175pt]{0.413pt}{0.350pt}}
\put(369,467){\rule[-0.175pt]{0.413pt}{0.350pt}}
\put(371,468){\rule[-0.175pt]{0.442pt}{0.350pt}}
\put(372,469){\rule[-0.175pt]{0.442pt}{0.350pt}}
\put(374,470){\rule[-0.175pt]{0.442pt}{0.350pt}}
\put(376,471){\rule[-0.175pt]{0.442pt}{0.350pt}}
\put(378,472){\rule[-0.175pt]{0.442pt}{0.350pt}}
\put(380,473){\rule[-0.175pt]{0.442pt}{0.350pt}}
\put(382,474){\rule[-0.175pt]{0.482pt}{0.350pt}}
\put(384,475){\rule[-0.175pt]{0.482pt}{0.350pt}}
\put(386,476){\rule[-0.175pt]{0.482pt}{0.350pt}}
\put(388,477){\rule[-0.175pt]{0.482pt}{0.350pt}}
\put(390,478){\rule[-0.175pt]{0.482pt}{0.350pt}}
\put(392,479){\rule[-0.175pt]{0.482pt}{0.350pt}}
\put(394,480){\rule[-0.175pt]{0.578pt}{0.350pt}}
\put(396,481){\rule[-0.175pt]{0.578pt}{0.350pt}}
\put(398,482){\rule[-0.175pt]{0.578pt}{0.350pt}}
\put(401,483){\rule[-0.175pt]{0.578pt}{0.350pt}}
\put(403,484){\rule[-0.175pt]{0.578pt}{0.350pt}}
\put(405,485){\rule[-0.175pt]{0.578pt}{0.350pt}}
\put(408,486){\rule[-0.175pt]{0.578pt}{0.350pt}}
\put(410,487){\rule[-0.175pt]{0.578pt}{0.350pt}}
\put(413,488){\rule[-0.175pt]{0.578pt}{0.350pt}}
\put(415,489){\rule[-0.175pt]{0.578pt}{0.350pt}}
\put(417,490){\rule[-0.175pt]{0.578pt}{0.350pt}}
\put(420,491){\rule[-0.175pt]{0.578pt}{0.350pt}}
\put(422,492){\rule[-0.175pt]{0.578pt}{0.350pt}}
\put(425,493){\rule[-0.175pt]{0.578pt}{0.350pt}}
\put(427,494){\rule[-0.175pt]{0.578pt}{0.350pt}}
\put(429,495){\rule[-0.175pt]{0.723pt}{0.350pt}}
\put(433,496){\rule[-0.175pt]{0.723pt}{0.350pt}}
\put(436,497){\rule[-0.175pt]{0.723pt}{0.350pt}}
\put(439,498){\rule[-0.175pt]{0.723pt}{0.350pt}}
\put(442,499){\rule[-0.175pt]{0.662pt}{0.350pt}}
\put(444,500){\rule[-0.175pt]{0.662pt}{0.350pt}}
\put(447,501){\rule[-0.175pt]{0.662pt}{0.350pt}}
\put(450,502){\rule[-0.175pt]{0.662pt}{0.350pt}}
\put(453,503){\rule[-0.175pt]{0.723pt}{0.350pt}}
\put(456,504){\rule[-0.175pt]{0.723pt}{0.350pt}}
\put(459,505){\rule[-0.175pt]{0.723pt}{0.350pt}}
\put(462,506){\rule[-0.175pt]{0.723pt}{0.350pt}}
\put(465,507){\rule[-0.175pt]{0.964pt}{0.350pt}}
\put(469,508){\rule[-0.175pt]{0.964pt}{0.350pt}}
\put(473,509){\rule[-0.175pt]{0.964pt}{0.350pt}}
\put(477,510){\rule[-0.175pt]{0.964pt}{0.350pt}}
\put(481,511){\rule[-0.175pt]{0.964pt}{0.350pt}}
\put(485,512){\rule[-0.175pt]{0.964pt}{0.350pt}}
\put(489,513){\rule[-0.175pt]{1.445pt}{0.350pt}}
\put(495,514){\rule[-0.175pt]{1.445pt}{0.350pt}}
\put(501,515){\rule[-0.175pt]{1.445pt}{0.350pt}}
\put(507,516){\rule[-0.175pt]{1.445pt}{0.350pt}}
\put(513,517){\rule[-0.175pt]{1.325pt}{0.350pt}}
\put(518,518){\rule[-0.175pt]{1.325pt}{0.350pt}}
\put(524,519){\rule[-0.175pt]{1.445pt}{0.350pt}}
\put(530,520){\rule[-0.175pt]{1.445pt}{0.350pt}}
\put(536,521){\rule[-0.175pt]{2.891pt}{0.350pt}}
\put(548,522){\rule[-0.175pt]{2.891pt}{0.350pt}}
\put(560,523){\rule[-0.175pt]{2.891pt}{0.350pt}}
\put(572,524){\rule[-0.175pt]{5.541pt}{0.350pt}}
\put(595,525){\rule[-0.175pt]{8.672pt}{0.350pt}}
\put(631,524){\rule[-0.175pt]{5.782pt}{0.350pt}}
\put(655,523){\rule[-0.175pt]{5.541pt}{0.350pt}}
\put(678,522){\rule[-0.175pt]{2.891pt}{0.350pt}}
\put(690,521){\rule[-0.175pt]{2.891pt}{0.350pt}}
\put(702,520){\rule[-0.175pt]{2.891pt}{0.350pt}}
\put(714,519){\rule[-0.175pt]{1.445pt}{0.350pt}}
\put(720,518){\rule[-0.175pt]{1.445pt}{0.350pt}}
\put(726,517){\rule[-0.175pt]{2.891pt}{0.350pt}}
\put(738,516){\rule[-0.175pt]{1.325pt}{0.350pt}}
\put(743,515){\rule[-0.175pt]{1.325pt}{0.350pt}}
\put(749,514){\rule[-0.175pt]{2.891pt}{0.350pt}}
\put(761,513){\rule[-0.175pt]{1.445pt}{0.350pt}}
\put(767,512){\rule[-0.175pt]{1.445pt}{0.350pt}}
\put(773,511){\rule[-0.175pt]{1.445pt}{0.350pt}}
\put(779,510){\rule[-0.175pt]{1.445pt}{0.350pt}}
\put(785,509){\rule[-0.175pt]{2.891pt}{0.350pt}}
\put(797,508){\rule[-0.175pt]{1.445pt}{0.350pt}}
\put(803,507){\rule[-0.175pt]{1.445pt}{0.350pt}}
\put(809,506){\rule[-0.175pt]{1.325pt}{0.350pt}}
\put(814,505){\rule[-0.175pt]{1.325pt}{0.350pt}}
\put(820,504){\rule[-0.175pt]{1.445pt}{0.350pt}}
\put(826,503){\rule[-0.175pt]{1.445pt}{0.350pt}}
\put(832,502){\rule[-0.175pt]{1.445pt}{0.350pt}}
\put(838,501){\rule[-0.175pt]{1.445pt}{0.350pt}}
\put(844,500){\rule[-0.175pt]{1.445pt}{0.350pt}}
\put(850,499){\rule[-0.175pt]{1.445pt}{0.350pt}}
\put(856,498){\rule[-0.175pt]{1.445pt}{0.350pt}}
\put(862,497){\rule[-0.175pt]{1.445pt}{0.350pt}}
\put(868,496){\rule[-0.175pt]{1.445pt}{0.350pt}}
\put(874,495){\rule[-0.175pt]{1.445pt}{0.350pt}}
\put(880,494){\rule[-0.175pt]{1.325pt}{0.350pt}}
\put(885,493){\rule[-0.175pt]{1.325pt}{0.350pt}}
\put(891,492){\rule[-0.175pt]{1.445pt}{0.350pt}}
\put(897,491){\rule[-0.175pt]{1.445pt}{0.350pt}}
\put(903,490){\rule[-0.175pt]{0.964pt}{0.350pt}}
\put(907,489){\rule[-0.175pt]{0.964pt}{0.350pt}}
\put(911,488){\rule[-0.175pt]{0.964pt}{0.350pt}}
\put(915,487){\rule[-0.175pt]{1.445pt}{0.350pt}}
\put(921,486){\rule[-0.175pt]{1.445pt}{0.350pt}}
\put(927,485){\rule[-0.175pt]{1.445pt}{0.350pt}}
\put(933,484){\rule[-0.175pt]{1.445pt}{0.350pt}}
\put(939,483){\rule[-0.175pt]{1.445pt}{0.350pt}}
\put(945,482){\rule[-0.175pt]{1.445pt}{0.350pt}}
\put(951,481){\rule[-0.175pt]{1.325pt}{0.350pt}}
\put(956,480){\rule[-0.175pt]{1.325pt}{0.350pt}}
\put(962,479){\rule[-0.175pt]{1.445pt}{0.350pt}}
\put(968,478){\rule[-0.175pt]{1.445pt}{0.350pt}}
\put(974,477){\rule[-0.175pt]{1.445pt}{0.350pt}}
\put(980,476){\rule[-0.175pt]{1.445pt}{0.350pt}}
\put(986,475){\rule[-0.175pt]{0.964pt}{0.350pt}}
\put(990,474){\rule[-0.175pt]{0.964pt}{0.350pt}}
\put(994,473){\rule[-0.175pt]{0.964pt}{0.350pt}}
\put(998,472){\rule[-0.175pt]{1.445pt}{0.350pt}}
\put(1004,471){\rule[-0.175pt]{1.445pt}{0.350pt}}
\put(1010,470){\rule[-0.175pt]{1.445pt}{0.350pt}}
\put(1016,469){\rule[-0.175pt]{1.445pt}{0.350pt}}
\put(1022,468){\rule[-0.175pt]{1.325pt}{0.350pt}}
\put(1027,467){\rule[-0.175pt]{1.325pt}{0.350pt}}
\put(1033,466){\rule[-0.175pt]{1.445pt}{0.350pt}}
\put(1039,465){\rule[-0.175pt]{1.445pt}{0.350pt}}
\put(1045,464){\rule[-0.175pt]{1.445pt}{0.350pt}}
\put(1051,463){\rule[-0.175pt]{1.445pt}{0.350pt}}
\put(1057,462){\rule[-0.175pt]{1.445pt}{0.350pt}}
\put(1063,461){\rule[-0.175pt]{1.445pt}{0.350pt}}
\put(1069,460){\rule[-0.175pt]{1.445pt}{0.350pt}}
\put(1075,459){\rule[-0.175pt]{1.445pt}{0.350pt}}
\put(1081,458){\rule[-0.175pt]{1.445pt}{0.350pt}}
\put(1087,457){\rule[-0.175pt]{1.445pt}{0.350pt}}
\put(1093,456){\rule[-0.175pt]{1.445pt}{0.350pt}}
\put(1099,455){\rule[-0.175pt]{1.445pt}{0.350pt}}
\put(1105,454){\rule[-0.175pt]{1.325pt}{0.350pt}}
\put(1110,453){\rule[-0.175pt]{1.325pt}{0.350pt}}
\put(1116,452){\rule[-0.175pt]{1.445pt}{0.350pt}}
\put(1122,451){\rule[-0.175pt]{1.445pt}{0.350pt}}
\put(1128,450){\rule[-0.175pt]{1.445pt}{0.350pt}}
\put(1134,449){\rule[-0.175pt]{1.445pt}{0.350pt}}
\put(1140,448){\rule[-0.175pt]{1.445pt}{0.350pt}}
\put(1146,447){\rule[-0.175pt]{1.445pt}{0.350pt}}
\put(1152,446){\rule[-0.175pt]{1.445pt}{0.350pt}}
\put(1158,445){\rule[-0.175pt]{1.445pt}{0.350pt}}
\put(1164,444){\rule[-0.175pt]{1.445pt}{0.350pt}}
\put(1170,443){\rule[-0.175pt]{1.445pt}{0.350pt}}
\put(1176,442){\rule[-0.175pt]{1.325pt}{0.350pt}}
\put(1181,441){\rule[-0.175pt]{1.325pt}{0.350pt}}
\put(1187,440){\rule[-0.175pt]{1.445pt}{0.350pt}}
\put(1193,439){\rule[-0.175pt]{1.445pt}{0.350pt}}
\put(1199,438){\rule[-0.175pt]{1.445pt}{0.350pt}}
\put(1205,437){\rule[-0.175pt]{1.445pt}{0.350pt}}
\put(1211,436){\rule[-0.175pt]{2.891pt}{0.350pt}}
\put(1223,435){\rule[-0.175pt]{1.445pt}{0.350pt}}
\put(1229,434){\rule[-0.175pt]{1.445pt}{0.350pt}}
\put(1235,433){\rule[-0.175pt]{1.445pt}{0.350pt}}
\put(1241,432){\rule[-0.175pt]{1.445pt}{0.350pt}}
\put(1247,431){\rule[-0.175pt]{1.325pt}{0.350pt}}
\put(1252,430){\rule[-0.175pt]{1.325pt}{0.350pt}}
\put(1258,429){\rule[-0.175pt]{2.891pt}{0.350pt}}
\put(1270,428){\rule[-0.175pt]{1.445pt}{0.350pt}}
\put(1276,427){\rule[-0.175pt]{1.445pt}{0.350pt}}
\put(1282,426){\rule[-0.175pt]{1.445pt}{0.350pt}}
\put(1288,425){\rule[-0.175pt]{1.445pt}{0.350pt}}
\put(1294,424){\rule[-0.175pt]{2.891pt}{0.350pt}}
\put(1306,423){\rule[-0.175pt]{1.445pt}{0.350pt}}
\put(1312,422){\rule[-0.175pt]{1.445pt}{0.350pt}}
\put(1318,421){\rule[-0.175pt]{2.650pt}{0.350pt}}
\put(1329,420){\rule[-0.175pt]{1.445pt}{0.350pt}}
\put(1335,419){\rule[-0.175pt]{1.445pt}{0.350pt}}
\put(1341,418){\rule[-0.175pt]{1.445pt}{0.350pt}}
\put(1347,417){\rule[-0.175pt]{1.445pt}{0.350pt}}
\put(1353,416){\rule[-0.175pt]{2.891pt}{0.350pt}}
\put(1365,415){\rule[-0.175pt]{1.445pt}{0.350pt}}
\put(1371,414){\rule[-0.175pt]{1.445pt}{0.350pt}}
\put(1377,413){\rule[-0.175pt]{2.891pt}{0.350pt}}
\put(1389,412){\rule[-0.175pt]{2.650pt}{0.350pt}}
\put(1400,411){\rule[-0.175pt]{1.445pt}{0.350pt}}
\put(1406,410){\rule[-0.175pt]{1.445pt}{0.350pt}}
\put(1412,409){\rule[-0.175pt]{2.891pt}{0.350pt}}
\put(1424,408){\rule[-0.175pt]{1.445pt}{0.350pt}}
\put(1430,407){\rule[-0.175pt]{1.445pt}{0.350pt}}
\sbox{\plotpoint}{\rule[-0.250pt]{0.500pt}{0.500pt}}%
\put(1600,676){\makebox(0,0)[r]{$\tilde{c}=0.5$}}
\put(1622,676){\usebox{\plotpoint}}
\put(1642,676){\usebox{\plotpoint}}
\put(1663,676){\usebox{\plotpoint}}
\put(1684,676){\usebox{\plotpoint}}
\put(1688,676){\usebox{\plotpoint}}
\put(264,325){\usebox{\plotpoint}}
\put(264,325){\usebox{\plotpoint}}
\put(281,336){\usebox{\plotpoint}}
\put(299,346){\usebox{\plotpoint}}
\put(317,357){\usebox{\plotpoint}}
\put(335,366){\usebox{\plotpoint}}
\put(354,375){\usebox{\plotpoint}}
\put(373,383){\usebox{\plotpoint}}
\put(393,389){\usebox{\plotpoint}}
\put(413,395){\usebox{\plotpoint}}
\put(433,400){\usebox{\plotpoint}}
\put(453,405){\usebox{\plotpoint}}
\put(474,408){\usebox{\plotpoint}}
\put(494,411){\usebox{\plotpoint}}
\put(515,413){\usebox{\plotpoint}}
\put(536,415){\usebox{\plotpoint}}
\put(556,416){\usebox{\plotpoint}}
\put(577,416){\usebox{\plotpoint}}
\put(598,416){\usebox{\plotpoint}}
\put(619,415){\usebox{\plotpoint}}
\put(639,414){\usebox{\plotpoint}}
\put(660,413){\usebox{\plotpoint}}
\put(681,411){\usebox{\plotpoint}}
\put(701,410){\usebox{\plotpoint}}
\put(722,407){\usebox{\plotpoint}}
\put(743,405){\usebox{\plotpoint}}
\put(763,402){\usebox{\plotpoint}}
\put(784,400){\usebox{\plotpoint}}
\put(804,397){\usebox{\plotpoint}}
\put(825,394){\usebox{\plotpoint}}
\put(845,391){\usebox{\plotpoint}}
\put(866,388){\usebox{\plotpoint}}
\put(886,385){\usebox{\plotpoint}}
\put(907,382){\usebox{\plotpoint}}
\put(927,378){\usebox{\plotpoint}}
\put(948,376){\usebox{\plotpoint}}
\put(968,372){\usebox{\plotpoint}}
\put(989,369){\usebox{\plotpoint}}
\put(1009,367){\usebox{\plotpoint}}
\put(1030,363){\usebox{\plotpoint}}
\put(1050,360){\usebox{\plotpoint}}
\put(1071,357){\usebox{\plotpoint}}
\put(1091,354){\usebox{\plotpoint}}
\put(1112,351){\usebox{\plotpoint}}
\put(1133,348){\usebox{\plotpoint}}
\put(1153,345){\usebox{\plotpoint}}
\put(1174,342){\usebox{\plotpoint}}
\put(1194,339){\usebox{\plotpoint}}
\put(1215,337){\usebox{\plotpoint}}
\put(1235,334){\usebox{\plotpoint}}
\put(1256,332){\usebox{\plotpoint}}
\put(1276,329){\usebox{\plotpoint}}
\put(1297,326){\usebox{\plotpoint}}
\put(1318,324){\usebox{\plotpoint}}
\put(1338,322){\usebox{\plotpoint}}
\put(1359,319){\usebox{\plotpoint}}
\put(1379,317){\usebox{\plotpoint}}
\put(1400,315){\usebox{\plotpoint}}
\put(1421,313){\usebox{\plotpoint}}
\put(1436,312){\usebox{\plotpoint}}
\sbox{\plotpoint}{\rule[-0.175pt]{0.350pt}{0.350pt}}%
\put(311,340){\circle*{12}}
\put(381,349){\circle*{12}}
\put(428,422){\circle*{12}}
\put(475,426){\circle*{12}}
\put(522,596){\circle*{12}}
\put(569,384){\circle*{12}}
\put(616,460){\circle*{12}}
\put(662,355){\circle*{12}}
\put(709,328){\circle*{12}}
\put(756,537){\circle*{12}}
\put(803,473){\circle*{12}}
\put(850,584){\circle*{12}}
\put(897,483){\circle*{12}}
\put(944,468){\circle*{12}}
\put(991,473){\circle*{12}}
\put(1038,466){\circle*{12}}
\put(1084,464){\circle*{12}}
\put(1131,617){\circle*{12}}
\put(1178,558){\circle*{12}}
\put(311,319){\rule[-0.175pt]{0.350pt}{10.118pt}}
\put(301,319){\rule[-0.175pt]{4.818pt}{0.350pt}}
\put(301,361){\rule[-0.175pt]{4.818pt}{0.350pt}}
\put(381,305){\rule[-0.175pt]{0.350pt}{21.199pt}}
\put(371,305){\rule[-0.175pt]{4.818pt}{0.350pt}}
\put(371,393){\rule[-0.175pt]{4.818pt}{0.350pt}}
\put(428,368){\rule[-0.175pt]{0.350pt}{26.258pt}}
\put(418,368){\rule[-0.175pt]{4.818pt}{0.350pt}}
\put(418,477){\rule[-0.175pt]{4.818pt}{0.350pt}}
\put(475,372){\rule[-0.175pt]{0.350pt}{26.258pt}}
\put(465,372){\rule[-0.175pt]{4.818pt}{0.350pt}}
\put(465,481){\rule[-0.175pt]{4.818pt}{0.350pt}}
\put(522,466){\rule[-0.175pt]{0.350pt}{62.634pt}}
\put(512,466){\rule[-0.175pt]{4.818pt}{0.350pt}}
\put(512,726){\rule[-0.175pt]{4.818pt}{0.350pt}}
\put(569,282){\rule[-0.175pt]{0.350pt}{49.384pt}}
\put(559,282){\rule[-0.175pt]{4.818pt}{0.350pt}}
\put(559,487){\rule[-0.175pt]{4.818pt}{0.350pt}}
\put(616,355){\rule[-0.175pt]{0.350pt}{50.589pt}}
\put(606,355){\rule[-0.175pt]{4.818pt}{0.350pt}}
\put(606,565){\rule[-0.175pt]{4.818pt}{0.350pt}}
\put(662,265){\rule[-0.175pt]{0.350pt}{43.362pt}}
\put(652,265){\rule[-0.175pt]{4.818pt}{0.350pt}}
\put(652,445){\rule[-0.175pt]{4.818pt}{0.350pt}}
\put(709,244){\rule[-0.175pt]{0.350pt}{40.471pt}}
\put(699,244){\rule[-0.175pt]{4.818pt}{0.350pt}}
\put(699,412){\rule[-0.175pt]{4.818pt}{0.350pt}}
\put(756,460){\rule[-0.175pt]{0.350pt}{37.339pt}}
\put(746,460){\rule[-0.175pt]{4.818pt}{0.350pt}}
\put(746,615){\rule[-0.175pt]{4.818pt}{0.350pt}}
\put(803,370){\rule[-0.175pt]{0.350pt}{49.384pt}}
\put(793,370){\rule[-0.175pt]{4.818pt}{0.350pt}}
\put(793,575){\rule[-0.175pt]{4.818pt}{0.350pt}}
\put(850,449){\rule[-0.175pt]{0.350pt}{64.802pt}}
\put(840,449){\rule[-0.175pt]{4.818pt}{0.350pt}}
\put(840,718){\rule[-0.175pt]{4.818pt}{0.350pt}}
\put(897,370){\rule[-0.175pt]{0.350pt}{54.443pt}}
\put(887,370){\rule[-0.175pt]{4.818pt}{0.350pt}}
\put(887,596){\rule[-0.175pt]{4.818pt}{0.350pt}}
\put(944,405){\rule[-0.175pt]{0.350pt}{30.353pt}}
\put(934,405){\rule[-0.175pt]{4.818pt}{0.350pt}}
\put(934,531){\rule[-0.175pt]{4.818pt}{0.350pt}}
\put(991,403){\rule[-0.175pt]{0.350pt}{33.485pt}}
\put(981,403){\rule[-0.175pt]{4.818pt}{0.350pt}}
\put(981,542){\rule[-0.175pt]{4.818pt}{0.350pt}}
\put(1038,387){\rule[-0.175pt]{0.350pt}{38.303pt}}
\put(1028,387){\rule[-0.175pt]{4.818pt}{0.350pt}}
\put(1028,546){\rule[-0.175pt]{4.818pt}{0.350pt}}
\put(1084,393){\rule[-0.175pt]{0.350pt}{34.208pt}}
\put(1074,393){\rule[-0.175pt]{4.818pt}{0.350pt}}
\put(1074,535){\rule[-0.175pt]{4.818pt}{0.350pt}}
\put(1131,535){\rule[-0.175pt]{0.350pt}{39.508pt}}
\put(1121,535){\rule[-0.175pt]{4.818pt}{0.350pt}}
\put(1121,699){\rule[-0.175pt]{4.818pt}{0.350pt}}
\put(1178,475){\rule[-0.175pt]{0.350pt}{40.230pt}}
\put(1168,475){\rule[-0.175pt]{4.818pt}{0.350pt}}
\put(1168,642){\rule[-0.175pt]{4.818pt}{0.350pt}}
\end{picture}

%% file: f_eewp1.tex
\setlength{\unitlength}{0.240900pt}
\ifx\plotpoint\undefined\newsavebox{\plotpoint}\fi
\begin{picture}(1720,900)(0,0)
\tenrm
\sbox{\plotpoint}{\rule[-0.175pt]{0.350pt}{0.350pt}}%
\put(264,158){\rule[-0.175pt]{282.335pt}{0.350pt}}
\put(264,158){\rule[-0.175pt]{4.818pt}{0.350pt}}
\put(242,158){\makebox(0,0)[r]{0}}
\put(1416,158){\rule[-0.175pt]{4.818pt}{0.350pt}}
\put(264,263){\rule[-0.175pt]{4.818pt}{0.350pt}}
\put(242,263){\makebox(0,0)[r]{5}}
\put(1416,263){\rule[-0.175pt]{4.818pt}{0.350pt}}
\put(264,368){\rule[-0.175pt]{4.818pt}{0.350pt}}
\put(242,368){\makebox(0,0)[r]{10}}
\put(1416,368){\rule[-0.175pt]{4.818pt}{0.350pt}}
\put(264,473){\rule[-0.175pt]{4.818pt}{0.350pt}}
\put(242,473){\makebox(0,0)[r]{15}}
\put(1416,473){\rule[-0.175pt]{4.818pt}{0.350pt}}
\put(264,577){\rule[-0.175pt]{4.818pt}{0.350pt}}
\put(242,577){\makebox(0,0)[r]{20}}
\put(1416,577){\rule[-0.175pt]{4.818pt}{0.350pt}}
\put(264,682){\rule[-0.175pt]{4.818pt}{0.350pt}}
\put(242,682){\makebox(0,0)[r]{25}}
\put(1416,682){\rule[-0.175pt]{4.818pt}{0.350pt}}
\put(264,787){\rule[-0.175pt]{4.818pt}{0.350pt}}
\put(242,787){\makebox(0,0)[r]{30}}
\put(1416,787){\rule[-0.175pt]{4.818pt}{0.350pt}}
\put(264,158){\rule[-0.175pt]{0.350pt}{4.818pt}}
\put(264,113){\makebox(0,0){1}}
\put(264,767){\rule[-0.175pt]{0.350pt}{4.818pt}}
\put(381,158){\rule[-0.175pt]{0.350pt}{4.818pt}}
\put(381,113){\makebox(0,0){1.05}}
\put(381,767){\rule[-0.175pt]{0.350pt}{4.818pt}}
\put(498,158){\rule[-0.175pt]{0.350pt}{4.818pt}}
\put(498,113){\makebox(0,0){1.1}}
\put(498,767){\rule[-0.175pt]{0.350pt}{4.818pt}}
\put(616,158){\rule[-0.175pt]{0.350pt}{4.818pt}}
\put(616,113){\makebox(0,0){1.15}}
\put(616,767){\rule[-0.175pt]{0.350pt}{4.818pt}}
\put(733,158){\rule[-0.175pt]{0.350pt}{4.818pt}}
\put(733,113){\makebox(0,0){1.2}}
\put(733,767){\rule[-0.175pt]{0.350pt}{4.818pt}}
\put(850,158){\rule[-0.175pt]{0.350pt}{4.818pt}}
\put(850,113){\makebox(0,0){1.25}}
\put(850,767){\rule[-0.175pt]{0.350pt}{4.818pt}}
\put(967,158){\rule[-0.175pt]{0.350pt}{4.818pt}}
\put(967,113){\makebox(0,0){1.3}}
\put(967,767){\rule[-0.175pt]{0.350pt}{4.818pt}}
\put(1084,158){\rule[-0.175pt]{0.350pt}{4.818pt}}
\put(1084,113){\makebox(0,0){1.35}}
\put(1084,767){\rule[-0.175pt]{0.350pt}{4.818pt}}
\put(1202,158){\rule[-0.175pt]{0.350pt}{4.818pt}}
\put(1202,113){\makebox(0,0){1.4}}
\put(1202,767){\rule[-0.175pt]{0.350pt}{4.818pt}}
\put(1319,158){\rule[-0.175pt]{0.350pt}{4.818pt}}
\put(1319,113){\makebox(0,0){1.45}}
\put(1319,767){\rule[-0.175pt]{0.350pt}{4.818pt}}
\put(1436,158){\rule[-0.175pt]{0.350pt}{4.818pt}}
\put(1436,113){\makebox(0,0){1.5}}
\put(1436,767){\rule[-0.175pt]{0.350pt}{4.818pt}}
\put(264,158){\rule[-0.175pt]{282.335pt}{0.350pt}}
\put(1436,158){\rule[-0.175pt]{0.350pt}{151.526pt}}
\put(264,787){\rule[-0.175pt]{282.335pt}{0.350pt}}
\put(199,877){\makebox(0,0)[l]{\shortstack{$\sigma$ (nb)}}}
\put(850,23){\makebox(0,0){$\sqrt{q^2}$ (GeV)}}
\put(264,158){\rule[-0.175pt]{0.350pt}{151.526pt}}
\sbox{\plotpoint}{\rule[-0.250pt]{0.500pt}{0.500pt}}%
\put(1624,766){\makebox(0,0)[r]{$\tilde{c}=0.5~\,$}}
\put(1646,766){\usebox{\plotpoint}}
\put(1666,766){\usebox{\plotpoint}}
\put(1687,766){\usebox{\plotpoint}}
\put(1708,766){\usebox{\plotpoint}}
\put(1712,766){\usebox{\plotpoint}}
\put(264,376){\usebox{\plotpoint}}
\put(264,376){\usebox{\plotpoint}}
\put(278,390){\usebox{\plotpoint}}
\put(293,405){\usebox{\plotpoint}}
\put(308,419){\usebox{\plotpoint}}
\put(324,432){\usebox{\plotpoint}}
\put(340,445){\usebox{\plotpoint}}
\put(357,458){\usebox{\plotpoint}}
\put(373,470){\usebox{\plotpoint}}
\put(390,482){\usebox{\plotpoint}}
\put(408,494){\usebox{\plotpoint}}
\put(426,504){\usebox{\plotpoint}}
\put(443,515){\usebox{\plotpoint}}
\put(462,524){\usebox{\plotpoint}}
\put(480,533){\usebox{\plotpoint}}
\put(499,542){\usebox{\plotpoint}}
\put(518,550){\usebox{\plotpoint}}
\put(537,558){\usebox{\plotpoint}}
\put(557,566){\usebox{\plotpoint}}
\put(577,572){\usebox{\plotpoint}}
\put(596,579){\usebox{\plotpoint}}
\put(616,585){\usebox{\plotpoint}}
\put(636,590){\usebox{\plotpoint}}
\put(656,595){\usebox{\plotpoint}}
\put(677,599){\usebox{\plotpoint}}
\put(697,604){\usebox{\plotpoint}}
\put(717,607){\usebox{\plotpoint}}
\put(738,612){\usebox{\plotpoint}}
\put(758,614){\usebox{\plotpoint}}
\put(779,618){\usebox{\plotpoint}}
\put(799,620){\usebox{\plotpoint}}
\put(820,623){\usebox{\plotpoint}}
\put(840,625){\usebox{\plotpoint}}
\put(861,627){\usebox{\plotpoint}}
\put(882,629){\usebox{\plotpoint}}
\put(903,630){\usebox{\plotpoint}}
\put(923,631){\usebox{\plotpoint}}
\put(944,633){\usebox{\plotpoint}}
\put(965,634){\usebox{\plotpoint}}
\put(985,634){\usebox{\plotpoint}}
\put(1006,635){\usebox{\plotpoint}}
\put(1027,636){\usebox{\plotpoint}}
\put(1048,637){\usebox{\plotpoint}}
\put(1068,637){\usebox{\plotpoint}}
\put(1089,637){\usebox{\plotpoint}}
\put(1110,638){\usebox{\plotpoint}}
\put(1130,638){\usebox{\plotpoint}}
\put(1151,638){\usebox{\plotpoint}}
\put(1172,638){\usebox{\plotpoint}}
\put(1193,637){\usebox{\plotpoint}}
\put(1213,637){\usebox{\plotpoint}}
\put(1234,637){\usebox{\plotpoint}}
\put(1255,636){\usebox{\plotpoint}}
\put(1276,636){\usebox{\plotpoint}}
\put(1296,635){\usebox{\plotpoint}}
\put(1317,635){\usebox{\plotpoint}}
\put(1338,634){\usebox{\plotpoint}}
\put(1359,634){\usebox{\plotpoint}}
\put(1379,633){\usebox{\plotpoint}}
\put(1400,632){\usebox{\plotpoint}}
\put(1421,632){\usebox{\plotpoint}}
\put(1436,631){\usebox{\plotpoint}}
\sbox{\plotpoint}{\rule[-0.500pt]{1.000pt}{1.000pt}}%
\put(1624,721){\makebox(0,0)[r]{$\tilde{c}=0.4~\,$}}
\put(1646,721){\rule[-0.500pt]{15.899pt}{1.000pt}}
\put(264,338){\usebox{\plotpoint}}
\put(264,338){\usebox{\plotpoint}}
\put(265,339){\usebox{\plotpoint}}
\put(266,340){\usebox{\plotpoint}}
\put(268,341){\usebox{\plotpoint}}
\put(269,342){\usebox{\plotpoint}}
\put(270,343){\usebox{\plotpoint}}
\put(272,344){\usebox{\plotpoint}}
\put(273,345){\usebox{\plotpoint}}
\put(274,346){\usebox{\plotpoint}}
\put(276,347){\usebox{\plotpoint}}
\put(277,348){\usebox{\plotpoint}}
\put(278,349){\usebox{\plotpoint}}
\put(279,350){\usebox{\plotpoint}}
\put(280,351){\usebox{\plotpoint}}
\put(282,352){\usebox{\plotpoint}}
\put(283,353){\usebox{\plotpoint}}
\put(284,354){\usebox{\plotpoint}}
\put(285,355){\usebox{\plotpoint}}
\put(286,356){\usebox{\plotpoint}}
\put(288,357){\usebox{\plotpoint}}
\put(289,358){\usebox{\plotpoint}}
\put(291,359){\usebox{\plotpoint}}
\put(292,360){\usebox{\plotpoint}}
\put(294,361){\usebox{\plotpoint}}
\put(295,362){\usebox{\plotpoint}}
\put(297,363){\usebox{\plotpoint}}
\put(298,364){\usebox{\plotpoint}}
\put(300,365){\usebox{\plotpoint}}
\put(301,366){\usebox{\plotpoint}}
\put(302,367){\usebox{\plotpoint}}
\put(303,368){\usebox{\plotpoint}}
\put(304,369){\usebox{\plotpoint}}
\put(306,370){\usebox{\plotpoint}}
\put(307,371){\usebox{\plotpoint}}
\put(308,372){\usebox{\plotpoint}}
\put(309,373){\usebox{\plotpoint}}
\put(311,374){\usebox{\plotpoint}}
\put(312,375){\usebox{\plotpoint}}
\put(314,376){\usebox{\plotpoint}}
\put(315,377){\usebox{\plotpoint}}
\put(317,378){\usebox{\plotpoint}}
\put(318,379){\usebox{\plotpoint}}
\put(320,380){\usebox{\plotpoint}}
\put(321,381){\usebox{\plotpoint}}
\put(323,382){\usebox{\plotpoint}}
\put(324,383){\usebox{\plotpoint}}
\put(326,384){\usebox{\plotpoint}}
\put(327,385){\usebox{\plotpoint}}
\put(329,386){\usebox{\plotpoint}}
\put(330,387){\usebox{\plotpoint}}
\put(332,388){\usebox{\plotpoint}}
\put(333,389){\usebox{\plotpoint}}
\put(335,390){\usebox{\plotpoint}}
\put(336,391){\usebox{\plotpoint}}
\put(338,392){\usebox{\plotpoint}}
\put(340,393){\usebox{\plotpoint}}
\put(341,394){\usebox{\plotpoint}}
\put(343,395){\usebox{\plotpoint}}
\put(345,396){\usebox{\plotpoint}}
\put(347,397){\usebox{\plotpoint}}
\put(348,398){\usebox{\plotpoint}}
\put(350,399){\usebox{\plotpoint}}
\put(351,400){\usebox{\plotpoint}}
\put(353,401){\usebox{\plotpoint}}
\put(354,402){\usebox{\plotpoint}}
\put(356,403){\usebox{\plotpoint}}
\put(357,404){\usebox{\plotpoint}}
\put(359,405){\usebox{\plotpoint}}
\put(361,406){\usebox{\plotpoint}}
\put(363,407){\usebox{\plotpoint}}
\put(365,408){\usebox{\plotpoint}}
\put(367,409){\usebox{\plotpoint}}
\put(369,410){\usebox{\plotpoint}}
\put(371,411){\usebox{\plotpoint}}
\put(372,412){\usebox{\plotpoint}}
\put(374,413){\usebox{\plotpoint}}
\put(375,414){\usebox{\plotpoint}}
\put(377,415){\usebox{\plotpoint}}
\put(378,416){\usebox{\plotpoint}}
\put(380,417){\usebox{\plotpoint}}
\put(382,418){\usebox{\plotpoint}}
\put(384,419){\usebox{\plotpoint}}
\put(386,420){\usebox{\plotpoint}}
\put(388,421){\usebox{\plotpoint}}
\put(390,422){\usebox{\plotpoint}}
\put(392,423){\usebox{\plotpoint}}
\put(394,424){\usebox{\plotpoint}}
\put(396,425){\usebox{\plotpoint}}
\put(398,426){\usebox{\plotpoint}}
\put(400,427){\usebox{\plotpoint}}
\put(402,428){\usebox{\plotpoint}}
\put(404,429){\usebox{\plotpoint}}
\put(406,430){\usebox{\plotpoint}}
\put(408,431){\usebox{\plotpoint}}
\put(410,432){\usebox{\plotpoint}}
\put(412,433){\usebox{\plotpoint}}
\put(414,434){\usebox{\plotpoint}}
\put(416,435){\usebox{\plotpoint}}
\put(418,436){\usebox{\plotpoint}}
\put(420,437){\usebox{\plotpoint}}
\put(422,438){\usebox{\plotpoint}}
\put(425,439){\usebox{\plotpoint}}
\put(427,440){\usebox{\plotpoint}}
\put(429,441){\usebox{\plotpoint}}
\put(432,442){\usebox{\plotpoint}}
\put(434,443){\usebox{\plotpoint}}
\put(437,444){\usebox{\plotpoint}}
\put(439,445){\usebox{\plotpoint}}
\put(441,446){\usebox{\plotpoint}}
\put(444,447){\usebox{\plotpoint}}
\put(446,448){\usebox{\plotpoint}}
\put(448,449){\usebox{\plotpoint}}
\put(450,450){\usebox{\plotpoint}}
\put(453,451){\usebox{\plotpoint}}
\put(455,452){\usebox{\plotpoint}}
\put(457,453){\usebox{\plotpoint}}
\put(460,454){\usebox{\plotpoint}}
\put(462,455){\usebox{\plotpoint}}
\put(464,456){\usebox{\plotpoint}}
\put(467,457){\usebox{\plotpoint}}
\put(469,458){\usebox{\plotpoint}}
\put(472,459){\usebox{\plotpoint}}
\put(474,460){\usebox{\plotpoint}}
\put(476,461){\usebox{\plotpoint}}
\put(480,462){\usebox{\plotpoint}}
\put(483,463){\usebox{\plotpoint}}
\put(486,464){\usebox{\plotpoint}}
\put(489,465){\usebox{\plotpoint}}
\put(492,466){\usebox{\plotpoint}}
\put(495,467){\usebox{\plotpoint}}
\put(498,468){\usebox{\plotpoint}}
\put(501,469){\usebox{\plotpoint}}
\put(504,470){\usebox{\plotpoint}}
\put(507,471){\usebox{\plotpoint}}
\put(510,472){\usebox{\plotpoint}}
\put(513,473){\usebox{\plotpoint}}
\put(516,474){\usebox{\plotpoint}}
\put(520,475){\usebox{\plotpoint}}
\put(524,476){\usebox{\plotpoint}}
\put(527,477){\usebox{\plotpoint}}
\put(530,478){\usebox{\plotpoint}}
\put(533,479){\usebox{\plotpoint}}
\put(536,480){\usebox{\plotpoint}}
\put(540,481){\usebox{\plotpoint}}
\put(544,482){\usebox{\plotpoint}}
\put(548,483){\usebox{\plotpoint}}
\put(552,484){\usebox{\plotpoint}}
\put(556,485){\usebox{\plotpoint}}
\put(560,486){\usebox{\plotpoint}}
\put(564,487){\usebox{\plotpoint}}
\put(568,488){\usebox{\plotpoint}}
\put(572,489){\usebox{\plotpoint}}
\put(576,490){\usebox{\plotpoint}}
\put(580,491){\usebox{\plotpoint}}
\put(584,492){\usebox{\plotpoint}}
\put(587,493){\usebox{\plotpoint}}
\put(591,494){\usebox{\plotpoint}}
\put(595,495){\rule[-0.500pt]{1.445pt}{1.000pt}}
\put(601,496){\rule[-0.500pt]{1.445pt}{1.000pt}}
\put(607,497){\usebox{\plotpoint}}
\put(611,498){\usebox{\plotpoint}}
\put(615,499){\usebox{\plotpoint}}
\put(619,500){\rule[-0.500pt]{1.445pt}{1.000pt}}
\put(625,501){\rule[-0.500pt]{1.445pt}{1.000pt}}
\put(631,502){\rule[-0.500pt]{1.445pt}{1.000pt}}
\put(637,503){\rule[-0.500pt]{1.445pt}{1.000pt}}
\put(643,504){\rule[-0.500pt]{1.445pt}{1.000pt}}
\put(649,505){\rule[-0.500pt]{1.445pt}{1.000pt}}
\put(655,506){\rule[-0.500pt]{1.445pt}{1.000pt}}
\put(661,507){\rule[-0.500pt]{1.445pt}{1.000pt}}
\put(667,508){\rule[-0.500pt]{1.325pt}{1.000pt}}
\put(672,509){\rule[-0.500pt]{1.325pt}{1.000pt}}
\put(678,510){\rule[-0.500pt]{2.891pt}{1.000pt}}
\put(690,511){\rule[-0.500pt]{1.445pt}{1.000pt}}
\put(696,512){\rule[-0.500pt]{1.445pt}{1.000pt}}
\put(702,513){\rule[-0.500pt]{2.891pt}{1.000pt}}
\put(714,514){\rule[-0.500pt]{1.445pt}{1.000pt}}
\put(720,515){\rule[-0.500pt]{1.445pt}{1.000pt}}
\put(726,516){\rule[-0.500pt]{2.891pt}{1.000pt}}
\put(738,517){\rule[-0.500pt]{2.650pt}{1.000pt}}
\put(749,518){\rule[-0.500pt]{2.891pt}{1.000pt}}
\put(761,519){\rule[-0.500pt]{2.891pt}{1.000pt}}
\put(773,520){\rule[-0.500pt]{2.891pt}{1.000pt}}
\put(785,521){\rule[-0.500pt]{2.891pt}{1.000pt}}
\put(797,522){\rule[-0.500pt]{2.891pt}{1.000pt}}
\put(809,523){\rule[-0.500pt]{2.650pt}{1.000pt}}
\put(820,524){\rule[-0.500pt]{5.782pt}{1.000pt}}
\put(844,525){\rule[-0.500pt]{5.782pt}{1.000pt}}
\put(868,526){\rule[-0.500pt]{5.541pt}{1.000pt}}
\put(891,527){\rule[-0.500pt]{8.672pt}{1.000pt}}
\put(927,528){\rule[-0.500pt]{42.880pt}{1.000pt}}
\put(1105,527){\rule[-0.500pt]{11.322pt}{1.000pt}}
\put(1152,526){\rule[-0.500pt]{8.431pt}{1.000pt}}
\put(1187,525){\rule[-0.500pt]{8.672pt}{1.000pt}}
\put(1223,524){\rule[-0.500pt]{5.782pt}{1.000pt}}
\put(1247,523){\rule[-0.500pt]{5.541pt}{1.000pt}}
\put(1270,522){\rule[-0.500pt]{5.782pt}{1.000pt}}
\put(1294,521){\rule[-0.500pt]{5.782pt}{1.000pt}}
\put(1318,520){\rule[-0.500pt]{5.541pt}{1.000pt}}
\put(1341,519){\rule[-0.500pt]{5.782pt}{1.000pt}}
\put(1365,518){\rule[-0.500pt]{5.782pt}{1.000pt}}
\put(1389,517){\rule[-0.500pt]{5.541pt}{1.000pt}}
\put(1412,516){\rule[-0.500pt]{5.782pt}{1.000pt}}
\sbox{\plotpoint}{\rule[-0.175pt]{0.350pt}{0.350pt}}%
\put(1624,676){\makebox(0,0)[r]{$\tilde{c}=0.25$}}
\put(1646,676){\rule[-0.175pt]{15.899pt}{0.350pt}}
\put(264,287){\usebox{\plotpoint}}
\put(264,287){\rule[-0.175pt]{0.413pt}{0.350pt}}
\put(265,288){\rule[-0.175pt]{0.413pt}{0.350pt}}
\put(267,289){\rule[-0.175pt]{0.413pt}{0.350pt}}
\put(269,290){\rule[-0.175pt]{0.413pt}{0.350pt}}
\put(270,291){\rule[-0.175pt]{0.413pt}{0.350pt}}
\put(272,292){\rule[-0.175pt]{0.413pt}{0.350pt}}
\put(274,293){\rule[-0.175pt]{0.413pt}{0.350pt}}
\put(276,294){\rule[-0.175pt]{0.482pt}{0.350pt}}
\put(278,295){\rule[-0.175pt]{0.482pt}{0.350pt}}
\put(280,296){\rule[-0.175pt]{0.482pt}{0.350pt}}
\put(282,297){\rule[-0.175pt]{0.482pt}{0.350pt}}
\put(284,298){\rule[-0.175pt]{0.482pt}{0.350pt}}
\put(286,299){\rule[-0.175pt]{0.482pt}{0.350pt}}
\put(288,300){\rule[-0.175pt]{0.482pt}{0.350pt}}
\put(290,301){\rule[-0.175pt]{0.482pt}{0.350pt}}
\put(292,302){\rule[-0.175pt]{0.482pt}{0.350pt}}
\put(294,303){\rule[-0.175pt]{0.482pt}{0.350pt}}
\put(296,304){\rule[-0.175pt]{0.482pt}{0.350pt}}
\put(298,305){\rule[-0.175pt]{0.482pt}{0.350pt}}
\put(300,306){\rule[-0.175pt]{0.442pt}{0.350pt}}
\put(301,307){\rule[-0.175pt]{0.442pt}{0.350pt}}
\put(303,308){\rule[-0.175pt]{0.442pt}{0.350pt}}
\put(305,309){\rule[-0.175pt]{0.442pt}{0.350pt}}
\put(307,310){\rule[-0.175pt]{0.442pt}{0.350pt}}
\put(309,311){\rule[-0.175pt]{0.442pt}{0.350pt}}
\put(311,312){\rule[-0.175pt]{0.578pt}{0.350pt}}
\put(313,313){\rule[-0.175pt]{0.578pt}{0.350pt}}
\put(315,314){\rule[-0.175pt]{0.578pt}{0.350pt}}
\put(318,315){\rule[-0.175pt]{0.578pt}{0.350pt}}
\put(320,316){\rule[-0.175pt]{0.578pt}{0.350pt}}
\put(322,317){\rule[-0.175pt]{0.482pt}{0.350pt}}
\put(325,318){\rule[-0.175pt]{0.482pt}{0.350pt}}
\put(327,319){\rule[-0.175pt]{0.482pt}{0.350pt}}
\put(329,320){\rule[-0.175pt]{0.482pt}{0.350pt}}
\put(331,321){\rule[-0.175pt]{0.482pt}{0.350pt}}
\put(333,322){\rule[-0.175pt]{0.482pt}{0.350pt}}
\put(335,323){\rule[-0.175pt]{0.578pt}{0.350pt}}
\put(337,324){\rule[-0.175pt]{0.578pt}{0.350pt}}
\put(339,325){\rule[-0.175pt]{0.578pt}{0.350pt}}
\put(342,326){\rule[-0.175pt]{0.578pt}{0.350pt}}
\put(344,327){\rule[-0.175pt]{0.578pt}{0.350pt}}
\put(346,328){\rule[-0.175pt]{0.723pt}{0.350pt}}
\put(350,329){\rule[-0.175pt]{0.723pt}{0.350pt}}
\put(353,330){\rule[-0.175pt]{0.723pt}{0.350pt}}
\put(356,331){\rule[-0.175pt]{0.723pt}{0.350pt}}
\put(359,332){\rule[-0.175pt]{0.578pt}{0.350pt}}
\put(361,333){\rule[-0.175pt]{0.578pt}{0.350pt}}
\put(363,334){\rule[-0.175pt]{0.578pt}{0.350pt}}
\put(366,335){\rule[-0.175pt]{0.578pt}{0.350pt}}
\put(368,336){\rule[-0.175pt]{0.578pt}{0.350pt}}
\put(370,337){\rule[-0.175pt]{0.662pt}{0.350pt}}
\put(373,338){\rule[-0.175pt]{0.662pt}{0.350pt}}
\put(376,339){\rule[-0.175pt]{0.662pt}{0.350pt}}
\put(379,340){\rule[-0.175pt]{0.662pt}{0.350pt}}
\put(382,341){\rule[-0.175pt]{0.723pt}{0.350pt}}
\put(385,342){\rule[-0.175pt]{0.723pt}{0.350pt}}
\put(388,343){\rule[-0.175pt]{0.723pt}{0.350pt}}
\put(391,344){\rule[-0.175pt]{0.723pt}{0.350pt}}
\put(394,345){\rule[-0.175pt]{0.723pt}{0.350pt}}
\put(397,346){\rule[-0.175pt]{0.723pt}{0.350pt}}
\put(400,347){\rule[-0.175pt]{0.723pt}{0.350pt}}
\put(403,348){\rule[-0.175pt]{0.723pt}{0.350pt}}
\put(406,349){\rule[-0.175pt]{0.964pt}{0.350pt}}
\put(410,350){\rule[-0.175pt]{0.964pt}{0.350pt}}
\put(414,351){\rule[-0.175pt]{0.964pt}{0.350pt}}
\put(418,352){\rule[-0.175pt]{0.723pt}{0.350pt}}
\put(421,353){\rule[-0.175pt]{0.723pt}{0.350pt}}
\put(424,354){\rule[-0.175pt]{0.723pt}{0.350pt}}
\put(427,355){\rule[-0.175pt]{0.723pt}{0.350pt}}
\put(430,356){\rule[-0.175pt]{0.964pt}{0.350pt}}
\put(434,357){\rule[-0.175pt]{0.964pt}{0.350pt}}
\put(438,358){\rule[-0.175pt]{0.964pt}{0.350pt}}
\put(442,359){\rule[-0.175pt]{0.883pt}{0.350pt}}
\put(445,360){\rule[-0.175pt]{0.883pt}{0.350pt}}
\put(449,361){\rule[-0.175pt]{0.883pt}{0.350pt}}
\put(452,362){\rule[-0.175pt]{0.964pt}{0.350pt}}
\put(457,363){\rule[-0.175pt]{0.964pt}{0.350pt}}
\put(461,364){\rule[-0.175pt]{0.964pt}{0.350pt}}
\put(465,365){\rule[-0.175pt]{1.445pt}{0.350pt}}
\put(471,366){\rule[-0.175pt]{1.445pt}{0.350pt}}
\put(477,367){\rule[-0.175pt]{0.964pt}{0.350pt}}
\put(481,368){\rule[-0.175pt]{0.964pt}{0.350pt}}
\put(485,369){\rule[-0.175pt]{0.964pt}{0.350pt}}
\put(489,370){\rule[-0.175pt]{1.445pt}{0.350pt}}
\put(495,371){\rule[-0.175pt]{1.445pt}{0.350pt}}
\put(501,372){\rule[-0.175pt]{1.445pt}{0.350pt}}
\put(507,373){\rule[-0.175pt]{1.445pt}{0.350pt}}
\put(513,374){\rule[-0.175pt]{0.883pt}{0.350pt}}
\put(516,375){\rule[-0.175pt]{0.883pt}{0.350pt}}
\put(520,376){\rule[-0.175pt]{0.883pt}{0.350pt}}
\put(524,377){\rule[-0.175pt]{2.891pt}{0.350pt}}
\put(536,378){\rule[-0.175pt]{1.445pt}{0.350pt}}
\put(542,379){\rule[-0.175pt]{1.445pt}{0.350pt}}
\put(548,380){\rule[-0.175pt]{1.445pt}{0.350pt}}
\put(554,381){\rule[-0.175pt]{1.445pt}{0.350pt}}
\put(560,382){\rule[-0.175pt]{1.445pt}{0.350pt}}
\put(566,383){\rule[-0.175pt]{1.445pt}{0.350pt}}
\put(572,384){\rule[-0.175pt]{2.891pt}{0.350pt}}
\put(584,385){\rule[-0.175pt]{2.650pt}{0.350pt}}
\put(595,386){\rule[-0.175pt]{1.445pt}{0.350pt}}
\put(601,387){\rule[-0.175pt]{1.445pt}{0.350pt}}
\put(607,388){\rule[-0.175pt]{2.891pt}{0.350pt}}
\put(619,389){\rule[-0.175pt]{2.891pt}{0.350pt}}
\put(631,390){\rule[-0.175pt]{2.891pt}{0.350pt}}
\put(643,391){\rule[-0.175pt]{2.891pt}{0.350pt}}
\put(655,392){\rule[-0.175pt]{5.541pt}{0.350pt}}
\put(678,393){\rule[-0.175pt]{2.891pt}{0.350pt}}
\put(690,394){\rule[-0.175pt]{5.782pt}{0.350pt}}
\put(714,395){\rule[-0.175pt]{5.782pt}{0.350pt}}
\put(738,396){\rule[-0.175pt]{8.431pt}{0.350pt}}
\put(773,397){\rule[-0.175pt]{28.426pt}{0.350pt}}
\put(891,396){\rule[-0.175pt]{11.563pt}{0.350pt}}
\put(939,395){\rule[-0.175pt]{8.431pt}{0.350pt}}
\put(974,394){\rule[-0.175pt]{8.672pt}{0.350pt}}
\put(1010,393){\rule[-0.175pt]{5.541pt}{0.350pt}}
\put(1033,392){\rule[-0.175pt]{5.782pt}{0.350pt}}
\put(1057,391){\rule[-0.175pt]{5.782pt}{0.350pt}}
\put(1081,390){\rule[-0.175pt]{5.782pt}{0.350pt}}
\put(1105,389){\rule[-0.175pt]{5.541pt}{0.350pt}}
\put(1128,388){\rule[-0.175pt]{2.891pt}{0.350pt}}
\put(1140,387){\rule[-0.175pt]{5.782pt}{0.350pt}}
\put(1164,386){\rule[-0.175pt]{5.541pt}{0.350pt}}
\put(1187,385){\rule[-0.175pt]{5.782pt}{0.350pt}}
\put(1211,384){\rule[-0.175pt]{2.891pt}{0.350pt}}
\put(1223,383){\rule[-0.175pt]{5.782pt}{0.350pt}}
\put(1247,382){\rule[-0.175pt]{2.650pt}{0.350pt}}
\put(1258,381){\rule[-0.175pt]{5.782pt}{0.350pt}}
\put(1282,380){\rule[-0.175pt]{2.891pt}{0.350pt}}
\put(1294,379){\rule[-0.175pt]{5.782pt}{0.350pt}}
\put(1318,378){\rule[-0.175pt]{2.650pt}{0.350pt}}
\put(1329,377){\rule[-0.175pt]{5.782pt}{0.350pt}}
\put(1353,376){\rule[-0.175pt]{2.891pt}{0.350pt}}
\put(1365,375){\rule[-0.175pt]{5.782pt}{0.350pt}}
\put(1389,374){\rule[-0.175pt]{2.650pt}{0.350pt}}
\put(1400,373){\rule[-0.175pt]{5.782pt}{0.350pt}}
\put(1424,372){\rule[-0.175pt]{2.891pt}{0.350pt}}
\put(311,340){\circle*{12}}
\put(381,349){\circle*{12}}
\put(428,422){\circle*{12}}
\put(475,426){\circle*{12}}
\put(522,596){\circle*{12}}
\put(569,384){\circle*{12}}
\put(616,460){\circle*{12}}
\put(662,355){\circle*{12}}
\put(709,328){\circle*{12}}
\put(756,537){\circle*{12}}
\put(803,473){\circle*{12}}
\put(850,584){\circle*{12}}
\put(897,483){\circle*{12}}
\put(944,468){\circle*{12}}
\put(991,473){\circle*{12}}
\put(1038,466){\circle*{12}}
\put(1084,464){\circle*{12}}
\put(1131,617){\circle*{12}}
\put(1178,558){\circle*{12}}
\put(311,319){\rule[-0.175pt]{0.350pt}{10.118pt}}
\put(301,319){\rule[-0.175pt]{4.818pt}{0.350pt}}
\put(301,361){\rule[-0.175pt]{4.818pt}{0.350pt}}
\put(381,305){\rule[-0.175pt]{0.350pt}{21.199pt}}
\put(371,305){\rule[-0.175pt]{4.818pt}{0.350pt}}
\put(371,393){\rule[-0.175pt]{4.818pt}{0.350pt}}
\put(428,368){\rule[-0.175pt]{0.350pt}{26.258pt}}
\put(418,368){\rule[-0.175pt]{4.818pt}{0.350pt}}
\put(418,477){\rule[-0.175pt]{4.818pt}{0.350pt}}
\put(475,372){\rule[-0.175pt]{0.350pt}{26.258pt}}
\put(465,372){\rule[-0.175pt]{4.818pt}{0.350pt}}
\put(465,481){\rule[-0.175pt]{4.818pt}{0.350pt}}
\put(522,466){\rule[-0.175pt]{0.350pt}{62.634pt}}
\put(512,466){\rule[-0.175pt]{4.818pt}{0.350pt}}
\put(512,726){\rule[-0.175pt]{4.818pt}{0.350pt}}
\put(569,282){\rule[-0.175pt]{0.350pt}{49.384pt}}
\put(559,282){\rule[-0.175pt]{4.818pt}{0.350pt}}
\put(559,487){\rule[-0.175pt]{4.818pt}{0.350pt}}
\put(616,355){\rule[-0.175pt]{0.350pt}{50.589pt}}
\put(606,355){\rule[-0.175pt]{4.818pt}{0.350pt}}
\put(606,565){\rule[-0.175pt]{4.818pt}{0.350pt}}
\put(662,265){\rule[-0.175pt]{0.350pt}{43.362pt}}
\put(652,265){\rule[-0.175pt]{4.818pt}{0.350pt}}
\put(652,445){\rule[-0.175pt]{4.818pt}{0.350pt}}
\put(709,244){\rule[-0.175pt]{0.350pt}{40.471pt}}
\put(699,244){\rule[-0.175pt]{4.818pt}{0.350pt}}
\put(699,412){\rule[-0.175pt]{4.818pt}{0.350pt}}
\put(756,460){\rule[-0.175pt]{0.350pt}{37.339pt}}
\put(746,460){\rule[-0.175pt]{4.818pt}{0.350pt}}
\put(746,615){\rule[-0.175pt]{4.818pt}{0.350pt}}
\put(803,370){\rule[-0.175pt]{0.350pt}{49.384pt}}
\put(793,370){\rule[-0.175pt]{4.818pt}{0.350pt}}
\put(793,575){\rule[-0.175pt]{4.818pt}{0.350pt}}
\put(850,449){\rule[-0.175pt]{0.350pt}{64.802pt}}
\put(840,449){\rule[-0.175pt]{4.818pt}{0.350pt}}
\put(840,718){\rule[-0.175pt]{4.818pt}{0.350pt}}
\put(897,370){\rule[-0.175pt]{0.350pt}{54.443pt}}
\put(887,370){\rule[-0.175pt]{4.818pt}{0.350pt}}
\put(887,596){\rule[-0.175pt]{4.818pt}{0.350pt}}
\put(944,405){\rule[-0.175pt]{0.350pt}{30.353pt}}
\put(934,405){\rule[-0.175pt]{4.818pt}{0.350pt}}
\put(934,531){\rule[-0.175pt]{4.818pt}{0.350pt}}
\put(991,403){\rule[-0.175pt]{0.350pt}{33.485pt}}
\put(981,403){\rule[-0.175pt]{4.818pt}{0.350pt}}
\put(981,542){\rule[-0.175pt]{4.818pt}{0.350pt}}
\put(1038,387){\rule[-0.175pt]{0.350pt}{38.303pt}}
\put(1028,387){\rule[-0.175pt]{4.818pt}{0.350pt}}
\put(1028,546){\rule[-0.175pt]{4.818pt}{0.350pt}}
\put(1084,393){\rule[-0.175pt]{0.350pt}{34.208pt}}
\put(1074,393){\rule[-0.175pt]{4.818pt}{0.350pt}}
\put(1074,535){\rule[-0.175pt]{4.818pt}{0.350pt}}
\put(1131,535){\rule[-0.175pt]{0.350pt}{39.508pt}}
\put(1121,535){\rule[-0.175pt]{4.818pt}{0.350pt}}
\put(1121,699){\rule[-0.175pt]{4.818pt}{0.350pt}}
\put(1178,475){\rule[-0.175pt]{0.350pt}{40.230pt}}
\put(1168,475){\rule[-0.175pt]{4.818pt}{0.350pt}}
\put(1168,642){\rule[-0.175pt]{4.818pt}{0.350pt}}
\end{picture}